\documentclass[11pt]{article}

\usepackage{pst-all}
\usepackage[dvips]{graphicx}

\def\eqcm{\: ,}           
\def\eqpt{\: .}

\newcommand{\lsim}{\raisebox{-4pt}{$
\,\stackrel{\textstyle <}{\sim}\,$}}
\newcommand{\gsim}{\raisebox{-4pt}{$
\,\stackrel{\textstyle >}{\sim}\,$}}

\def\d{{\rm d}}

\newcommand{\tvec}[1]{\mbox{\boldmath{$#1$}}}
\newcommand{\svec}[1]{\mbox{\boldmath{$\scriptstyle #1$}}}

\newcommand{\bsq}[2]{\langle \tvec{b}^2 \rangle^{#1}_{#2}}  

\newcommand{\dof}{\mathrm{d.o.f.}}

\def\gev{\,{\rm GeV}}

\begin{document}

\begin{flushright}
DESY 04-146 \\
CERN-PH-TH/04-154 \\
WUB 04-08 \\
hep-ph/0408173 \\
\end{flushright}

\renewcommand{\thefootnote}{\fnsymbol{footnote}}

\begin{center}
\vskip 3.5\baselineskip
{\LARGE \bf
Generalized parton distributions \\[0.5em]
from nucleon form factor data
}

\vskip 2.5\baselineskip
M.~Diehl$^{1}$,
Th.~Feldmann$^{2}$,
R.~Jakob$^{3}$ and P.~Kroll$^{3}$
\vskip \baselineskip
{\it
1. Deutsches Elektronen-Synchroton DESY, 22603 Hamburg, Germany \\ 
2. CERN, Dept.\ of Physics, Theory Division, 1211 Geneva, Switzerland\\
3. Fachbereich Physik, Universit\"at Wuppertal, 42097 Wuppertal,
   Germany
}
\vskip 3\baselineskip
\textbf{Abstract}\\[0.5\baselineskip]
\parbox{0.9\textwidth}{\small We present a simple empirical
parameterization of the $x$- and $t$-dependence of generalized parton
distributions at zero skewness, using forward parton distributions as
input.  A fit to experimental data for the Dirac, Pauli and axial form
factors of the nucleon allows us to discuss quantitatively the
interplay between longitudinal and transverse partonic degrees of
freedom in the nucleon (``nucleon tomography'').  In particular we
obtain the transverse distribution of valence quarks at given momentum
fraction~$x$.  We calculate various moments of the distributions,
including the form factors that appear in the handbag approximation to
wide-angle Compton scattering.  This allows us to estimate the minimal
momentum transfer required for reliable predictions in that approach
to be around $|t| \simeq 3 \gev^2$.  We also evaluate the valence
contributions to the energy-momentum form factors entering Ji's sum
rule.}
\vskip 1.5\baselineskip
\end{center}

\renewcommand{\thefootnote}{\arabic{footnote}}
\setcounter{footnote}{0}


\newpage

\tableofcontents

\clearpage

\section{Introduction}
\label{sec:introduction}
In recent years hard exclusive reactions have found increased
attention because of new theoretical developments. For a number of
such reactions, for instance deeply virtual \cite{ji:96,rad:97,cf:98}
or wide-angle \cite{rad:98,DFJK1} Compton scattering off the nucleon,
the scattering amplitudes factorize into partonic subprocesses and
soft hadronic matrix elements, called generalized parton distributions
(GPDs) \cite{ji:96,rad:97,mul:94,Blumlein:1997pi}.  While the partonic
subprocesses can be evaluated in perturbation theory, calculation of
GPDs requires non-perturbative methods.  As an approach starting from
first principles, lattice QCD is suitable for evaluating $x$-moments
of GPDs.  Interesting results on the first few moments have recently
been obtained \cite{goe:03,hae:03a,hae:03b,neg:04}, although such
calculations still come with important systematic uncertainties.

At present one is therefore unable to get along without models for the
GPDs in order to describe data on hard exclusive reactions. These
models are frequently simple parameterizations, constrained by general
symmetry properties, by reduction formulas which express that certain
GPDs become the usual parton densities in the forward limit of
vanishing momentum transfer, and by the sum rules which express that
the integrals of quark GPDs over $x$ give the contributions of these
quarks to the elastic form factors of the nucleon.  Other attempts to
model GPDs are for instance based on effective quark-model Lagrangians
\cite{pet:97,pen:99} or on the concept of constituent quarks
\cite{Scopetta:2003et,Pasquini:2004gc}.  Another approach constructs
GPDs from the partonic degrees of freedom through the overlap of
light-cone wave functions \cite{DFJK1,bro:01,DFJK3}.  For large $x$ or
for large momentum transfer, where essentially only the valence Fock
state matters, this concept provides reasonable results as comparison
with the measured parton densities and form factors reveals
\cite{DFJK1}.  Recent reviews of the general properties of GPDs and
efforts to model them can be found in \cite{goe:01,die:03}.

A long-term goal is the (almost) model-independent extraction of GPDs
from experimental data, in analogy to the determination of the usual
parton densities from hard inclusive processes performed for instance
in \cite{CTEQ,blum:03}.  For GPDs this is admittedly a very
challenging task since they are functions of three kinematic
variables: the average momentum fraction $x$ of the partons, the
skewness $\xi$, and the invariant momentum transfer $t$.  Furthermore,
compared to inclusive processes, we will for quite some time have much
less data on exclusive reactions at our disposal.  In the near future
we can aim at suitable parameterizations of the GPDs, with a few free
parameters adjusted to data.  An attempt of such a parameterization
will be presented in this work.  The proposed parameterization is
physically motivated on one hand by Regge phenomenology in the limit
$x\to 0$ and, on the other hand, by the physical intuition gained in
the impact parameter representation of GPDs.  The free parameters of
this ansatz are fitted to the experimental data on the Dirac and Pauli
form factors of the nucleon, exploiting the sum rules mentioned above.
Only the valence quark combinations of parton distributions are
constrained by these observables, but the combination of proton and
neutron form factors allows for a separation of $u$ and $d$ quark
distributions in a certain kinematical range.  Similarly, the GPDs for
polarized quarks can be constrained by the axial-vector form factor.
The kinematic dependence of GPDs involves two aspects: the interplay
between the two independent longitudinal momentum fractions $x$ and
$\xi$, and the interplay between the longitudinal variables and $t$.
As a consequence of Lorentz invariance, the $\xi$-dependence of GPDs
drops out in the sum rules for the form factors of the quark vector
and axial-vector currents \cite{ji:96}.  We therefore restrict our
study to GPDs at $\xi=0$ and concentrate on the correlation between
the variables $x$ and $t$.  This can be done in a physically very
intuitive representation: after a Fourier transform to impact
parameter space, GPDs at zero skewness $\xi$ describe the joint
density to find a parton at a given longitudinal momentum fraction $x$
and a given transverse separation from the center of the nucleon in
the infinite-momentum frame~\cite{bur:00}.

For the valence part of the GPDs $H$ and $\widetilde{H}$ we will use
their forward limits $q(x)$ and $\Delta q(x)$ as an input to our
parameterization, and our fit to form factor data will allow us to
extract the average impact parameter $\bsq{}{x}$ of valence quarks
with a given momentum fraction.  For the proton-helicity dependent GPD
$E$ our task is more difficult since we need to make an ansatz for
both its forward limit $e(x)$ and for its dependence on $t$ or on
impact parameter.  Lack of sufficient data on the pseudoscalar nucleon
form factor prevents a similar study for the GPD $\widetilde{E}$ at
present.

With the GPDs at hand we are in the position to compute various
moments and compare them for instance with lattice QCD results.  We
can in particular evaluate the contribution of valence quarks to Ji's
angular momentum sum rule \cite{ji:96}.  In impact parameter space our
results can be turned into ``tomographic'' images of the proton as
suggested in \cite{bur:00,ral:02,bur:02}.  We shall also discuss
wide-angle Compton scattering in some detail.  The soft hadronic
matrix elements appearing in the soft handbag description of this
process are new form factors, which are expressed as $1/x$-moments of
GPDs \cite{rad:98,DFJK1} and can also be evaluated from our results.
Comparison of the corresponding observables with precision data that
have been taken at Jefferson Lab and will be published shortly
\cite{bogdan}, will subsequently allow for an examination of our
theoretical understanding of wide-angle Compton scattering.

This paper is organized as follows: In Sect.\ \ref{sec:GPD} we recall
some properties of GPD at zero skewness. In Sect.\ \ref{sec:ansatz} we
present the physical motivation of our parameterization and analyze
the GPD $H$. The corresponding analyses of the GPDs $\widetilde{H}$
and $E$ are described in Sects.\ \ref{sec:axial} and \ref{sec:gpd-e},
respectively.  Various properties of our GPDs are shown in Sect.\
\ref{sec:results}.  In Sect.\ \ref{sec:compton} we discuss wide-angle
Compton scattering in the soft handbag approach and investigate the
corresponding form factors. The paper ends with our conclusions in
Sect.\ \ref{sec:summary}.  In two appendices we provide details on the
nucleon form factor data we have used (App.\ \ref{app:A}) and collect
all fit results (App.\ \ref{app:B}).

\section{Generalized parton distributions at zero skewness}
\label{sec:GPD}

Let us start by recalling some properties of generalized parton
distributions at $\xi=0$.  We use Ji's definitions of GPDs and their
arguments \cite{ji:96} and for simplicity suppress the argument $\xi$,
writing $H(x,t)$ instead of $H(x,\xi=0,t)$, $\widetilde{H}(x,t)$
instead of $\widetilde{H}(x,\xi=0,t)$, etc.

Let us first concentrate on the combination
\begin{equation}
H_v^q(x,t,\mu^2) = H^q(x,t,\mu^2) + H^q(-x,t,\mu^2)
\label{def-val}
\end{equation}
of the quark helicity averaged GPDs for flavor $q$ in the proton.
This is the combination entering the proton and neutron Dirac form
factors as
\begin{eqnarray}
        \label{dirac-sum-rules}
F_1^p(t) &=& \int_0^1 \d x\,
  \Big( {\textstyle\frac{2}{3}} H_v^u(x,t,\mu^2) 
      - {\textstyle\frac{1}{3}} H_v^d(x,t,\mu^2) \Big) \eqcm
\nonumber \\
F_1^n(t) &=& \int_0^1 \d x\,
  \Big( {\textstyle\frac{2}{3}} H_v^d(x,t,\mu^2) 
      - {\textstyle\frac{1}{3}} H_v^u(x,t,\mu^2) \Big) \eqpt
\end{eqnarray}
In the forward limit $t=0$ the distribution $H_v^q(x,0)$ becomes the
valence quark density $q_v(x) = q(x) - \bar{q}(x)$.  In both $F_1^p$
and $F_1^n$ we have neglected the contribution $e_s H_v^s$ from
strange quarks: the difference of strange and antistrange
distributions in the nucleon is not large \cite{Olness:2003wz} and the
strange contribution to nucleon form factors at low $t$ is seen to be
small in neutral-current elastic scattering \cite{Beck:2001yx}.  In
(\ref{def-val}) and (\ref{dirac-sum-rules}) we have displayed the
dependence of the GPDs on the factorization scale~$\mu$, which we will
often omit for ease of notation.  We will also use the notation
$F_1^q(t) = \int_0^1 \d x\, H_v^q(x,t)$ for the individual quark
flavor contributions to the Dirac form factor.

As shown by Burkardt \cite{bur:00,bur:02}, a density interpretation of
GPDs at $\xi=0$ is obtained in the mixed representation of
longitudinal momentum and transverse position in the infinite-momentum
frame.  In particular,
\begin{equation}        
        \label{impact-gpd}
q_v(x,\tvec{b}) = \int \frac{\d^2 \tvec{\Delta}}{(2\pi)^2}\,
e^{-i \svec{b} \svec{\Delta}}\, H_v^q(x,t=-\tvec{\Delta}^2)
\end{equation}
gives the probability to find a quark with longitudinal momentum
fraction $x$ and impact parameter $\tvec{b}$ minus the corresponding
probability to find an antiquark, where we reserve boldface notation
for two-dimensional vectors in the transverse plane.  The average
impact parameter of this distribution at given $x$ is
\begin{equation}
  \label{b-def}
\bsq{q}{x} = \frac{\int \d^2\tvec{b}\, \tvec{b}^2\, q_v(x,\tvec{b})}{
               \int \d^2\tvec{b}\, q_v(x,\tvec{b})}
= 4 \frac{\partial}{\partial t} \log H_v^q(x,t) \Bigg|_{t=0}  \eqpt
\end{equation}
Since $q_v(x,\tvec{b})$ is a difference of probabilities, $\bsq{q}{x}$
is not an average in the strict sense.  It gives however the typical
value of $\tvec{b}^2$ in $q_v(x,\tvec{b})$ as long as this
distribution is positive (which is the case for the parameterizations
we will use, and which is generally expected when $x$ is sufficiently
large to neglect antiquarks compared with quarks).

GPDs can be written as the overlap of light-cone wave functions.  In
impact parameter space this representation has an especially simple
form:
\begin{eqnarray}
  \label{impact-overlap}
q_v(x,\tvec{b}) &=& \sum_{N,\beta}(4\pi)^{N-1}
 \int \prod_{i} \d x_i \, \delta\Big(1 - \sum_i x_i\Big) 
 \int \prod_{i} \d^2 \tvec{b}_i \, 
      \delta^{(2)}\Big(\sum_i x_i \tvec{b}_i\Big)
\nonumber \\
&& \sum_j \, \eta_j \delta(x-x_j)\, \delta^{(2)}(\tvec{b} - \tvec{b}_j)\,
 \Big| \tilde{\psi}_{N\beta}^{+}(x_i, \tvec{b}_i) \Big|^2  \eqpt
\end{eqnarray}
The index $i$ runs over the $N$ partons in a given Fock state, whose
quantum numbers are collectively denoted by the index $\beta$, and
$\tilde{\psi}_{N\beta}^{+}$ is the light-cone wave function of this
Fock state in a proton with positive helicity.  This impact-parameter
wave function is obtained from the wave function in momentum space by
a Fourier transform as given in \cite{die:02}.  The index $j$ singles
out the struck parton and runs over all quarks or antiquarks with
flavor $q$, with $\eta_j=1$ for quarks and $\eta_j=-1$ for antiquarks.

\begin{figure}
\begin{center}
\leavevmode
\includegraphics[width=0.4\textwidth]{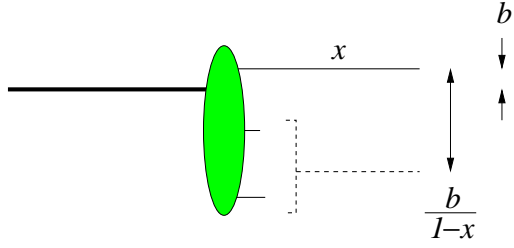}
\end{center}
\caption{\label{geometry} A three-quark configuration in the mixed
representation of definite transverse position and definite
plus-momentum.  $x$ denotes the plus-momentum fraction of the upper
quark with respect to the nucleon.  The dashed line indicates the
center of momentum of the two lower quarks and the thick solid line
the center of momentum of the proton.}
\end{figure}

As explained in \cite{bur:02} the impact parameter $\tvec{b}$ in
$q_v(x,\tvec{b})$ is the transverse distance between the struck parton
and the center of momentum of the hadron (see Fig.~\ref{geometry}).
The latter is the average transverse position of the partons in the
hadron with weights given by the parton momentum fractions.  It was
chosen to be the origin in (\ref{impact-overlap}), so that the
transverse positions $\tvec{b}_i$ and momentum fractions $x_i$ of the
partons satisfy $\sum_i x_i \tvec{b}_i =0$.  The center of momentum of
the spectator partons is easily identified as $-\tvec{b} x /(1-x)$.
The relative distance $\tvec{b} /(1-x)$ between the struck parton and
the spectator system provides an estimate of the size of the hadron as
a whole, and we denote its average square by
\begin{equation}
  \label{r-def}
d_q^2(x) = \frac{\bsq{q}{x}}{(1-x)^2} \eqpt
\end{equation}
It does however not account for the spatial extension of the spectator
system itself, which remains unaccessible in quantities like GPDs at
zero skewness, where only one single parton within a hadron is probed.
{}From Fig.~\ref{geometry} one readily sees that $d_q$ provides a
lower limit on the transverse size of the hadron.  This quantity has
also been considered in recent work on color transparency
\cite{Burkardt:2003mb,Liuti:2004hd}.

Just as the usual quark densities, GPDs depend on the factorization
scale $\mu$ at which the partons are resolved.  For $\xi=0$ the
evolution in $\mu$ is described by the usual DGLAP equation, which for
the valence combination $H_v^q$ reads
\begin{equation}
        \label{dglap-H}
\mu^2 \frac{\d}{\d\mu^2} H_v^q(x,t,\mu^2) 
= \int_x^1 \frac{\d z}{z}\,
  \Big[ P\Big(\frac{x}{z}\Big) \Big]_+ H_v^q(z,t,\mu^2) \eqcm
\end{equation}
where $[~]_+$ denotes the usual plus-distribution, and where the
kernel reads $P(z) = \frac{\alpha_s}{2\pi}\, C_F \frac{1+z^2}{1-z}$ at
leading order in the strong coupling.  We note that the situation for
$\mu^2 \ll -t$ is rather subtle; it will be discussed in some detail
in Sect.~\ref{sec:compton-nlo}.  Dividing (\ref{dglap-H}) by
$H_v^q(x,t)$ and subsequently taking the derivative in $t$ at $t=0$ we
obtain an evolution equation for the average impact parameter:
\begin{equation}
        \label{dglap-bsq}
\mu^2 \frac{\d}{\d\mu^2} \bsq{q}{x} = - \frac{1}{q_v(x)}
 \int_x^1 \frac{\d z}{z}\,
  P\Big(\frac{x}{z}\Big)\, q_v(z) \Big[ \bsq{q}{x} - \bsq{q}{z} \Big] ,
\end{equation}
where the plus-prescription is no longer needed since the term in
square brackets vanishes at $z=x$.  We see that the average impact
parameter decreases with $\mu$ for all $x$, provided that $\bsq{q}{x}$
is a decreasing function of $x$.

Notice that the right-hand sides of (\ref{dirac-sum-rules}) must be
evaluated at a particular resolution scale $\mu$, whereas the
left-hand sides are the form factors of a conserved current and hence
independent of the scale $\mu$ where the current is renormalized.
Physically speaking, the transverse distribution of quarks of a given
momentum fraction $x$ is modified by the parton splitting processes
that underly DGLAP evolution.  It hence depends on the spatial
resolution $1/\mu$ at which quarks are probed, whereas the transverse
distribution of charge described by the electromagnetic form factors
does not \cite{die:02}.

For the quark helicity dependent GPDs we define the valence
combination
\begin{equation}
\widetilde{H}_v^q(x,t,\mu^2) = 
   \widetilde{H}^q(x,t,\mu^2) - \widetilde{H}^q(-x,t,\mu^2) \eqcm
\end{equation}
whose forward limit is $\Delta q_v(x) = \Delta q(x) -
\Delta\bar{q}(x)$.  The impact parameter distribution $\Delta
q_v(x,\tvec{b})$ is then defined in analogy to (\ref{impact-gpd}) and
again can be interpreted as a difference of probability densities in
$x$ and $\tvec{b}$ space.  It has a wave function representation akin
to (\ref{impact-overlap}), with an extra minus sign in front of the
squared wave function if the struck quark or antiquark has negative
helicity.  The evolution of $\widetilde{H}_v^q$ in the scale $\mu$ is
described by a DGLAP equation as in (\ref{dglap-H}), with an evolution
kernel that is identical to the one for $H_v^q$ at leading order in
$\alpha_s$.  The properties of the proton helicity flip GPD $E$ will
be discussed in Sect.~\ref{sec:gpd-e}.

\section{A parameterization for the unpolarized GPD $H_v(x,t)$}
\label{sec:ansatz}

\subsection{Physical motivation}
\label{sec:phys-mot}

We now develop a parameterization for $H_v^u(x,t)$ and $H_v^d(x,t)$.
For its functional form we will use theoretical guidance in the
regions of very small and very large $x$ and then attempt a suitable
interpolation for intermediate~$x$.  We will fit this parameterization
to the data on the nucleon Dirac form factors $F_1^p(t)$ and
$F_1^n(t)$.

For small $t$ and very small $x$ one can expect Regge behavior of
$H_v(x,t)$, employing the same argument as in the well-known case of
forward parton densities \cite{lan:71}.  The simplest form of Regge
behavior is the dominance of a single Regge pole,
\begin{equation}
        \label{regge-form}
H_v(x,t) \sim c(t)\, \Big( \frac{x_0}{x} \Big)^{\alpha(t)} =
  \Big( \frac{x_0}{x} \Big)^{\alpha(0)}\, \exp \Bigg[ 
     \Big( \alpha' \log\frac{x_0}{x} + b_0 \Big)\, t\, \Bigg] ,
\end{equation}
where in the second step we have used a simple parameterization $c(t)
= \exp(b_0\, t)$ for the small-$t$ region and a linear form of the
Regge trajectory $\alpha(t) = \alpha(0) + \alpha' t$.  The leading
Regge trajectories with the quantum numbers of $H_v$ are those of the
$\omega$ and the $\rho$.  They can be phenomenologically determined
from suitable hadronic cross sections and from the Chew-Frautschi plot
(showing the spin of a meson versus its squared mass, which we take
from \cite{PDG}).  {}From the masses of $\omega(782)$ and
$\omega_3(1670)$ one obtains a linear trajectory $\alpha_\omega(t) =
0.42 + 0.95 \gev^{-2}\, t$, whose intercept at $t=0$ agrees well with
the value extracted from $\sigma_{\rm{tot}}(pp) -
\sigma_{\rm{tot}}(p\bar{p})$ in \cite{Igi:1988vp}.  The masses of
$\rho(770)$ and $\rho_3(1690)$ give a linear trajectory
$\alpha_\rho(t) = 0.48 + 0.88 \gev^{-2}\, t$, in good agreement with
the intercept from $\sigma_{\rm{tot}}(\pi^- p) -
\sigma_{\rm{tot}}(\pi^+ p)$ and with the trajectory extracted from the
data on $d\sigma/dt (\pi^- p\to \pi^0 n)$ up to about $|t| \approx
0.3 \gev^2$ \cite{Barnes:1976ek}.

We emphasize that (\ref{regge-form}) is not a prediction of Regge
theory, but rather corresponds to a simple form of Regge
phenomenology: on one hand one expects subleading Regge trajectories
to become important if $x$ is not sufficiently small, and on the other
hand the importance of Regge cuts, which lead to a more complicated
behavior on $x$ and $t$, is notoriously difficult to determine without
further assumptions.  To assess how well the ansatz (\ref{regge-form})
fares at $t=0$ we have investigated the CTEQ6M parton densities
\cite{CTEQ} at $\mu=2 \gev$ and found that for $10^{-5} < x < 10^{-2}$
one has $u_v + d_v \approx x^{-0.435}$ and $(u_v - d_v) /(u_v + d_v)
\approx x^{-0.07}$, both within $1\%$ accuracy.  Scanning the 40 sets
of parton densities given by CTEQ as error estimates, we found an
exponent in the power-law for $u_v+d_v$ between $-0.38$ and $-0.495$
and a corresponding exponent for $(u_v-d_v)/(u_v+d_v)$ between
$-0.165$ and $0.13$.  Similar values are found when taking the
distributions at scales $\mu=1 \gev$ or $\mu=4 \gev$.  We conclude
that a simple Regge pole ansatz with an intercept taken from the
phenomenology of soft hadronic interactions is in fair agreement with
valence quark distributions at low factorization scale, and assume in
the following that this description generalizes to small finite $-t$.
Note that the form (\ref{regge-form}) translates into an average
impact parameter $\bsq{}{x}$ diverging like $\log(1/x)$ at small $x$.
A physical mechanism that gives such a behavior is Gribov diffusion,
the generation of small-$x$ partons through a cascade of branching
processes \cite{gri:73}.

As $x$ increases, the struck parton takes more and more weight in the
center of momentum $\sum_i x_i \tvec{b}_i$ of all partons, so that the
distribution in $\tvec{b}$ should become more and more narrow
\cite{bur:01}.  This means that the $t$-dependence of GPDs should
become less steep with increasing $x$.  If the average distance
$d_q(x)$ between the struck quark and the center of momentum of the
spectators is to remain finite, which one may expect for a system
subject to confinement, then $\bsq{}{x}$ must vanish at least like
$(1-x)^2$ in the limit $x\to 1$ \cite{bur:04}.  The actual limiting
behavior of $\bsq{}{x}$ in QCD remains unknown.  Certainly the impact
parameter dependence of GPDs at large $x$ contains interesting
information about the dynamics of confinement, and we shall see how
much information on this dependence can be extracted from existing
form factor data.

In our parameterization we will make an exponential ansatz for the
$t$-dependence: 
\begin{equation}
  \label{master-ansatz}
H_v^q(x,t) = q_v(x) \exp{[t f_q(x)]} \eqpt
\end{equation}
The function $f_q(x)$ parameterizes how the profile of the quark
distribution in the impact parameter plane changes with $x$, as is
readily seen from
\begin{equation}
  \label{master-impact}
q_v(x,\tvec{b}) = \frac{1}{4\pi}\, \frac{q_v(x)}{f_q(x)}\, 
  \exp\Bigg[ -\frac{\tvec{b}^2}{4 f_q(x)} \Bigg] \eqcm
\qquad \qquad
\bsq{q}{x} = 4 f_q(x) \eqpt
\end{equation}
Apart from being suitable for analytic calculations, an exponential
$t$-dependence of $H_v^q(x,t)$ guarantees that $q_v(x,\tvec{b})$ is
positive.  It ensures a rapid falloff at any $x$ as $-t$ becomes
large, and it readily matches with the Regge form (\ref{regge-form})
for small $x$ and $-t$ if we impose
\begin{equation}
  \label{regge-constraint}
f_q(x) \to \alpha' \log\frac{1}{x} + B_q \qquad\qquad 
\mbox{for~} x\to 0 \eqpt
\end{equation}
The $t$-slope at $x=x_0$ is obtained as $b_0 = \alpha' \log(1/x_0) +
B_q$.  In our fits we will explore a possible flavor dependence of
$B_q$ but keep $\alpha'$ flavor independent, as suggested by Regge
phenomenology.  We are aware that our exponential ansatz
(\ref{master-ansatz}) has no rigorous theoretical backing, and we
shall explore alternative forms of the $t$-dependence in
Sect.~\ref{sec:power-law-t}.  Let us emphasize already here that we
cannot trust the detailed form of our extracted GPDs in the region $-t
\gg 1/f_q(x)$, where according to (\ref{master-ansatz}) they are
exponentially small and thus give only a tiny contribution to the form
factor integrals (\ref{dirac-sum-rules}).  In particular we claim no
validity of our ansatz at $-t$ of several $\gev^2$ and very small $x$,
where its motivation from Regge phenomenology does indeed not apply.

For large $x$ one can expect that the overlap representation
(\ref{impact-overlap}) is dominated by Fock states with few partons.
In \cite{DFJK1} we have evaluated GPDs from model wave functions for
the lowest Fock states, whose dependence on the transverse parton
momenta or impact parameters was taken as Gaussian,
\begin{equation}
  \label{gauss-wfs}
\psi \propto \exp\Bigg[ -a^2 
     \sum_i \frac{\tvec{k}_i^2}{x_i} \, \Bigg] \eqcm
\qquad\qquad
\tilde{\psi} \propto \exp\Bigg[ -\frac{1}{4a^2} 
             \sum_i x_i^{\phantom{.}} \tvec{b}_i^2 \, \Bigg] \eqcm
\end{equation}
a form going back to \cite{Brodsky:1981aa} and explored in detail for
the nucleon in \cite{bol:96}.  With a parameter $a^2 = 0.72 \gev^{-2}$
or somewhat larger, this model allowed a fair description of
unpolarized and polarized $u$ and $d$ quark densities for $x\gsim 0.6
\mbox{~to~} 0.7$ and of $F_1^p(t)$ for $-t \gsim 4 \gev^2$.  The
resulting GPDs at $\xi=0$ take the form given in (\ref{master-ansatz})
and (\ref{master-impact}) with $\bsq{}{x} = 2 a^2 (1-x) /x$.  The
average distance between the struck quark and the spectators hence
diverges like $d_q(x) \sim (1-x)^{-1/2}$ in the limit $x\to 1$.
Indeed the impact parameter form of the model wave functions
(\ref{gauss-wfs}) allows $\tvec{b}_i^2$ to grow like $1/x_i$ when the
spectators become soft.  In the limit $x\to 1$ such a behavior is
difficult to reconcile with confinement as was pointed out in
\cite{bur:04}, and one aim of our study here is to explore
quantitatively at which $x$ the behavior of such wave functions
becomes physically suspect.  In our ansatz (\ref{master-ansatz}) for
the valence GPDs we will impose the constraint
\begin{equation}
\label{large-x-constraint}
f_q(x) \to A_q (1-x)^n  \qquad\qquad
\mbox{for~} x\to 1 \eqcm
\end{equation}
either with $n=1$ as in the model just discussed, or with $n=2$, which
corresponds to $d_q(x)$ tending to a constant at $x=1$.

The ansatz (\ref{master-ansatz}) must be made at a particular
factorization scale $\mu$ and may work better for some scales than for
others.  Let us see that the limiting behavior we take for $\bsq{}{x}$
at small and at large $x$ retains its form under leading-order DGLAP
evolution.  To be more precise, let us first assume that $\bsq{}{x}
\approx 4 \alpha' \log(1/x) + 4 B$ and $q_v(x) \approx c x^{-\alpha}$
with $\alpha>0$ at small $x$ and for a given $\mu$.  We need not take
the mathematical limit of $x\to 0$ but only require these forms to be
good approximations in a range of $x$ where the small-$x$
approximations of the following arguments are numerically adequate.
With the evolution equation (\ref{dglap-bsq}) for $\bsq{}{x}$ and the
leading-order evolution kernel we have
\begin{equation}
  \label{lo-dglap-bsq}
\mu^2 \frac{\d}{\d \mu^2} \bsq{}{x} \approx \frac{\alpha_s C_F}{2\pi}
   \int_x^1 \d z \Bigg[ 1 + \Big(\frac{x}{z}\Big)^2 \Bigg] 
   \frac{q_v(z)}{q_v(x)} \,
        \frac{\bsq{}{x} - \bsq{}{z}}{x-z} \eqpt
\end{equation}
Let $\delta$ be a fixed value of $x$ below which $q_v(x)$ and
$\bsq{q}{x}$ can be approximated as stated above.  For $z>\delta$ the
integrand behaves like $x^{\alpha} \log x$ for $x\to 0$, so that the
integral over $z$ from $\delta$ to 1 gives a vanishing contribution to
the right-hand side.  We can hence approximate
\begin{eqnarray}
  \label{dglap-small-x}
\mu^2  \frac{\d}{\d \mu^2} \bsq{}{x} 
 &\approx & - \frac{\alpha_s C_F}{2\pi}
   \int_x^\delta \d z \Bigg[ 1 + \Big(\frac{x}{z}\Big)^2 \Bigg] 
   \Big( \frac{x}{z} \Big)^{\alpha} \,
        \frac{4 \alpha' \log(z/x)}{z-x}
\nonumber \\
 &=& - 4\alpha' \, \frac{\alpha_s C_F}{2\pi} \int_1^{\delta /x} \d u\,
    ( 1 + u^{-2}\, )\, u^{-\alpha}\, \frac{\log u}{u-1} \eqcm
\end{eqnarray}
which tends to a negative constant for $x\to 0$.  The divergent part
$4 \alpha' \log(1/x)$ of $\bsq{}{x}$ is hence $\mu$ independent,
whereas the constant $4 B$ will decrease with $\mu$.  Our argument can
be generalized to other forms of $q_v(x)$ at small $x$, for instance
to a sum $\sum_k c_k x^{-\alpha_k}$ with $\alpha_k>0$.

Concerning the large-$x$ behavior, one can show that a form $\bsq{}{x}
\approx 4 A (1-x)^n$ with $n>0$ is stable under leading-order DGLAP
evolution, provided that the forward densities at a given $\mu$ behave
as $q_v(x) \sim (1-x)^\beta$.  More precisely, the coefficient $A$
decreases with $\mu$, whereas the power $n$ remains stable.  To see
this one starts with the evolution equation (\ref{lo-dglap-bsq}),
replaces $q_v$ and $\bsq{}{}$ on the right-hand side with the
approximations just given, and Taylor expands the evolution kernel to
leading order in $(1-x)$.  The result is
\begin{equation}
        \label{large-x-dglap}
\mu^2  \frac{\d}{\d \mu^2} \bsq{}{x} \approx
  - 4 A (1-x)^n \, \frac{\alpha_s C_F}{\pi}
     \int_0^1 \d v\, v^\beta\, \frac{1-v^n}{1-v} \eqpt
\end{equation}
The leading $x$-dependence is hence in $(1-x)^n$ on both sides, and
one obtains an equation for the evolution of its coefficient $4 A$
with $\mu$.  Our finding is in line with a numerical study of pion
GPDs by Vogt \cite{Vogt:2000ku}, who found that in a finite interval
of large $x$ the form (\ref{master-ansatz}) with $f(x) = \frac{1}{2}
a^2 (1-x)/x$ is approximately stable under DGLAP evolution, with a
moderate decrease with $\mu$ of the parameter $a^2(\mu)$.  For the
Gaussian model wave functions giving this form of GPDs, a decrease of
$a^2$ entails a decreasing probability of the corresponding lowest
Fock states.  This is in agreement with physical intuition: at higher
resolution scale $\mu$ one resolves more and more partons in the
hadron, and to find a configuration with only a few partons becomes
less likely.

The exponential $t$-dependence (\ref{master-ansatz}) of GPDs is
generally not stable under DGLAP evolution.  To see this let us
consider the Taylor expansion
\begin{equation}
  \label{taylor-t}
\log H_v^q(x,t) = \log q_v(x) 
  + t \Bigg[ \frac{\partial}{\partial t} \log H_v^q(x,t) \Bigg]_{t=0}
  + \frac{1}{2} t^2 
    \Bigg[ \frac{\partial^2}{\partial t^2} \log H_v^q(x,t) \Bigg]_{t=0}
  + \ldots \eqcm
\end{equation}
which ends after the linear term if the $t$-dependence of $H_v^q$ is
exponential.  Dividing (\ref{dglap-H}) by $H_v^q(x,t)$ and taking the
second derivative in $t$ we obtain the scale dependence of the
quadratic term in (\ref{taylor-t}) as
\begin{eqnarray}
  \label{dglap-second-deriv}
\lefteqn{
\mu^2 \frac{\d}{\d\mu^2}\, \Bigg[ \frac{\partial^2}{\partial t^2}
  \log H_v^q(x,t) \Bigg]_{t=0}
= \frac{1}{q_v(x)} \int_x^1 \frac{\d z}{z} \,
  P\Big(\frac{x}{z}\Big)\, q_v(z) 
}
\nonumber \\
&& {}\times \Bigg[
  \Big( \frac{\partial}{\partial t} \log H_v^q(z,t)
      - \frac{\partial}{\partial t} \log H_v^q(x,t) \Big)^2
  + \frac{\partial^2}{\partial t^2} 
       \Big( \log H_v^q(z,t) - \log H_v^q(x,t) \Big) \Bigg]_{t=0} .
\end{eqnarray}
If at a given scale $\mu$ the $t$-dependence is exponential, then the
right-hand side of (\ref{dglap-second-deriv}) is positive so that the
quadratic term in (\ref{taylor-t}) becomes positive as one evolves to
a higher scale.

In the small-$x$ limit we do however find approximate stability under
evolution.  With the same assumptions and approximations that led to
(\ref{dglap-small-x}) we get
\begin{equation}
\mu^2 \frac{\d}{\d\mu^2}\, \Bigg[ \frac{\partial^2}{\partial t^2}
  \log H_v^q(x) \Bigg]_{t=0} \approx 
(\alpha')^2 \, \frac{\alpha_s C_F}{2\pi} \int_1^{\delta /x} \d u\,
    ( 1 + u^{-2}\, )\, u^{-\alpha}\, \frac{(\log u)^2}{u-1}
\end{equation}
if the second derivative in $t$ of $\log H_v^2$ vanishes at the scale
$\mu$.  The change with $\mu$ of the quadratic term in the Taylor
expansion (\ref{taylor-t}) is then of order $(t \alpha')^2$.  For
moderate values of $t \alpha'$ this is small compared with the linear
term $t \alpha' \log(1/x)$.

In the large-$x$ limit we get in analogy to (\ref{large-x-dglap})
\begin{equation}
  \label{large-x-second}
\mu^2 \frac{\d}{\d\mu^2}\, \Bigg[ \frac{\partial^2}{\partial t^2}
  \log H_v^q(x) \Bigg]_{t=0} \approx
  A^2 (1-x)^{2n} \frac{\alpha_s C_F}{\pi}
     \int_0^1 \d v\, v^\beta\, \frac{(1-v^n)^2}{1-v}
\end{equation}
if at the scale $\mu$ the $t$-dependence of $H_v^q$ is exponential.
The change with $\mu$ of the quadratic term in the Taylor expansion
(\ref{taylor-t}) is then parametrically of order $(t A)^2 (1-x)^{2n}$.
This is not small compared with the linear term $t A (1-x)^n$ if the
latter is of order $1$.  Numerically however the $v$-integral in
(\ref{large-x-second}) is rather small, namely $0.05$ for $n=1$ and
$0.14$ for $n=2$ if $\beta=3$.  For large $x$ we thus find that the
departure from the exponential behavior (\ref{master-ansatz}) under
evolution should not be too strong in the $t$-region where the
exponent $t f(x)$ does not take too large negative values.  This is
again in agreement with the numerical study of Vogt \cite{Vogt:2000ku}
mentioned above.

So far we have considered the valence combination $H_v$ of quark GPDs.
Let us briefly comment on what one would expect for the ``sea quark''
distributions $H(x,t)$ at $x<0$.  At large $x$, the wave function
overlap picture suggests a different impact parameter or
$t$-dependence than for the valence distribution, because sea quark
distributions require Fock states with at least one $q\bar{q}$ pair in
addition to the minimal three-quark configuration.  In the small-$x$
limit one may expect a form as in (\ref{regge-form}) at low scale
$\mu$, given that the leading $\rho$, $\omega$, $a_1$ and $f_1$ Regge
trajectories all have approximately the same $\alpha(t)$.  The
situation for sea quarks is however more complicated because the
singlet combination $\sum_q [ H^q(x,t) - H^q(-x,t) ]$ mixes under
DGLAP evolution with the gluon GPD $H^g(x,t)$, whose small-$x$
behavior is dominated by Pomeron exchange.  It is well-known that for
the forward quark densities this leads to a drastic modification of
the small-$x$ behavior as one increases the scale $\mu$ even to
moderate values of a few GeV, and one cannot exclude similarly strong
modifications for the parameter $\alpha'$ of sea quarks.  There is no
data for form factors which might constrain the sea quark GPDs at
$\xi=0$ in a similar fashion as the electromagnetic form factors
constrain $H_v^q$.  To investigate the sea quark sector one will
rather rely on measurements of exclusive processes like deeply virtual
Compton scattering or meson electroproduction at small $t$, where GPDs
at nonzero $\xi$ are accessible.

\subsection{Selecting a profile function}
\label{sec:select}

The assumption of an exponential $t$-dependence and the
parameterization for the profile function $f_q(x)$ in
(\ref{master-ansatz}) represent a source of theoretical bias, which
translates into a systematic error in the determination of GPDs from
experimental data.  To gain a feeling for this error we will carefully
compare different parameterizations.  Our criteria for a good
parameterization are:
\begin{itemize}
\item simplicity,
\item consistency with theoretical and phenomenological constraints,
\item easy physical interpretation of parameters, if possible,
\item stability with respect to variations of the forward parton
densities.
\end{itemize} 

In this section we discuss a few examples, including the default
parameterization we will use in the remainder of the paper.  For each
parameterization we fix the free parameters by a $\chi^2$ fit to the
experimental data on $F_1^p$ and $F_1^n$.  For the forward densities
$q_v(x)$ we will use the CTEQ6M distributions \cite{CTEQ} at $\mu=2
\gev$ as a default, where the choice of scale is a compromise between
being large enough for $q_v(x)$ to be rather directly fixed by data
and small enough to make contact with soft physics like conventional
Regge phenomenology.  Tables with the results of our fits are
collected in App.~\ref{app:B}, and details of our data selection and
error treatment are given in App.~\ref{app:A}.

The simplest form of the profile function $f_q(x)$ satisfying our
constraints (\ref{regge-constraint}) and (\ref{large-x-constraint})
with $n=1$ is actually $f_q(x) = \alpha' \log(1/x)$ itself.  Such a
Regge behavior of the GPDs has already been mentioned in \cite{pen:99}
and \cite{bur:01} and was explored in some detail in \cite{goe:01}.
One can however not expect this simple form, where one and the same
parameter describes physics at small and large $x$, to work beyond a
rough accuracy.  Note that even in the small-$x$ limit this form is
special since it fixes the parameter $b_0$ in~(\ref{regge-form}) to be
$\alpha' \log(1/x_0)$.  As a simple extension of this ansatz one may
try
\begin{equation}
f_q(x) = \alpha' \log\frac{1}{x} + 
        ( A_q - \alpha' ) (1-x) \eqcm
\label{fit:var1}
\end{equation}
which is still in agreement with (\ref{regge-constraint}) and
(\ref{large-x-constraint}).  A very good fit ($\chi^2/\dof=1.34$) of
the nucleon Dirac form factors can then be obtained with three free
parameters, $\alpha'$, $A_u$ and $A_d$, see Table~\ref{tab:old-fits}.
The fitted value $\alpha' = (1.38\pm 0.01) \gev^{-2}$ is however
significantly larger than what Regge phenomenology would lead one to
expect.  This disagreement becomes even stronger if we take the
forward parton densities at $\mu=1 \gev$ instead of $2 \gev$.  The fit
then gives $\alpha' = (1.65\pm 0.02) \gev^{-2}$ with
$\chi^2/\dof=1.14$.

\begin{figure}[t]
\begin{center}
\leavevmode
\includegraphics[width=0.45\textwidth,
  bb=60 320 410 565]{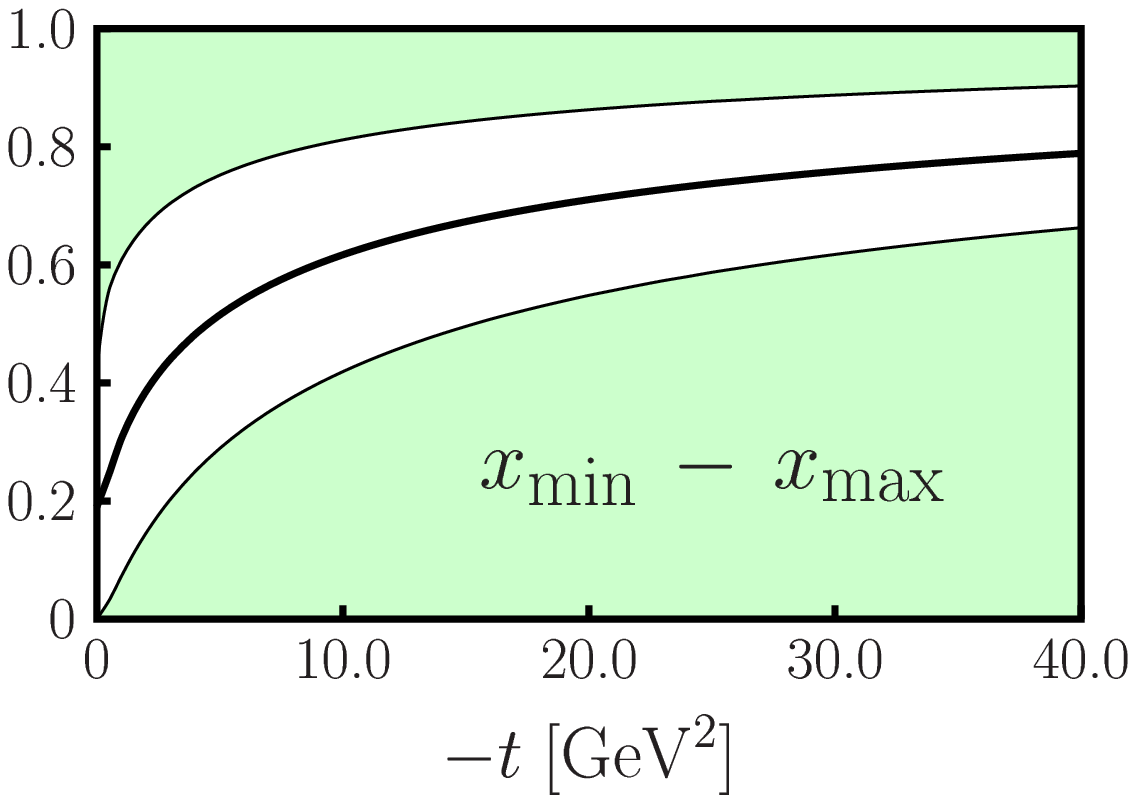}
\end{center}
\caption{\label{fig:var1_xminmax} Region of $x$ (white region) which
accounts for $90\%$ of $F_1^p(t)$ in the best fit to
(\protect\ref{fit:var1}) at $\mu=2\gev$.  The upper and lower shaded
$x$-regions each account for $5\%$ of $F_1^p(t)$, see
(\protect\ref{eq:xminmax}).  The thick line shows the average $\langle
x\rangle_t$ defined in (\protect\ref{eq:xavg}).}
\end{figure}

To better understand the situation we first determine the region of
$x$ in (\ref{dirac-sum-rules}) to which our fit is actually sensitive.
To this end we consider the minimal and maximal value of $x$ which is
needed to account for $95\%$ of the form factor in the sum rule,
\begin{eqnarray}
  \int\limits_{x_{\rm min}(t)}^1 \d x \, \sum_q e_q \, H^q_{v}(x,t) 
  &=& \mbox{95\%} \cdot F_1^p(t) \eqcm
\nonumber \\[0.2em]
  \int\limits^{x_{\rm max}(t)}_0 \d x \, \sum_q e_q \, H^q_{v}(x,t) 
  &=& \mbox{95\%} \cdot F_1^p(t) \eqcm
\label{eq:xminmax}
\end{eqnarray}
where $e_u= \frac{2}{3}$ and $e_d= -\frac{1}{3}$.  We concentrate on
the proton form factor for this purpose, which is the most important
input to our fit given the available data.  A related quantity is the
average value of $x$ in the form factor integral, given by
\begin{eqnarray}
\langle x\rangle_t &=&
  \frac{\int_0^1 \d x \, \sum_q e_q \, x \, H^q_{v}(x,t)} 
       {\int_0^1 \d x \, \sum_q e_q \, H^q_{v}(x,t)} \ .
\label{eq:xavg}
\end{eqnarray}
In Fig.~\ref{fig:var1_xminmax} we show $x_{\rm min}$, $x_{\rm max}$
and $\langle x\rangle_t$ obtained for the best fit to (\ref{fit:var1})
as a function of $|t|$.  The relevant $x$-range moves towards higher
values for increasing momentum transfer.  With the existing data on
$F_1^p$ going up to $-t=31.2 \gev^2$, the region of $x$ where the
profile function can be constrained by our fit goes up to about 0.8 or
0.9.  On the other extreme we have $x_{\rm min} \approx 1.4 \times
10^{-3}$ at $t=0$.  Clearly, the form factor integrals are only
sensitive to the small-$x$ behavior of $H_v^q$ if $t$ is also small,
as we anticipated below (\ref{regge-constraint}).

\begin{figure}[t]
\begin{center}
\leavevmode
\includegraphics[width=0.45\textwidth,
  bb=45 340 400 630]{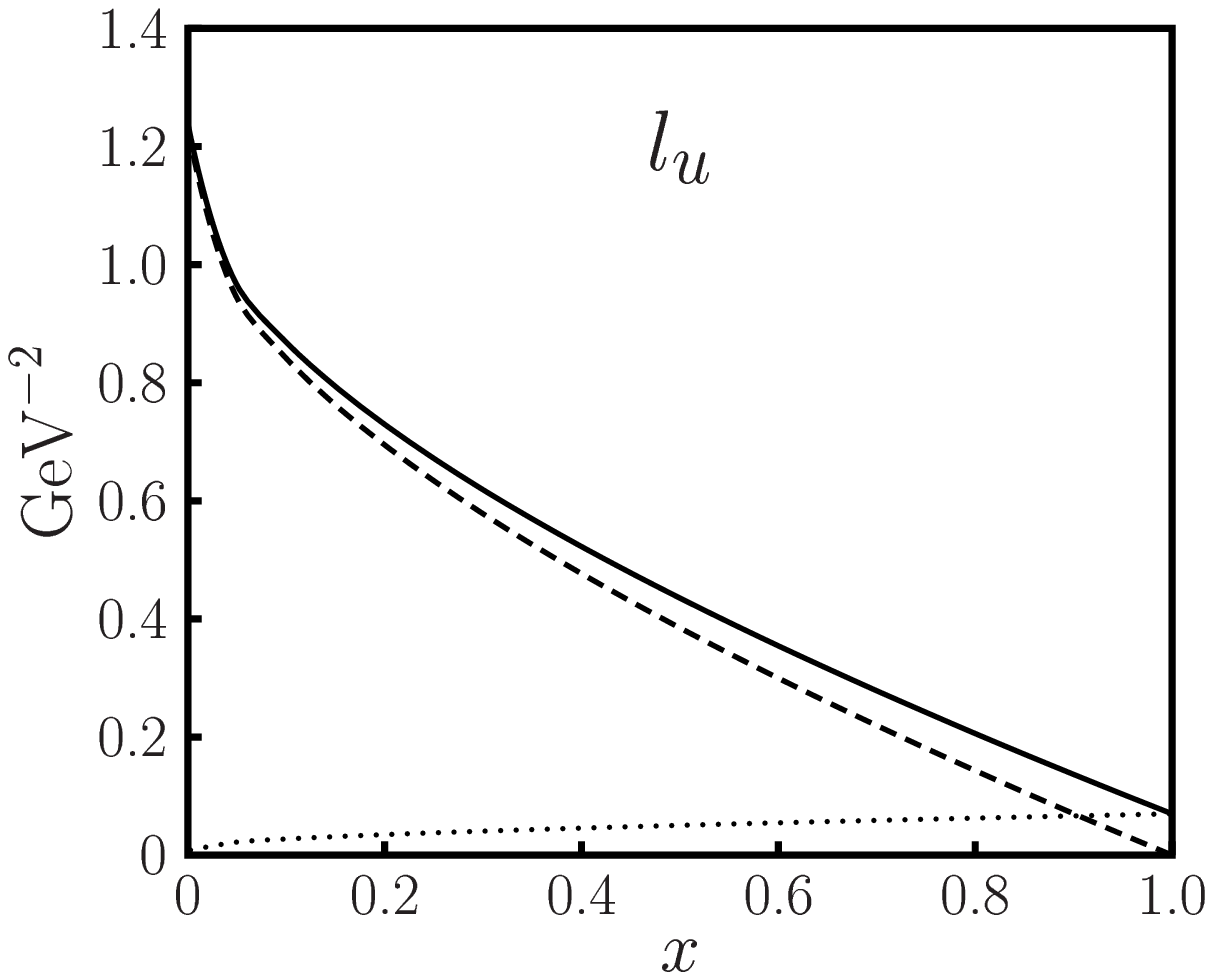}
\hspace{3em}
\includegraphics[width=0.45\textwidth,
  bb=45 340 400 630]{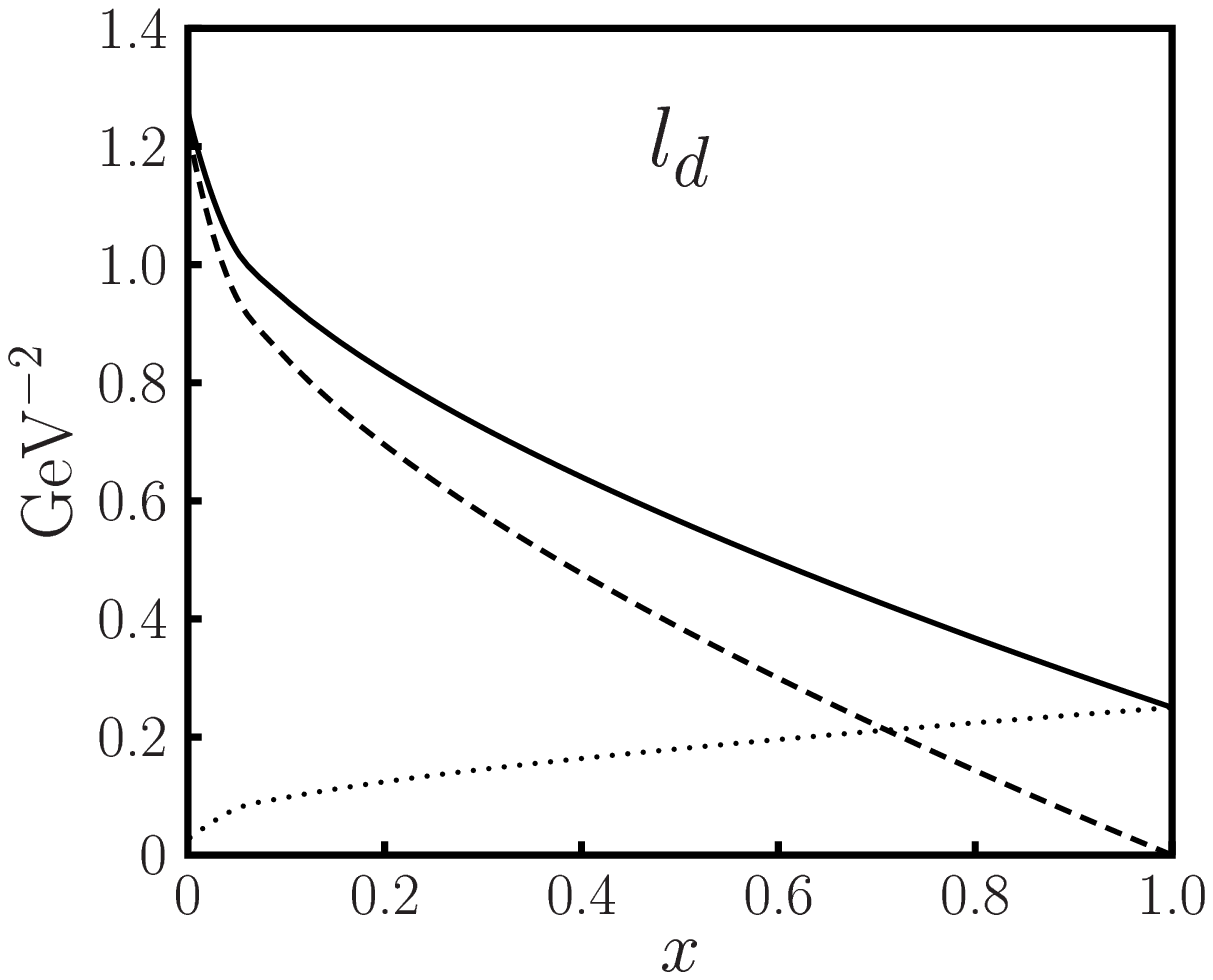}
\end{center}
\caption{\label{fig:var1_leff} The function $l_q(x)$ from
(\protect\ref{l-def}) for the best fit to (\protect\ref{fit:var1})
with $\mu=2\gev$.  The contributions from terms going with $\alpha'$
(dashed line) and $A_q$ (dotted line) are shown separately.}
\end{figure}

Let us now see which of the parameters in our fit is most important in
the profile function at given~$x$.  In Fig.~\ref{fig:var1_leff} we
show the quantities
\begin{equation}
  \label{l-def}
l_q(x) = \frac{f_q(x)}{\log\frac{1}{x}}
\end{equation}
resulting from our fit, where we have divided out the factor
$\log(1/x)$ in order to have a finite quantity in the limit $x\to 0$.
We also show the individual contributions to $l_q$ from the terms
$\alpha' [\log(1/x) - (1-x)]$ and $A_q (1-x)$ in $f_q$ and see that
the value of $\alpha'$ controls the profile function in almost the
entire $x$ range for $u$ quarks and in a substantial region for $d$
quarks.  The fitted value of $\alpha'$ thus reflects dynamics at both
small and large $x$ (in the fit it has to find a compromise between
these regions).  We can hardly expect it to give a good representation
of the physics in the region where Regge phenomenology is relevant,
say for $x \lsim 10^{-2}$.

The simplest profile function satisfying the constraints
(\ref{regge-constraint}) and (\ref{large-x-constraint}) with $n=2$ is
$f_q(x) = \alpha' (1-x) \log(1/x)$, which has been proposed in
\cite{bur:01} and used for numerical studies in \cite{bur:02}.  An
obvious extension of this ansatz is
\begin{equation}
f_q(x) = \alpha' (1-x) \log\frac{1}{x} + 
        ( A_q - \alpha' ) (1-x)^2 \eqpt
\label{fit:var2}
\end{equation}
A fit with free parameters $\alpha'$, $A_u$ and $A_d$ does not give a
good description of the form factor data, see
Table~\ref{tab:old-fits}.  Having an overall $\chi^2/\dof=5.25$ it
systematically overshoots the $F_1^p$ data at $|t|$ above $10 \gev^2$,
namely by 15\% to 20\% for $|t| \gsim 20 \gev^2$.  Comparing with the
good fit we obtained with (\ref{fit:var1}) one might conclude that the
data prefers a behavior $f_q(x) \sim (1-x)$ over $f_q(x) \sim (1-x)^2$
at $x\to 1$, but this would be mistaken as we shall see below.

In search of a more adequate profile function we demand that
\begin{itemize}
\item the low-$x$ behavior of $f_q(x)$ should match the form
      (\ref{regge-constraint}), where we now impose the value
      $\alpha'=0.9$~GeV$^2$ from Regge phenomenology,
\item the high-$x$ behavior should be controlled by the parameter
      $A_q$ in (\ref{large-x-constraint}) and not by $\alpha'$,
\item the intermediate $x$-region should smoothly interpolate between
      the two limits, with a few additional parameters providing
      enough flexibility to enable a good fit to the form factor data.
\end{itemize}
We found  these requirements to be satisfied by the forms
\begin{equation}
        \label{fit:n1}
f_q(x) = \alpha' (1-x)^{2} \log\frac{1}{x} + B_q (1-x)^{2} 
        + A_q \, x (1-x) 
\end{equation}
and
\begin{equation}
        \label{fit:n2}
f_q(x) = \alpha' (1-x)^{3} \log\frac{1}{x} + B_q (1-x)^{3} 
        + A_q \, x (1-x)^{2} \eqcm
\end{equation}
which respectively correspond to $n=1$ and $n=2$ in
(\ref{large-x-constraint}).  At large $x$, the individual terms behave
like $\alpha' (1-x)^{n+2}$, $B_q (1-x)^{n+1}$ and $A_q (1-x)^n$, which
in particular prevents the term with $\alpha'$ from being too
important in the high-$x$ region.

\begin{figure}
\begin{center}
\leavevmode 
\includegraphics[width=0.47\textwidth,
  bb=45 340 400 630]{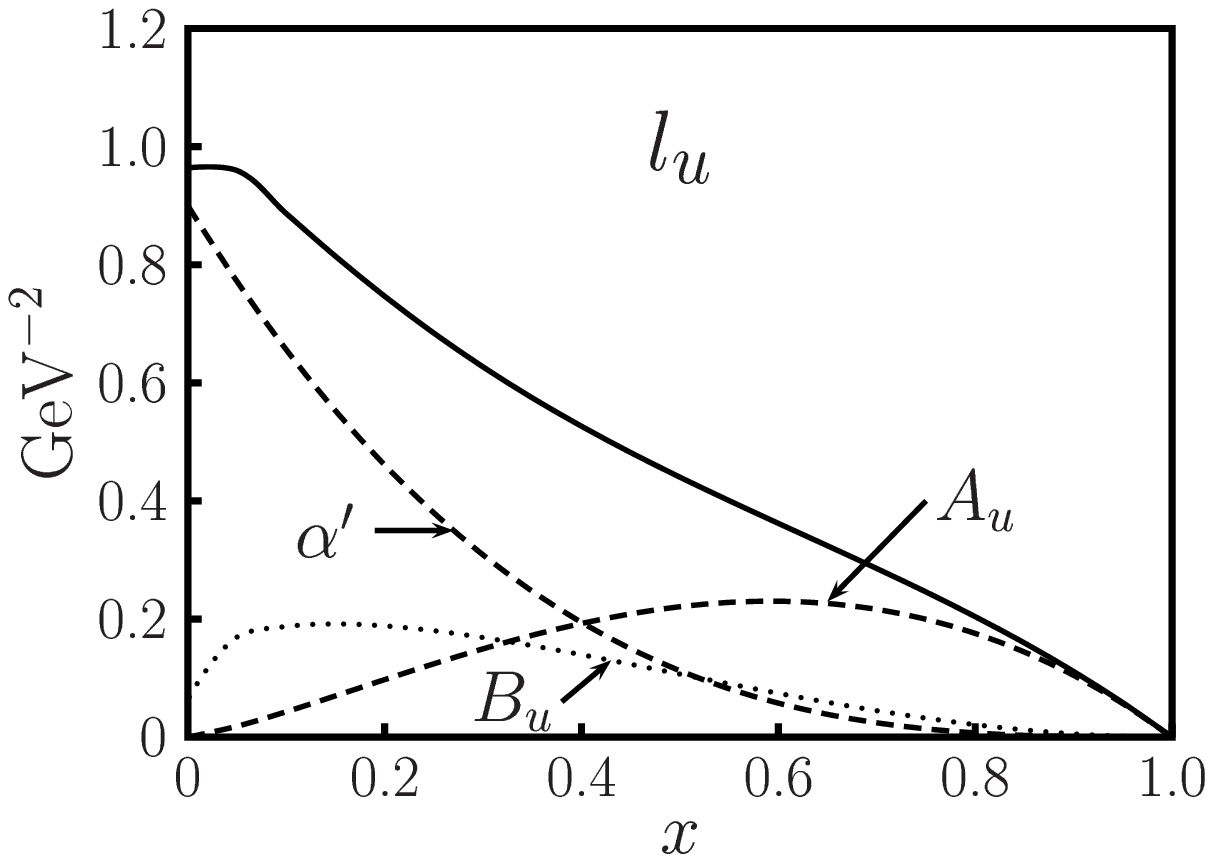}
\hspace{2em}
\includegraphics[width=0.47\textwidth,
  bb=45 340 400 630]{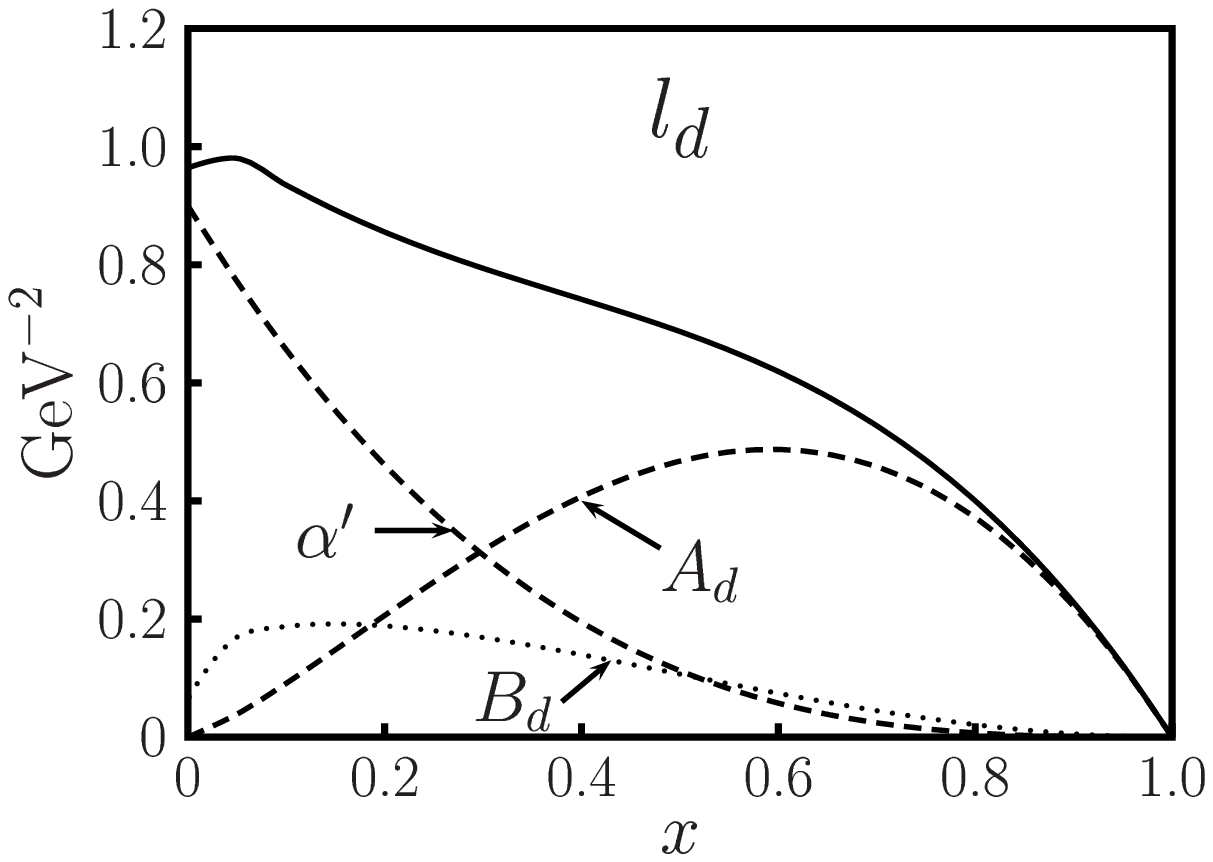}
\end{center}
\caption{\label{fig:luld_n2} The function $l_q(x)$ from
(\protect\ref{l-def}) for the best fit to (\protect\ref{fit:n2}) with
$\mu=2\gev$.  The contributions from terms going with $\alpha'$, $B_q$
and $A_q$ are shown separately.}
\end{figure}

As we see in Tables~\ref{tab:n1-fits} and \ref{tab:n2-fits}, a fit to
either (\ref{fit:n1}) or (\ref{fit:n2}) with $A_u$, $A_d$ and $B_u =
B_d$ as free parameters provides a good description of the form factor
data.  We will comment on setting $B_u = B_d$ in
Sect.~\ref{sec:varia}.  In Fig.~\ref{fig:luld_n2} we show the profile
functions divided by $\log(1/x)$ obtained in the fit to
(\ref{fit:n2}), as well as the individual contributions from the terms
with $\alpha'$, $B_q$ and $A_q$.  Our criterion that the profile
function should be controlled by $\alpha'$ at low $x$ and by $A_q$ at
high $x$ is now well satisfied.  To a lesser degree this also holds
for the fit to (\ref{fit:n1}), where the contribution of the $\alpha'$
term to $f_u(x)$ is about $30\%$ at $x=0.7$ and $15\%$ at $x=0.8$.
The sensitive region of $x$ in both fits is essentially the same as
for the fit to (\ref{fit:var1}), with the values of $x_{\rm max}(t)$
and $\langle x\rangle_t$ differing by less than 2\% from those shown
in Fig.~\ref{fig:var1_xminmax}.  The difference of $x_{\rm min}(t)$ in
the different fits is more pronounced at small $t$ but below 5\% for
$-t$ above $1\gev^2$.

We see from Tables~\ref{tab:n1-fits} and \ref{tab:n2-fits} that
$f_d(x)$ has clearly larger errors than $f_u(x)$.  This is a feature
of all our fits (except when we force $f_u(x)$ and $f_d(x)$ to be
equal) and can readily be understood from the sum rules
(\ref{dirac-sum-rules}).  Due to the charge factors, the integrand
giving $F_1^p$ is dominated by the contribution from $u$ quarks.  This
trend is enhanced by the fact that $u$ quarks are more abundant in the
proton than $d$ quarks.  For large $x$ the ratio $d_v/u_v$ of parton
densities becomes very small indeed (see Fig.~\ref{fig:ud-ratio}), and
one can expect this trend to persist for $H_v^d/H_v^u$ at least over
some range in $t$.  The combination of $F_1^p$ and $F_1^n$ provides
sensitivity to $d$ quarks, but data on both form factors is only
available in a relatively small interval of $|t|$.  Improved data on
$F_1^n$ in a wider range of $t$ would be highly welcome in this
context.

\begin{figure}[t]
\begin{center}
\leavevmode
\includegraphics[width=0.45\textwidth,
  bb=45 210 405 545]{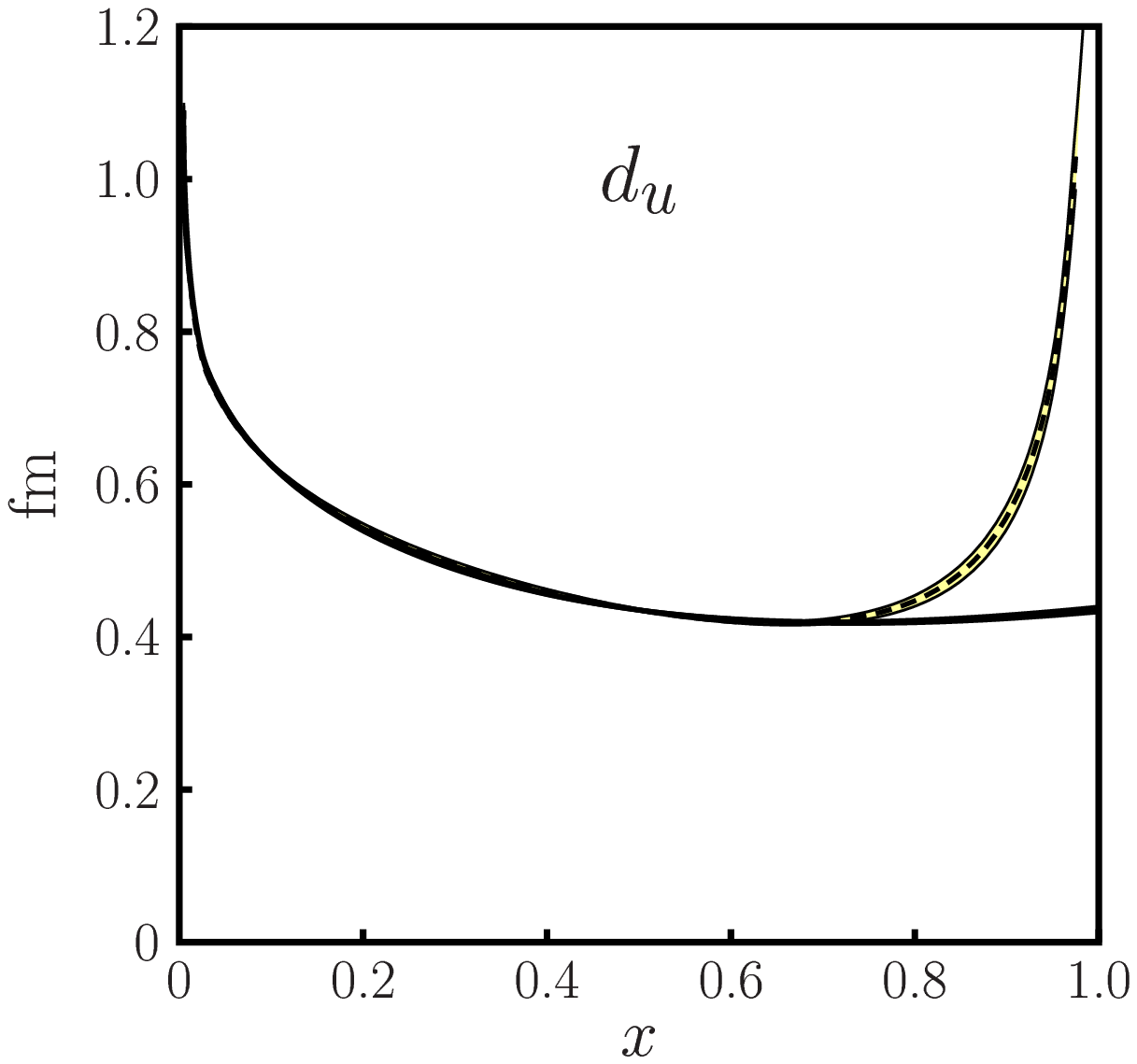}
\hspace{3em}
\includegraphics[width=0.45\textwidth,
  bb=45 210 405 545]{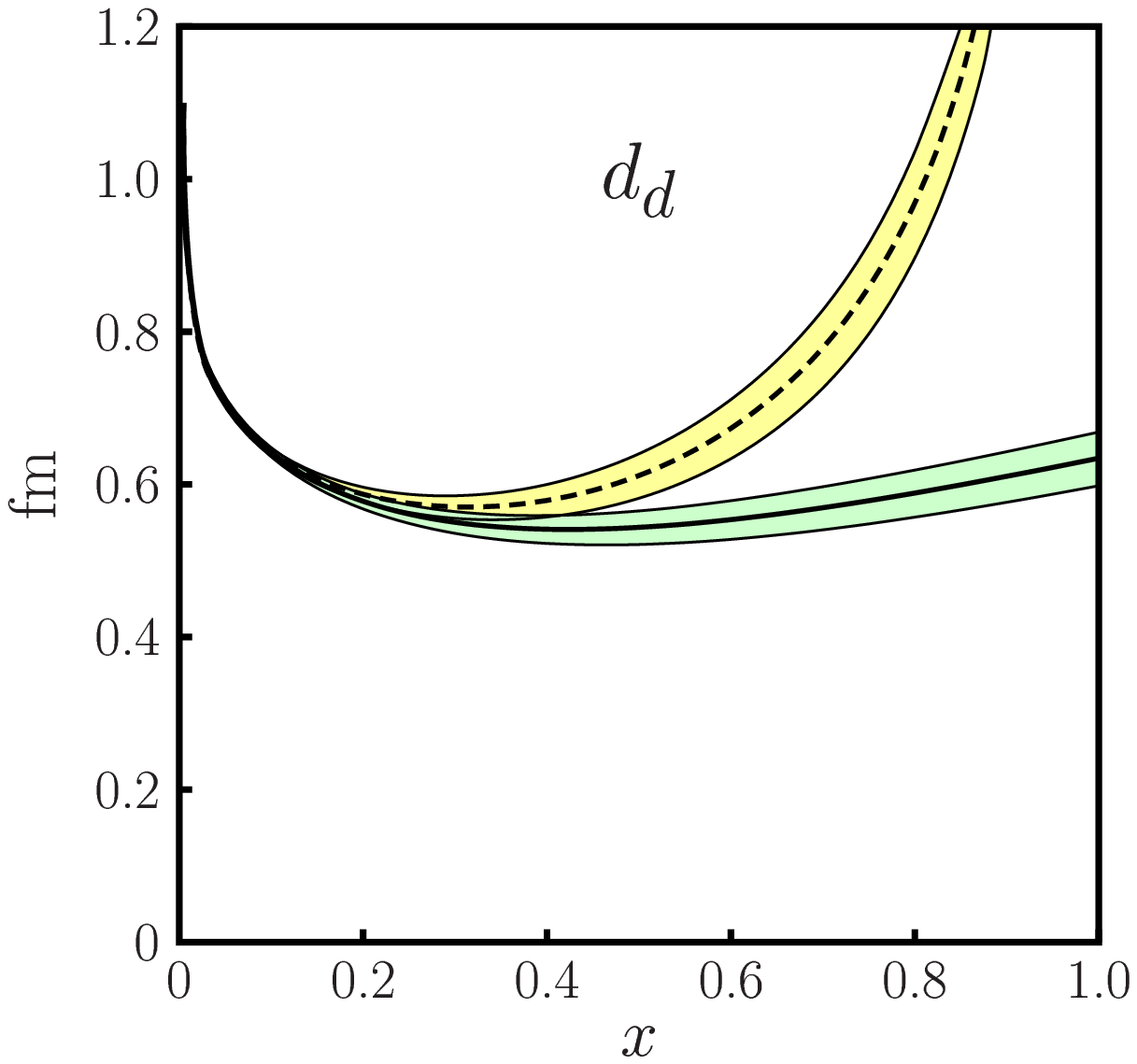}
\end{center}
\caption{\label{fig:n2_rq} The distance $d_q(x)$ between struck quark
and spectators, evaluated for the best fits to (\protect\ref{fit:n1})
(dashed line) and (\protect\ref{fit:n2}) (solid line) with
$\mu=2\gev$.  The smallest $x$ value plotted is $5 \cdot 10^{-3}$.
Shaded bands indicate the $1\, \sigma$ uncertainties of the fits as
explained in App.~\ref{app:B}.}
\end{figure}

In Fig.~\ref{fig:n2_rq} we show the distance $d_q(x)$ obtained with
our fits to (\ref{fit:n1}) and to (\ref{fit:n2}).  For $u$ quarks the
results of the two fits are fully compatible within their errors up to
about $x=0.75$.  The $x$-region where they differ significantly is
outside the range where our fit to the form factors can constrain
them.  For $d$ quarks the results start to differ at lower values of
$x$, but their errors are significantly bigger as well.  We conclude
that the limiting behavior of $d_q(x)$ for $x\to 1$ cannot be
determined by data on $F_1^p$ up to $|t|$ around $30 \gev^2$.  We note
that the description of $F_1^p$ at high $|t|$ is slightly better for
the fit to (\ref{fit:n2}) than for the one to (\ref{fit:n1}), where
$F_1^p(t)$ falls off a bit too fast.  One should however not interpret
this as a preference of the data for $n=2$ rather than $n=1$ in the
power-law falloff (\ref{large-x-constraint}) since the situation is
opposite for the fits to (\ref{fit:var1}) and (\ref{fit:var2}), see
Tables~\ref{tab:old-fits}, \ref{tab:n1-fits} and \ref{tab:n2-fits}.
Which value of $n$ a fit prefers thus depends on the remaining
functional dependence of $d_q(x)$.  Without data constraining $d_q(x)$
for $x$ above $0.8$ we cannot determine its behavior around $x=1$.

We also see in Fig.~\ref{fig:n2_rq} that $d_u$ from the fit with $n=1$
takes values one may suspect to be unphysically large only for $x$
above $0.9$ or so, where even the forward parton densities are barely
known.  Similarly, for models obtained with the Gaussian wave
functions (\ref{gauss-wfs}) and the parameters used in
\cite{bol:96,DFJK1}, the distance $d_q$ stays below 1~fm for $x \lsim
0.9$.  In the kinematic range where these models have been used to
describe or predict data we hence do not find them physically
inconsistent.  In the following we will however take the form with
$n=2$ in (\ref{large-x-constraint}), since its limiting behavior for
$x\to 1$ is more plausible than the one with $n=1$.  An exponent $n$
above $2$, which results in a vanishing $d_q$ for $x\to 1$, may also
be possible \cite{bur:04}.  Since the form factor data cannot
determine $n$ we refrain however from further investigation of this
point.  We henceforth refer to the fit to (\ref{master-ansatz}) and
(\ref{fit:n2}) at $\mu=2\gev$ as our ``default fit''.  Its parameters
are
\begin{eqnarray}
  \label{params:n2}
A_u  &=&  (1.22 \pm 0.02) \gev^{-2}, \qquad\qquad
A_d \:=\: (2.59 \pm 0.29) \gev^{-2}, 
\nonumber \\
B_u &=& B_d \:=\: (0.59 \pm 0.03) \gev^{-2},
\end{eqnarray}
and $\alpha' = 0.9 \gev^{-2}$, with full details given in
App.~\ref{app:B}.  Note that in this fit the parameter $A_q$ has a
simple physical interpretation as the limit of $d_q^{\,2} /4$ for
$x\to 1$.  To a good approximation it also gives the value of this
quantity over a finite range of large $x$.

\subsection{Features of the default fit}
\label{sec:feat}

\begin{figure}[p]
\leavevmode
\begin{center}
\includegraphics[width=.47\textwidth,
  bb=120 235 500 590]{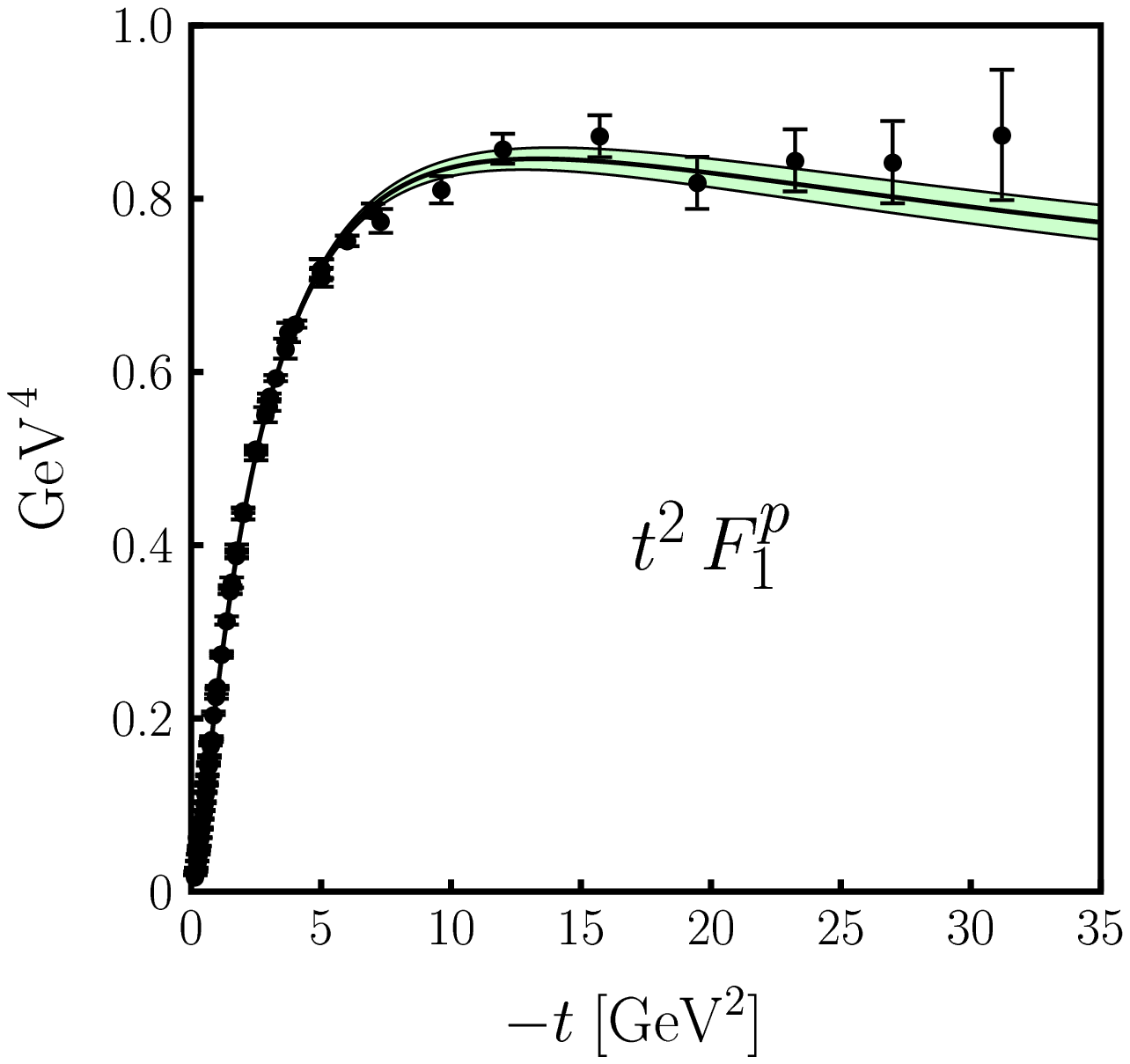}
\hspace{2em}
\includegraphics[width=.47\textwidth,
  bb=110 337 490 690]{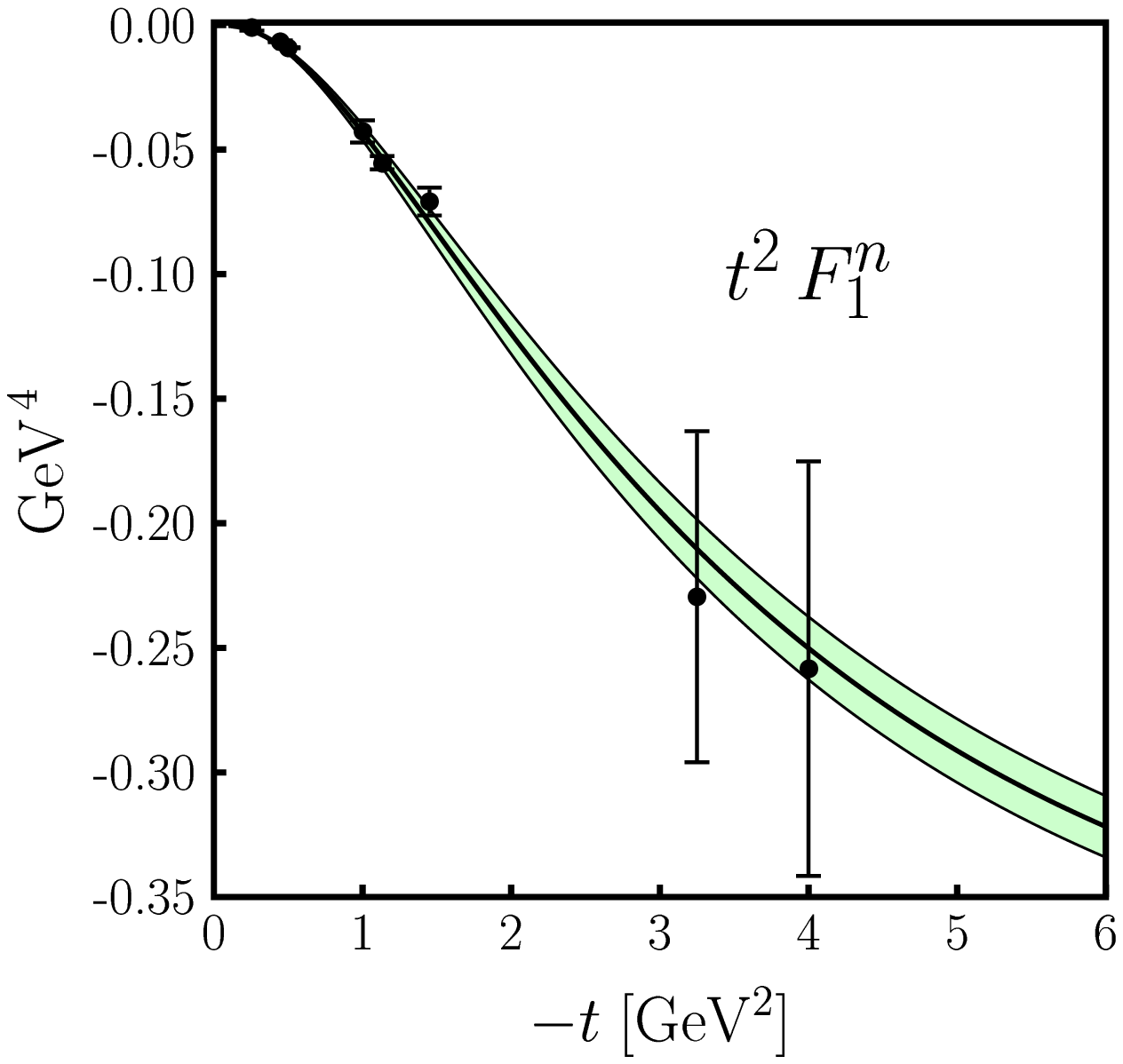}
\end{center}
\caption{\label{fig:n2} Results for the Dirac form factor of proton
and neutron with our default fit, defined by
(\protect\ref{master-ansatz}) and (\protect\ref{fit:n2}) with the
parameters (\protect\ref{params:n2}) and the valence quark densities
evaluated at scale $\mu=2 \gev$.  The solid lines represent the best
fit, and the error bands represent the $1\, \sigma$ uncertainties of
the fit as explained in App.~\protect\ref{app:B}.  Details on the form
factor data and their errors are given in App.~\protect\ref{app:A}.}
\end{figure}

\begin{figure}[p]
\begin{center}
\includegraphics[width=.382\textwidth,
  bb=146 284 443 565]{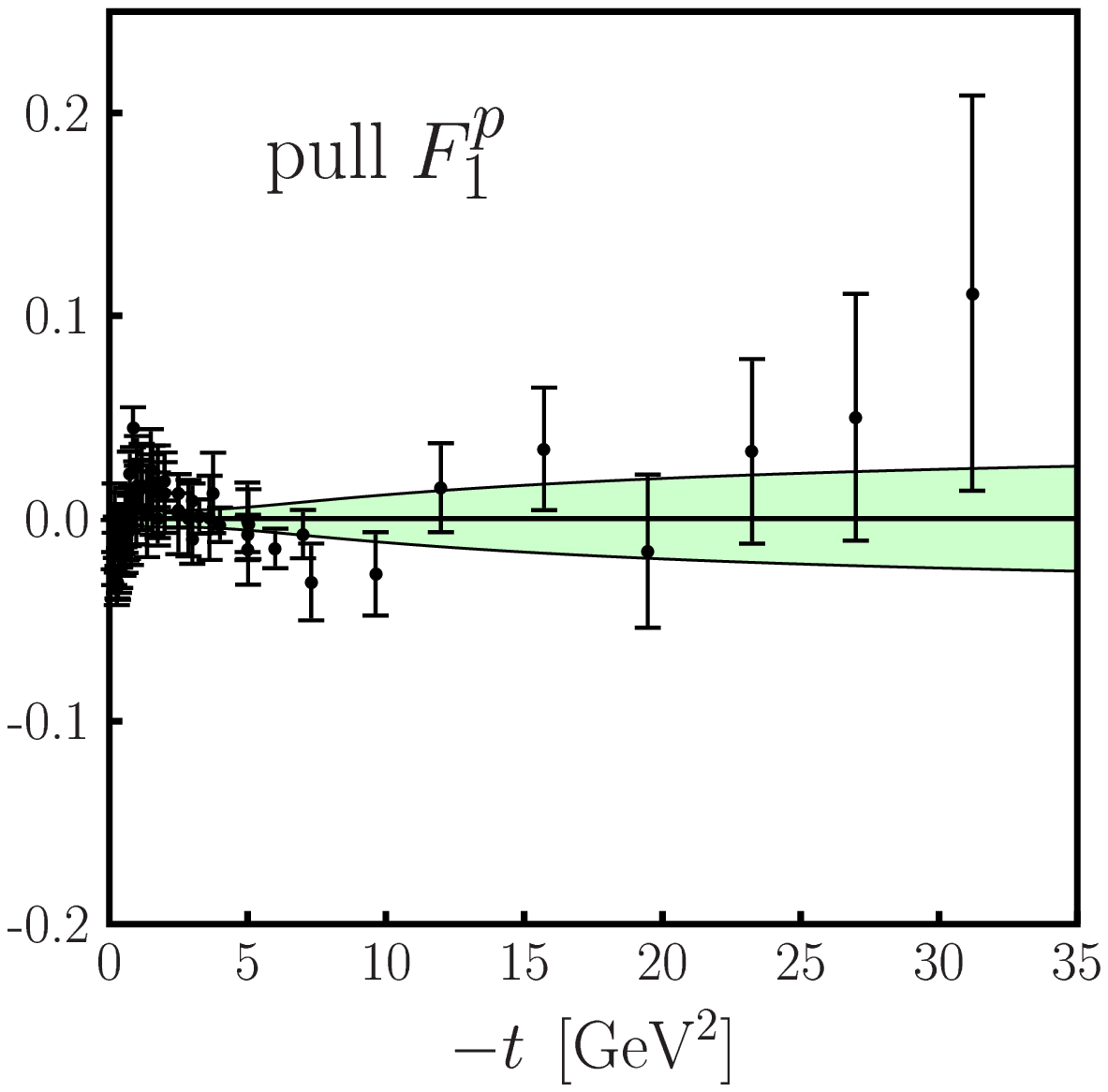}
\hspace{6em}
\includegraphics[width=.36\textwidth,
  bb=151 416 434 701]{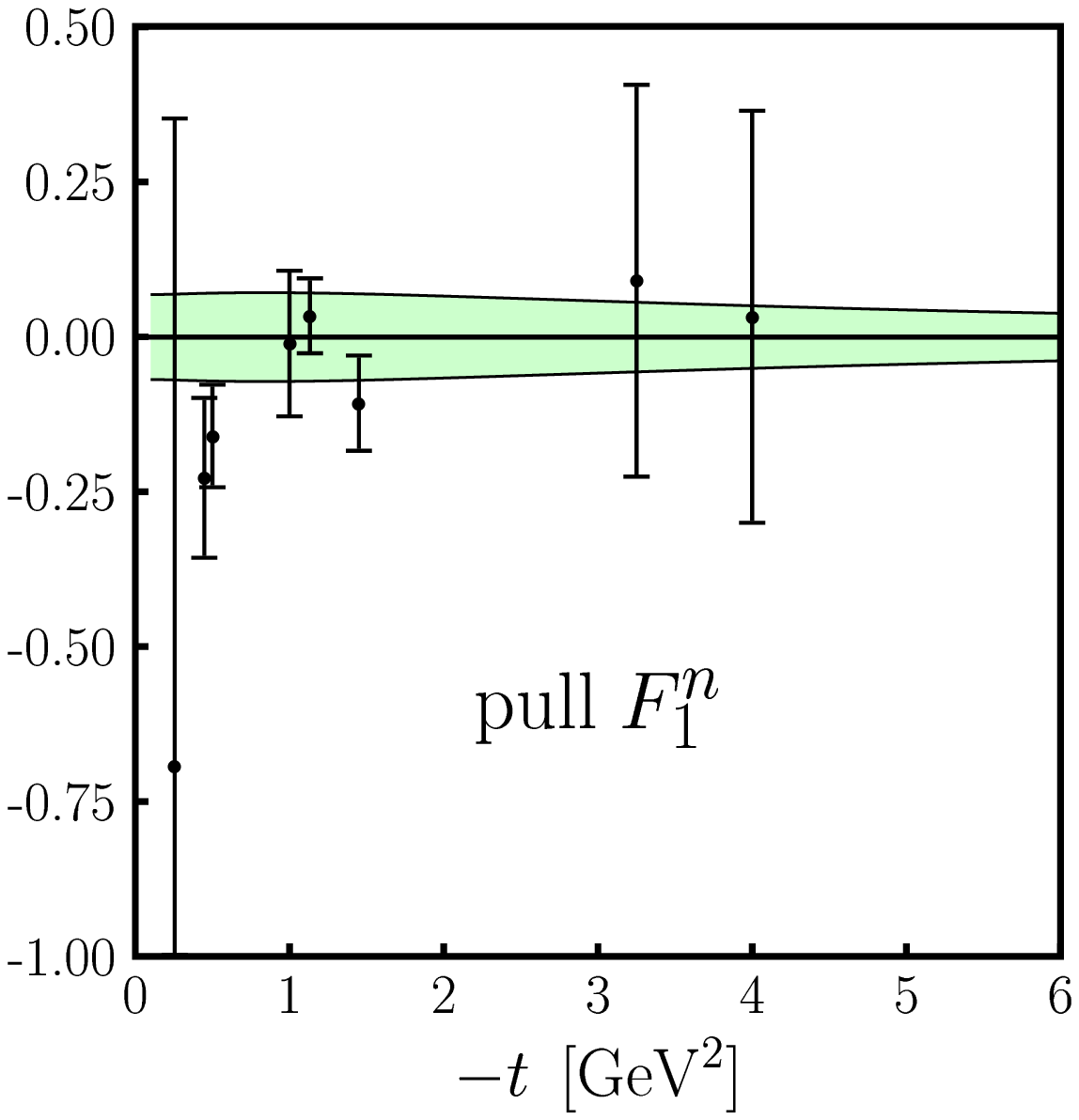}
\\[4em]
\end{center}
\caption{\label{fig:n2pull} The pull $[F_1({\rm data})/F_1({\rm fit})
    - 1]$ for our default fit to the Dirac form factors of the
  nucleon.  The error bands represent the $1\, \sigma$ fit
  uncertainties relative to the best fit.}
\end{figure}

In Fig.~\ref{fig:n2} we compare the form factor data with the result
of our default fit, whose parametric uncertainties are shown by the
shaded bands.  To have a clear representation of the data at large $t$
we have scaled the form factors with $t^2$.  The quality of the fit at
smaller $t$ is better seen from the ``pull'', defined as $[F_1({\rm
data})/F_1({\rm fit}) - 1]$ and shown in Fig.~\ref{fig:n2pull}.  Our
fitted GPDs describe $F_1^p$ within 5\% for $-t$ up to $27 \gev^2$,
only the last data point at $-t = 31.2 \gev^2$ has a larger pull.
Apart from the data point at $-t = 0.255 \gev^2$ with its huge errors,
our fit describes $F_1^n$ within $25\%$.  A detailed inspection
reveals that a large contribution to $\chi^2$ is due to five data
points in the sample of \cite{Janssens:1966}, with $0.18\gev^2 \le -t
\le 0.86\gev^2$.  The relative errors for these data are only about
$1\%$, which is below the accuracy we are aiming at.  We are hence not
worried by the comparatively high $\chi^2$ value found in our fit.

The distance $d_q(x)$ between the struck quark and the spectators
shown in Fig.~\ref{fig:n2_rq} is one of our main physics results.  Our
analysis provides the first data-driven determination of this
important quantity.  It exhibits a significant decrease when going
from small to intermediate $x$.  We will comment on the increase of
$d_d(x)$ for $x \gsim 0.4$ at the end of Sect.~\ref{sec:varia}.

\begin{figure}[t]
\begin{center}
\leavevmode
\includegraphics[width=0.47\textwidth,
  bb=105 350 455 690]{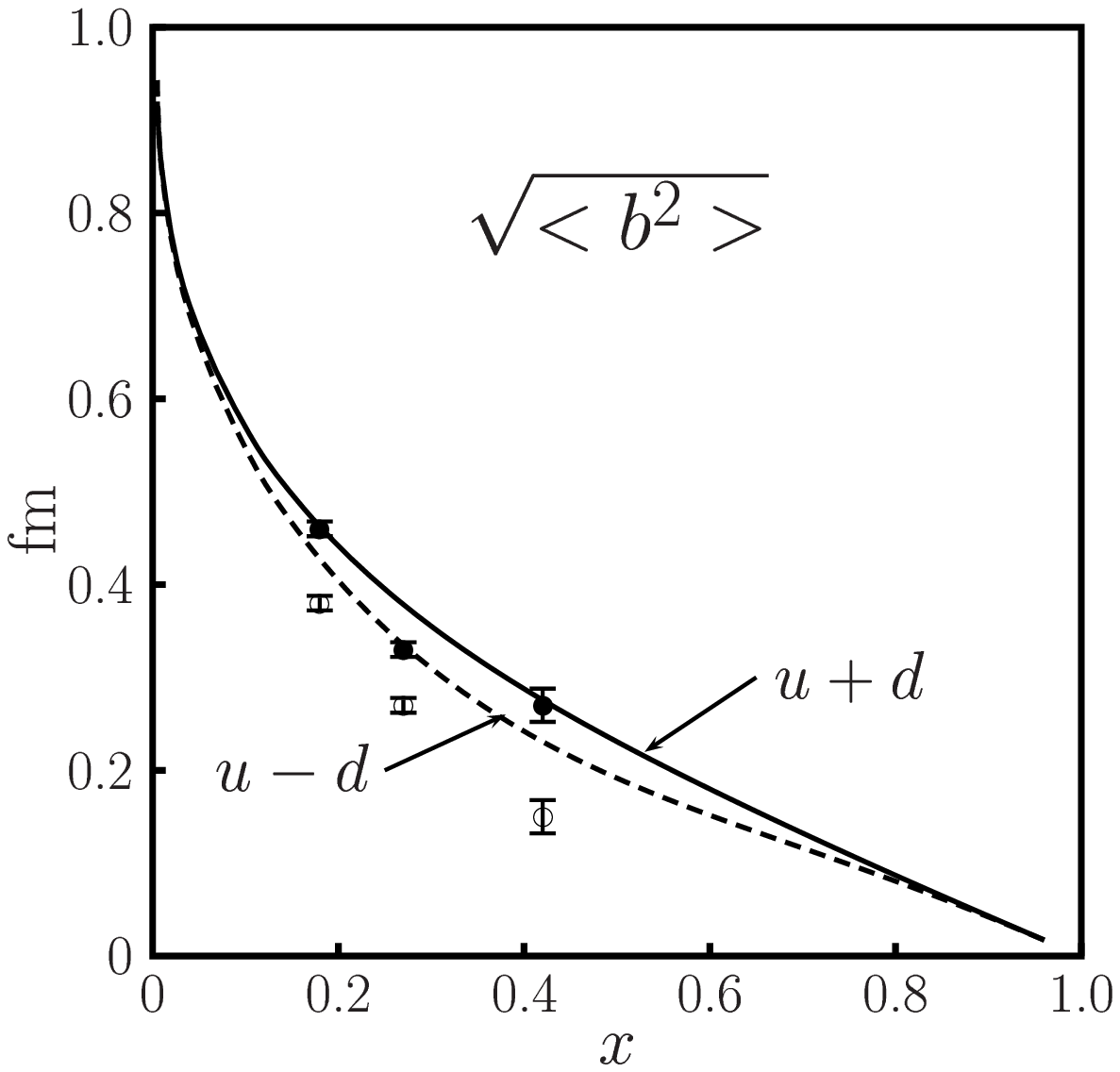}
\hspace{2em}
\includegraphics[width=0.47\textwidth,
  bb=105 380 455 720]{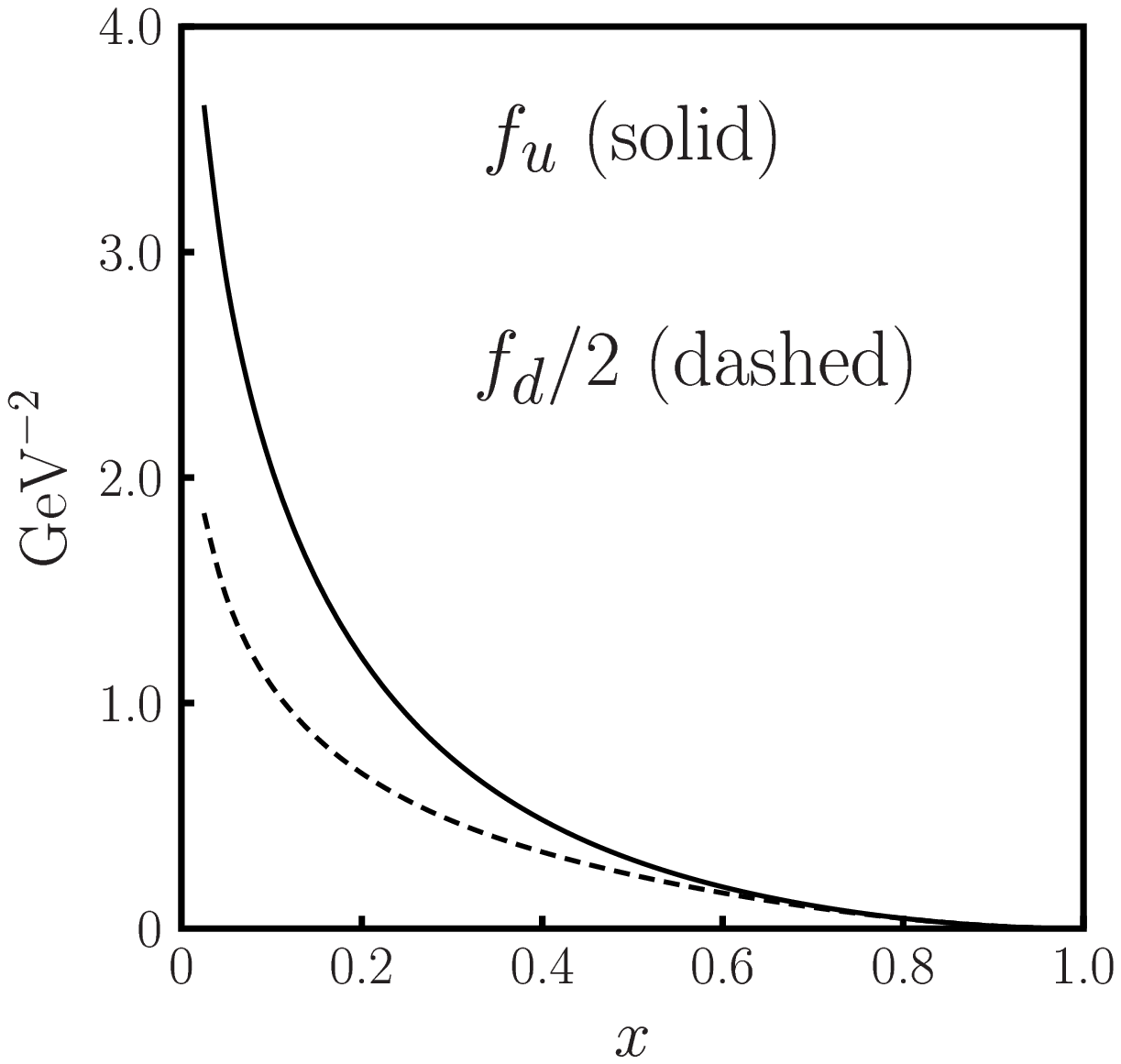}
\end{center}
\caption{\label{fig:n2_bf} Left: Square root of the average impact
parameters $\bsq{}{x}$ for the sum and the difference of $u$ and $d$
quark distributions obtained in our default fit, compared with lattice
QCD results from~\protect\cite{neg:04}.  The smallest $x$-value
plotted is $5 \cdot 10^{-3}$.  Right: The profile functions $f_u$ and
$f_d$ obtained in our default fit (\protect\ref{fit:n2}) (solid) with
$\mu=2\gev$.  For better visibility $f_d$ has been scaled by a factor
$1/2$.  The smallest $x$-value plotted is $2.5 \cdot 10^{-2}$.}
\end{figure}

In Fig.~\ref{fig:n2_bf} we show the square root of the average impact
parameters $\bsq{}{x}$ for the sum and the difference of $u$ and $d$
quark distributions, defined as in (\ref{b-def}) with $H_v^q$ replaced
by $H_v^u \pm H_v^d$.  That $\bsq{}{x}$ comes out somewhat bigger for
$u+d$ than for $u-d$ reflects that $\bsq{u}{x} < \bsq{d}{x}$ in our
fit.  The points shown in the figure are results from an evaluation of
moments $\int_{-1}^1 \d x\, x^m [ H^u(x,t) \pm H^d(x,t) ]$ in lattice
QCD 
\cite{neg:04}.  A quantitative comparison of the two results must be
made with due caution.  For one thing, the values of $x$ in the
lattice calculation have been estimated from the ratios of successive
moments in $x$ at $t=0$.  We could of course avoid this problem by
directly comparing our results for $x$ moments with those obtained on
the lattice.  More importantly, however, the lattice calculation was
performed for a pion mass of 870~MeV, and an extrapolation to the
physical pion mass has not been attempted in \cite{neg:04}.  Indeed,
the falloff in $t$ of $(F_1^u + F_1^d)$ and $(F_1^u - F_1^d)$ obtained
in that calculation \cite{hae:03b,neg:04} is too slow to correctly
describe the data for $F_1^p$ and $F_1^n$.  Together with the
uncertainties inherent in our phenomenological extraction, we
nevertheless find the overall agreement of the results in
Fig.~\ref{fig:n2_bf} remarkable.

In Fig.~\ref{fig:n2_bf} we also show the profile functions $f_u$ and
$f_d$ themselves.  The scale controlling the decrease of $H_v^q(x,t)$
with $t$ is seen to depend very strongly on $x$, not only in the
regions of very large or very small $x$ but also when going from, say,
$x=0.1$ to $x=0.3$.


\subsection{Variations of the fit}
\label{sec:varia}

The $1\, \sigma$ errors quoted in our tables and the associated error
bands in the figures reflect the uncertainty on the free parameters in
a \emph{given} parameterization of the GPDs, but not the uncertainty
due to the \emph{choice} of parameterization.  In order to obtain a
better feeling how significant various features of our default fit
are, we have tested their stability under several modifications of the
fit.

To investigate the difference between the profile functions for $u$
and $d$ quarks we have fitted the data to the form (\ref{fit:n2}) with
or without setting $B_u=B_d$ and $A_u=A_d$, see
Table~\ref{tab:n2-fits}.  The fit with all four parameters free leads
only to a slightly better description of the data than our default fit
with $B_u=B_d$.  The fitted parameters are compatible with those in
the default fit at $2\, \sigma$ level, but the errors on $A_d$ and
$B_d$ are much larger.  We find that the presently available data on
$F_2^n$ do not warrant two free parameters for $d$ quarks.  Note that
although $B_u > B_d$ in this fit, the resulting function $f_u(x)$ only
becomes larger than $f_d(x)$ for $x<0.1$, where the difference between
the two functions is at most 5\%.  Already at $x=0.3$ the ratio
$f_d/f_u$ has grown to a value of 1.4, to be compared with 1.3 in our
default fit.  A fit where we impose both $A_u=A_d$ and $B_u=B_d$
cannot adequately describe the neutron data.  The $\chi^2$ in this
subsample is 74 for 8 data points, and the fit result for $|F_1^n|$
undershoots the data by at least $30\%$ for $-t \ge 1 \gev^2$.  The
same happens if we take the profile function (\ref{fit:n1}) with
$f_u(x)=f_d(x)$, see Table~\ref{tab:n1-fits}.  A three-parameter fit
to (\ref{fit:n2}) with the constraint $A_u=A_d$ finds $B_u < B_d$.  It
still undershoots the data on $|F_1^n|$ for $-t \ge 1 \gev^2$,
although not as badly as the fit where both $A_u=A_d$ and $B_u=B_d$.

We conclude that if we insist on having a good description of both
proton and neutron form factors, we need $f_d(x) > f_u(x)$ at moderate
to large values of $x$.  This implies that the suppression of $d$
compared with $u$ quarks seen in the forward parton densities at high
$x$ becomes even stronger for $H_v^d$ and $H_v^u$ as $|t|$ increases.
Note that the observed rise of the form factor ratio $-F_1^n(t)
/F_1^p(t)$ implies that the flavor contribution $F_1^d(t)$ decreases
faster with $|t|$ than $F_1^u(t)$.  This is seen by writing
\begin{equation}
  \label{ud-ratio-F1}
R_1(t) = - \frac{2 F_1^n(t)}{F_1^p(t)} = 
       \frac{1 - r_1(t)}{1 - \frac{1}{4}\, r_1(t)} \, ,
\qquad \qquad \qquad
r_1(t) = \frac{2 F_1^d(t)}{F_1^u(t)} \, .
\end{equation}
With the data giving $R_1 \approx 0.37$ at $-t= 1 \gev^2$ one has $r_1
\approx 0.7$, which is clearly different from the value $r_1 = 1$ at
$t=0$.  To have $r_1(t)$ decreasing sufficiently fast, the damping
factor $f_q$ in the $t$ dependence must be bigger for $H_v^d$ than for
$H_v^u$ when we take the exponential form (\ref{master-ansatz}).  The
same trend is observed with the power-law dependence on $t$ we
investigate in Sect.~\ref{sec:power-law-t}, see
Table~\ref{tab:power-fits}.

\begin{figure}
\begin{center}
\leavevmode
\includegraphics[width=0.60\textwidth]{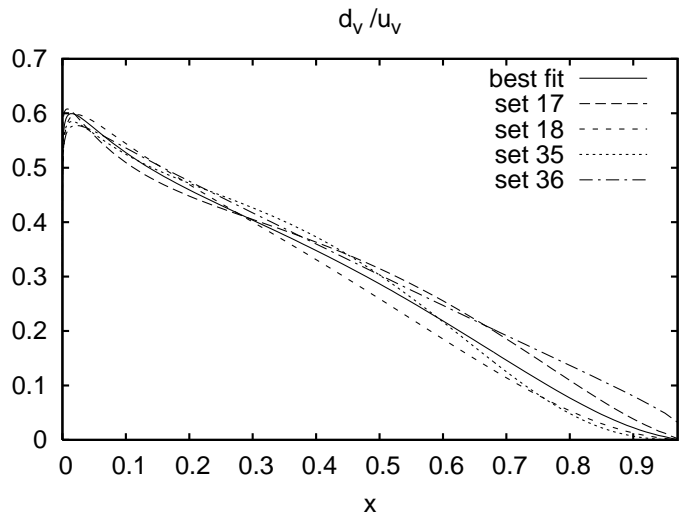}

\vspace{-1em}

\end{center}
\caption{\label{fig:ud-ratio} The ratio $d_v(x) /u_v(x)$ obtained with
different sets of CTEQ6M parton distributions~\protect\cite{CTEQ} at
scale $\mu=2 \gev$.  The largest plotted value of $x$ is $0.97$.}
\end{figure}

It is well known that as $x$ becomes larger, the parton densities
extracted from data become more and more uncertain.  In their analysis
\cite{CTEQ} CTEQ provide 40 sets of parton densities which reflect
variations from their best fit results that are allowed within errors.
Among these we find sets 17, 18 and 35, 36 to provide the largest
deviations from the best fit parton densities at large $x$, reaching
20\% for $u_v$ at $x=0.9$ and for $d_v$ at $x=0.7$ and growing well
beyond $100\%$ for $d$ quarks at higher~$x$.  In
Fig.~\ref{fig:ud-ratio} we show the corresponding ratio $d_v(x)
/u_v(x)$, which is especially important for the simultaneous
description of the form factors $F_1^p$ and $F_1^n$ in our fit.
Repeating our default fit with these error distributions as input, we
find good stability of the obtained GPDs, see
Table~\ref{tab:error-fits}.  The profile functions $f_u$ for the error
distributions deviate by at most 3\% from the one for the CTEQ best
fit, as well as $f_d$ for sets 17 and 18.  With both sets 35 and 36
the ratio of $f_d$ obtained for the error distribution and for the
CTEQ best fit grows from 1 to about 1.1 as $x$ rises from 0 to 1.  The
uncertainties on the forward parton densities thus hardly affect our
extraction of the impact parameter profile of parton distributions as
a function of $x$.

\begin{figure}
\begin{center}
\leavevmode
\includegraphics[width=0.4\textwidth, height=0.35\textwidth,
  bb =119 333 466 671,clip=true]{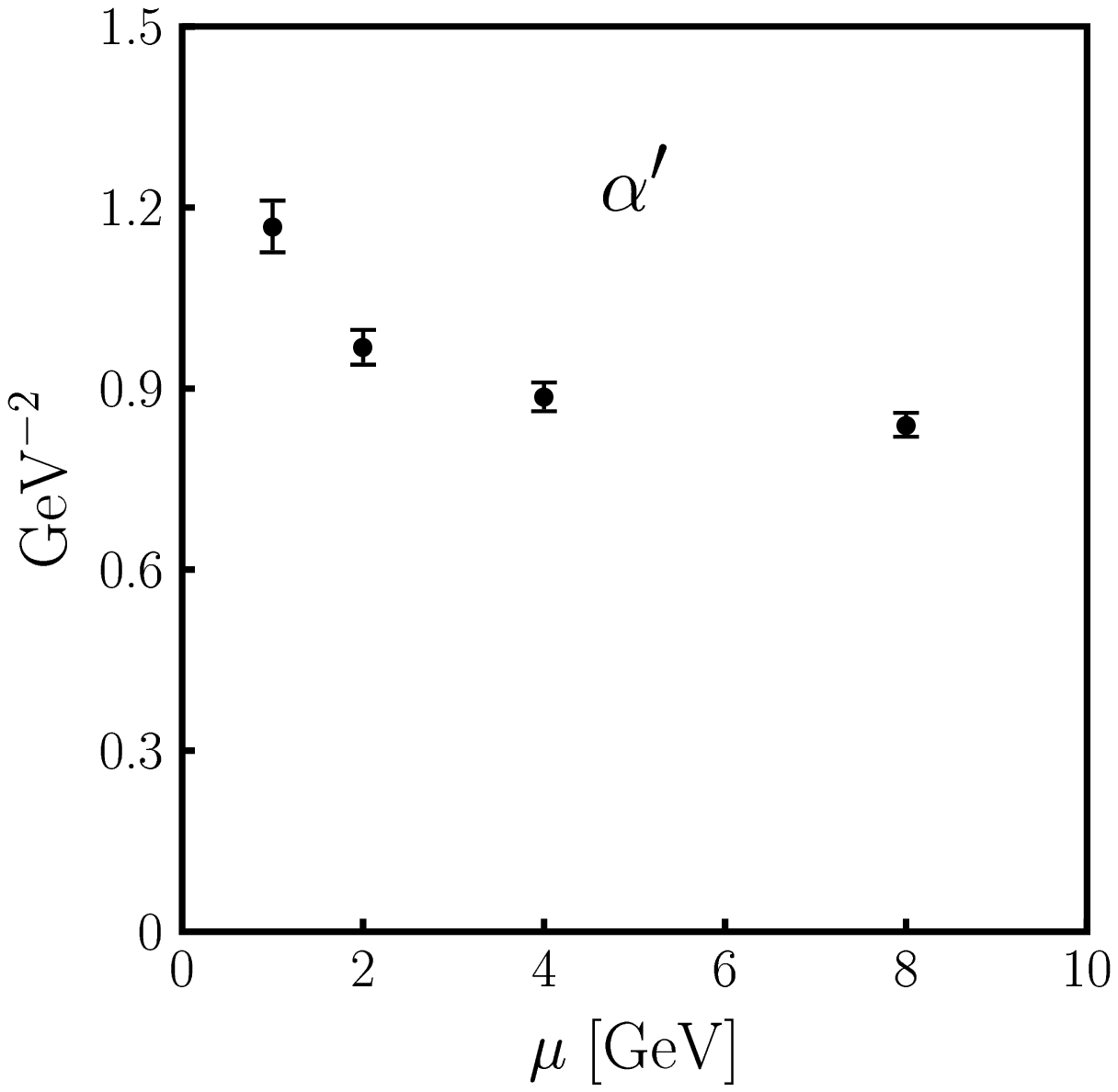}
\hspace{3em}
\includegraphics[width=0.4\textwidth, height=0.345\textwidth,
  bb= 103 383 453 722,clip=true]{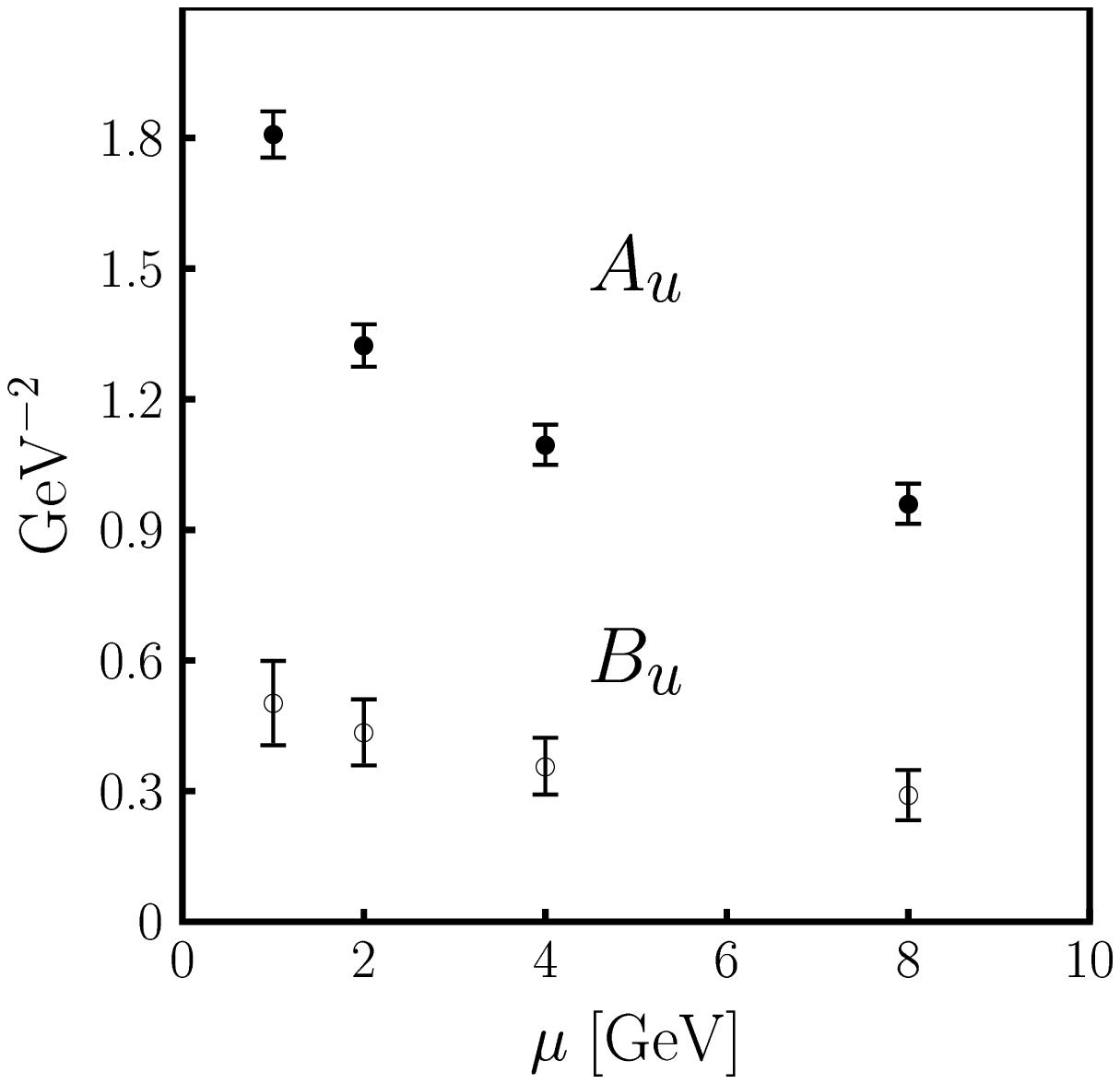}
\end{center}
\caption{\label{fig:evolv} Parameters and their errors obtained in a
fit to (\protect\ref{master-ansatz}) and (\protect\ref{fit:n2}) with
$B_u=B_d$ and forward parton distributions taken at different scales
$\mu$.  The corresponding parameters are given in
Table~\protect\ref{tab:alpha-mu-fits}.}
\end{figure}

We have finally allowed the value of $\alpha'$ in (\ref{fit:n2}) to be
selected by the fit, and in addition we have taken the forward
distributions in our ansatz (\ref{master-ansatz}) at different scales
$\mu$.  The results are given in Table~\ref{tab:alpha-mu-fits}.  In
all cases we obtain a rather good description of the form factor data,
although there is a tendency for the fits to become worse for larger
$\mu$.  We see that our parameterization of GPDs is reasonably
flexible to cover a range of factorization scales.  This also
validates the analytical considerations in Sect.~\ref{sec:phys-mot},
which showed that in selected regions of $x$ and $t$ our functional
form of GPDs is approximately stable under a change of $\mu$.  The
profile functions $f_u$ and $f_d$ obtained in our fits decrease with
$\mu$, precisely as we expect from the evolution equation
(\ref{dglap-bsq}) for $\bsq{}{x}$.  In Fig.~\ref{fig:evolv} we show
the change of parameters $\alpha'$, $A_u$ and $B_u=B_d$ with $\mu$.
The uncertainty on $A_d$ is too large to observe a clear evolution
effect.  The mild $\mu$ dependence in the central values of $\alpha'$
does not contradict our general analysis in Sect.~\ref{sec:phys-mot},
where $\alpha'$ describes the behavior of the profile function at very
small~$x$, whereas the parameter $\alpha'$ in our ansatz for $f_q(x)$
is relevant in a finite $x$ interval (see Fig.~\ref{fig:luld_n2}).
Given this and the uncertainties on $\alpha'$ within Regge
phenomenology, we conclude that the fitted values of this parameter
confirm our assumption that the valence GPDs at small $x$ and $t$ can
be described by the leading Regge trajectories known from the
phenomenology of soft hadronic interactions.  The results of this
exercise also justifies the choice of fixing $\alpha'$ in our default
fit, although it is clear that its parametric errors underestimate the
uncertainty of the profile function in the small-$x$ limit.

\begin{figure}
\begin{center}
\leavevmode
\includegraphics[width=0.45\textwidth,
  bb=45 400 400 730]{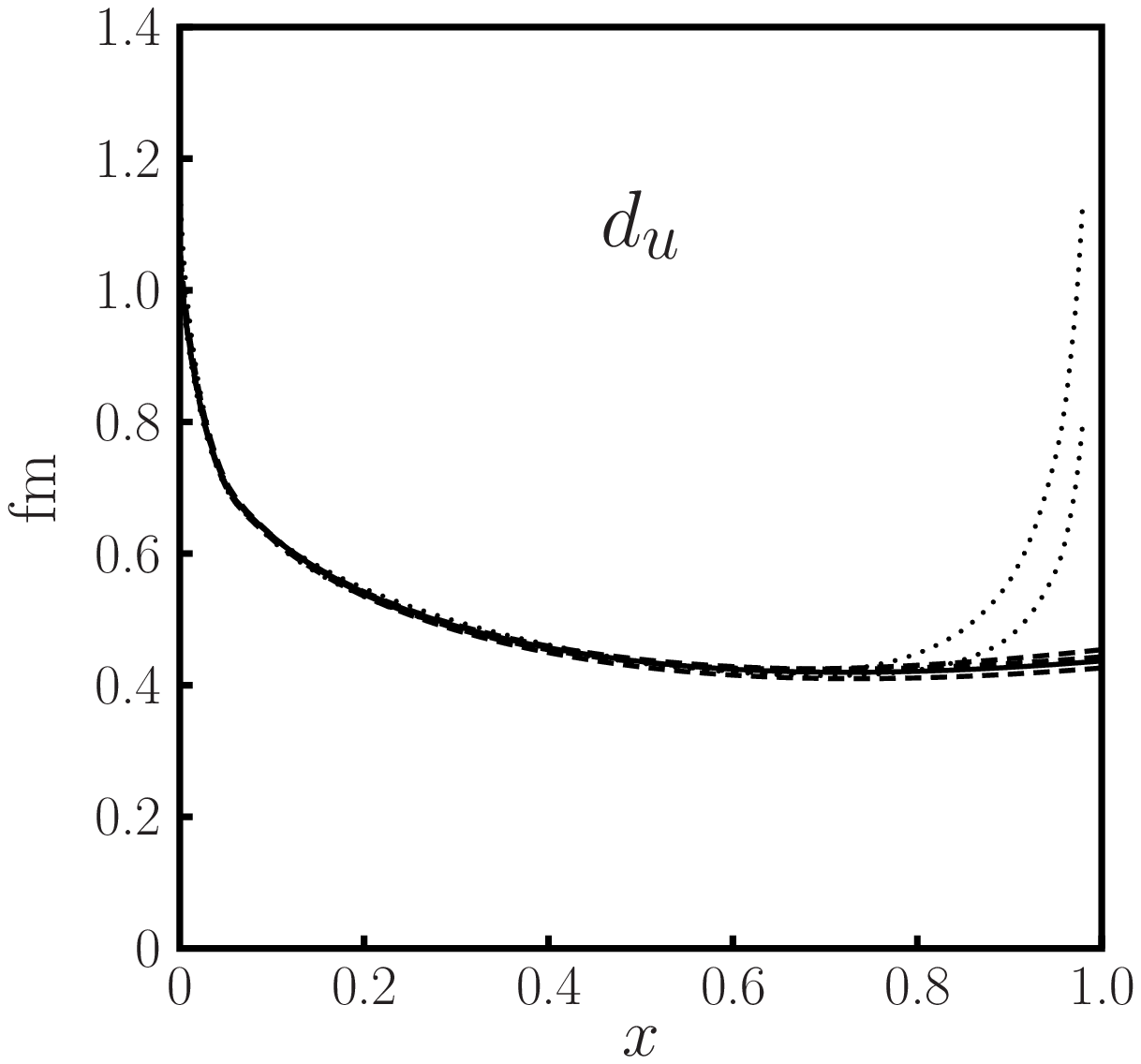}
\hspace{3em}
\includegraphics[width=0.45\textwidth,
  bb=45 400 400 730]{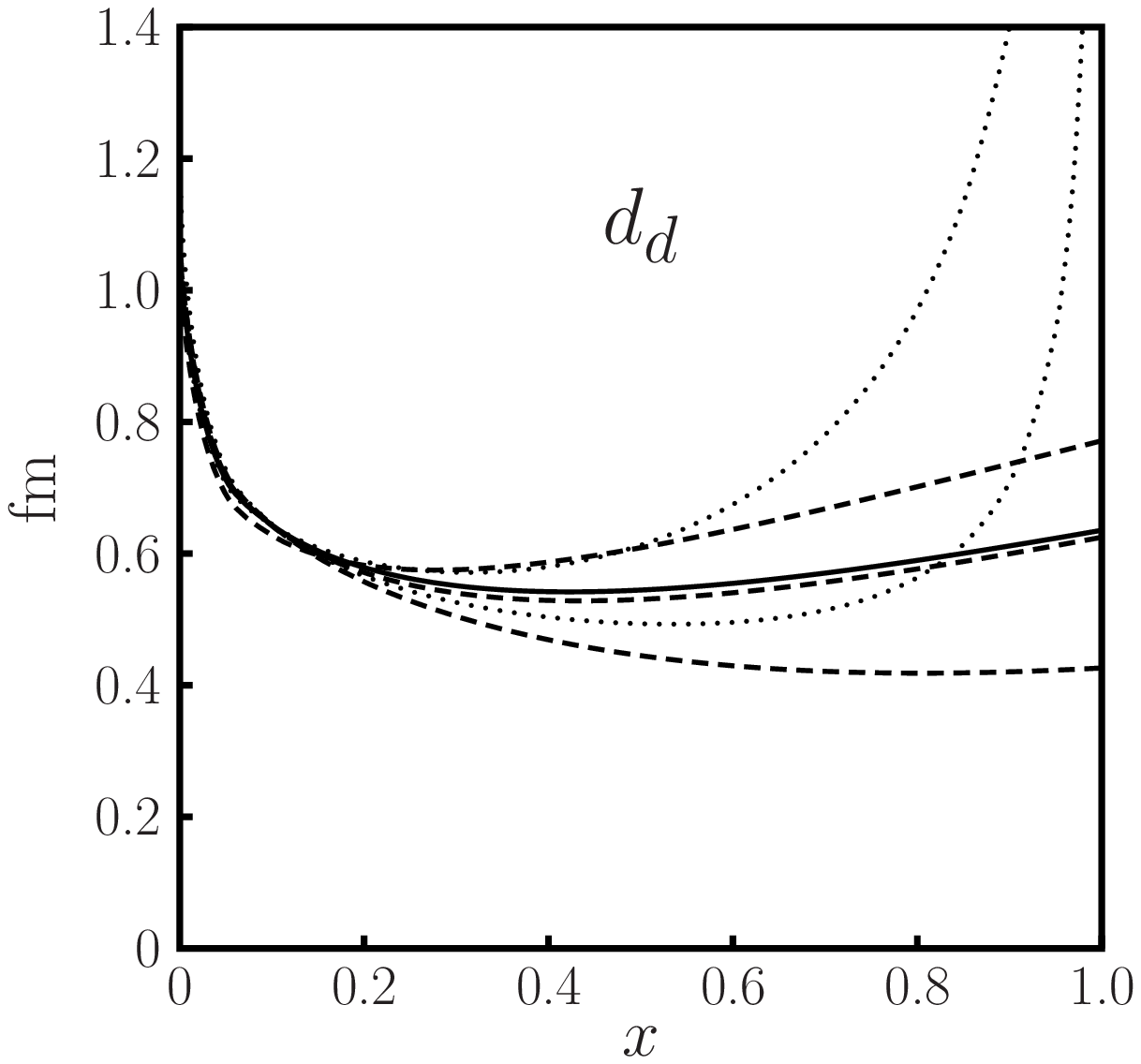}
\end{center}
\caption{\label{fig:H-syst} The average distance $d_q(x)$ for $u$ and
$d$ quarks for selected fits.  The solid lines show our default fit,
and dashed lines the fits in the second and third rows of
Table~\protect\ref{tab:n2-fits} and in the second row of
Table~\ref{tab:alpha-mu-fits}.  Dotted lines show the fits in the
first row of Table~\protect\ref{tab:old-fits} and the first row of
Table~\protect\ref{tab:n1-fits}, where $d_q(x) \sim (1-x)^{-1/2}$ for
$x\to 1$.}
\end{figure}

To illustrate the uncertainties of our fit results due to the choice
of parameterization, we show in Fig.~\ref{fig:H-syst} the average
distances $d_u(x)$ and $d_d(x)$ obtained with selected fits, which
provide good descriptions for both proton and neutron data.  We see
that for $u$ quarks there is little influence of the parameterization
up to $x \sim 0.8$. For $d$ quarks the uncertainties are larger, due
to the lack of good neutron data at higher values of $t$.  The curve
with the lowest values of $d_d(x)$ in the figure belongs to the fit in
Table~\ref{tab:n2-fits} where $A_u=A_d$ but $B_u \neq B_d$, which
provides only a moderately good description of the neutron data.  This
is the only fit among those shown where $d_u$ and $d_d$ differ by less
than $10\%$ over the entire $x$ range.  In all other cases we observe
in particular that $d_d(x)$ rises for $x$ above a certain moderate
value, in order to accommodate a rise of the ratio $f_d/f_u$.

\subsection{Large $t$ and the Feynman mechanism}
\label{sec:feynman}

\begin{figure}[b]
\begin{center}
\leavevmode
\includegraphics[width=0.95\textwidth]{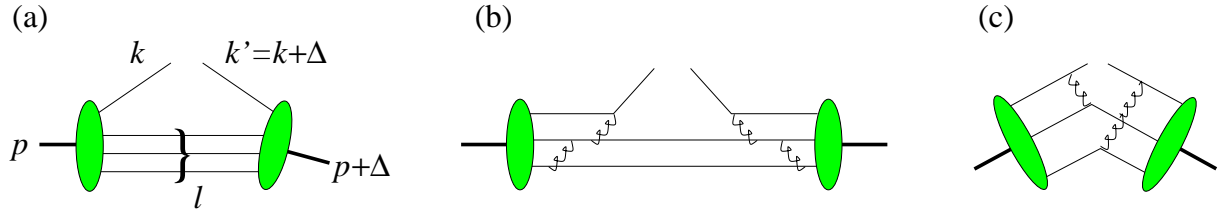}
\end{center}
\caption{\label{fig:feynman} (a) Momenta of nucleon, active quark and
spectator system in a GPD.  (b) Perturbative mechanism for large $x$.
(c) Hard-scattering mechanism \protect\cite{Lepage:1980fj} for large
$t$.}
\end{figure}

In this subsection we will see that our parameterization of GPDs
enables the Feynman mechanism at large $t$, where the struck quark
carries most of the nucleon momentum and thus avoids large internal
virtualities of order $t$.  Let us first consider $H_v^q(x,t)$ in the
limit of large $x$ and take a closer look at the scaling of momenta,
which we denote as shown in Fig.~\ref{fig:feynman}a.  Defining
light-cone coordinates, $v^{\pm} = (v^0 \pm v^3) /\sqrt{2}$ and
$\tvec{v}_\perp = (v^1, v^2)$ for any four-vector $v$, we have $x= k^+
/p^+$.  We choose a reference frame where $\Delta^+ = 0$ (i.e.\
$\xi=0$) and $\tvec{p}_\perp = \tvec{0}$.  Then $t = -
\tvec{\Delta}_\perp^2 \,$, and the virtuality of the active quark
before it is struck can be written as
\begin{equation}
  \label{active-before}
k^2 = x m^2 + l^2 - \frac{l^2 + \tvec{l}_\perp^2}{1-x} ,
\end{equation}
where $m$ is the nucleon mass.  For small $(1-x)$ we distinguish two
momentum regions according to the virtuality and transverse momentum
of the spectator system:
\begin{equation}
\begin{array}[b]{lll}
  \label{soft-classif-x}
\textit{soft region:} \qquad &
|l^2| \sim \tvec{l}_\perp^2 \sim \Lambda^2 , \qquad &
|k^2| \sim \Lambda^2 /(1-x) ,
\nonumber \\[0.4em]
\textit{ultrasoft region:} &
|l^2| \sim \tvec{l}_\perp^2 \sim (1-x) \Lambda^2 , \qquad &
|k^2| \sim \Lambda^2 ,
\end{array}
\end{equation}
where $\Lambda$ is a typical scale of soft interactions.  Our
nomenclature is similar to the one in recent work on soft-collinear
effective theory \cite{Beneke:2003pa}.  Note that in the ``soft
region'' the spectator partons are soft, but the struck quark is far
off-shell (in the parlance of \cite{Beneke:2003pa} it would be
identified as ``hard-collinear'').  In the ``ultrasoft region'' the
spectator system has virtuality and squared transverse momentum much
smaller than $\Lambda^2$.  Such momentum regions do appear in the
perturbative analysis of graphs if quarks are treated as massless (see
e.g.~\cite{Collins:1996fb,Becher:2003qh}), but one may suspect that
due to confinement they cannot be important in physical matrix
elements.

An analogous classification holds with respect to the virtuality of
the active quark after it is struck, with
\begin{equation}
  \label{active-after}
k'^2 = x m^2 + l^2 - \frac{l^2 + (\tvec{l}_\perp 
  - (1-x) \tvec{\Delta}_\perp)^2}{1-x} .
\end{equation}
Note that with our choice of frame, $\tvec{l}_\perp - (1-x)
\tvec{\Delta}_\perp$ is the intrinsic transverse momentum of the
spectator system in the outgoing nucleon (see e.g.~\cite{DFJK1}).

That the struck quark in the soft region is far off-shell is the basis
of a perturbative analysis, which has long ago been given for the $x
\to 1$ limit of parton distributions or deeply inelastic structure
functions (see e.g.~\cite{Mueller:1981sg}) and recently also for the
$x\to 1$ limit of GPDs at arbitrary fixed $t$ \cite{Yuan:2003fs}.  It
is based on graphs as the one shown in Fig.~\ref{fig:feynman}b, where
a configuration of three quarks with momentum fractions of order
$\frac{1}{3}$ and virtualities of order $\Lambda^2$ turns into a
configuration with a fast off-shell quark and two soft quarks by
successive emission of gluons, which need to be off-shell too.
Standard perturbative power counting for these graphs in the momentum
region just stated gives a behavior $H_v^q(x,t) \sim (1-x)^3$ for
$x\to 1$.  Their actual calculation in perturbation theory leads
however to severe divergences $\d \tvec{l}^2 /\tvec{l}^4$ in the
infrared unless the quark mass is kept finite, which indicates that
standard hard-scattering factorization \cite{Lepage:1980fj} does not
provide an adequate separation of short-and long-distance physics for
the mechanism represented by the graphs.  While the general
power-counting argument might still give the correct answer, further
details such as the overall normalization are currently beyond
theoretical control.  Resummation of radiative corrections into
Sudakov form factors can give a stronger suppression than
the power-law $(1-x)^3$ obtained from fixed-order graphs, but given
the above difficulties the detailed form of these corrections (let
alone their quantitative impact) is unknown.  We recall that the
single logarithms resummed by DGLAP evolution also modify the power of
$(1-x)$, see e.g.~\cite{Peterman:1978tb}.

Phenomenologically, the limiting behavior of parton densities for
$x\to 1$ is poorly known.  Their extraction from hard processes is
increasingly difficult in this limit due to higher-twist
contributions, and leading-twist analyses can use data only for values
of $x$ where such contributions are under control.  It is then
difficult to infer a power behavior for $x\to 1$, as our attempt to
extract the large-$x$ asymptotics of the profile function in
Sect.~\ref{sec:select} has taught us.  The powers $(1-x)^\beta$
appearing in many parameterizations of parton densities are to be seen
as parameters describing the approximate behavior of these functions
over a certain range of large $x$.  The CTEQ6M distributions we use in
our analysis have powers $\beta_u\approx 2.9$ for $u_v$ and
$\beta_d\approx 5.0$ for $d_v$ in the parameterization at the starting
scale 
$\mu= 1.3 \gev$.  We find that at $\mu= 2\gev$ these distributions are
described by $u_v(x) \sim (1-x)^{3.4}$ and $d_v(x) \sim (1-x)^{5.0}$
within 5\% for $0.5\le x \le 0.9$.  Taking different $x$-intervals,
these parameters slightly change.

In form factors at large $t$ the relevant $x$-values of the
corresponding GPDs are selected by the dynamics.  Demanding $k^2$ and
$k'^2$ to be of the same order, the soft and ultrasoft regions are
identified as
\begin{equation}
\begin{array}[b]{llll}
  \label{soft-classif-t}
\textit{soft region:} \qquad &
1-x \sim \Lambda\, /\sqrt{|t|} , \qquad  &
|k^2|, |k'^2| \sim \Lambda \sqrt{|t|} , \qquad &
l^+, l^-, \tvec{l}_\perp \sim \Lambda ,
\nonumber \\[0.4em]
\textit{ultrasoft region:} &
1-x \sim \Lambda^2 /\, |t| , \qquad  &
|k^2|, |k'^2| \sim \Lambda^2 , \qquad &
l^+, l^-, \tvec{l}_\perp \sim \Lambda^2 /\sqrt{|t|} .
\end{array}
\end{equation}
The contribution from the soft region has been analyzed in
perturbation theory in the same way as for parton distributions
\cite{Lepage:1980fj,Mueller:1981sg}.  Power counting for the graph in
Fig.~\ref{fig:feynman}b gives $F_1(t) \sim t^{-2}$.  This is the same
power as obtained with the standard hard-scattering mechanism shown in
Fig.~\ref{fig:feynman}c, where parton virtualities are of order $|t|$.
Again one may expect a further damping in the soft region by Sudakov
corrections, leaving the hard-scattering mechanism as the dominant
contribution at asymptotically large $-t$.

Let us now investigate the large-$t$ behavior of the form factors
obtained with our parameterizations of GPDs.  We see in
Fig.~\ref{fig:var1_xminmax} that at large $t$ the dominant
contribution to $F_1^p$ comes from a rather narrow region of large
$x$.  To simplify the analysis we take the large-$x$ approximations
$q_v(x) \sim (1-x)^{\beta_q}$ and $f_q(x) \sim A_q (1-x)^n$.  At
sufficiently large
$t$ the integral $F_1^q(t) = \int \d x\, H_v^q(x,t)$ can then be
evaluated in the saddle point approximation, obtained by minimizing
the exponent in $H_v^q(x,t) = \exp[\, \beta_q \log(1-x) + t f_q(x) ]$
with respect to $x$.  We then find
\begin{equation}
  \label{saddle}
F_1^q(t) \sim |t|^{-(1+\beta_q)/n} \; , \qquad \qquad
1 - x_s = \Bigg( \frac{n}{\beta_q}\, A_q |t| \Bigg)^{-1/n} ,
\end{equation}
where $x_s$ is the position of the saddle point.  We see that for our
default fit with $n=2$ the dominant values of $x$ in the form factor
are in the soft region.  As observed in \cite{Burkardt:2003ck}, the
power behavior in (\ref{saddle}) with $n=2$ corresponds to the
Drell-Yan relation \cite{Drell:1969km} between the large-$t$ behavior
of form factors and the large-$x$ behavior of deep inelastic structure
functions.  Indeed, the kinematical assumptions made in the work of
Drell and Yan correspond to the dominance of the soft region.  Putting
$\beta_q=3$ as obtained from dimensional counting we recover the
$t^{-2}$ behavior mentioned above.  With the phenomenological values
$\beta_u=3.4$ and $\beta_d=5.0$ for our input parton distributions at
large $x$ one finds that for large $t$ the form factor $F_1^u$
obtained in our fit should fall slightly faster than $t^{-2}$, whereas
$F_1^d$ should decrease much more strongly.  At large $t$ both proton
and neutron form factor should then be dominated by $F_1^u$.  Our fit
result for $F_1^p$ in Fig.~\ref{fig:n2} shows the large-$t$ behavior
expected from these arguments, which means that the above
approximations are indeed applicable in kinematics where there is
data.  We will show $F_1^u$ and $F_1^d$ separately in
Sect.~\ref{sec:moments}.

Taking $n=1$ for the large-$x$ behavior of the profile function, the
dominant values of $x$ in the saddle point approximation are from the
ultrasoft region and give $F_1^q(t) \sim |t|^{-1-\beta_q}$.  Our fits
to (\ref{fit:var1}) and (\ref{fit:n1}) give a good description of the
data for $F_1^p$, which are clearly incompatible with such a strong
decrease in $t$.  This means that the asymptotic behavior has not yet
set in at $-t \approx 30 \gev^2$ for these parameterizations of GPDs.
To understand this, we notice that the approximation $f^q(x) \approx
A_q (1-x)^n$ works rather well in the interval $0.6 \le x \le 0.9$ for
our default fit, but not so well for our fits with $n=1$.  Any
inaccuracy in $f_q(x)$ appears however exponentiated in $H_v^q$ and
thus in the form factors.  The validity of asymptotic expansions like
(\ref{saddle}) must hence be carefully investigated on a case-by-case
basis.

\begin{figure}
\begin{center}
\includegraphics[width=.45\textwidth, height=.42\textwidth,
  bb=70 210 420 550]{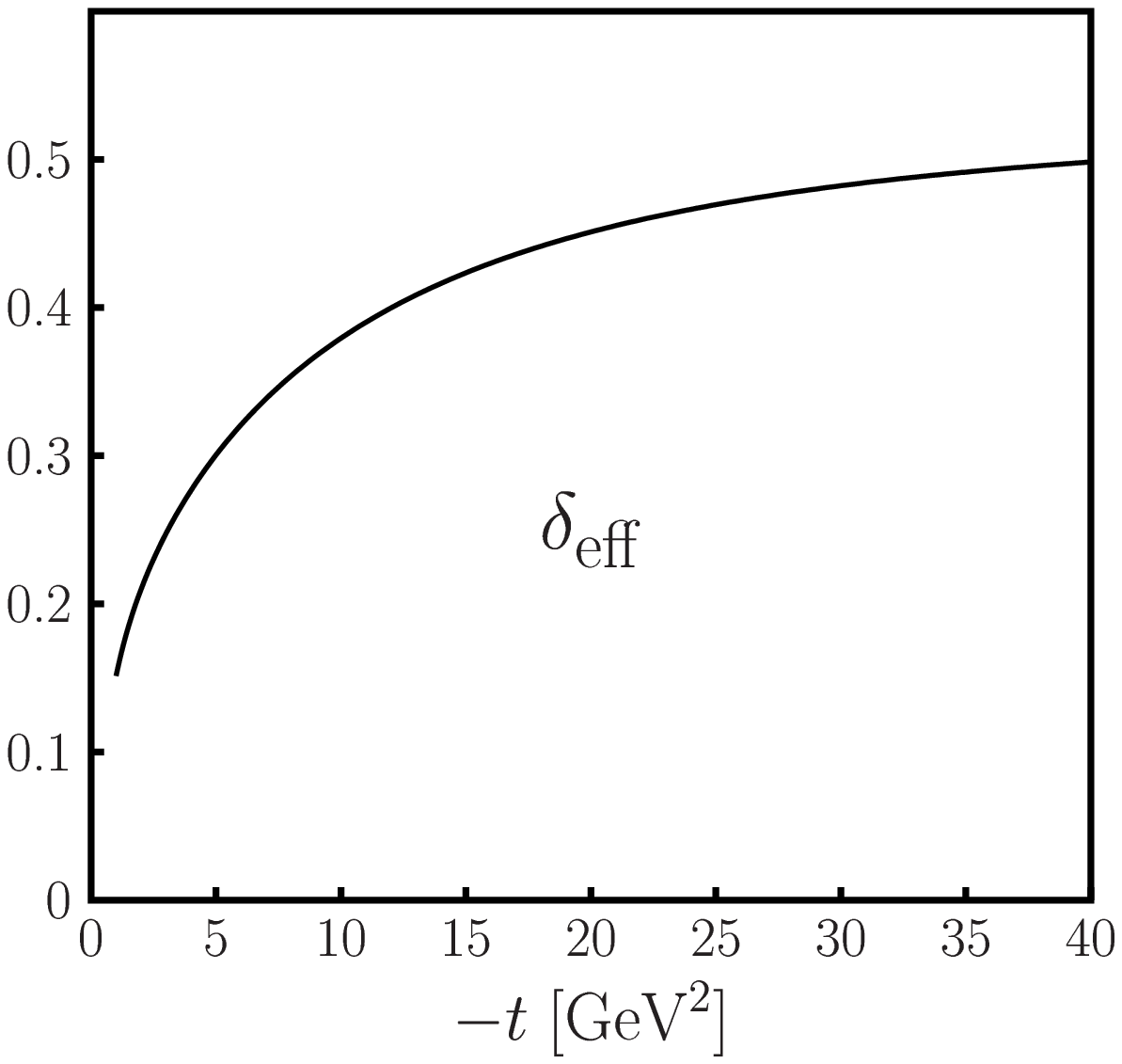}
\hspace{3em}
\includegraphics[width=.45\textwidth, height=.415\textwidth,
  bb=70 480 420 820]{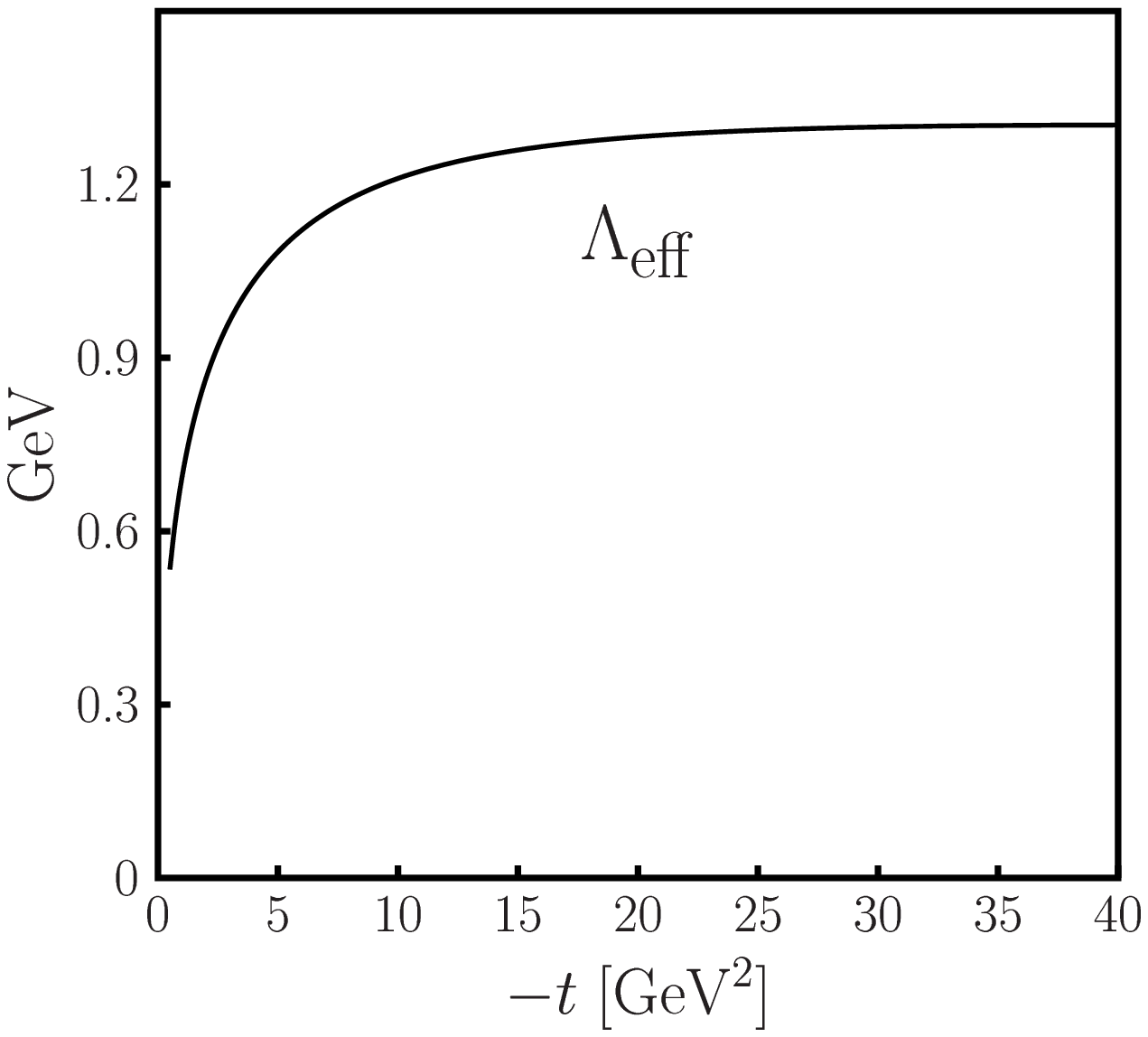}
\end{center}
\caption{\label{fig:neff} Dynamical interpretation of our default fit.
Left: The effective power $\delta_{\rm eff}$, which describes the
scaling of $(1-x)$ as a function of $|t|$.  Right: The effective soft
scale $\Lambda_{\rm eff}$ as a function of $|t|$.}
\end{figure}

To quantify how close our parameterizations are from the scaling laws
in (\ref{soft-classif-t}), we start with $\langle x\rangle_t$ defined
in (\ref{eq:xavg}) and introduce the quantity
\begin{equation}
\delta_{\rm eff}(t) = t\, \frac{\d}{\d t} \, 
   \log[1-\langle x\rangle_t] \, ,
\end{equation}
which is $\frac{1}{2}$ in the soft and $1$ in the ultrasoft region.
Using the saddle point approximation for both $\int \d x\, H_v^q(x,t)$
and $\int \d x\, (1-x) H_v^q(x,t)$, one readily finds that
$\delta_{\rm eff}(t)$ tends to $n^{-1}$ at asymptotic values of $t$.
In Fig.~\ref{fig:neff} we show $\delta_{\rm eff}(t)$ for our default
fit and find that for $-t$ above $5$ to $10 \gev^2$ the scaling of the
relevant $x$ values in the form factor integral is indeed the one for
the soft region.  In the same figure we show
\begin{equation}
\Lambda_{\rm eff}(t) = [1- \langle x\rangle_t]\, \sqrt{|t|} , 
\end{equation}
which according to (\ref{soft-classif-t}) should be of typical
hadronic size for $t$ where the soft scaling law applies.  This is
indeed the case for our fit.  We remark that with the saddle point
approximation we find $\Lambda_{\rm eff} = 1.26 \gev$ when neglecting
the $d$ quark contribution to $F_1^p$ at large $t$ and taking the
parameters $A_u=1.22 \gev^{-2}$ and $\beta_u= 3.4$ for $u$ quarks.
This shows once more the relevance of asymptotic considerations for
our default fit in kinematics where data is available.

The values of $\delta_{\rm eff}$ for our fit to (\ref{fit:n1}), where
$n=1$, differ from those shown in Fig.~\ref{fig:neff} by at most 8\%
and the values of $\Lambda_{\rm eff}$ by at most 4\%.  In the
large-$t$ region of present data the soft contribution hence also
dominates for this fit, where we find that ultrasoft behavior with
$\delta_{\rm eff} = 1$ only sets in for $|t|$ well above $100
\gev^{2}$.  We remark that in our previous work \cite{DFJK1} we used
GPDs obtained from Gaussian wave functions~(\ref{gauss-wfs}), which
had a profile function decreasing like $(1-x)$ at large $x$ and gave
an asymptotic behavior $F_1(t) \sim t^{-4}$.  The power counting for
the ``soft overlap mechanism'' we set up in that work corresponds to
the ultrasoft region in the parlance of our present paper.  The
considerations of this section make it clear that this (physically
suspect) region is not the dominant one in the kinematics where the
model of \cite{DFJK1} has been used for phenomenology.

We conclude that the description of $F_1^p$ provided by our fitted
GPDs supports the hypothesis that for $-t$ from about $5\gev^2$ to
several $10 \gev^2$ the dynamics is dominated by the Feynman mechanism
in the soft kinematics specified above.  We emphasize that by itself
our result cannot exclude the dominance of the standard
hard-scattering mechanism at large $t$.  Applied to GPDs
\cite{Hoodbhoy:2003uu} this mechanism results in a behavior
\begin{equation}
  \label{hsp-for-gpd}
H_v^q(x,t) \sim t^{-2} h(x) \qquad \qquad 
\mbox{for $-t \to \infty$ at fixed $x$}
\end{equation}
up to logarithms in $t$.  The function $h(x)$ diverges for $x\to 1$,
signaling the breakdown of the factorization scheme in that limit, see
\cite{Diehl:1999ek}.  Our parameterization (\ref{master-ansatz}) does
not tend to the factorized form~(\ref{hsp-for-gpd}) in the large-$t$
limit and thus does not incorporate the physics of the hard-scattering
mechanism.  The dominance of this mechanism in $F_1^p$ at
experimentally accessible $t$ is very doubtful: to be close to the
data one requires proton distribution amplitudes for which a
substantial fraction of the form factor comes from configurations
where partons are soft and the approximations of leading-twist
hard-scattering factorization are inadequate.  References can e.g.\ be
found in Sect.~10 of \cite{die:03}.  In our present work we make the
assumption that in the $t$ region we consider, the hard-scattering
mechanism is not dominant and that the Feynman mechanism controls form
factors at large $t$, despite its possible Sudakov suppression in the
asymptotic limit.


\subsection{The $t$ dependence}
\label{sec:power-law-t}

In this subsection we explore an ansatz for the $t$ dependence of
$H_v^q$ that is different from the exponential form we have used so
far.  To cover a range of possibilities we take
\begin{equation}
  \label{power-law-H}
H_v^q(x,t) = q_v(x) \Bigg( 1 - \frac{t f_q(x)}{p} \Bigg)^{-p} ,
\end{equation}
with different powers $p$.  In the limit $p\to \infty$ we recover the
exponential (\ref{master-ansatz}).  The corresponding impact parameter
distribution $q_v(x,\tvec{b}^2)$ can be expressed in terms of the
modified Bessel function $K_{p-1}$ and satisfies positivity.  At fixed
$x$ the form (\ref{power-law-H}) gives a power-law falloff $H_v^q(x,t)
\sim |t|^{-p}$ for $|t| \to\infty$.

Before proceeding to fits let us investigate some general properties
of this ansatz.  For $p \neq 1$ the form (\ref{power-law-H}) is not
stable under DGLAP evolution.  To see this we observe that it
satisfies
\begin{equation}
  \label{pow-law-relation}
\Bigg( \frac{\partial^2}{\partial t^2} \log H_v^q(x,t) \Bigg) 
\Bigg( \frac{\partial}{\partial t} \log H_v^q(x,t) \Bigg)^{-2}
 = \frac{1}{p}
\end{equation}
for all $x$ and $t$, and starting from (\ref{dglap-H}) calculate the
evolution of the left-hand-side of this relation.  If at a given scale
$\mu$ the GPD satisfies (\ref{pow-law-relation}) then this evolution
equation simplifies to
\begin{eqnarray}
\lefteqn{
\mu^2 \frac{\d}{\d\mu^2}\, \Bigg[ 
\Bigg( \frac{\partial^2}{\partial t^2} \log H_v^q(x) \Bigg) 
\Bigg( \frac{\partial}{\partial t} \log H_v^q(x) \Bigg)^{-2} \, \Bigg]
= \Bigg( 1 - \frac{1}{p} \Bigg) 
  \Bigg( \frac{\partial}{\partial t} \log H_v^q(x) \Bigg)^{-2}
} \hspace{6em}
\nonumber \\[0.3em]
&& {}\times
  \int_x^1 \frac{\d z}{z} \, P\Big(\frac{x}{z}\Big)\, 
  \frac{H_v^q(z)}{H_v^q(x)}
  \Bigg( \frac{\partial}{\partial t} \log H_v^q(z)
      - \frac{\partial}{\partial t} \log H_v^q(x) \Bigg)^2 ,
\end{eqnarray}
where for better legibility we have omitted the arguments $t$ and
$\mu$ in the GPDs.  Except for trivial functions $H_v^q(x)$ the
right-hand-side of this expression is positive and furthermore depends
on $x$, so that relation (\ref{pow-law-relation}) is destroyed by
evolution.

At very large $t$ the form factors obtained with (\ref{power-law-H})
are again dominated by large $x$ provided that $p n > \beta_q$, where
as before we assume the large-$x$ behaviors $q_v(x) \sim
(1-x)^{\beta_q}$ and $f_q(x) \approx A_q (1-x)^n$.  For $u$ quarks and
$n=2$ this condition is fulfilled even when $p=2$.  One can then
again use the saddle point approximation and finds
\begin{equation}
  \label{saddle-p}
F_1^q(t) \sim |t|^{-(1+\beta_q)/n} , \qquad \qquad
1 - x_s = \Bigg( \Big(\frac{n}{\beta_q}\, - \frac{1}{p}\Big)
          A_q |t| \Bigg)^{-1/n} ,
\end{equation}
where remarkably the large-$t$ behavior of $F_1^q$ is independent of
$p$.  The dominant $x$ in the form factor integral are in the region
of the soft Feynman mechanism discussed in the previous subsection.
Note that in this region $t f_q(x)$ is of order 1, so that the
approximation giving a power-law $H_v^q(x,t) \sim |t|^{-p}$ is not
valid.

We note that the form (\ref{power-law-H}) does not have the Regge
behavior (\ref{regge-form}) at small $x$ and $t$.  Depending on $p$,
this can be significant in kinematics appropriate for Regge
phenomenology.  Taking for example $x=10^{-3}$ and $-t= 0.4 \gev^2$,
we have $t f(x) \approx \alpha' t \log(1/x) \approx -2.5$ with
$\alpha' = 0.9 \gev^{-2}$, and $(1+2.5/p)^{-p}$ is about twice as
large as $\exp(-2.5)$ if $p=3$.  Having lost the connection with Regge
phenomenology, we will not impose a fixed value of $\alpha'$ in our
fits.  A logarithmic increase of $\bsq{}{x}$ at small $x$ seems
however plausible from a more general point of view, and we keep the
analytic form of our profile function as in our default fit.  One
could of course construct parameterizations which interpolate between
an exponential $t$ dependence at very small $x$ and a power-law when
$x$ becomes larger.  Since the small-$x$ region does however not
affect our fit to nucleon form factors for $-t$ above $1 \gev^2$ or
so, we shall not pursue this possibility here.

The results of our fits to (\ref{power-law-H}) and (\ref{fit:n2}) are
shown in Table~\ref{tab:power-fits} for selected values of $p$, where
for easier comparison we have also given the result of our exponential
fit with free $\alpha'$ discussed in Sect.~\ref{sec:varia}.  As $p$
decreases from $\infty$, the quality of the fit first improves (except
for the neutron data).  The smallest $\chi^2$ is attained for $p=2.5$,
and further decrease of $p$ makes the fit worse.  With $p=2$ no good
description of the large-$t$ data can be achieved: the fit result for
$F_1^p$ systematically overshoots the data above $-t= 9 \gev^2$, by
about $15\%$ when $-t> 19 \gev^2$.  We find the same qualitative
picture when fixing $\alpha' = 0.9 \gev^{-2}$ in the fit.  The lowest
$\chi^2$ is then obtained at $p=3$, and for $p=2$ the description of
the large-$t$ data is again very bad.  It is important to realize
that, despite a clear decrease in $\chi^2$, the description of the
data for $F_1^p$ does not improve dramatically when going from
$p=\infty$ to $p=2.5$.  Except for the two data points with the
largest values of $-t$ and the data point with $-t= 0.86\gev^2$ (which
has the highest $-t$ in the sample of \cite{Janssens:1966}), the pull
for $F_1^p$ of our fit with $p=2.5$ is at most $3\%$ in magnitude,
whereas it is at most $3.5\%$ with our default fit.

\begin{figure}
\begin{center}
\includegraphics[width=.45\textwidth,
  bb=130 375 485 710]{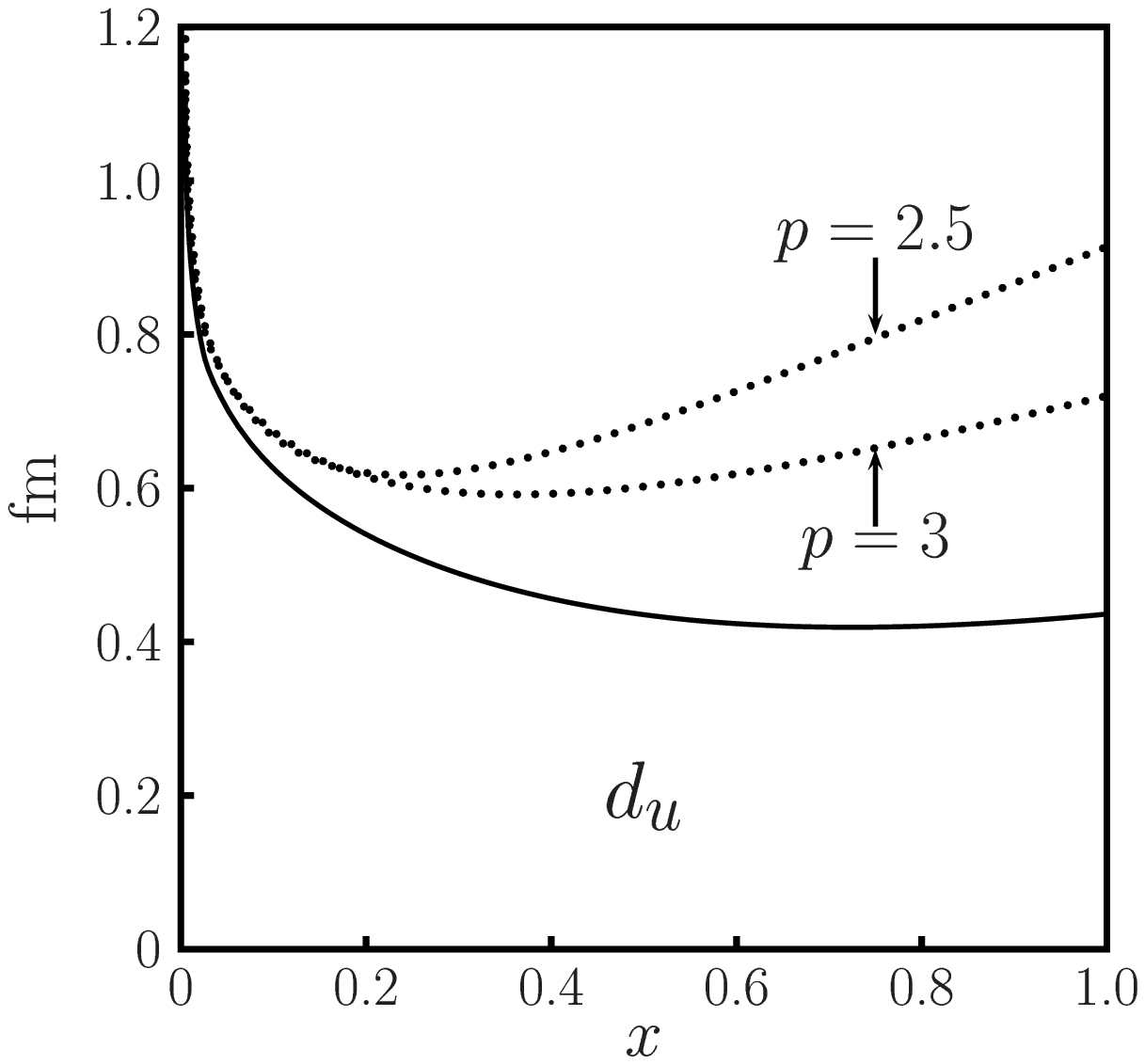}
\hspace{3em}
\includegraphics[width=.45\textwidth,
  bb=65 485 427 820]{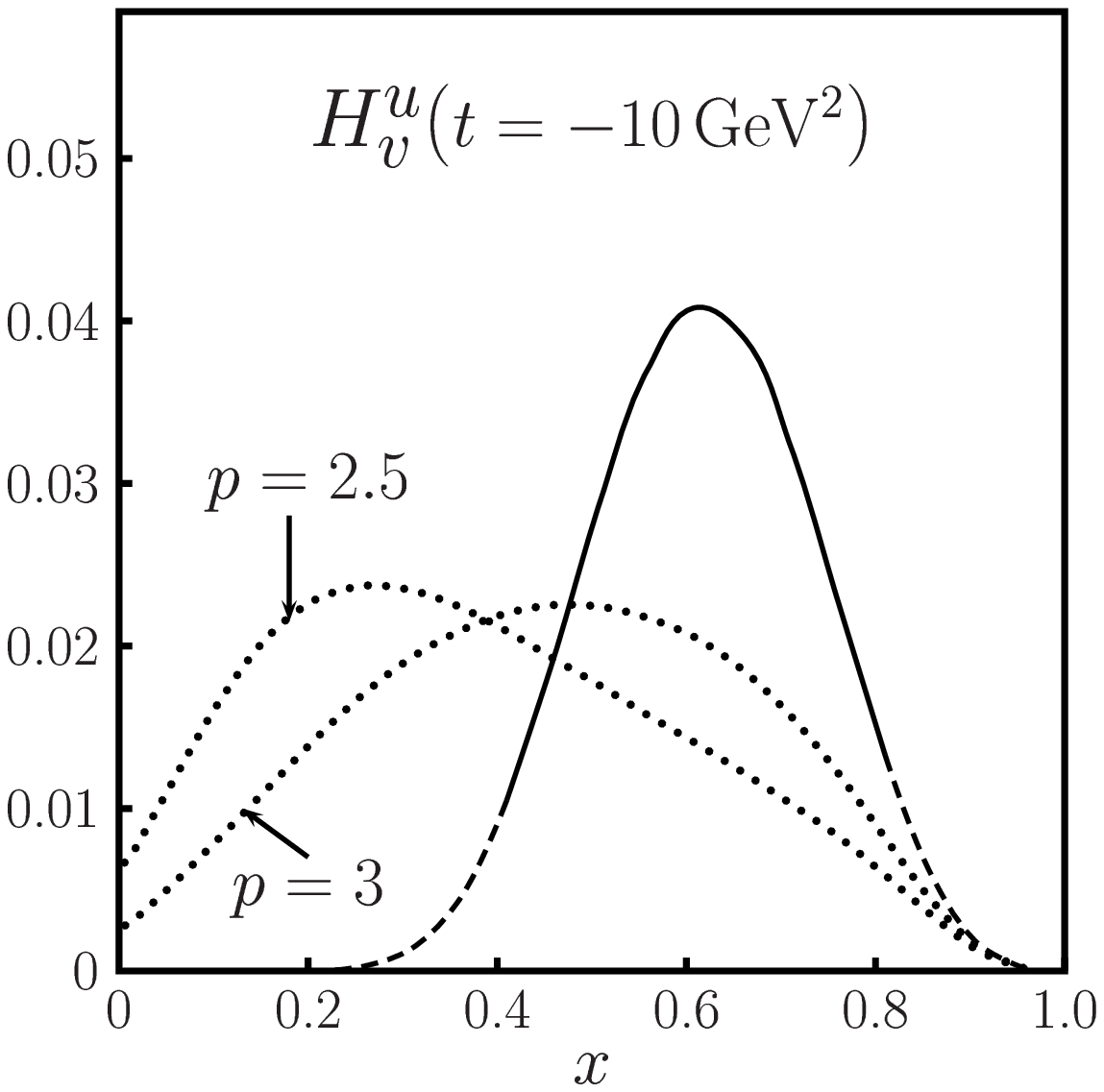}
\end{center}
\caption{\label{fig:power-fit} Left: The distance $d_u(x)$ between
struck quark and spectators obtained with fits to the power-law
(\protect\ref{power-law-H}) with $p=2.5$ and $p=3$, compared with the
result of our default fit, which corresponds to $p\to\infty$ and
differs only slightly from the corresponding fit with free $\alpha'$.
Right: The results for $H_v^u(x,t)$ at $-t= 10\gev^2$ from the same
fits.}
\end{figure}

With decreasing $p$ the profile functions $f_q(x)$ increase
significantly, which is shown in Fig.~\ref{fig:power-fit} for $u$
quarks.  This can be readily understood: the GPDs in
(\ref{power-law-H}) decrease much smaller in the variable $|t| f_q(x)$
for small $p$ than for large $p$, so that small $p$ requires a larger
$f_q(x)$ in order to not overshoot the form factors at large $|t|$.

In the figure we also show the $x$ dependence of $H_v^u$ at large $t$.
Its shape is much broader for the fits with smaller $p$ and reaches
down to smaller $x$.  This reflects that when $p$ becomes smaller it
takes significantly larger $t$ to suppress $H_v^q$ at a given $x$.
For small $p$ the asymptotic behavior (\ref{saddle-p}) obtained
with the saddle-point approximation should hence set in at much larger
$t$.  Taking $-t= 40\gev^2$ we find indeed that $\langle
x\rangle_t$ is only 0.6 for $p=3$ and 0.5 for $p=2.5$, and that
$\delta_{\rm eff}$ only becomes 0.3 for $p=3$ and 0.15 for $p=2.5$, to
be compared with the corresponding values $\langle x\rangle_t= 0.8$
and $\delta_{\rm eff} = 0.5$ for the default fit.

As is seen in Fig.~\ref{fig:power-fit}, the GPDs with power-law form
(\ref{power-law-H}) do not vanish for $x\to 0$ at large $t$, so that
small $x$ also contribute in the high-$t$ form factors to some extent.
This is in contrast to the GPDs with exponential $t$ dependence.
Using that $q_v(x) \sim x^{-\alpha}$ for $x\to 0$, one readily finds
that $H_v^q(x,t) \sim x^{-\alpha - \alpha' t}$ vanishes in this limit
as soon as $-t \gsim \alpha /\alpha'$, i.e., already for $-t$ above
$0.7 \gev^{2}$ or~so.

The fits we have shown make it clear that a significant ambiguity
remains when one tries to determine the correlated dependence on $x$
and $t$ of GPDs from experimental knowledge of their integrals over
$x$ and of their limits at $t=0$.  Without data on observables
depending on both variables we need a certain theoretical bias in our
extraction of GPDs.  We regard the connection of our default fit with
Regge phenomenology at small $x$ and $t$ and with the dynamics of the
soft Feynman mechanism at large $t$ as physically attractive features
and retain the exponential form (\ref{master-ansatz}) for the $t$
dependence of $H_v^q(x,t)$.

\section{The polarized distribution $\widetilde{H}_v(x,t)$}
\label{sec:axial}

In this section we investigate the distribution of polarized quarks
and its connection with the isovector axial form factor $F_A(t)$ of
the nucleon.  The relevant sum rule reads
\begin{equation}
F_A(t) = \int_{-1}^1 \d x \,
\Big( \widetilde{H}^u(x,t,\mu^2) - \widetilde{H}^d(x,t,\mu^2) \Big) .
\label{sumrule-FA-a}
\end{equation}
Its value at $t=0$ is given by the axial charge $F_A(0)$ and well
known from $\beta$ decay experiments.  The measurement of $F_A(t)$
covering the largest $t$ range  \cite{kita:83} has been performed in
charged current scattering $\nu_\mu\, n\to \mu^- p$.  The result has
been given in the form of a dipole parameterization
\begin{equation}
F_A(t) = \frac{F_A(0)}{(1 - t /M_A^2)^{2}_{\phantom{A}}} 
\label{dipole-FA}
\end{equation}
with $F_A(0)=1.23\pm 0.01$ and $M_A= (1.05\,^{+0.12}_{-0.16}) \gev$
for a measured range $0.1 \gev^2 \le -t \le 3 \gev^2$.  The resulting
$t$ dependence of $F_A(t)$ is not too different from the one of the
proton Dirac form factor: we find that $F_1^p(t)$ can be approximated
to 9\% accuracy for $-t \le 3 \gev^2$ by $(1 - t/M_D^2)^{-2}$ with
$M_D = 0.98 \gev$.  Note that the well-known dipole mass of $0.84
\gev$ does not refer to $F_1^p$ but to a dipole parameterization of
the magnetic form factors of proton and neutron, $G_M^p$ and $G_M^n$.

Many other measurements of $F_A(t)$ in either charged current
scattering or $ep \to e \pi^+ n$ have also been presented in terms of
a dipole mass, with a considerable spread of results for $-t \le 1
\gev^2$ (see e.g.~\cite{Bernard:2001rs}).  We remark that while a
dipole form (\ref{dipole-FA}) provides a simple and compact
parameterization, it is not well suited for a description of data
beyond a certain accuracy.  It is instructive in this respect to
perform dipole fits to the data on $F_1^p$ or on $G_M^p$ in different
ranges of $t$ and to observe the shift in the dipole mass.

Given the data situation for $F_A(t)$ we do not attempt a fit of GPDs
to the sum rule (\ref{sumrule-FA-a}), but rather test the simple
ansatz
\begin{equation}
  \label{Htilde-ansatz}
\widetilde{H}^q_v(x,t) 
  = \Delta q_v(x)\, \exp{[t \tilde{f}_q(x)]}
\end{equation}
with the profile functions $\tilde{f}_q(x)$ taken equal to $f_q(x)$
obtained in our default fit for the unpolarized distributions $H_v^q$.
In physical terms this ansatz assumes that the distribution of quarks
minus antiquarks in the transverse plane does not depend on the quark
or antiquark helicity relative to the helicity of the proton.  For our
evaluation we take the polarized parton densities $\Delta q_v(x)$ from
\cite{blum:03}, more specifically the NLO distributions in their
scenario 1 at $\mu= 2\gev$.

Since the axial form factor has positive charge parity, it is not
directly connected with the valence quark distributions.  Instead we
have
\begin{equation}
F_A(t) = \int_{0}^1 \d x \,
\Big( \widetilde{H}_v^u(x,t) - \widetilde{H}_v^d(x,t) \Big) 
+ 2 \int_{0}^1 \d x \,
\Big( \widetilde{H}^{\bar{u}}(x,t) - \widetilde{H}^{\bar{d}}(x,t) \Big) ,
\label{sumrule-FA-b}
\end{equation}
where the generalized antiquark distributions are given by
$\widetilde{H}^{\bar{q}}(x,t) = \widetilde{H}^{q}(-x,t)$.  The flavor
non\-singlet combination $\Delta\bar{u}(x) - \Delta\bar{d}(x)$ of
forward densities is poorly known, and at present there is no
experimental evidence that it might be large \cite{Airapetian:2003ct}.
To make a motivated ansatz for its analog at finite $t$ is beyond the
scope of this work.  For a very crude estimate we have taken
$\widetilde{H}^{\bar{q}}(x,t) = \Delta\bar{q}(x) \exp[ t f_q(x) ]$
with the polarized antiquark distributions from \cite{blum:03}, where
$\Delta\bar{u}(x) =\Delta\bar{d}(x)$.  The resulting contribution from
antiquark GPDs in (\ref{sumrule-FA-b}) is below a percent, both for
$f_q(x)$ from our default fit and from the corresponding
four-parameter fit in Table~\ref{tab:n2-fits}, where $B_u \neq B_d$,
This is because the profile functions for $u$ and $d$ quarks differ
mostly at larger $x$, where antiquarks do not abound.

The present uncertainties on polarized quark distributions are
significantly larger than those on their unpolarized counterparts.  We
have calculated the resulting error on $F_A(t)$ using the covariance
matrix on the parameters in the parton densities provided in
\cite{blum:03}.  This error is at least a factor of 5 larger than the
error resulting from the uncertainty in $f_q(x)$.  For estimating the
parametric uncertainties of $F_A(t)$ obtained with our ansatz, we have
added the errors from the two sources in quadrature.

\begin{figure}
\begin{center}
\includegraphics[width=.45\textwidth,
  bb=105 355 470 705]{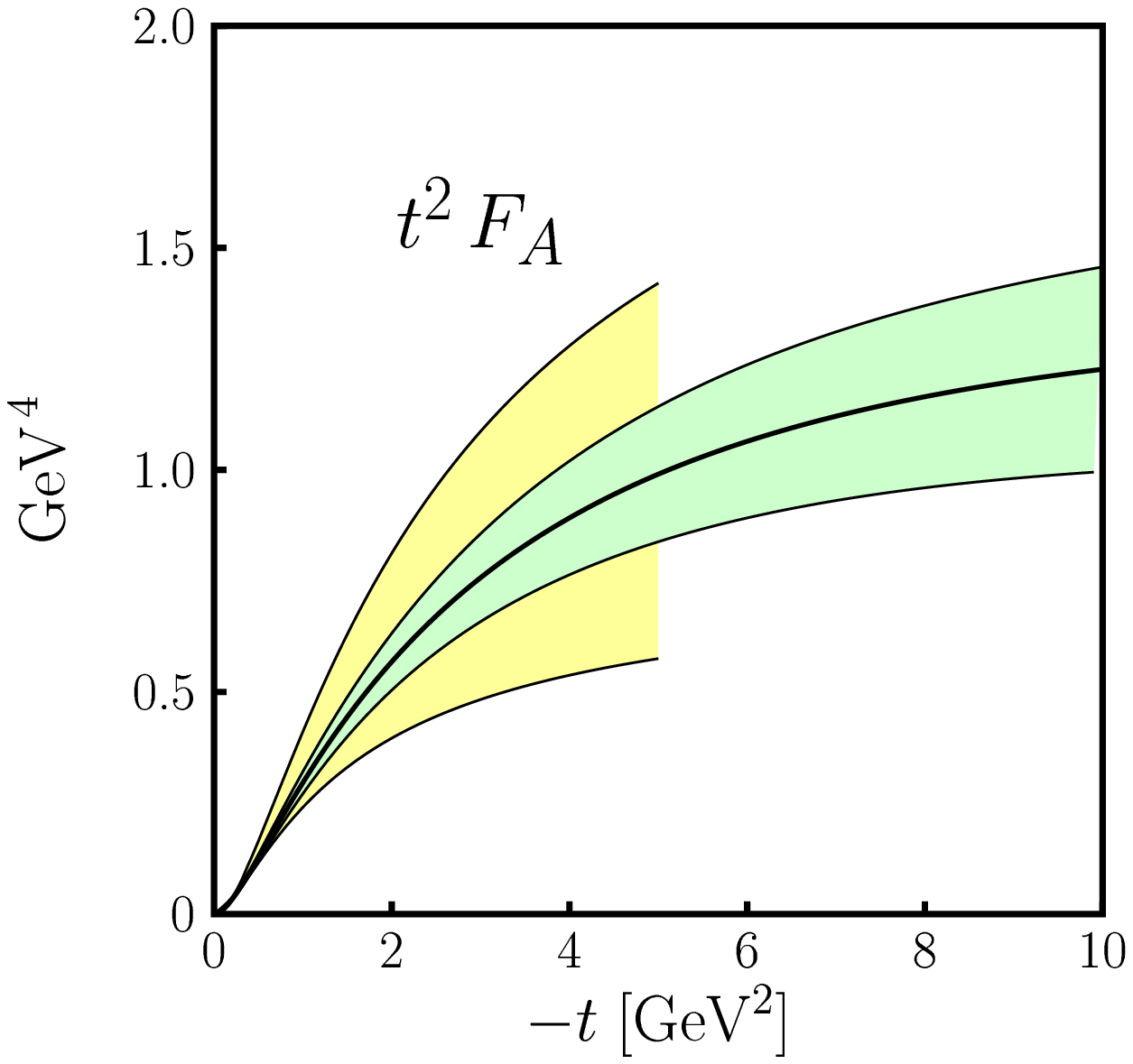}
\end{center}
\caption{\label{fig:Q4FA} Axial form factor $F_A(t)$ obtained from the
valence distributions (\protect\ref{Htilde-ansatz}) with polarized
densities from \protect\cite{blum:03} at $\mu=2 \gev^2$ and the result
of our default fit for $f_q(x)$ specified in
Sect.~\protect\ref{sec:select}.  The contribution from sea quarks has
been neglected.  The error band gives the $1\, \sigma$ uncertainties of
the fit to $f_q(x)$ added in quadrature to those of $\Delta q(x)$.
The curve and band shown up to $-t = 5 \gev^2$ represents the dipole
parameterization from \protect\cite{kita:83} with $F_A(0)=1.23\pm 0.01$
and $M_A= (1.05\,^{+0.12}_{-0.16}) \gev$.}
\end{figure}

In Fig.~\ref{fig:Q4FA} we show the contribution to $F_A(t)$ from the
valence quark GPDs specified above.  We compare this with the dipole
parameterization of \cite{kita:83}, which we have extrapolated up to
$-t = 5 \gev^2$.  Our result undershoots the central value of that
parameterization by at most 13\%, with the largest discrepancy for $-t$
between $0.5$ and $1 \gev^2$.  It is however consistent within the
errors of the parameterization in the full range $0.1 \gev^2 \le -t \le
3 \gev^2$ of the measurement in \cite{kita:83}.  We note that one
obtains a larger form factor with the ansatz (\ref{Htilde-ansatz})
when taking a smaller profile function $\tilde{f}_q(x)$.  This is
indeed physically allowed, whereas taking $\tilde{f}_q(x) > f_q(x)$
would violate the positivity of parton densities in impact parameter
space (see Sect.~\ref{sec:ansatz-e}).

In conclusion, we find that the present data on the axial form factor
is consistent with valence quark dominance in the sum rule
(\ref{sumrule-FA-b}) and with only a weak helicity dependence in the
transverse distribution of valence quarks at not too large values of
$x$.  We note that $x_{\rm max}(t)$ for $F_A(t)$, defined in analogy
to (\ref{eq:xminmax}), equals $0.73$ at $-t= 3\gev^2$.

\section{The helicity-flip distribution $E$ from the Pauli form
  factors}
\label{sec:gpd-e}

\subsection{General properties}
\label{sec:general-e}

The GPDs $E^q$ are related to the Pauli form factors of proton and
neutron through the sum rules
\begin{eqnarray}
F^p_2(t) &=& \int_0^1 \d x \, 
  \Big[ {\textstyle\frac{2}{3}} E_v^u(x,t,\mu^2) 
      - {\textstyle\frac{1}{3}} E_v^d(x,t,\mu) \Big]\, , 
\nonumber \\
F^n_2(t) &=& \int_0^1 \d x \, 
  \Big[ {\textstyle\frac{2}{3}} E_v^d(x,t,\mu^2) 
      - {\textstyle\frac{1}{3}} E_v^u(x,t,\mu) \Big]\, , 
\label{sumrule-e}
\end{eqnarray}
where in analogy to (\ref{def-val}) we have introduced valence
distributions
\begin{equation}
E_v^q(x,t,\mu^2) = E^q(x,t,\mu^2) + E^q(-x,t,\mu^2)
\end{equation}
for quarks of flavor $q$ in the proton.  Contributions from sea quarks
cancel in (\ref{sumrule-e}).  We have neglected in (\ref{sumrule-e})
the contribution from strange quarks, as we did for the Dirac form
factors.  The scale dependence of $E_v$ is described by the same DGLAP
equation as the scale dependence of $H_v$, since both distributions
belong to the same quark operator.

The distribution $E^q_v(x,t)$ describes proton helicity flip in a
frame where the proton moves fast (or more precisely light-cone
helicity flip, see e.g.\ Sect.~3.5 of \cite{die:03}).  Since quark
helicity is conserved by the vector current, proton helicity flip
requires orbital angular momentum between the struck quark and the
spectator system.  This becomes for instance manifest by writing
$E^q_v$ as the overlap of light-cone wave functions, whose orbital
angular momentum must differ by exactly one unit.  Another
manifestation is Ji's sum rule for the combination $H^q(x,t) +
E^q(x,t)$, see Section~\ref{sec:ji}.

$E^q_v$ admits a probability interpretation in impact parameter space
if one changes basis from longitudinal to transverse polarization
states of the proton \cite{bur:02}.  More precisely, one considers
states of definite proton transversity, which is the light-cone
analog of transverse polarization, see e.g.~\cite{Jaffe:1996zw}.  The
distribution
\begin{equation}
  \label{impact-pol}
q_{v}^{X}(x,\tvec{b}) = q_v(x,\tvec{b}) - \frac{b^y}{m}\,
  \frac{\partial}{\partial \tvec{b}^2}\, e_v^q(x,\tvec{b})
\end{equation}
gives the probability to find an unpolarized quark with momentum
fraction $x$ and impact parameter $\tvec{b} = (b^x, b^y)$ in a proton
polarized along the $x$ direction, minus the corresponding probability
to find an antiquark.  Here we have introduced the Fourier transform
\begin{equation}        
        \label{fourier-e}
e_v^q(x,\tvec{b}) = 
  \int \frac{\d^2 \tvec{\Delta}}{(2\pi)^2}\,
  e^{-i \svec{b} \svec{\Delta}}\, E_v^q(x,t=-\tvec{\Delta}^2) .
\end{equation}
Note that $e_v^q(x,\tvec{b})$ and $q_v(x,\tvec{b})$ depend on $b^x$
and $b^y$ only via $\tvec{b}^2$.  We see from (\ref{impact-pol}) that
target polarization along the $x$-axis induces a shift in the quark
distribution along the $y$-axis.  As explained in \cite{bur:02}, this
effect is consistent with the classical picture of the polarized
proton as a sphere rotating about the $x$-axis and moving in the
$z$-direction (see also \cite{Belitsky:2003nz}).  The average
displacement of this shift is
\begin{equation}
  \label{av-displ}
\langle b^y \rangle^{q}_{x} 
 = \frac{\int \d^2\tvec{b}\; b^y\, q_v^X(x,\tvec{b})}{
               \int \d^2\tvec{b}\, q_v^X(x,\tvec{b})}
 = \frac{1}{2m}\, \frac{E_v^q(x,0)}{H_v^q(x,0)} \, ,
\end{equation}
and its scale evolution is given by
\begin{equation}
        \label{dglap-by}
\mu^2 \frac{\d}{\d\mu^2} \langle b^y \rangle^{q}_{x} = 
 - \frac{1}{q_v(x)}  \int_x^1 \frac{\d z}{z}\,
  P\Big(\frac{x}{z}\Big)\, q_v(z) \Big[ \langle b^q \rangle^{q}_{x}
                                - \langle b^q \rangle^{q}_{z} \Big] ,
\end{equation}
in analogy to the evolution of $\bsq{q}{x}$.  The corresponding shift
for the distance between the struck quark and the spectator system is
\begin{equation}
  \label{sq-def}
s_q(x) = \frac{\langle b^y \rangle^{q}_{x}}{1-x} ,
\end{equation}
in analogy to the distance function $d_q(x)$ we introduced in
(\ref{r-def}).

The impact parameter space distributions satisfy inequalities which
insure that the quark densities for various combinations of proton and
quark spins are positive.  Using the methods of
\cite{Pobylitsa:2002iu} one finds in particular \cite{Burkardt:2003ck}
\begin{equation}
  \label{b-bound}
\frac{\tvec{b}^2}{m^2}  \Bigg(
  \frac{\partial}{\partial \tvec{b}^2}\, e^q(x,\tvec{b}) 
\Bigg)^2  \leq 
\Big[ q(x,\tvec{b}) + \Delta q(x,\tvec{b}) \Big]\, 
\Big[ q(x,\tvec{b}) - \Delta q(x,\tvec{b}) \Big]\, ,
\end{equation}
where $e^q(x,\tvec{b})$, $q(x,\tvec{b})$ and $\Delta q(x,\tvec{b})$
are the respective Fourier transforms of $E^q(x,t)$, $H^q(x,t)$ and
$\widetilde{H}^q(x,t)$, defined in analogy to (\ref{impact-gpd}) and
(\ref{fourier-e}).  The bound (\ref{b-bound}) is stable under
leading-order DGLAP evolution to higher scales, and a closer look at
its derivation shows that it should be valid when $\mu$ is large
enough for the application of leading-twist factorization theorems to
the exclusive processes where GPDs appear.  For a discussion and
references see \cite{die:03}.

Multiplying (\ref{b-bound}) with $\tvec{b}^2$ and integrating over
$\tvec{b}$, one obtains after a few steps an inequality for GPDs in
momentum space \cite{Burkardt:2003ck},
\begin{equation}
  \label{forward-bound}
\Big( E^q(x,0) \Big)^2 \leq  4 m^2 
\Big[ q(x) + \Delta q(x) \Big]\, \Big[ q(x) - \Delta q(x) \Big]\,
  \frac{\partial}{\partial t} 
     \ln \left[H^q(x,t)\pm \tilde H^q(x,t)\right]_{t=0} \ ,
\end{equation}
where either the sum or the difference of $H^q$ and $\widetilde{H}^q$
may be taken.  The $t$-derivative on the right-hand side is
$\frac{1}{4}$ times the average squared impact parameter of quarks
with positive or negative helicity.  According to our discussion in
Sect.~\ref{sec:phys-mot} this quantity may be expected to decrease at
least like $(1-x)^2$ in the limit $x\to 1$.  Since in addition the
longitudinal polarization $| \Delta q /q |$ of quarks is
phenomenologically seen to grow as $x$ becomes large, the inequality
(\ref{forward-bound}) severely limits the high-$x$ behavior of
$E^q(x,t=0)$ and will be an essential input in the following.

The positivity constraints (\ref{b-bound}) and (\ref{forward-bound})
hold for the distribution of quarks (i.e.\ for $x>0$) and have
analogs for antiquarks.  They do not hold for the valence
combinations we aim to determine in this work, which are
\emph{differences} of quark and antiquark distributions.  It is
however natural to neglect antiquarks at large enough $x$, which is in
fact the region where the bounds give the strongest constraints.  With
this proviso in mind we will in the following use (\ref{b-bound}) and
(\ref{forward-bound}) for the valence distributions.

Our discussion of the bound (\ref{forward-bound}) implies that the
average displacement $\langle b^y \rangle^{q}_{x}$ should vanish in
the limit of $x\to 1$.  At large enough $x$ it should hence be a
decreasing function, which according to (\ref{dglap-by}) implies that
$\langle b^y \rangle^{q}_{x}$ becomes smaller when evolving to higher
scales $\mu$.  We comment on the small-$x$ behavior of this
displacement in the next subsection.


\subsection{Ansatz for $E_v(x,t)$}
\label{sec:ansatz-e}

Our ansatz for $E_v$ is taken in analogy to the other GPDs as
\begin{equation}
  \label{master-e}
E_v^q(x,t) = e_v^q(x)\, \exp[ t g_q(x) ] ,
\end{equation}
where we use the notation
\begin{equation}
e_v^q(x) = E_v^q(x,t=0) 
\end{equation}
for the forward limit.  The normalization integrals
\begin{equation}
  \label{little-sumrules}
\int_0^1 \d x\, e_v^q(x,t=0,\mu) = \kappa_q 
\end{equation}
give the contribution of quark flavor $q$ to the anomalous magnetic
moment of the proton (up to quark charge factors).  Neglecting the
contribution from strange quarks one has according to
(\ref{sumrule-e})
\begin{equation}
   \kappa_u = 2 \kappa_p + \kappa_n \approx 1.67\,, 
\qquad \qquad
   \kappa_d = \kappa_p + 2 \kappa_n \approx -2.03\,.
\end{equation}
For the profile function in (\ref{master-e}) we take the same form as
in our default fit for $H_v^q$,
\begin{equation}
  \label{profile-e}
g_q(x) = \alpha' (1-x)^{3} \log\frac{1}{x} + D_q (1-x)^{3} 
        + C_q \, x (1-x)^{2} 
\hspace{4em}
\mbox{with~} \alpha' = 0.9 \gev^{-2}  \eqpt
\end{equation}
The motivation from Regge phenomenology for the behavior of $g_q$ at
very small $x$ applies in the same way as for $f_q$.  Indeed the Regge
exchanges contributing to $H_v^q$ and $E_v^q$ are the same, and only
their coupling strengths differ.  We therefore take the same fixed
value of $\alpha'$ as in our default fit for $H_v^q$.

The exponential $t$-dependence for $\widetilde{H}_v^q$ and $E_v^q$
taken in (\ref{Htilde-ansatz}) and (\ref{master-e}) gives
\begin{equation}
  \label{polar-impact}
\Delta{q}_v(x,\tvec{b}) = \frac{1}{4\pi}\, 
  \frac{\Delta q_v(x)}{\tilde{f}_q(x)}\, 
  \exp\Bigg[-\frac{\tvec{b}^2}{4 \tilde{f}_q(x)} \Bigg] \eqcm
\qquad \qquad
e_v(x,\tvec{b}) = \frac{1}{4\pi}\, 
  \frac{e_v(x)}{g_q(x)}\, 
  \exp\Bigg[ -\frac{\tvec{b}^2}{4 g_q(x)} \Bigg] 
\end{equation}
in analogy to the form (\ref{master-impact}) of $q_v(x,\tvec{b})$.
According to (\ref{b-bound}) one must have $| \Delta q_v(x,\tvec{b} |
\le q_v(x,\tvec{b})$ for all $\tvec{b}$ when antiquarks are
negligible, which implies $\tilde{f}_q(x) \le f_q(x)$ as anticipated
in Sect.~\ref{sec:axial}.  With our simplified ansatz $\tilde{f}_q(x)
= f_q(x)$, the bounds (\ref{b-bound}) and (\ref{forward-bound})
respectively read
\begin{equation}
  \label{b-bou-ans}
  \frac{\tvec{b}^2}{(4 m g_q)^2} \, 
  \Bigg( \frac{e_v^q}{g_q}  \Bigg)^2
         \exp\Bigg[-\frac{\tvec{b}^2}{2 g_q}\Bigg]
\leq \frac{(q_v)^2 - (\Delta q_v)^2}{(f_q)^2} \,
         \exp\Bigg[-\frac{\tvec{b}^2}{2 f_q}\Bigg] 
\end{equation}
and
\begin{equation}
  \label{forw-bou-ans}
(e_v^q)^2  \leq 4 m^2 f_q \, \Big[ (q_v)^2 - (\Delta q_v)^2 \Big]
\end{equation}
when antiquarks can be neglected, where for the sake of legibility we
have omitted the argument $x$ in all functions.  Because of the factor
$\tvec{b}^2$ on the left-hand side of (\ref{b-bou-ans}) we must have
\begin{equation}
  \label{profile-bound}
g_q(x) < f_q(x) 
\end{equation}
with strict inequality, otherwise the bound will be violated at
sufficiently large $\tvec{b}^2$.  Multiplying both sides of
(\ref{b-bou-ans}) with $\exp[\tvec{b}^2 /(2 f_q)]$ and then maximizing
the left-hand side, one finds that the bound is most stringent for
$\tvec{b}^2 = 2 g_q f_q /(f_q - g_q)$, where it reads
\begin{equation}
  \label{super-bound}
(e_v^q)^2  \leq 4 m^2\, e^{1+\log 2}\,
   \Bigg( \frac{g_q}{f_q} \Bigg)^3 (f_q - g_q)\, 
          \Big[ (q_v)^2 - (\Delta q_v)^2 \Big] .
\end{equation}
We note that the bound (\ref{forw-bou-ans}) is weaker than
(\ref{super-bound}) but has the practical advantage to be independent
of the profile function $g_q$.  If we require the distance $d_q(x)$
between struck quark and spectators in an unpolarized proton to stay
finite in the limit $x\to 1$, then this bound guarantees that the
shift $s_q(x)$ of this distance also remains finite, given that $e_v^q
= 2m\, (1-x) q_v\, s_q$ and $f_q = (1-x)^2\, d_q^2 /4$.

For the shape of $e_v^q$ we make the time-honored ansatz
\begin{equation}
  \label{good-old-ans}
e_v^q(x) = N_q \kappa_q \, x^{-\alpha}\, (1-x)^{\beta_q} \, ,
\end{equation}
whose analog for $q_v$ and $\Delta q_v$ gives a reasonable first
approximation of phenomenologically extracted parton densities.  The
factor
\begin{equation}
N_q = \frac{\Gamma(2-\alpha+\beta_q)}{\Gamma(1-\alpha)
      \Gamma(1+\beta_q)}
\end{equation}
ensures the proper normalization (\ref{little-sumrules}).  In the
small-$x$ limit $\alpha$ takes the role of a Regge intercept if one
assumes dominance of a single Regge pole, see
Sect.~\ref{sec:phys-mot}.  {}From Regge phenomenology and from
experience with $q_v$ one expects $\alpha \approx 0.5$ for both $u$
and $d$ quarks.  Note however that the form (\ref{good-old-ans}) is
sensitive to $\alpha$ over a finite interval of $x$, and we do not
have enough data to introduce further parameters which would make the
description of $e_v^q$ more flexible.  For the average displacement
$\langle b^y \rangle^q_x$ in (\ref{av-displ}) we expect a relatively
weak $x$-dependence at small $x$, say a power-law $x^{\delta}$, where
$\delta$ should be of order $0.1$ but may be positive or negative.  In
line with our treatment of $\alpha'$ we have taken a single parameter
$\alpha$ for $u$ and $d$ quarks in all our fits.  Trying to determine
a flavor dependence at the level of what is seen in the distributions
$q_v(x)$ is beyond the accuracy we can hope for in this study.


\subsection{Fit to the Pauli form factors}
\label{sec:fit-e}

\begin{figure}
\leavevmode
\begin{center}
\includegraphics[width=0.67\textwidth]{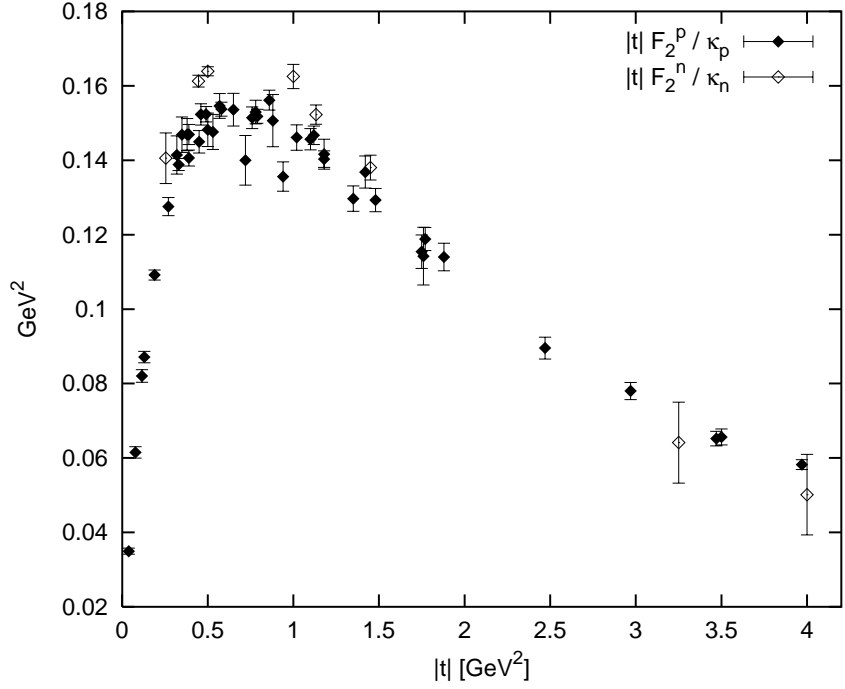}
\end{center}
\caption{\label{fig:F2data1} Data for the normalized form factors
$\kappa_p^{-1} F_2^p$ and $\kappa_n^{-1} F_2^n$ weighted with $|t|$.}
\end{figure}

Before proceeding to the fits of $E^q_v$ we would like to point out
two features of the data on the Pauli form factors.  As we saw in
Sect.~\ref{sec:varia}, a comparison of the Dirac form factors for
proton and neutron clearly shows that with growing $|t|$ the ratio of
$F_1^d$ and $F_1^u$ must become smaller than its value $\frac{1}{2}$
at 
$t=0$.  Let us investigate the evolution with $|t|$ for the analogous
ratio of individual flavor contributions $F_2^q(t) = \int_0^1 \d x\,
E_v^q(x,t)$ to the Pauli form factor.  Writing $r_2 = (\kappa_d^{-1}
F_2^d) / (\kappa_u^{-1} F_2^u)$ we obtain
\begin{equation}
  \label{ud-ratio-F2}
R_2 = \frac{\kappa_n^{-1} F_2^n}{\kappa_p^{-1} F_2^p} \approx
      \frac{1 + 0.71\, (r_2-1)}{1 + 0.38\, (r_2-1)} 
      \approx 1 + 0.33\, (r_2-1) ,
\end{equation}
where we have inserted the values of the anomalous magnetic moments
and in the last step expanded in the deviation of $r_2$ from its value
at $t=0$.  In Fig.~\ref{fig:F2data1} we show the normalized form
factors $\kappa_p^{-1} F_2^p$ and $\kappa_n^{-1} F_2^n$, weighted with
a factor $|t|$ to make the region where we have neutron data more
visible.  We observe that the five neutron data points with $-t$
between $0.25$ and $1.2 \gev^2$ have a tendency to be above the data
for the proton.  We note that these points are from three different
measurements \cite{Mar:1993hx,Eden:1994ji}, \cite{War:2003ma}
and \cite{Madey:2003av}, so that this effect is at least not due to a
normalization problem in a single experiment.  According to
(\ref{ud-ratio-F2}), a positive value of $R_2-1$ implies a positive
value of $r_2-1$ about three times as large, so that $\kappa_u^{-1}
F_2^u(t)$ must decrease faster than $\kappa_d^{-1} F_2^d(t)$ starting
from $t=0$.  This is the opposite of what is found for the Dirac form
factors, and it would be interesting to have better neutron data to
see if this trend is confirmed, and possibly reversed at larger $t$.
With the data and errors at our disposal, the trend has a clear effect
on our fit of $E_v^u$ and $E_v^d$.

\begin{figure}
\leavevmode
\begin{center}
\includegraphics[width=0.67\textwidth]{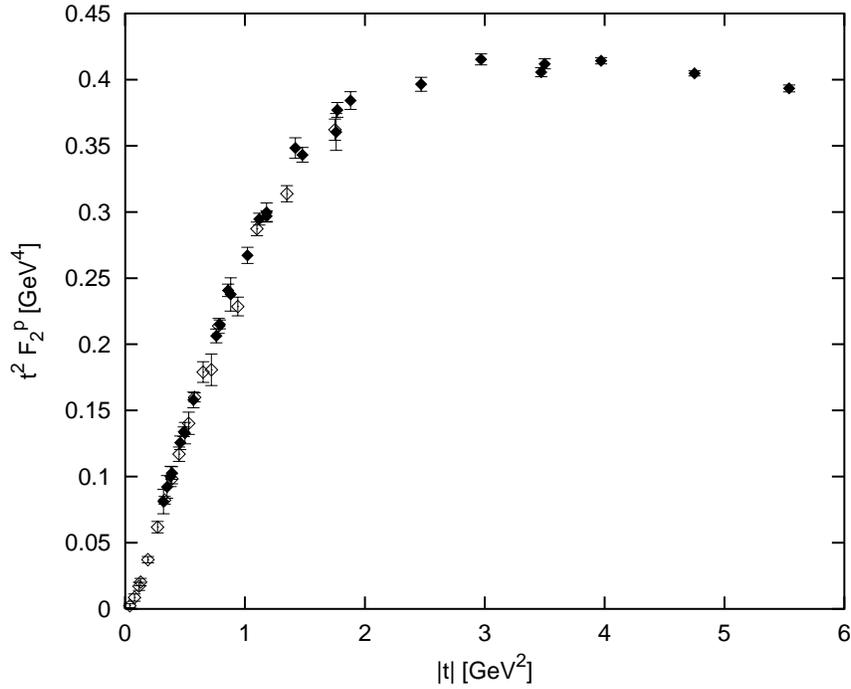}
\end{center}
\caption{\label{fig:F2data2} Data for the weighted form factor $t^2
F_2^p(t)$.  Full symbols: results obtained with the recoil
polarization method
\protect\cite{Jones:1999rz,Gayou:2001qd,Milbrath:1997de}.  Open
symbols: results from the Rosenbluth separation in
\protect\cite{Price:1971zk}.}
\end{figure}

In Fig.~\ref{fig:F2data2} we show $F_2^p$ weighted with a factor $t^2$
(as we did for the Dirac form factors) and find a striking plateau for
$-t$ between $2.5$ and $5.5 \gev^2$.  A behavior $F_2^p \sim t^{-2}$
is certainly not expected in the large-$t$ limit.  The plot in
Fig.~\ref{fig:F2data2} thus instructs us that observables may exhibit
an approximate power-law behavior in an \emph{intermediate} range of a
kinematical variable, which has little to do with the
\emph{asymptotic} behavior.

Apart from their connection with the Pauli form factors, we have at
present no phenomenological constraints on the shape of the GPDs
$E^q(x,\xi,t)$.  With data for $F_2$ going up to $-t = 5.5 \gev^2$ for
the proton and even less for the neutron, we must expect that a
significant range of functions $e_v^q(x)$ and $g_q(x)$ in
(\ref{master-e}) is able to describe the form factors.  In particular,
the fit can partially accommodate a decreased value of $\beta_q$ by
simultaneously increasing $g_q$, thus partially compensating a shift
of $e_v^q(x)$ to larger values of $x$ by a stronger suppression
through the exponential factor $\exp[ t g_q(x) ]$ in that region of
$x$.  The bounds (\ref{forw-bou-ans}) and (\ref{profile-bound})
provide lower limits on $\beta_q$ and upper limits on $g_q(x)$.  In
the opposite direction we have only the requirement that $g_q(x)$
must be positive for the exponential ansatz (\ref{master-e}) to make
sense.

Performing fits to the data on $F_2^p$ and $F_2^n$ where the
parameters $C_u$ and $C_d$ are left free, we find that the values
of $C_d$ and (depending on the details of the fit) also of $C_u$
are too large to fulfill the bound (\ref{profile-bound}), which
results in an exponentially strong violation of (\ref{b-bou-ans}) for
large enough $x$ and $\tvec{b}^2$.  We have therefore in the
minimum-$\chi^2$ fits imposed the bounds
\begin{equation}
  \label{par-bounds}
C_u \le A_u = 1.22 \gev^{-2} , \qquad \qquad
C_d \le A_d = 2.59 \gev^{-2} .
\end{equation}
The values
of $A_u$ and $A_d$ in (\ref{par-bounds}) are those obtained in our
default fit to the Dirac form factors (see Sect.~\ref{sec:select}).
Directly implementing (\ref{forw-bou-ans}) or (\ref{super-bound}) in
the fit would be more involved, and we have instead verified these
bounds only for the fit results.  We remark that with our input parton
densities at $\mu=2 \gev$ (CTEQ6M \cite{CTEQ} and the NLO
distributions in scenario 1 of \cite{blum:03}) the inequalities $|
\Delta q(x) | \le q(x)$ and $| \Delta q_v(x) | \le q_v(x)$ are
violated for the best fit values of the distributions at $x$ above 0.6
or 0.7, but are well satisfied within the $1\, \sigma$ error bands on
the polarized densities from \cite{blum:03}.  We therefore require the
bounds (\ref{b-bou-ans}), (\ref{forw-bou-ans}), (\ref{super-bound}) to
hold within these errors.

Notice that when fitting $E_v^q(x,t,\mu^2)$ to the Pauli form factors
alone, we cannot determine the scale $\mu$ since the form factor
integrals in (\ref{sumrule-e}) are scale independent.  We fix $\mu$
implicitly in our fit by requiring the positivity condition
(\ref{super-bound}) to hold for the parton densities at $\mu=2 \gev$
and the associated profile function $f_q(x)$ obtained in our default
fit of $H_v^q$.

To avoid dealing with too many free parameters, we will present fits
where we imposed $\alpha = 0.55$.  Leaving $\alpha$ free in those fits
which give a good description of the data, we obtain values between
$0.5$ and $0.6$.  Following the discussion at the end of the previous
subsection, this validates once more our assumption that simple Regge
phenomenology provides a guide for the behavior of GPDs at small $x$
and small $t$.  For those fits which fail to describe the data, making
$\alpha$ a free parameter does not help.  The conclusions we draw in
the following are hence not affected by having fixed the parameter
$\alpha$.

Even then we still have six parameters to determine, namely $\beta_q$,
$C_q$ and $D_q$ for each quark flavor.  To gain some insight into
the range of allowed parameters, we have performed a series of fits
with fixed values of $\beta_u$ and $\beta_d$, leaving the other four
parameters to be fitted.  The resulting values of $\chi^2 /\dof$ are
shown in Table~\ref{tab:scan-chi} and the results for a subset of
these fits in Table~\ref{tab:scan}.  Distributions with $\beta_u=3$ or
with $\beta_d=4$ violate the positivity condition (\ref{forw-bou-ans})
at large $x$ and have been discarded.  A number of observations can be
made in these fits:
\begin{enumerate}
\item Both $C_q$ and $D_q$ increase when $\beta_u$ decreases, and for
$\beta_u=4$ the fit selects the maximum allowed value $C_u= A_u$.  For
all fits shown in Table~\ref{tab:scan-chi} the fit selects the maximum
$C_d= A_d$, and only $D_d$ increases when $\beta_d$ becomes smaller.
\item The minimum $\chi^2$ is very flat as a function of $\beta_u$ and
even more of $\beta_d$.  In particular there is only a slight
preference to have $\beta_u < \beta_d$, with good fits being also
obtained for $\beta_u \ge \beta_d$.
\item The fitted parameters $D_u$ and $D_d$ differ significantly
within their errors---much more strongly than in our our fits to the
Dirac form factors (see Sect.~\ref{sec:varia}).
\end{enumerate}

\begin{table}[hb]
\caption{\label{tab:scan-chi} Values of $\chi^2 /\dof$ obtained in
fits to (\protect\ref{master-e}), (\protect\ref{profile-e}) and
(\protect\ref{good-old-ans}) with fixed $\alpha = 0.55$.  Free
parameters are $D_u$, $D_d$ and $C_u$, $C_d$ under the
constraints (\protect\ref{par-bounds}).}
\begin{center}
\renewcommand{\arraystretch}{1.2}
\begin{tabular}{ccccc} \hline\hline
          & \multicolumn{4}{c}{$\beta_u$} \\
          & 4    & 5    & 6    & 7  \\
\hline
$\beta_d$ & \multicolumn{4}{c}{$\chi^2 /\dof$} \\
   5      & 1.32 & 1.36 & 1.41 & 1.47 \\
   6      & 1.31 & 1.36 & 1.41 & 1.47 \\
   7      & 1.33 & 1.37 & 1.42 & 1.48 \\
   8      & 1.36 & 1.39 & 1.44 & 1.50 \\
   9      & 1.44 & 1.42 & 1.47 & 1.52 \\
\hline\hline
\end{tabular}
\end{center}
\end{table}

These fits with a grid of fixed values $\beta_u$ and $\beta_d$ are not
well suited for the propagation of correlated errors.  We have thus
performed a six-parameter fit with free $\beta_u$, $\beta_d$, $C_u$,
$C_d$, $D_u$ and $D_d$, fixing only $\alpha=0.55$.  The minimum
$\chi^2$ is achieved for $\beta_u=3.99$ and $\beta_d=5.59$, and the
parameters $C_u$ and $C_d$ take their maximum values given in
(\ref{par-bounds}).  To avoid the treatment of errors on parameters at
the boundary of their allowed range, we have fixed their values to
$C_u= 1.22 \gev^{-2}$ and $C_d= 2.59 \gev^{-2}$ and repeated the
fit.  The resulting four-parameter fit still has a very large error of
$2.67$ on $\beta_d$.  In addition, this error is strongly correlated
with the error on $D_d$, with a linear combination of the two
parameters being essentially undetermined.  The available data on the
Pauli form factors do not allow us to extract four independent
parameters within an acceptable accuracy.  As a simple solution one
might impose the constraint $\beta_u=\beta_d$, but this results in a
fitted value of $\beta_q = 4.21$, which is in clear violation of the
positivity bound (\ref{forw-bou-ans}).  As an alternative we have
fixed the difference $\beta_d-\beta_u$ to the value $1.6$ obtained in
the six- and four-parameter fits just discussed.  This value is then
to be regarded as an external input to a three-parameter fit with
\begin{equation}
  \label{par-def-e}
\renewcommand{\arraystretch}{1.3}
\renewcommand{\arraycolsep}{0.15ex}
\begin{array}[b]{llll}
 \beta_u & =\, 3.99 \pm 0.22 ,    & \beta_d & =\, \beta_u + 1.60 , \\
 C_u   & =\, 1.22 \gev^{-2} ,   & C_d & =\, 2.59 \gev^{-2} , \\
 D_u   & =\, (0.38 \pm 0.11) \gev^{-2} , ~~~~~~ &
 D_d   & =\, - (0.75 \pm 0.05) \gev^{-2} ,
\end{array}
\end{equation}
and $\alpha=0.55$.  Full details are given in App.~\ref{app:B}.  We
take this as our default fit for $E_v^q$ in the rest of this paper.
We remark that a four-parameter fit where we additionally leave
$\alpha$ free finds it to be $0.55 \pm 0.03$, with the remaining
parameters essentially as in (\ref{par-def-e}).

The profile function $g_u$ obtained with this fit is smaller than
$f_u$ by at most 10\% for all $x$.  In contrast, $g_d$ is
significantly below its counterpart $f_d$ at moderate values of $x$,
namely by as much as a factor of 2 for $x\sim 0.2$.  This can be
understood from the observation we made at the beginning of this
section.  The forward limit $e_d(x)$ in this fit is concentrated at
smaller values of $x$ than $e_u(x)$, which favors $|F_2^d|$ decreasing
faster with $|t|$ than $F_2^u$.  To obtain the reverse trend at small
$t$, which is favored by the data, the fit requires a rather weak
damping factor $g_d$ in the $t$ dependence of $E_v^d$ at the
relevant $x$ values.  In fact, one has $g_d < g_u$ up to $x
<0.45$.  For larger $x$ the hierarchy is reversed, so that eventually
$|F_2^d|$ will decrease faster than $F_2^u$ when $t$ becomes large.

\begin{figure}
\begin{center}
\leavevmode
\includegraphics[width=0.45\textwidth,
  bb=100 410 430 650]{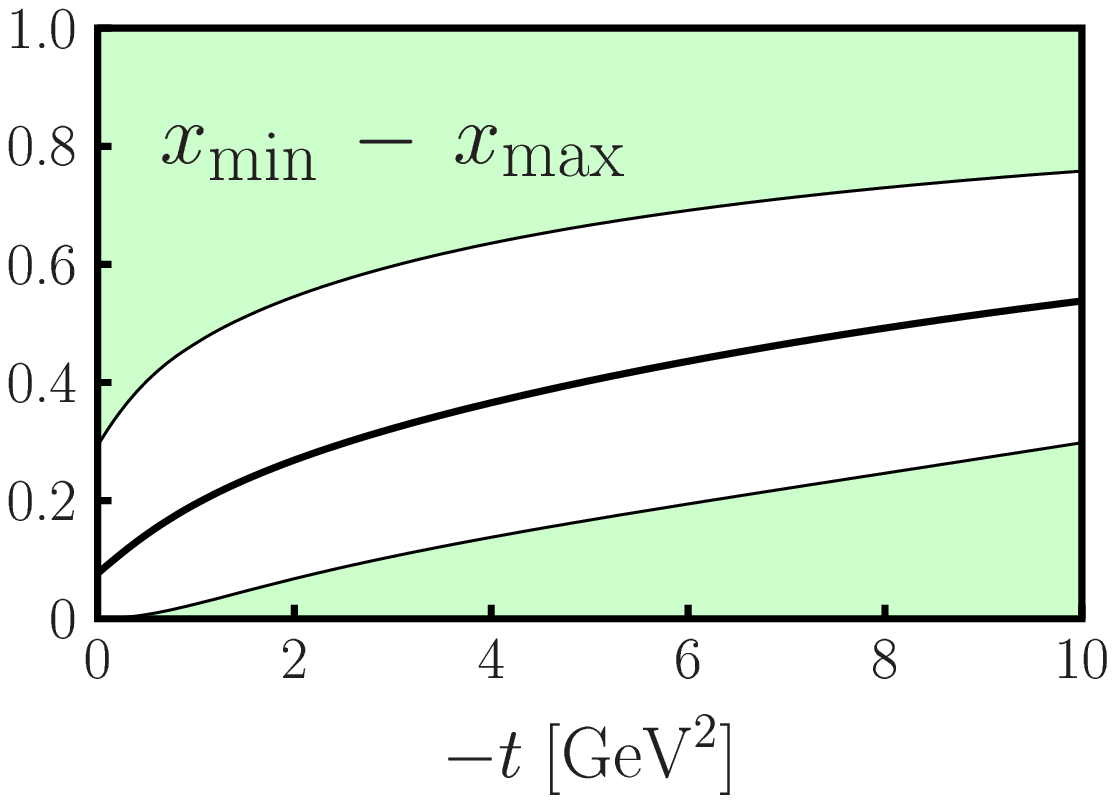}
\end{center}
\caption{\label{fig:F2_xminmax} Region of $x$ (white region) which
accounts for $90\%$ of $F_2^p(t)$ in our default fit for $E_v^q(x,t)$.
The upper and lower shaded $x$-regions each account for $5\%$ of
$F_2^p(t)$.  The thick line shows the average~$\langle x\rangle_t$ in
the form factor integral.}
\end{figure}

\begin{figure}[p]
\leavevmode
\begin{center}
\includegraphics[width=.47\textwidth,
  bb=100 355 470 705]{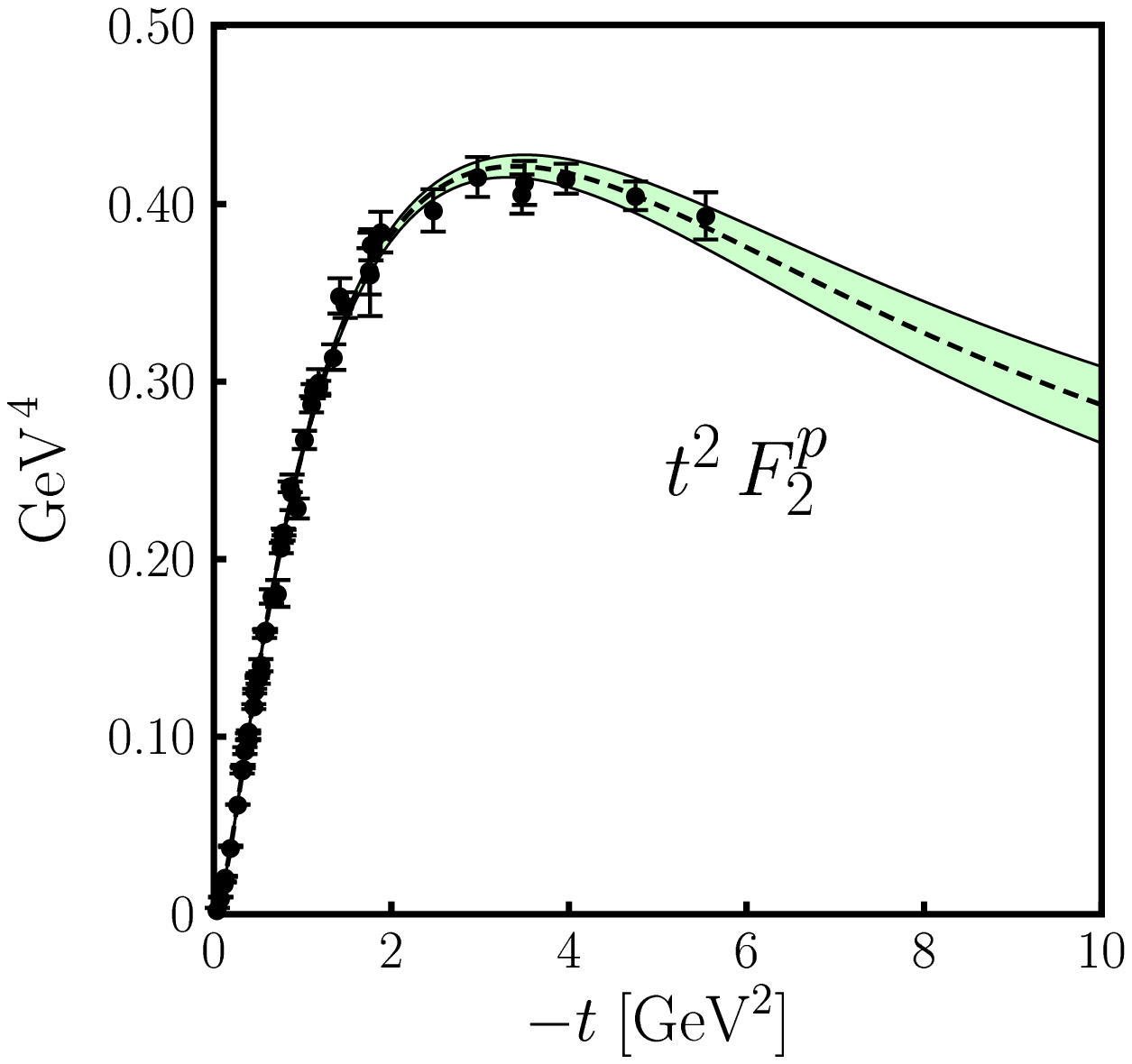}
\hspace{2em}
\includegraphics[width=.47\textwidth,
  bb=110 325 480 675]{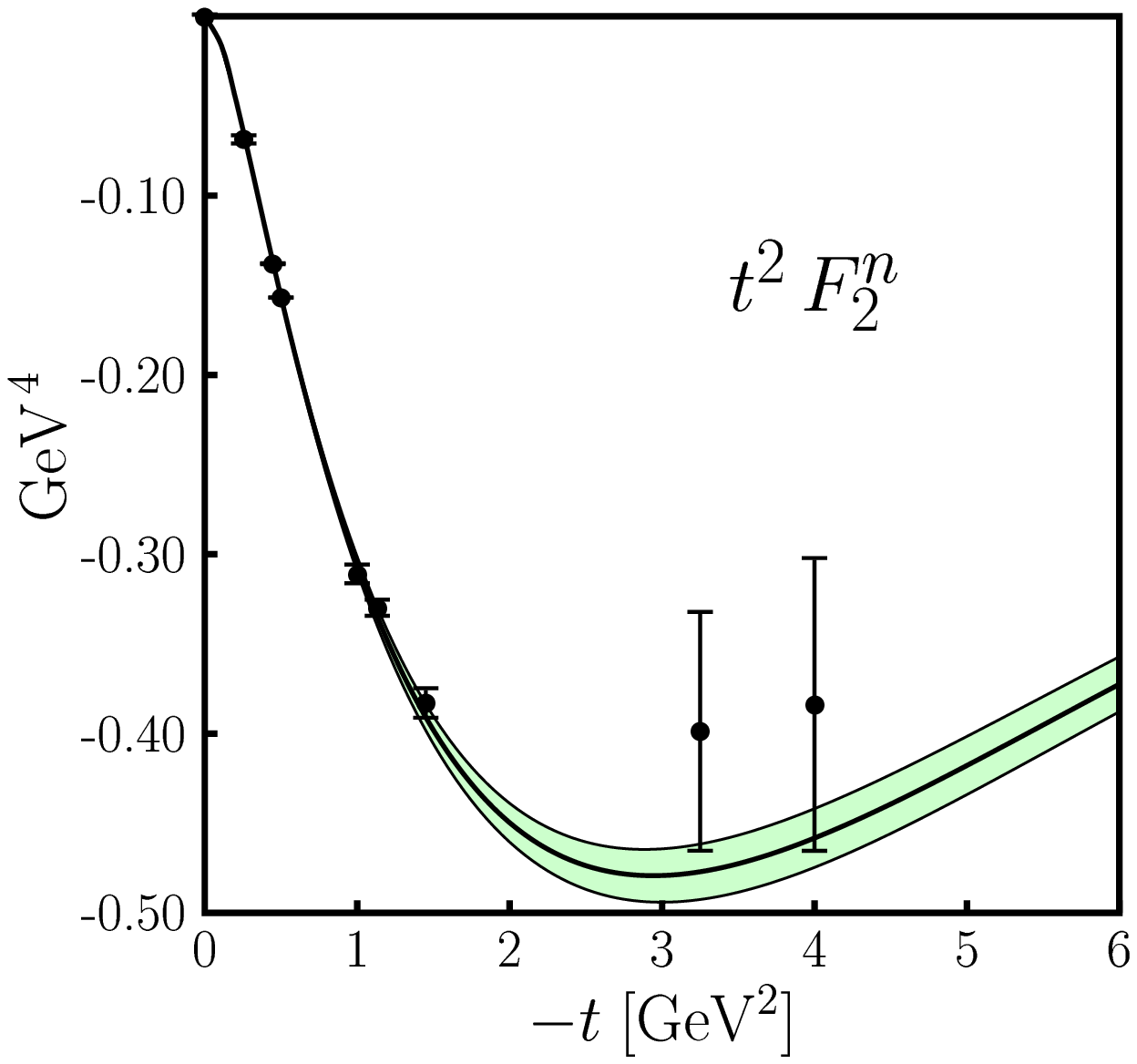}
\end{center}
\caption{\label{fig:defF2} Results for the Pauli form factor of proton
and neutron with our default fit defined in the text.  The error bands
represent the $1\, \sigma$ uncertainties of the fit as explained in
App.~\protect\ref{app:B}.}
\end{figure}
\begin{figure}[p]
\begin{center}
\includegraphics[width=.47\textwidth,
  bb=155 260 495 610]{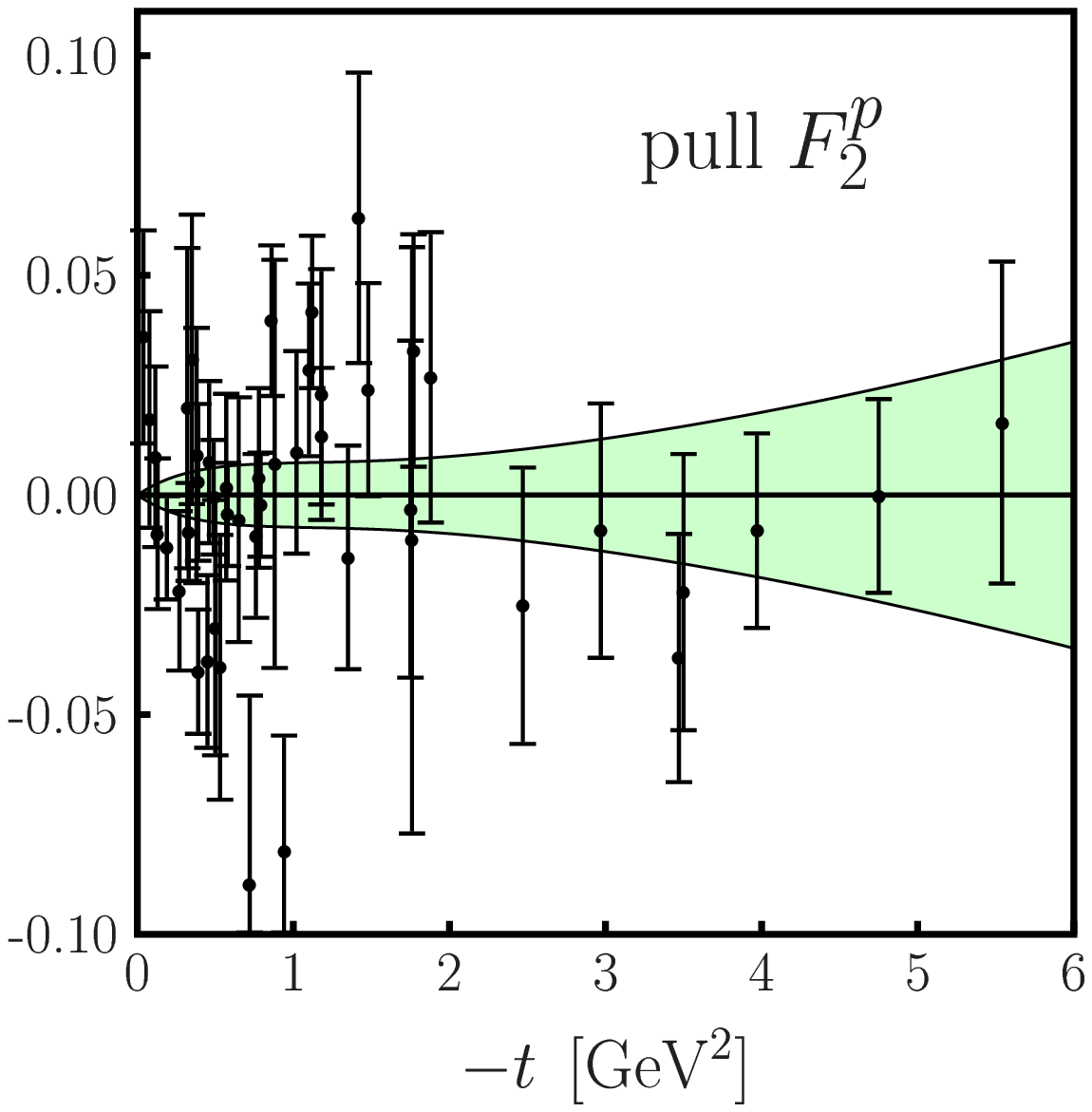}
\hspace{2em}
\includegraphics[width=.47\textwidth,
  bb=125 350 468 700]{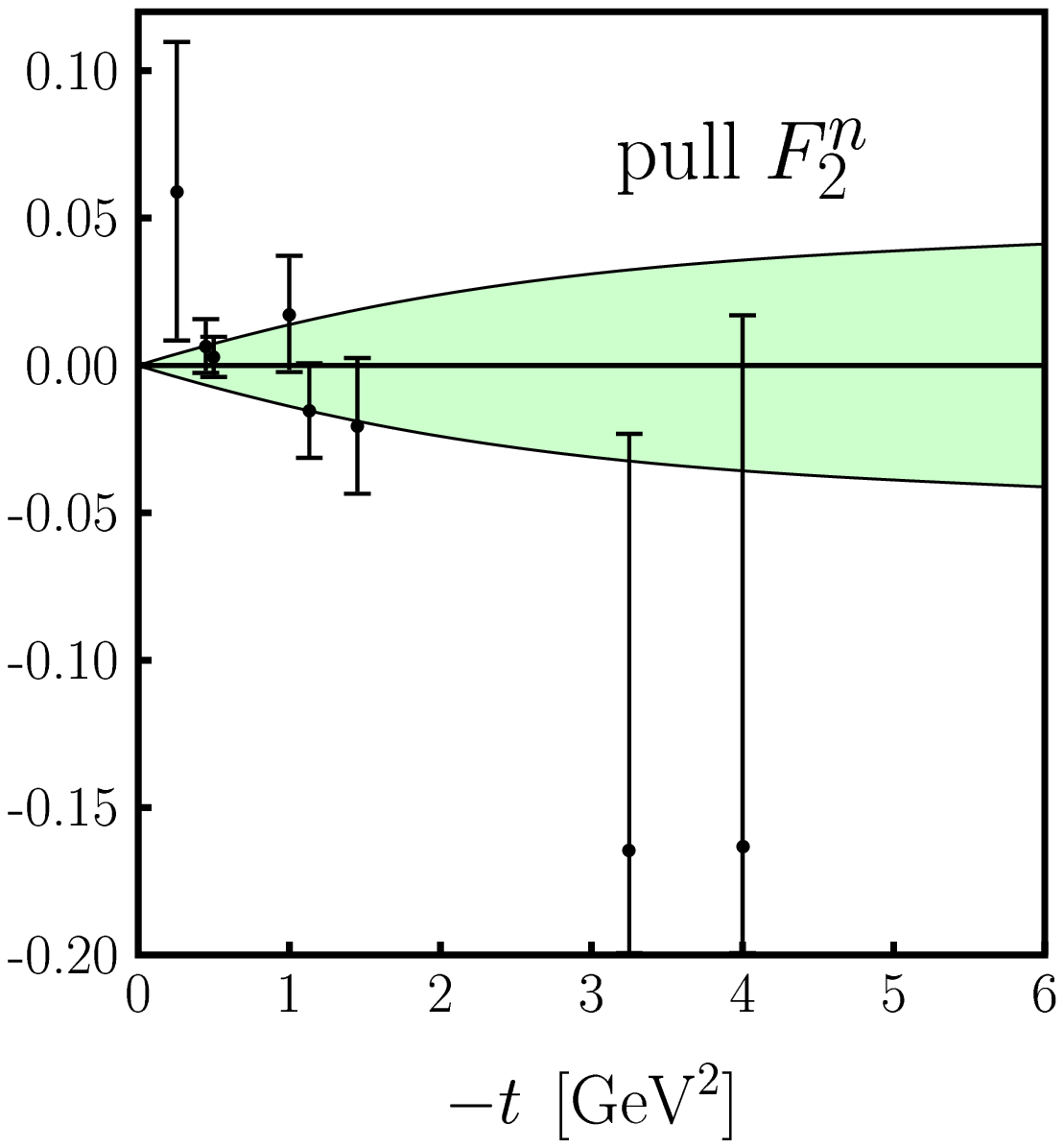}
\end{center}
\caption{\label{fig:F2pull} The pull $[F_2({\rm data})/F_2({\rm fit})
    - 1]$ for our default fit to the Pauli form factors of the
  nucleon.  The error bands represent the $1\, \sigma$ fit uncertainty
  relative to the best fit.}
\end{figure}

To quantify the sensitivity of our fit to the range of $x$ in the sum
rules (\ref{sumrule-e}) we plot in Fig.~\ref{fig:F2_xminmax} the
quantities $x_{\rm min}(t)$, $x_{\rm max}(t)$ and $\langle x\rangle_t$
for the Pauli form factor of the proton, defined in analogy to
(\ref{eq:xminmax}) and (\ref{eq:xavg}) by replacing $H_v^q$ with
$E_v^q$ and $F_1^p$ with $F_2^p$.  We find that we are sensitive to
values up to about $x\sim 0.7$ with the form factor data at hand.  In
Fig.~\ref{fig:defF2} we show the result of our fit in comparison with
the data, and in Fig.~\ref{fig:F2pull} the corresponding pull.  With
the exception of three (somewhat outlying) data points, our fit
describes $F_2^p$ within 5\% over the entire $t$ range.  The
description of $F_2^n$ is of similar quality, except for the two data
points with the highest $-t$, where the central value of the fit is
just compatible with the errors of the data.


\subsection{Flavor structure}
\label{sec:flavor-e}

The fit we have just described exhibits a clear difference in the
parameters of the $u$ and the $d$ quark distributions.  To test the
significance of this result, we have performed a fit with the same
$\beta_q$, $C_q$ and $D_q$ for both flavors, leaving also $\alpha$
as a free parameter in order not to be biased by its particular value.
The bounds (\ref{par-bounds}) imply of course $C_q \le 1.22
\gev^{-2}$ in this case.  Such a fit does not give a satisfactory
description of the neutron data, with a $\chi^2$ of 106 for the 8 data
points according to Table~\ref{tab:flavor-e}.  This fit undershoots
all data for $|F_2^n|$ with $-t$ below $1.5 \gev^2$ by 5\% to 10\%,
which is a large effects given the errors on the data in that region
(see App.~\ref{app:A}).  To see whether the positivity constraints are
responsible for this failure, we have repeated the fit without
restricting $C_q$.  The fit then selects parameters $C_q$ and
$\beta_q$ that badly violate the positivity constraints, but the
resulting $F_2^n$ is barely changed for $-t < 2\gev^2$ and thus
equally inadequate.

Having seen that the data and errors we use require different
$\kappa_u^{-1} E_v^u$ and $\kappa_d^{-1} E_v^d$, we would like to
explore for which of our parameters a flavor dependence is required
most.  {}From the fits with fixed $\beta_u$ and $\beta_d$ presented in
Tables~\ref{tab:scan-chi} and \ref{tab:scan} we have already seen that
$\kappa_u^{-1} e_v^u(x) = \kappa_d^{-1} e_v^d(x)$ for the forward
limit is compatible with existing data.  Concerning the profile
functions, one may ask whether the data can be described when taking
$D_u = D_d$, as was the case for the analogous fits of $H_v^q$.  A
fit with free $\beta_u$ and $\beta_d$ and this constraint selects
$\beta_d$ much too small to comply with the positivity bound
(\ref{forw-bou-ans}), so that instead we have performed a series of
fits with fixed $\beta_u$ and $\beta_d$, leaving free parameters
$D_u=D_d$ in addition to $C_u$ and $C_d$ subject to
(\ref{par-bounds}).  The resulting values of $\chi^2 /\dof$ are given
in Table~\ref{tab:constr-scan-chi} and details of selected fits in
Table~\ref{tab:constr-scan}.  We find that a good description of both
proton and neutron data is possible only if $\beta_u$ is significantly
larger than $\beta_d$.  In analogy to the discussion at the end of
Sect.~\ref{sec:fit-e}, this can again be understood from the necessity
of the fit to accommodate $\kappa_u^{-1} F_2^u < \kappa_d^{-1} F_2^d$
at small~$t$.  In Table~\ref{tab:constr-scan} we observe that with
increasing $\beta_u$ the fit requires smaller and smaller profile
functions $g_u(x)$, otherwise $F_2^u$ would decrease too fast to be
compatible with the proton data at larger values of $t$.  For the
largest $\beta_u$ in the table this does not seem a physically very
plausible scenario.  In the case $\beta_u=11$ and $\beta_d=5$ the
fitted values even lead to $g_u(x) <0$ for $0.5<x<0.8$, so that the
corresponding entry in Table~\ref{tab:constr-scan-chi} has been
omitted.

\begin{table}[ht]
\caption{\label{tab:constr-scan-chi} Values of $\chi^2 /\dof$ obtained
in fits as specified in Table~\protect\ref{tab:scan-chi} but with the
additional constraint $D_u=D_d$.}
\begin{center}
\renewcommand{\arraystretch}{1.2}
\begin{tabular}{ccccccccc} \hline\hline
          & \multicolumn{7}{c}{$\beta_u$} \\
          & 5    & 6    & 7    & 8    & 9    & 10   & 11 \\
\hline
$\beta_d$ & \multicolumn{7}{c}{$\chi^2 /\dof$} \\
    5     & 5.86 & 3.01 & 1.98 & 1.72 & 1.62 & 1.60 &      \\
    6     & 8.94 & 4.85 & 2.96 & 2.12 & 1.84 & 1.72 & 1.67 \\
\hline\hline
\end{tabular}
\end{center}
\end{table}

We have finally performed a series of fits where we imposed
$C_u=C_d$ with the constraint $C_q \le 1.22 \gev^{-2}$,
summarized in Table~\ref{tab:constr}.  Provided that we allow a flavor
dependence of either $\beta_q$ or $D_q$, a reasonably good
description of the data can be achieved, although the $\chi^2$ for the
neutron data is rather high.

We note that positivity restricts $E_v^u$ and $E_v^d$ in a rather
asymmetric way.  The bound (\ref{forw-bou-ans}) is stronger for $d$
quarks, given the stronger decrease of $d_v(x)$ with $x$.  With the
profile functions $f_q(x)$ from our default fit of $H_v^q$ one needs
at least $\beta_d \ge 5$, whereas $\beta_u$ may be as small as $3.5$.
For the parameters of the profile function, the constraint
(\ref{profile-bound}) is in turn more restrictive on $u$ quarks.  We
have seen that even with these constraints the flavor structure of the
forward limit and of the profile function in $E_v^q$ cannot be
uniquely determined by present data on the Pauli form factors.  The
neutron data can exclude a complete flavor independence of the
distributions (up to their normalization) and somewhat disfavors a
flavor independent profile function towards larger $x$.  Flavor
independence of $g_q$ at small to moderate $x$ cannot be excluded
from the data, but it requires a rather extreme behavior of $E_v^u$.
In view of these uncertainties, precise data on $F_2^n$ in a wider $t$
range would be of great help.

\begin{figure}
\begin{center}
\includegraphics[width=.45\textwidth,
  bb=45 400 400 730]{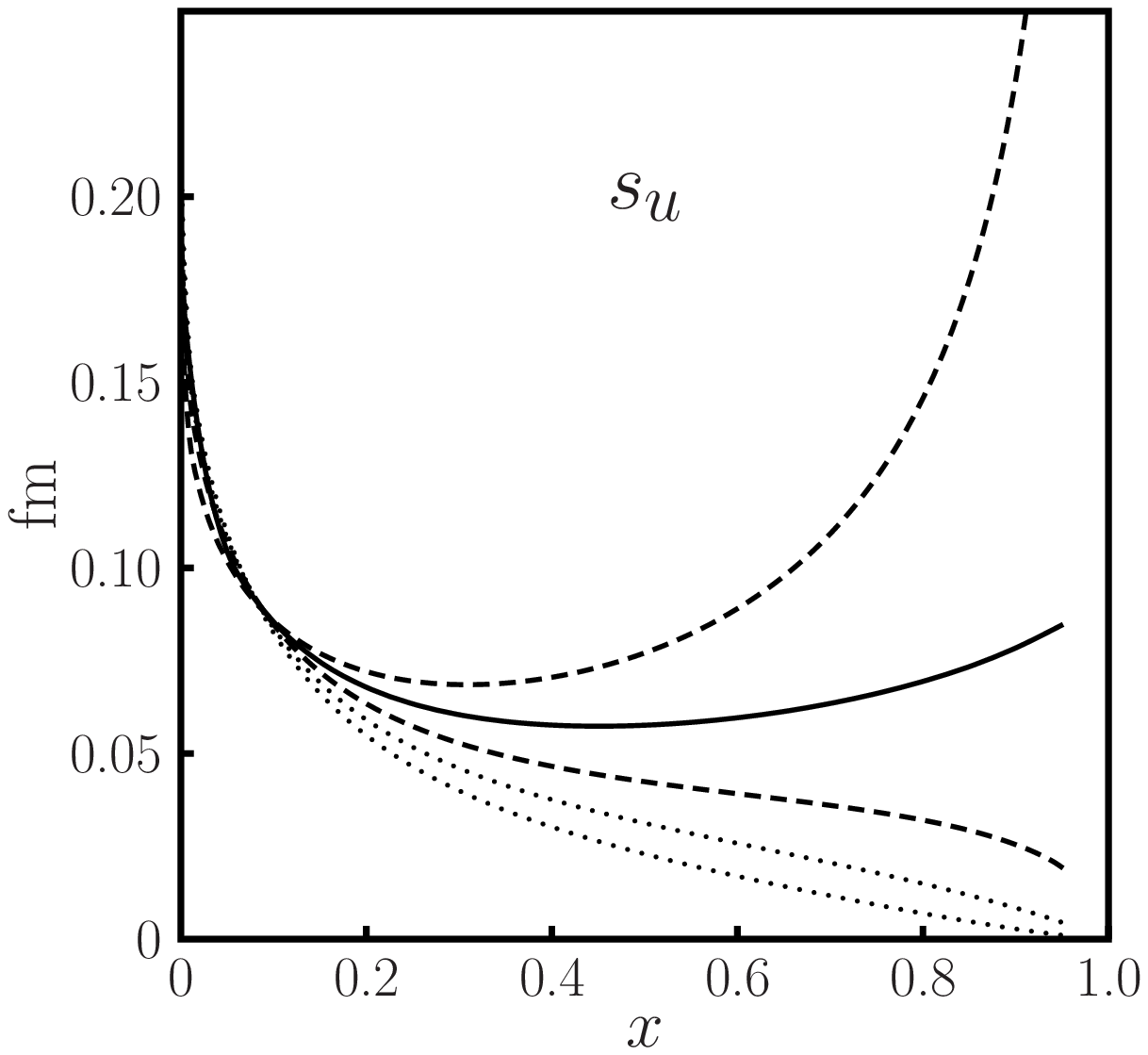}
\hspace{3em}
\includegraphics[width=.45\textwidth,
  bb=45 400 400 730]{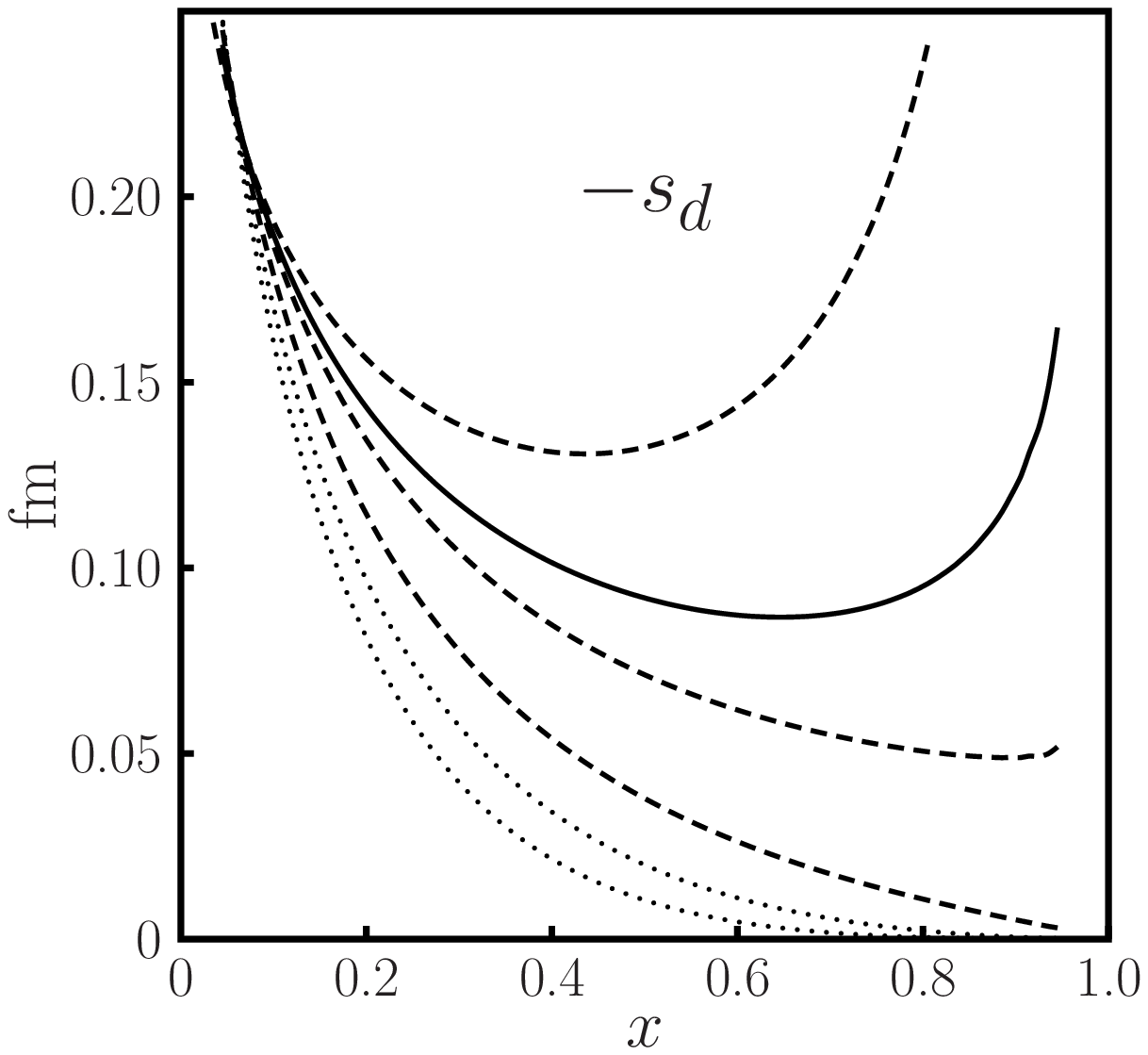}
\end{center}
\caption{\label{fig:s-syst} Average shift $s_q(x)$ of the distance
between struck quark and spectators in a polarized proton, defined in
(\protect\ref{sq-def}).  Solid curves correspond to the default fit
(\protect\ref{par-def-e}), dashed curves to $\beta_u=3.5, 4.5$ or
$\beta_d=5, 6, 7$, and dotted curves to $\beta_u=5, 5.5$ or
$\beta_d=8, 9$.  In all cases $\alpha=0.55$.}
\end{figure}

In Fig.~\ref{fig:s-syst} we plot the average sideways shift $s_q(x)$
of the distance between struck quark and spectators in a transversely
polarized proton, obtained with different values of $\beta_u$ and
$\beta_d$ for which satisfactory fits to the Pauli form factors can be
obtained.  Despite the considerable spread of possibilities,
comparison with Fig.~\ref{fig:n2_rq} shows that $|s_q(x)|$ is clearly
below $d_q(x)$ for all $x$ where we can trust our results.

\section{Results for valence GPDs and their moments}
\label{sec:results}

In this section we present the results for the valence GPDs at zero
skewness which follow from our default fits to the proton and neutron
form factors.

\subsection{GPDs as a function of $x$ and $t$}
\label{GPD-results}

\begin{figure}[p]
\begin{center}
\includegraphics[width=.35\textwidth, height=.35\textwidth,
  bb=101 492 389 779]{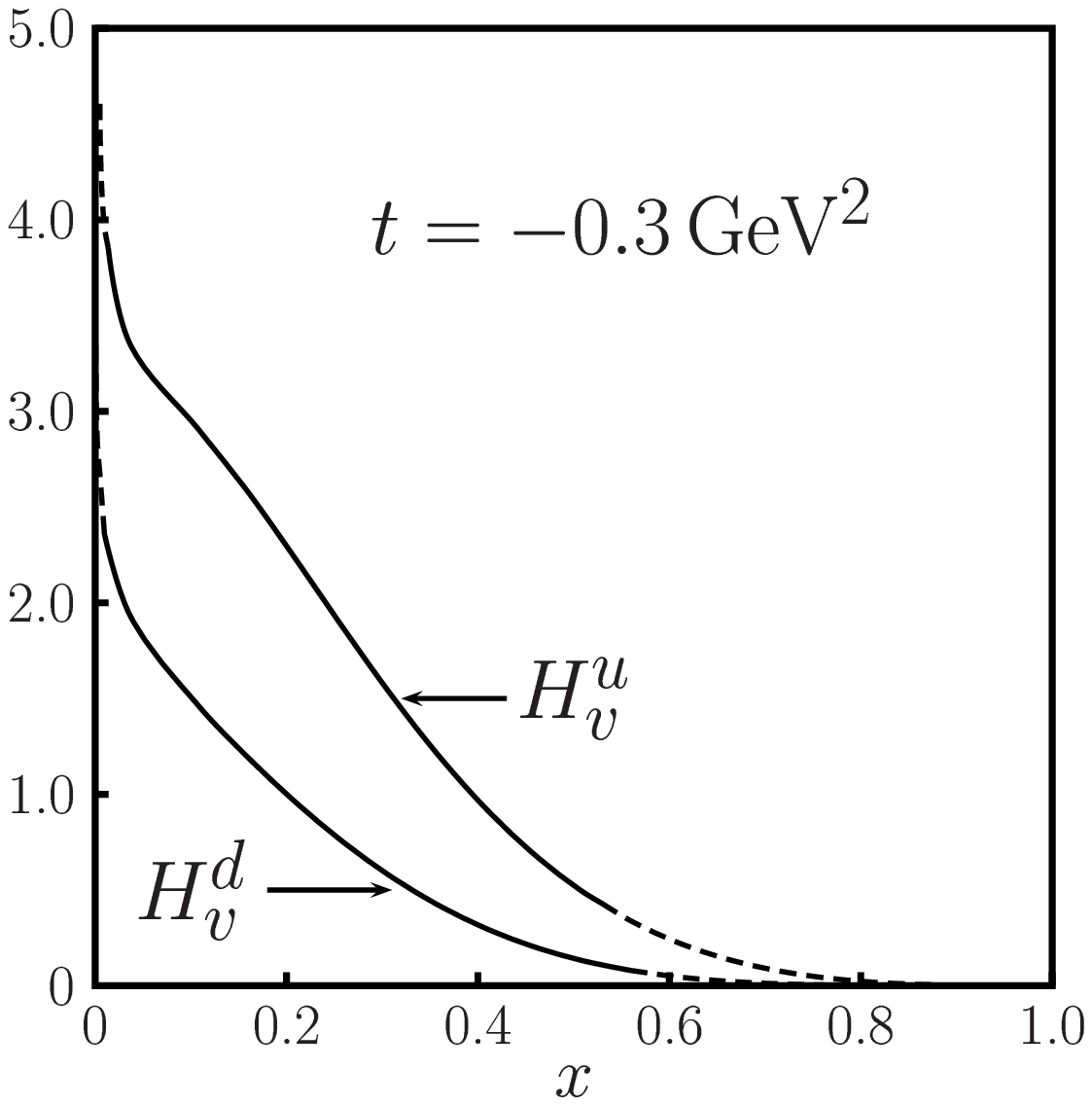}
\hspace{4em}
\includegraphics[width=.35\textwidth, height=.35\textwidth,
  bb=101 339 390 628]{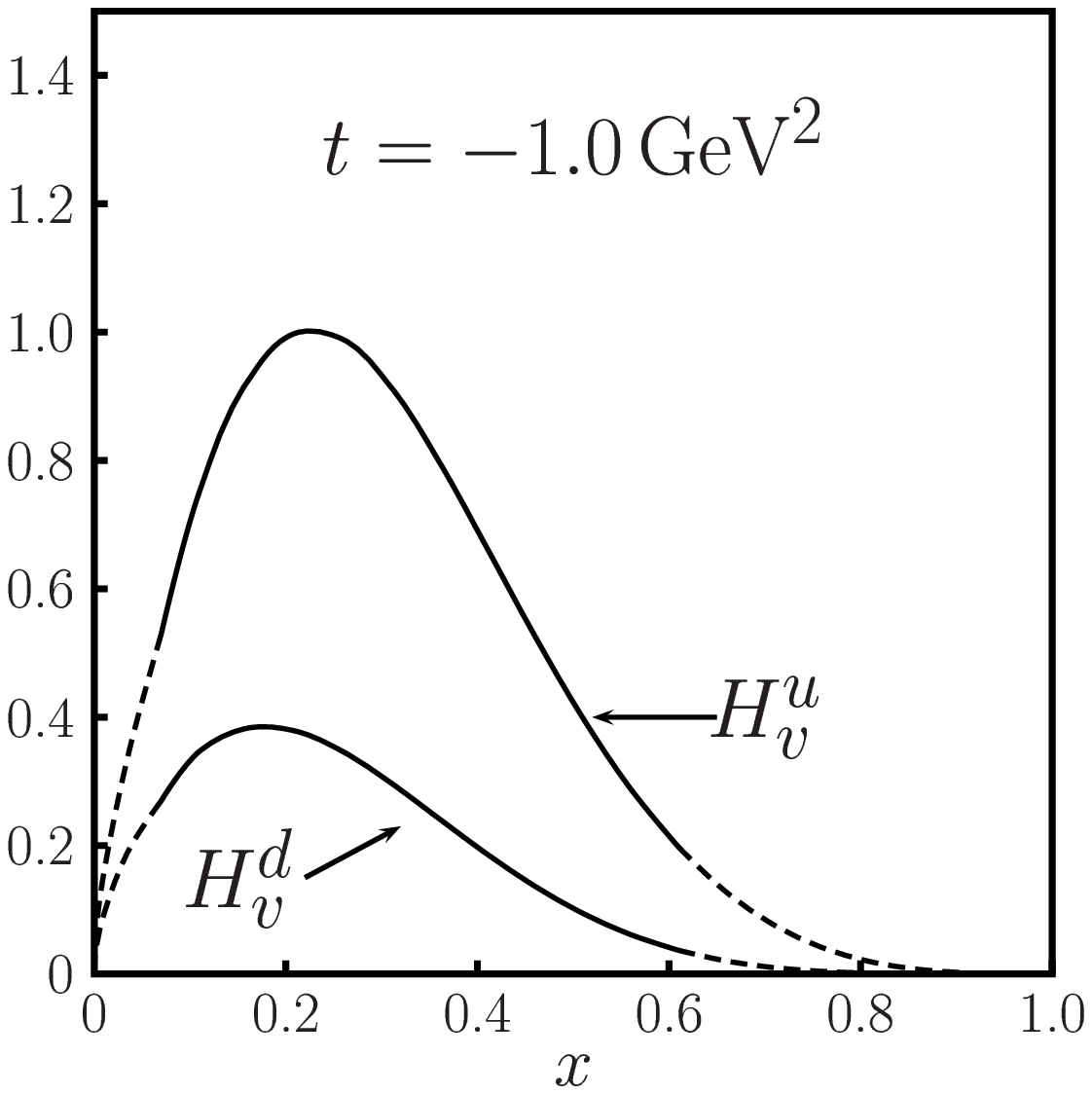} \\[4em]
\includegraphics[width=.35\textwidth, height=.35\textwidth,
  bb=101 401 390 683]{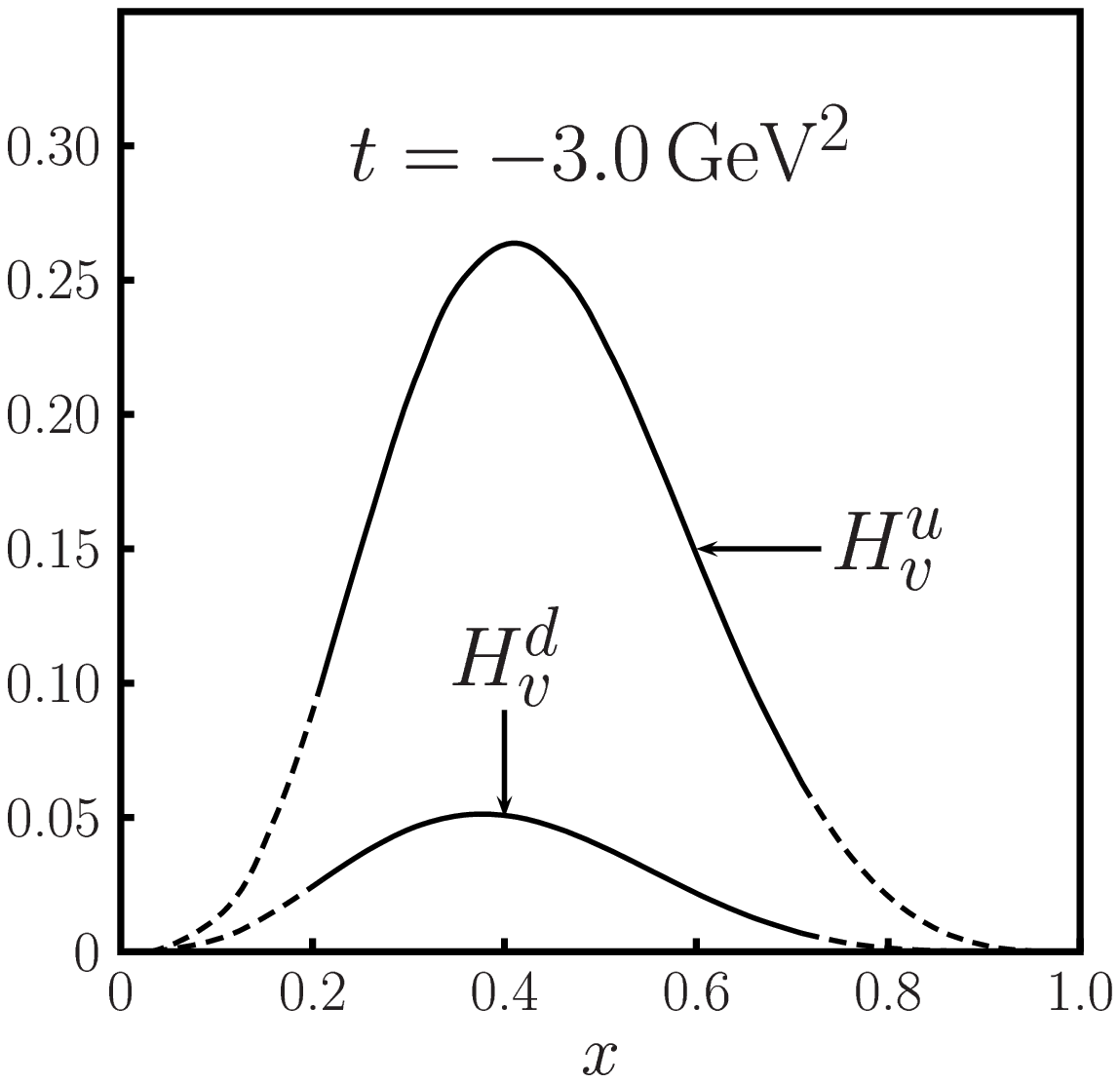}
\hspace{4em}
\includegraphics[width=.35\textwidth, height=.35\textwidth,
 bb=101 524 390 820 ]{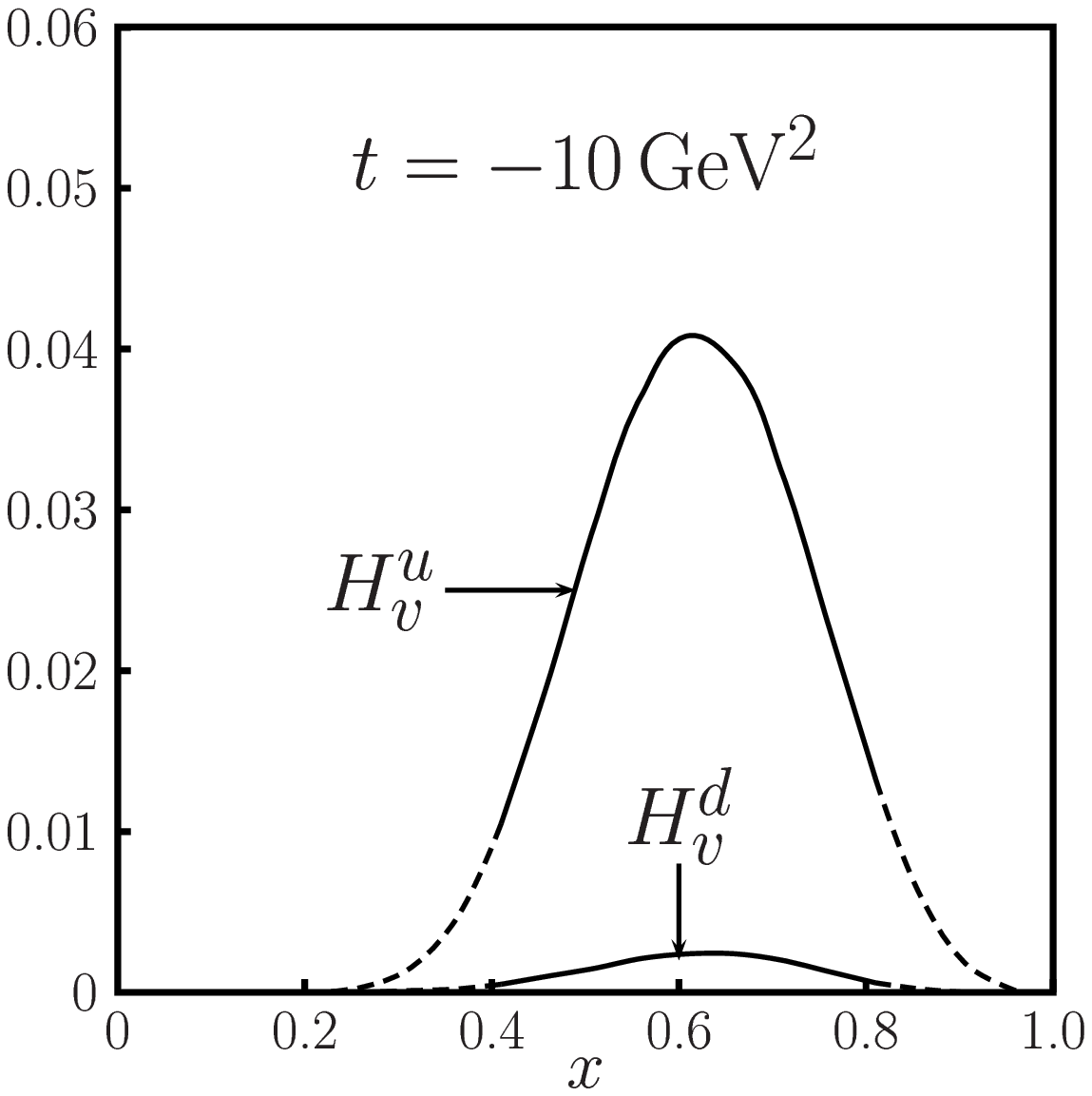}
\\[1em]
\ 
\end{center}
\caption{\label{fig:Hgpd} Result for the valence GPDs $H_v^q(x,t)$ at
$\mu=2 \gev$ obtained in default fit to $F_1^{p}(t)$ and $F_1^{n}(t)$.
Dashed lines indicate the regions where $x < x_{\rm min}(t)$ or $x >
x_{\rm max}(t)$, see (\protect\ref{eq:xminmax}) and
Fig.~\protect\ref{fig:var1_xminmax}.}
\end{figure}

In Fig.~\ref{fig:Hgpd} we have plotted $H_v^u(x,t)$ and $H_v^d(x,t)$
as functions of $x$ for some fixed values of momentum transfer $t$.
We recall that the distributions in our default fits refer to the
scale $\mu=2 \gev$.  Notice the qualitative change when going from
small to large values of $|t|$. For momentum transfer well below $1
\gev^2$ the GPDs resemble the forward parton distributions and in
particular show a divergent behavior at small $x$.  For increasing
momentum transfer, the GPDs become narrower and develop a pronounced
maximum, which shifts towards higher values of $x$ in accordance with
the behavior of $\langle x\rangle_t$ in Fig.~\ref{fig:var1_xminmax}.
The height of this maximum decreases with $|t|$, and this effect is
more dramatic for $d$ quarks than for $u$ quarks.  This can be traced
back to our fit result for the profile functions, where
$f_d(x,t)>f_u(x,t)$ at large $x$.
The dashed lines in Fig.~\ref{fig:Hgpd} indicate the values of $x$
where according to Fig.~\ref{fig:var1_xminmax} the sensitivity of the
form factor fits is not sufficient to constrain the GPDs, and
therefore the result should be viewed as an extrapolation which
follows from our particular ansatz for the functional form of the $x$
and $t$ dependence.

\begin{figure}[p]
\begin{center}
\includegraphics[width=.35\textwidth, height=.35\textwidth,
 bb=131 414 415 698]{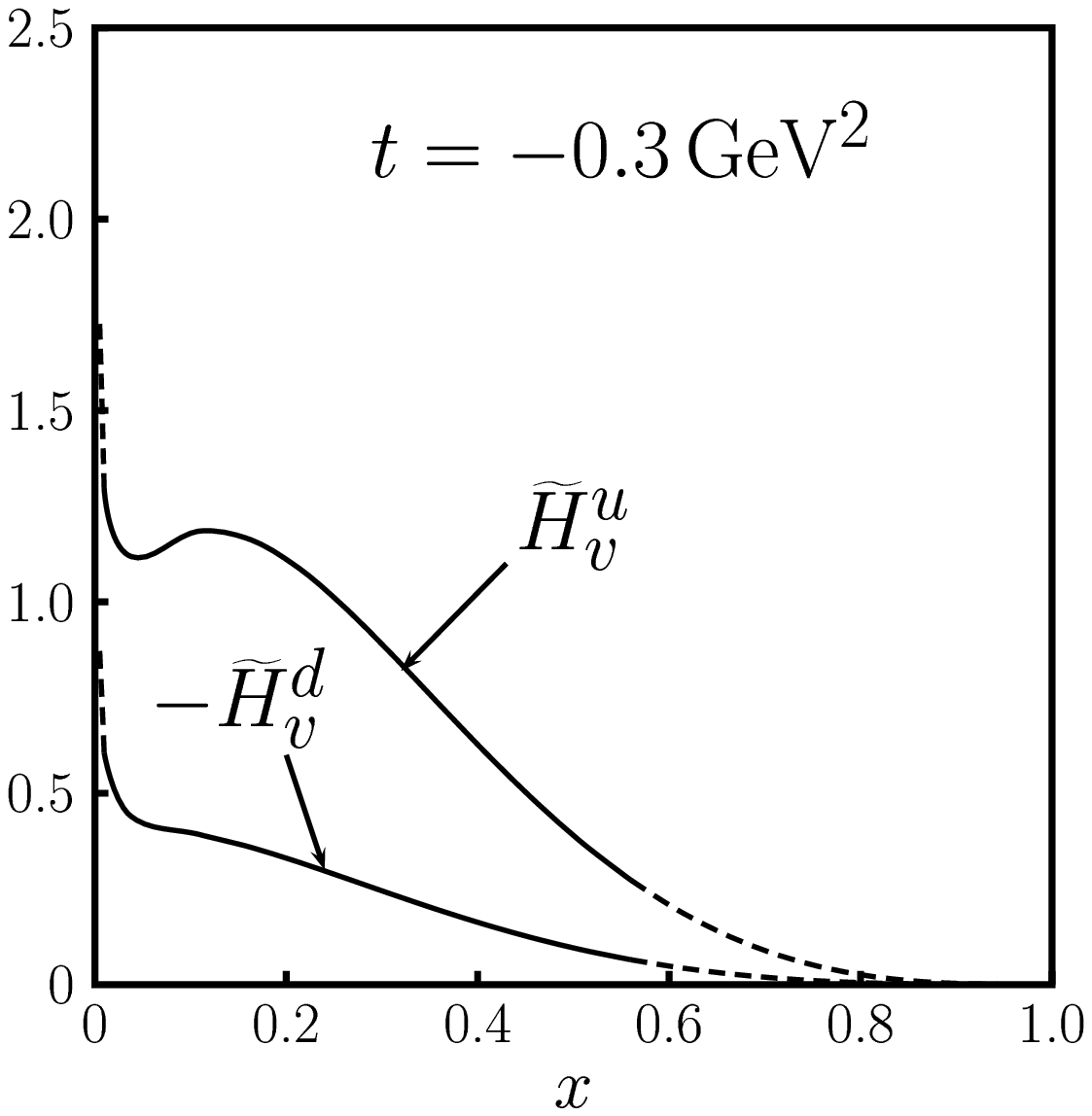}
\hspace{4em}
\includegraphics[width=.35\textwidth, height=.35\textwidth,
 bb=131 396 416 680]{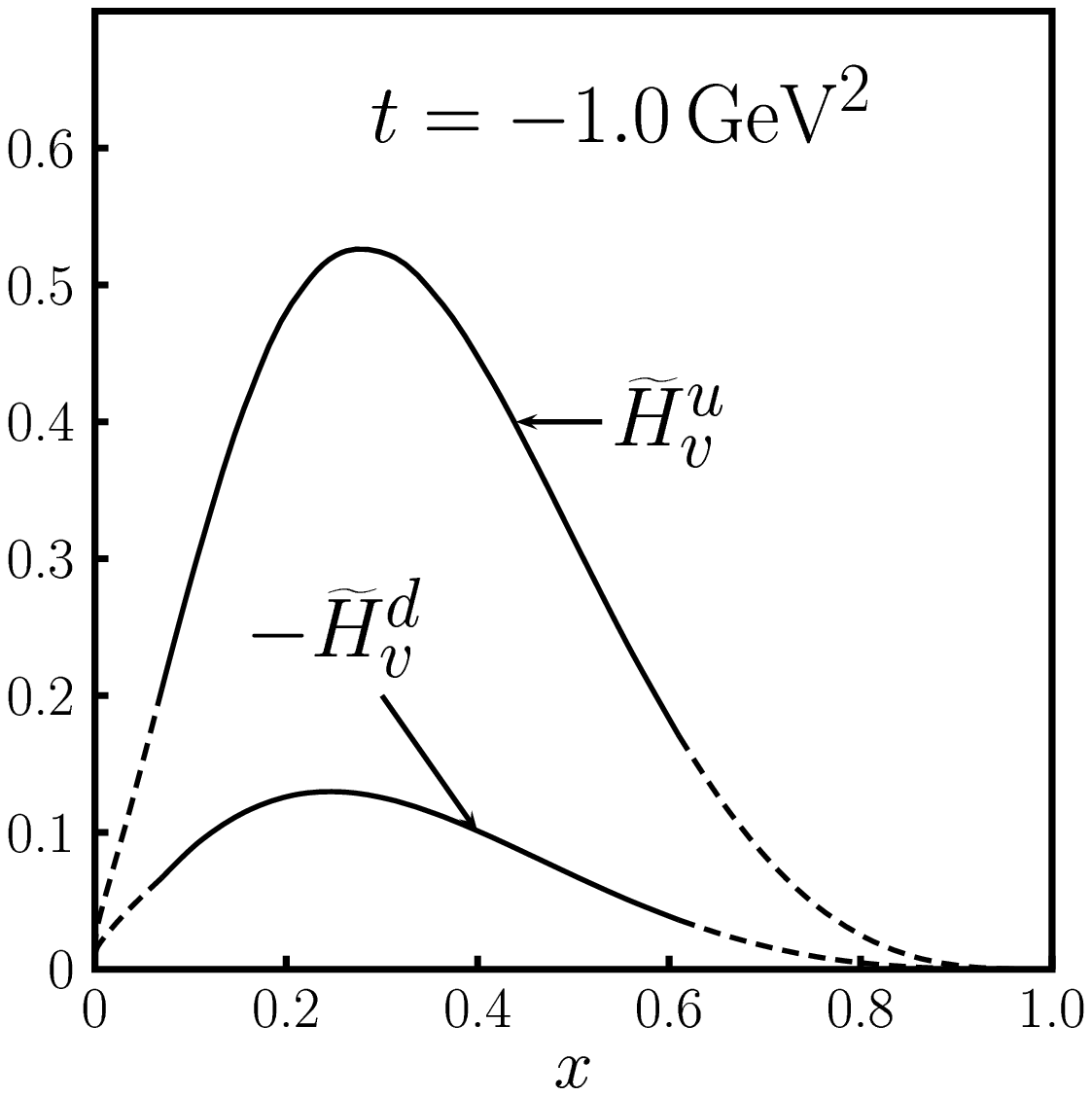} \\[4em]
\includegraphics[width=.35\textwidth, height=.35\textwidth,
 bb =139 403 424 687]{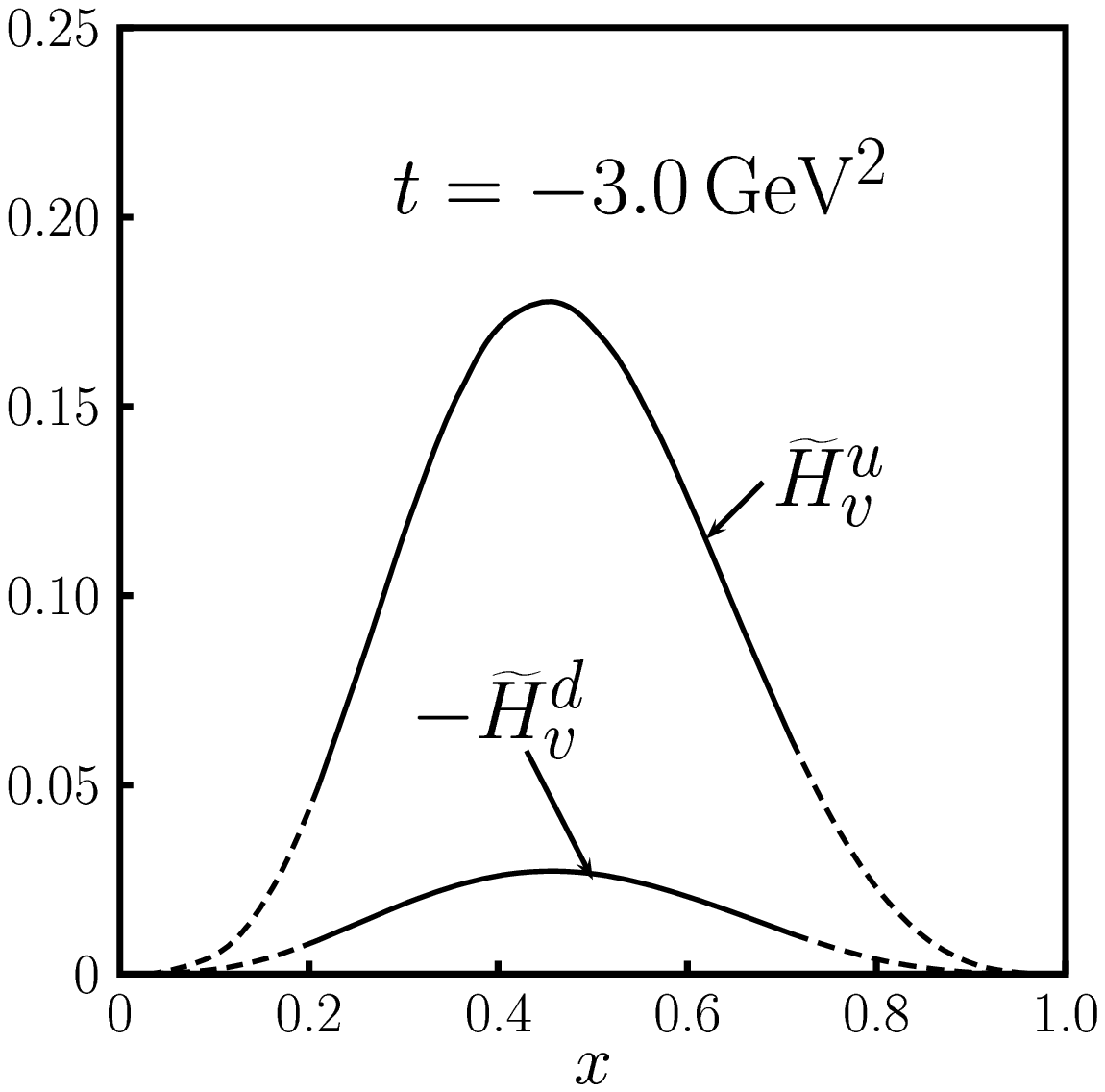}
\hspace{4em}
\includegraphics[width=.35\textwidth, height=.35\textwidth,
 bb =148 409 432 692]{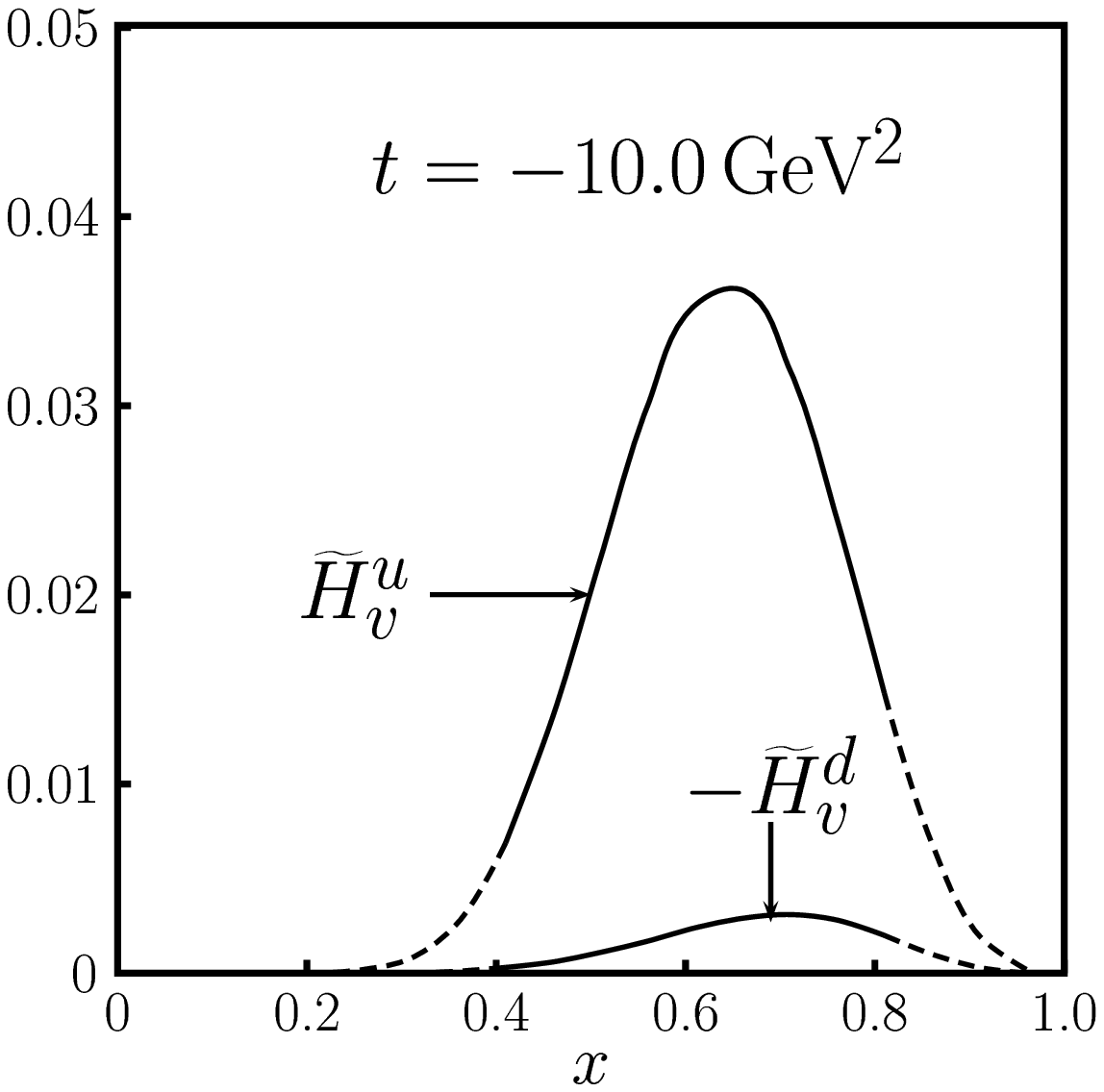}
\\[1em]
\
\end{center}
\caption{\label{fig:Htgpd} The same as Fig.~\protect\ref{fig:Hgpd} for
the polarized valence GPDs $\widetilde{H}_v^q$.}
\end{figure}

\begin{figure}[p]
\begin{center}
\includegraphics[width=.35\textwidth, height=.35\textwidth,
 bb = 158 399 446 687]{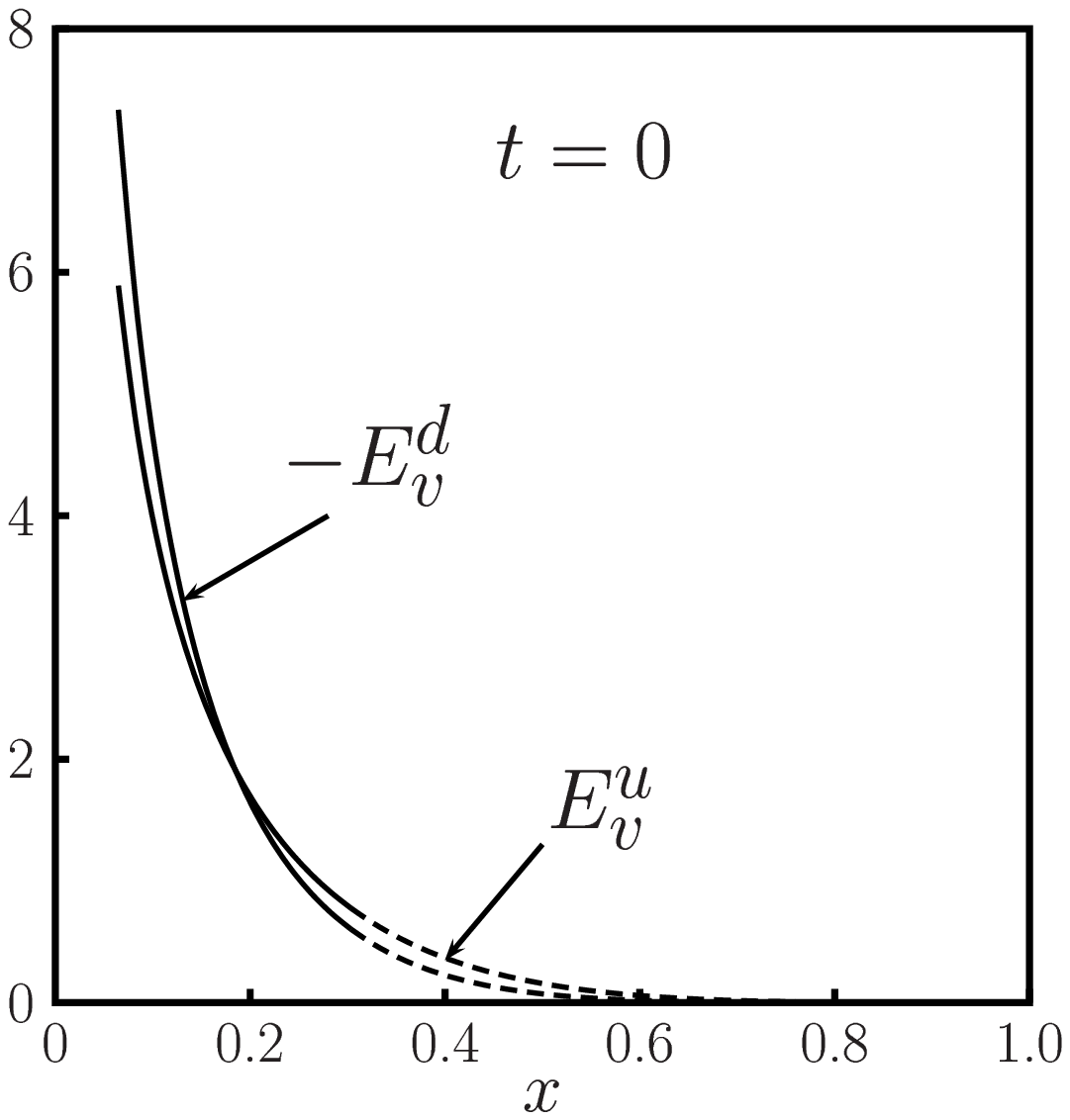} \\[3.5em]
\includegraphics[width=.35\textwidth, height=.35\textwidth,
 bb = 157 378 446 676]{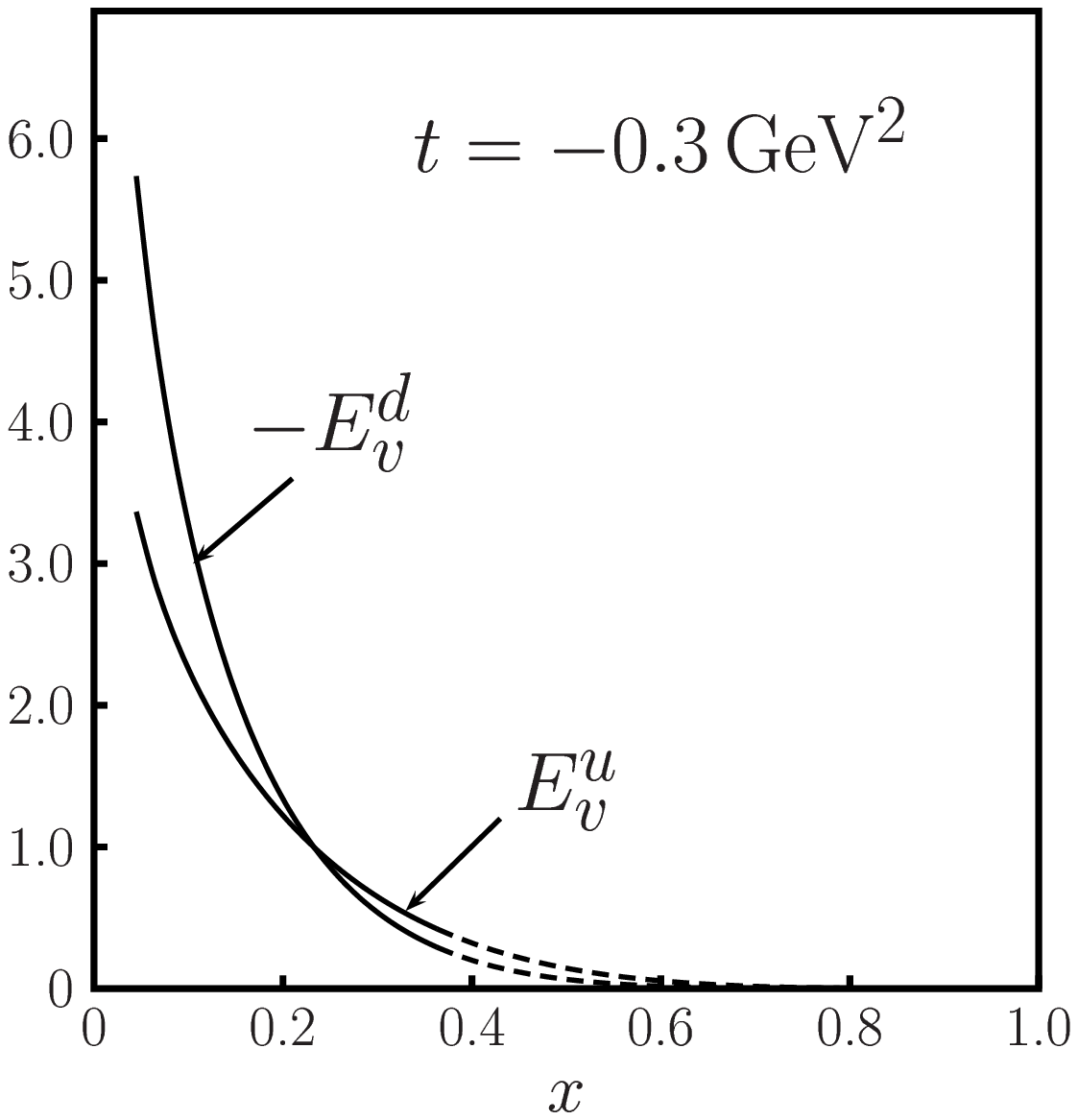}
\hspace{4em}
\includegraphics[width=.35\textwidth, height=.35\textwidth,
 bb = 157 401 446 675]{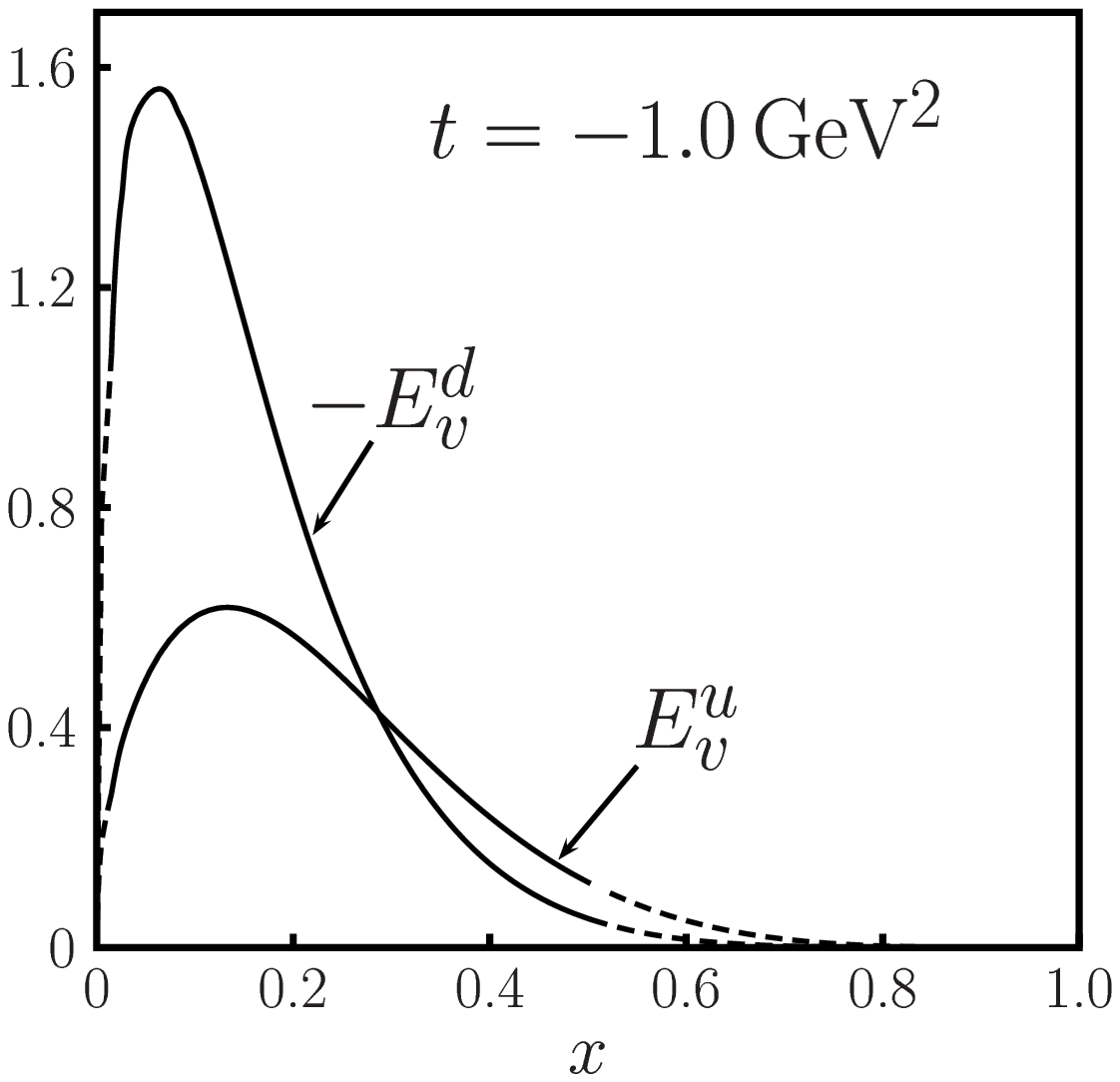} \\[3.5em]
\includegraphics[width=.35\textwidth, height=.35\textwidth,
 bb = 158 379 446 666]{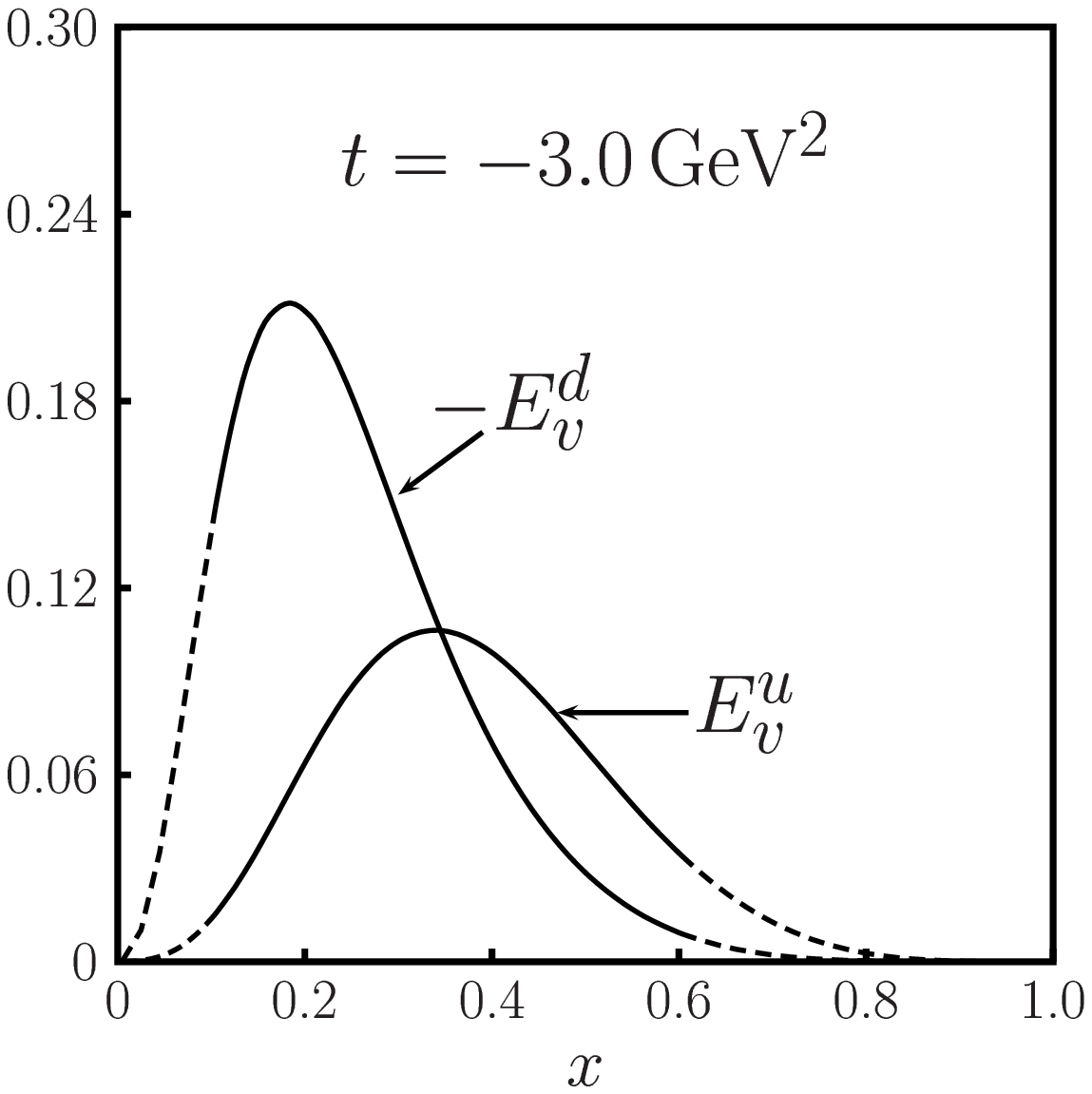}
\hspace{4em}
\includegraphics[width=.35\textwidth, height=.35\textwidth,
 bb = 158 417 446 691]{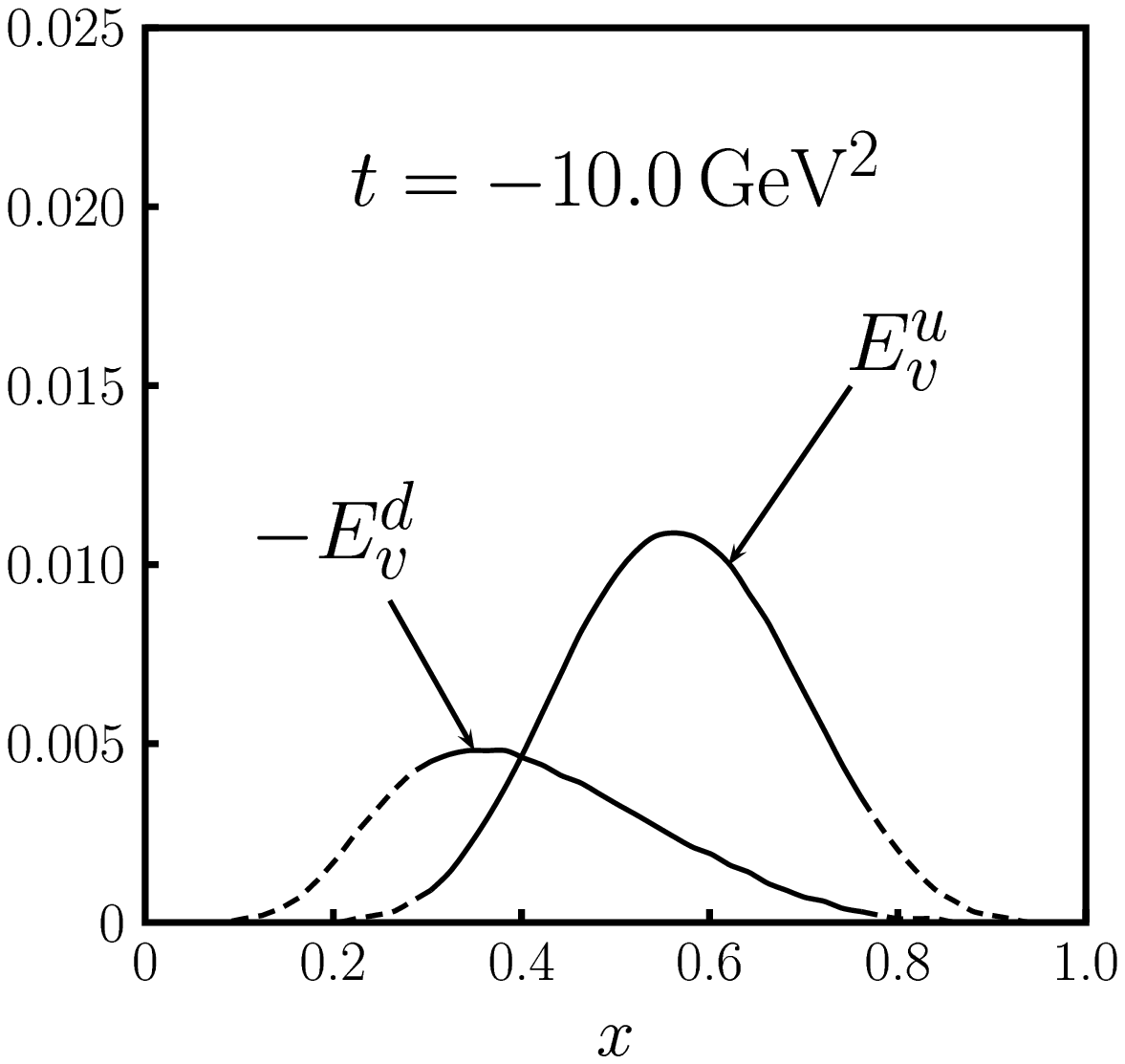}
\\
\
\end{center}
\caption{\label{fig:Egpd} The same as Fig.~\ref{fig:Hgpd} for the
valence GPDs $E_v^q$.  Dashed lines indicate the regions where $x <
x_{\rm min}(t)$ or $x > x_{\rm max}(t)$ according to
Fig.~\protect\ref{fig:F2_xminmax}.}
\end{figure}

The same exercise is repeated for $\widetilde H_v^{u}(x,t)$ and
$\widetilde H_v^{d}(x,t)$ in Fig.~\ref{fig:Htgpd}.  Since we take the
same profile functions for $H_v^q$ and $\widetilde H_v^q$, the
qualitative behavior with increasing $|t|$ is similar for both cases.
Remember that for $\widetilde H_v$ our imprecise knowledge of the
polarized parton densities represents a major source of uncertainty.

In Fig.~\ref{fig:Egpd} we show the corresponding plots for the
helicity-flip distributions $E_v^u(x,t)$ and $E_v^d(x,t)$ obtained in
our default fit to the Pauli form factors $F_2^{p}$ and $F_2^{n}$.
Again we observe a remarkably different behavior for $u$ and $d$
quarks.  At $t=0$ the distributions $e_v^u(x)$ and $-e_v^d(x)$ are not
too different, but for intermediate $|t|$ the distribution for $d$
quark develops a maximum which is more pronounced and located at
significantly smaller $x$ than in the $u$ quark distribution.  At
larger values of $|t|$ we observe again a faster decrease of the $d$
quark distribution.  This is because our default fits have equal
values for the parameters $A_q$ and $C_q$, which respectively govern
the large-$x$ behavior of the profile functions in $H_v^q$ and
$E_v^q$.


\subsection{Moments of GPDs}
\label{sec:moments} 

Important quantities obtained from GPDs are their moments in $x$.  For
zero skewness we define moments of valence GPDs as
\begin{eqnarray}
  h^q_{n,0}(t) &=& \int_{0}^1 \d x \, x^{n-1} \, H_v^q(x,t) \ , 
\nonumber \\
  \tilde h^q_{n,0}(t) &=& 
     \int_{0}^1 \d x \, x^{n-1} \, \widetilde H_v^q(x,t) \ , 
\nonumber \\
  e^q_{n,0}(t) &=& \int_{0}^1 \d x \, x^{n-1} \, E_v^q(x,t) \ . 
\end{eqnarray}
In terms of the notation for moments used in \cite{Ji:1998pc,die:03}
we have
\begin{eqnarray}
h^q_{n,0}(t) = A^q_{n,0}(t) & , &
e^q_{n,0}(t) = B^q_{n,0}(t)  \hspace{3em} \mbox{for odd $n$} , 
\nonumber \\
\tilde{h}^q_{n,0}(t) = \tilde{A}^q_{n,0}(t) &&
                             \hspace{10.1em} \mbox{for even $n$} .
\end{eqnarray}
For even $n$ the moments $h^q_{n,0}$ and $e^q_{n,0}$ are \emph{not}
form factors of local operators, but can rather be seen as the
valence contributions to the corresponding moments $A^q_{n,0}$ and
$B^q_{n,0}$, which involve both valence and sea quarks.  Likewise,
$\tilde{h}^q_{n,0}$ is the valence contribution to
$\tilde{A}^q_{n,0}$ when $n$ is odd.  Note that for the lowest moments
we just have $h^q_{1,0} = F_1^q$ and $e^q_{1,0} = F_2^q$.

\begin{figure}[p]
\begin{center}
\includegraphics[height=.35\textwidth,
 bb = 101 350 429 649]{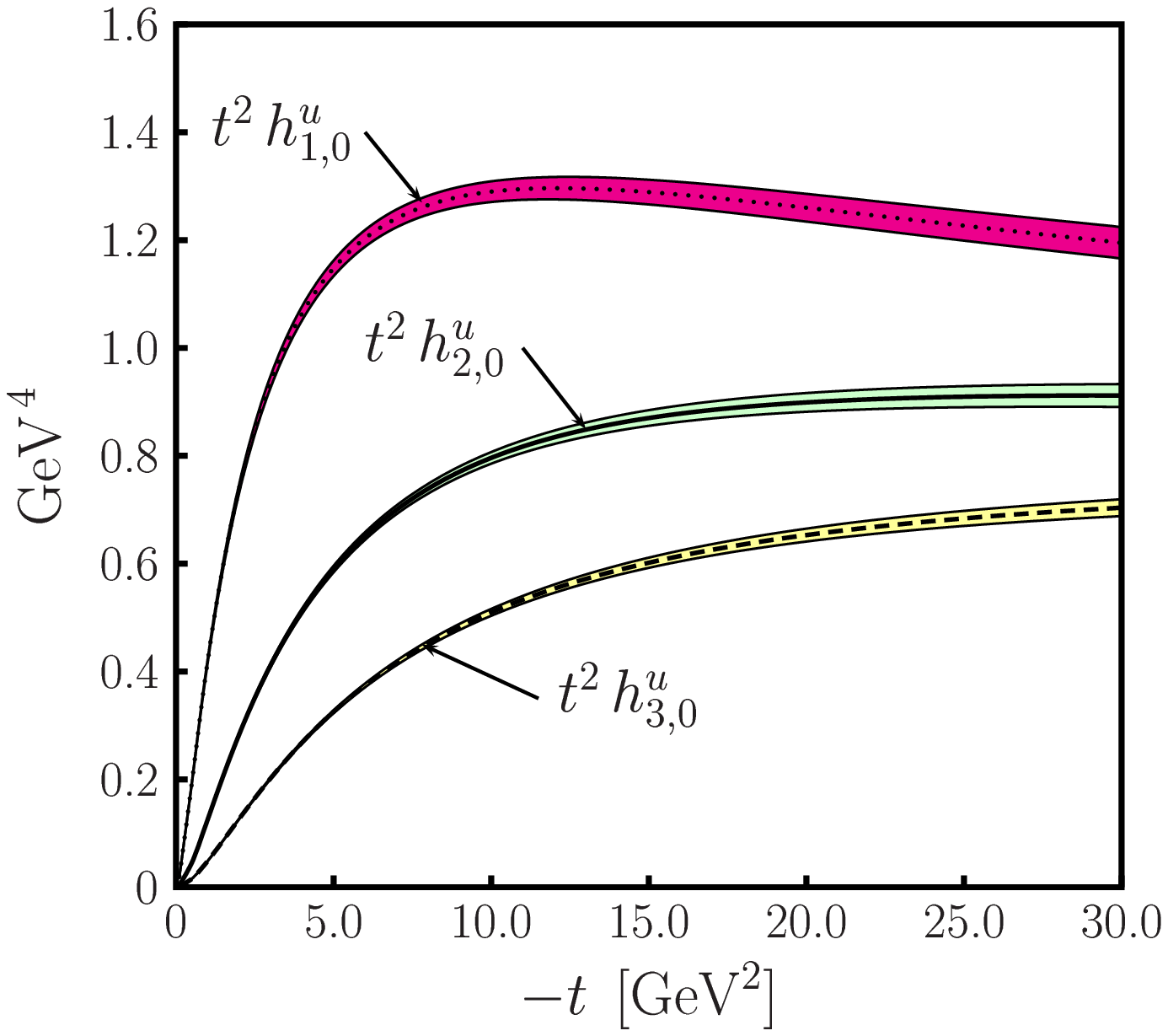}
\hspace{5em} 
\includegraphics[height=.35\textwidth,
 bb = 158 392 452 686]{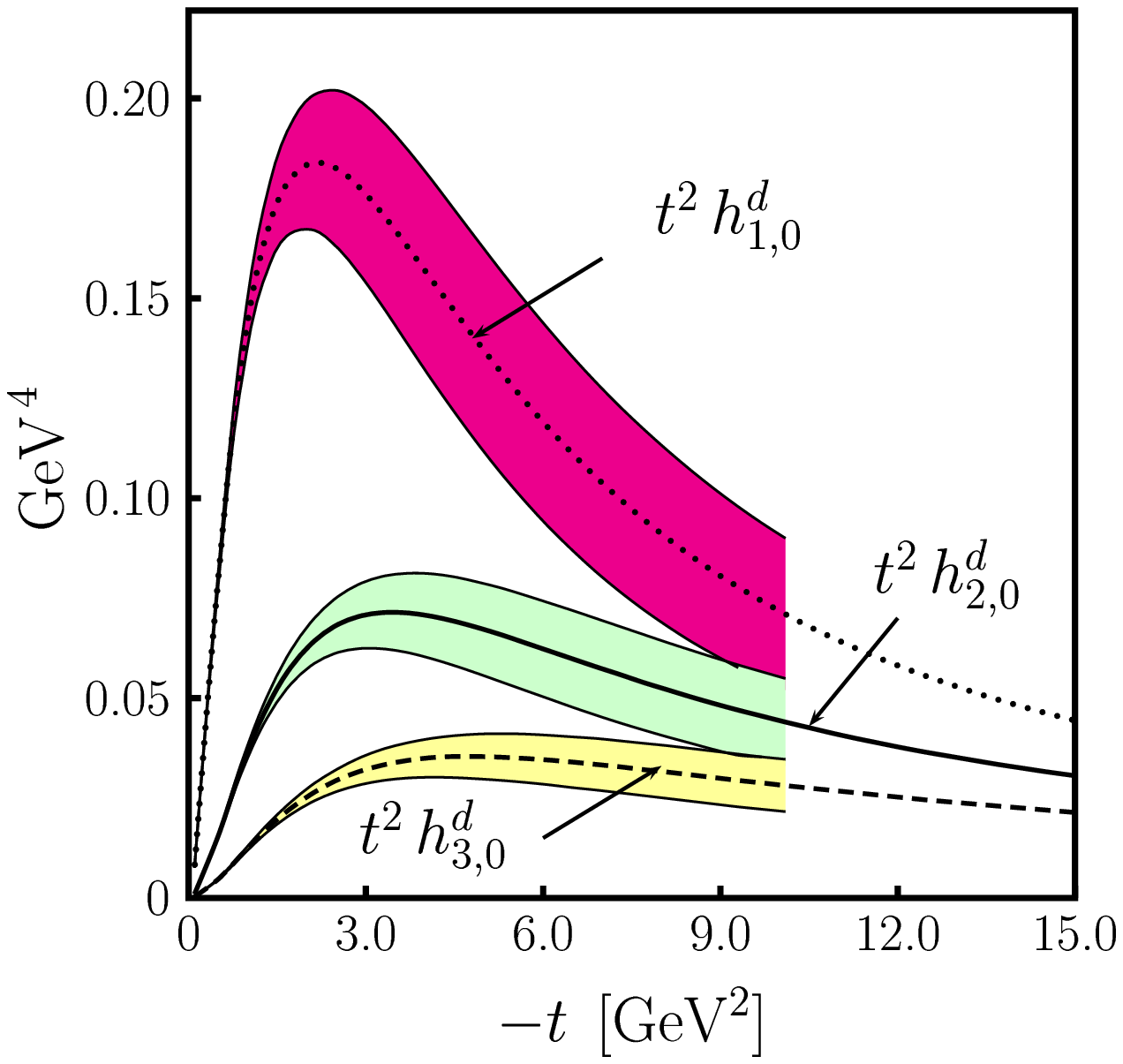}
\\[1em] \ 
\end{center}
\caption{\label{fig:Hmom} The first three moments of the valence GPDs
$H_v^u$ (left) and $H_v^d$ (right), scaled with $t^2$.  The error
bands denote the parametric uncertainty resulting from the fit to the
Dirac form factors $F_1^{p}$ and~$F_1^{n}$.}
\end{figure}

In Fig.~\ref{fig:Hmom} we have plotted the first three moments of
$H_v^u$ and $H_v^d$.  The scaled $u$ quark moments $t^2 h_{n,0}^u(t)$
show a similar behavior as the Dirac form factor of the proton, with
a smooth increase up to values of about $-t =5 \gev^2$ and a rather
flat plateau between $10 \gev^2 \leq -t \leq 30 \gev^2$.  On the other
hand, the lowest $d$ quark moments $t^2 h_{n,0}^d(t)$ have a
pronounced maximum around $-t =3 \gev^2$ and die out much faster for
higher values of $-t$.  This can be understood in terms of the soft
Feynman mechanism, or more precisely of the Drell-Yan relation, which
relates a large-$x$ behavior like $(1-x)^\beta$ of a forward parton
distribution with a large-$t$ behavior like $|t|^{-(1+\beta_q)/2}$ of
the associated form factor, see Sect.~\ref{sec:feynman}.  Since
$d_v(x)$ behaves approximately like $(1-x)^5$ at large $x$, we expect
that $t^2 h_{n,0}^d(t) \sim |t|^{-1}$ at intermediate values of $|t|$,
which is roughly what we observe in Fig.~\ref{fig:Hmom}.

\begin{figure}[p]
\begin{center}
\includegraphics[width=.45\textwidth,
  bb=45 385 400 730]{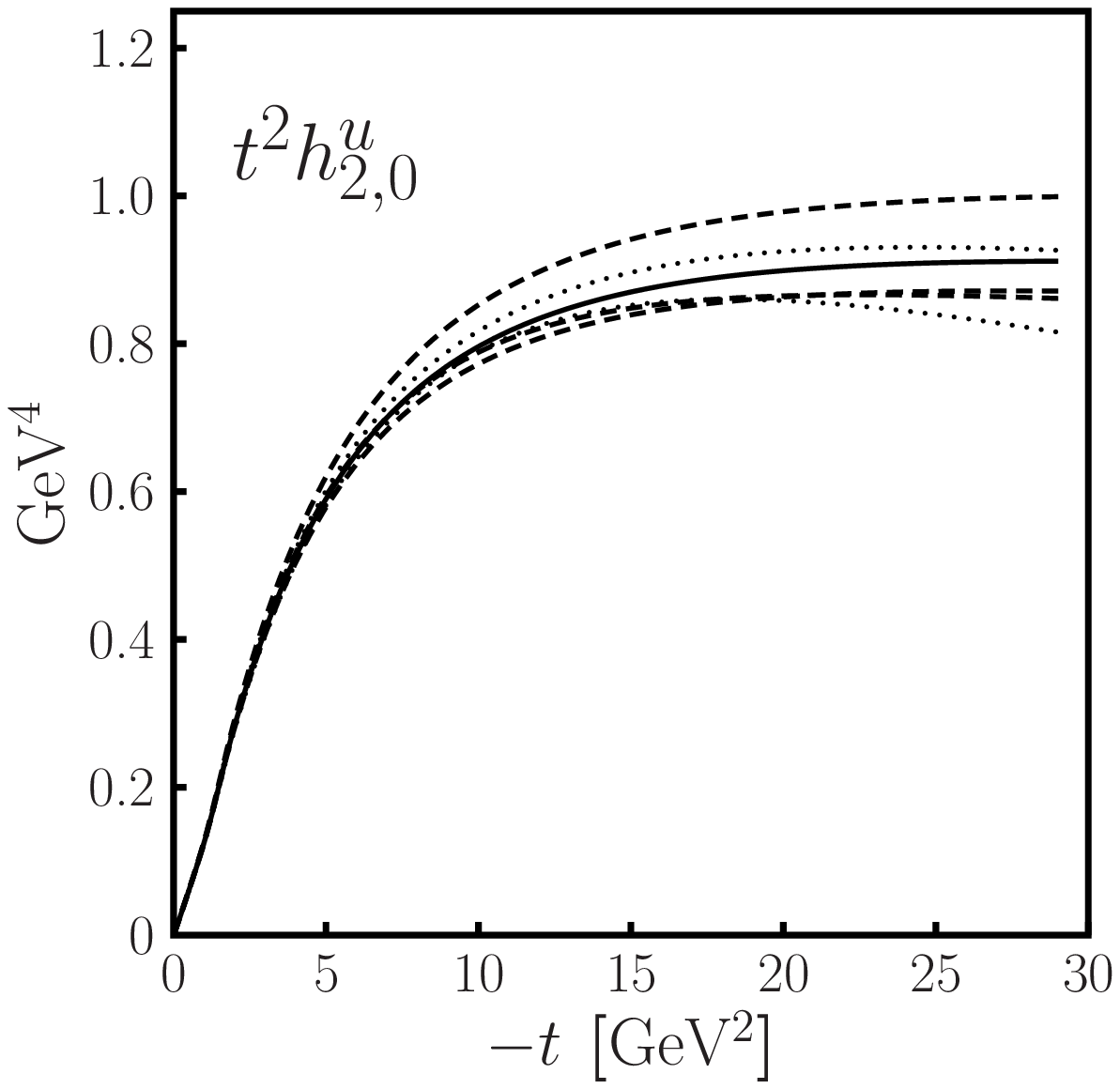}
\hspace{1em}
\includegraphics[width=.45\textwidth,
  bb=45 385 400 730]{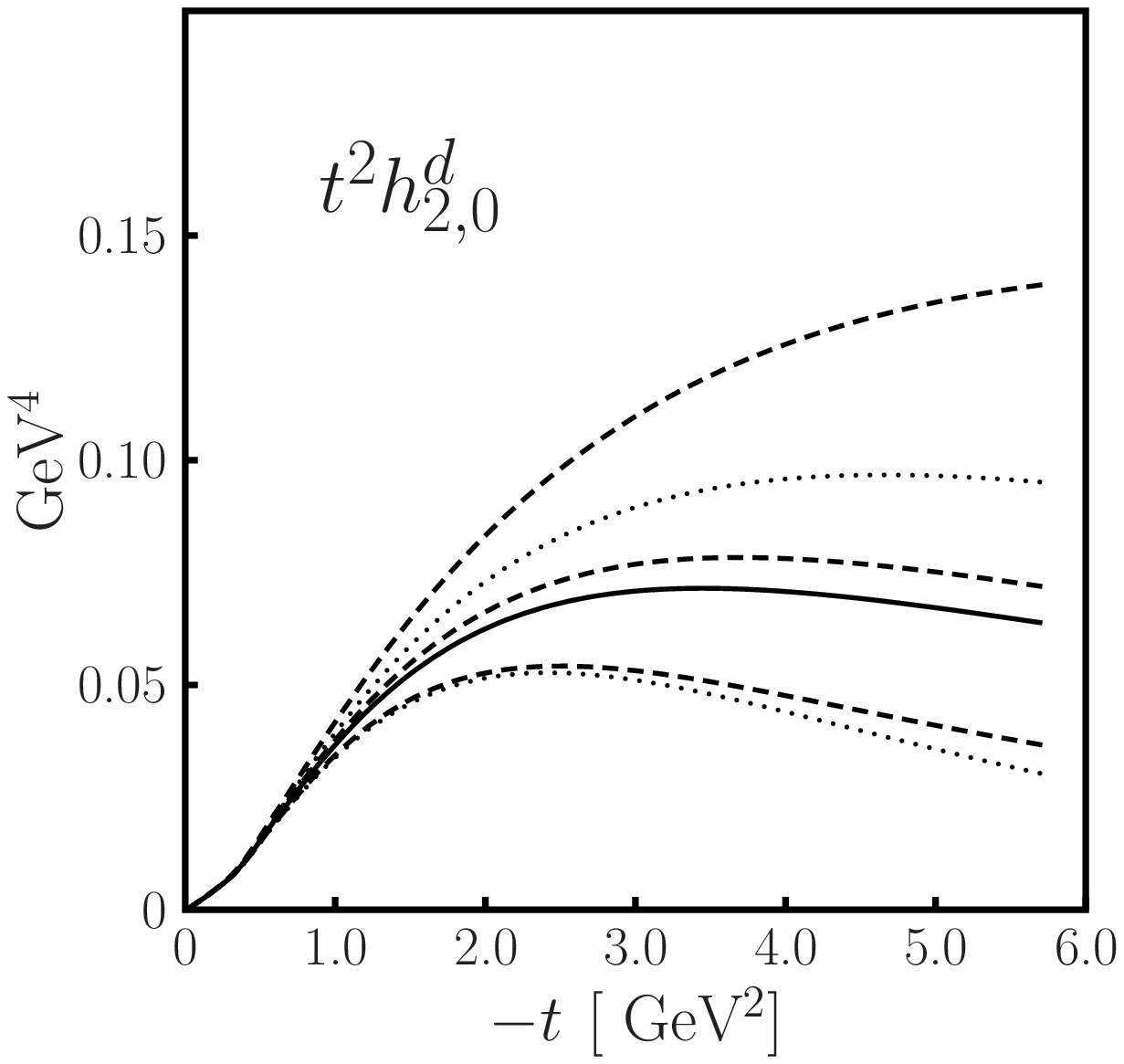}
\end{center}
\caption{\label{fig:Hmom_sys} Systematic uncertainties for $n=2$
moments of valence GPDs $H_v^q$.  The different line styles represent
alternative fits to the Dirac form factors, as specified in the
caption of Fig.~\protect\ref{fig:H-syst}.}
\end{figure}

\begin{figure}[p]
\begin{center}
\includegraphics[height=.35\textwidth, width=.35\textwidth,
 bb = 158 397 448 687]{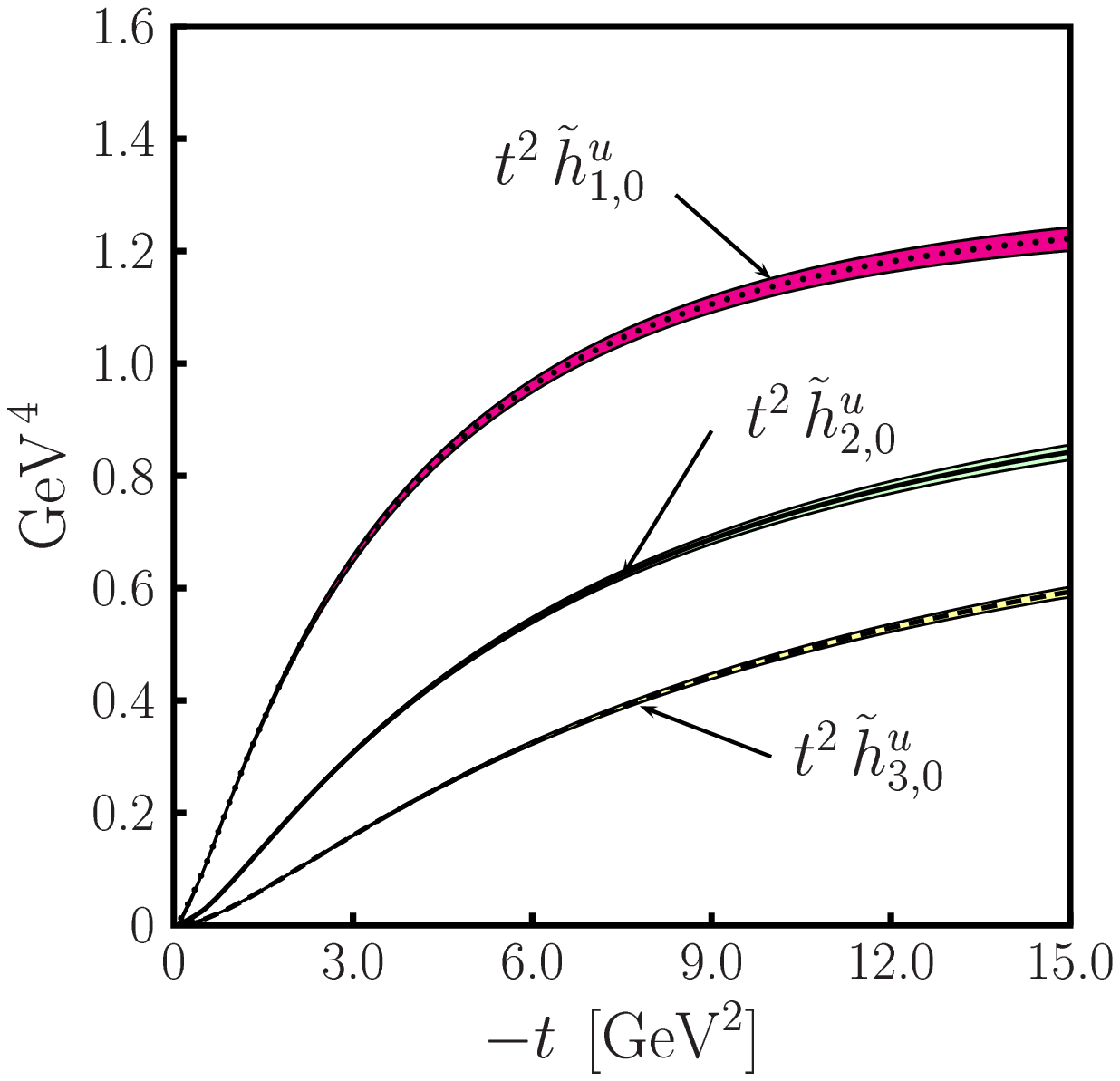}
\hspace{5em}
\includegraphics[height=.35\textwidth, width=.35\textwidth,
 bb = 159 394 440 679]{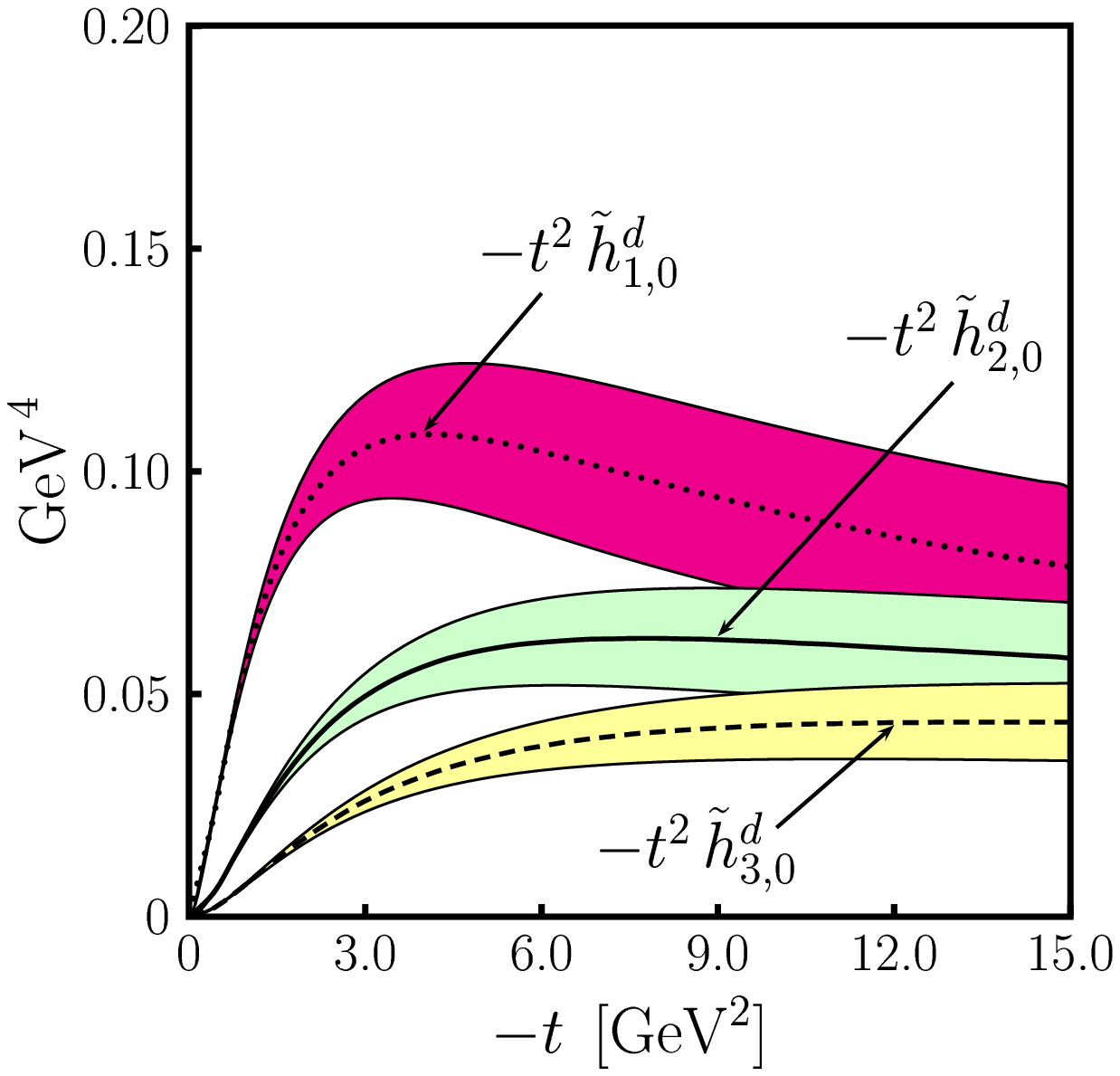}
\\[1em]\
\end{center}
\caption{\label{fig:Htmom} The first three moments of the valence GPDs
$\widetilde H_v^u$ (left) and $\widetilde H_v^d$ (right), scaled with
$t^2$.  The error bands denote the parametric uncertainty of our
default fit.}
\end{figure}

\begin{figure}[p]
\begin{center}
\includegraphics[height=.35\textwidth, width=.35\textwidth,
 bb = 101 435 389 723]{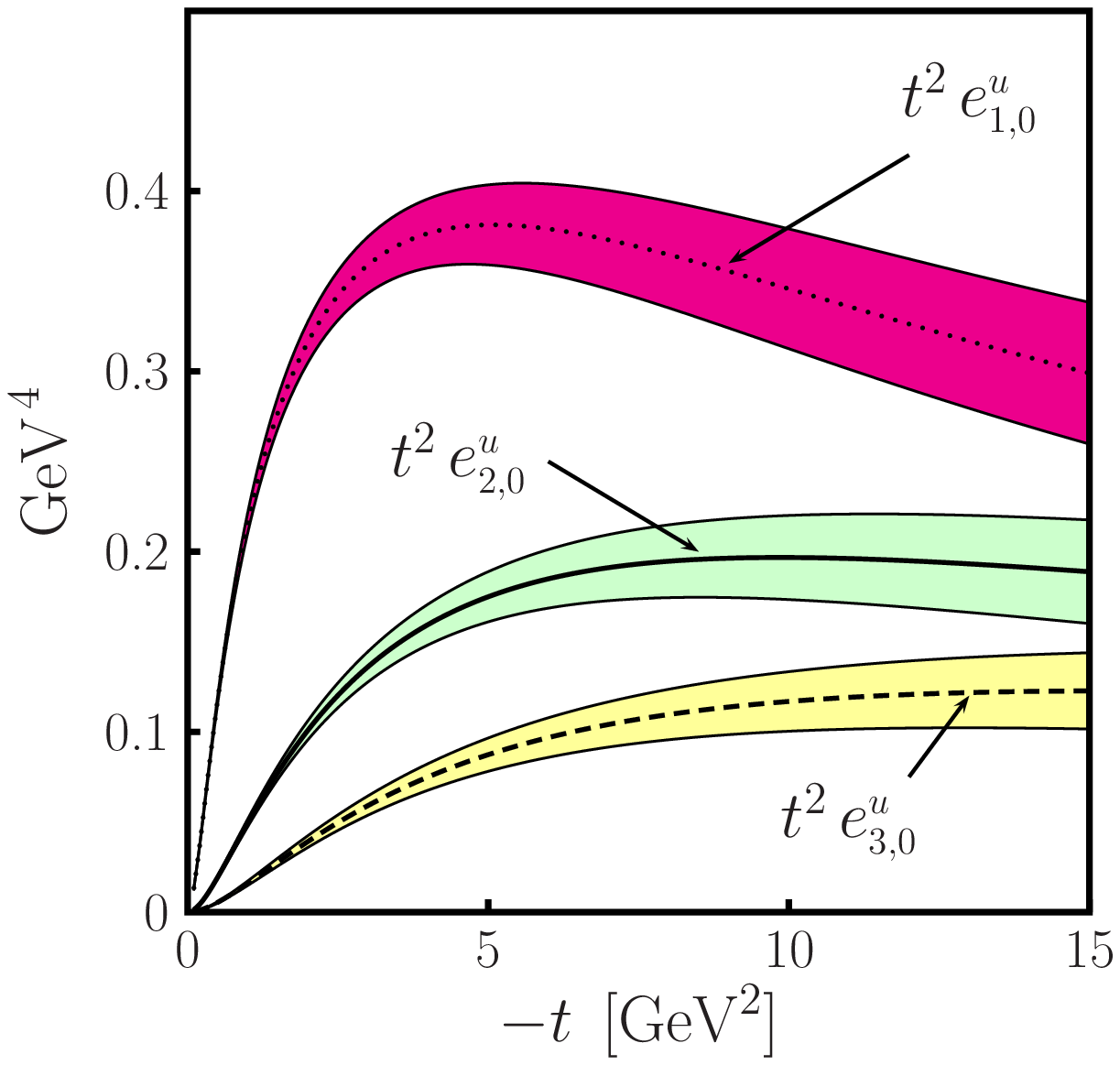} 
\hspace{5em}
\includegraphics[height=.35\textwidth, width=.35\textwidth,
 bb = 101 435 389 723]{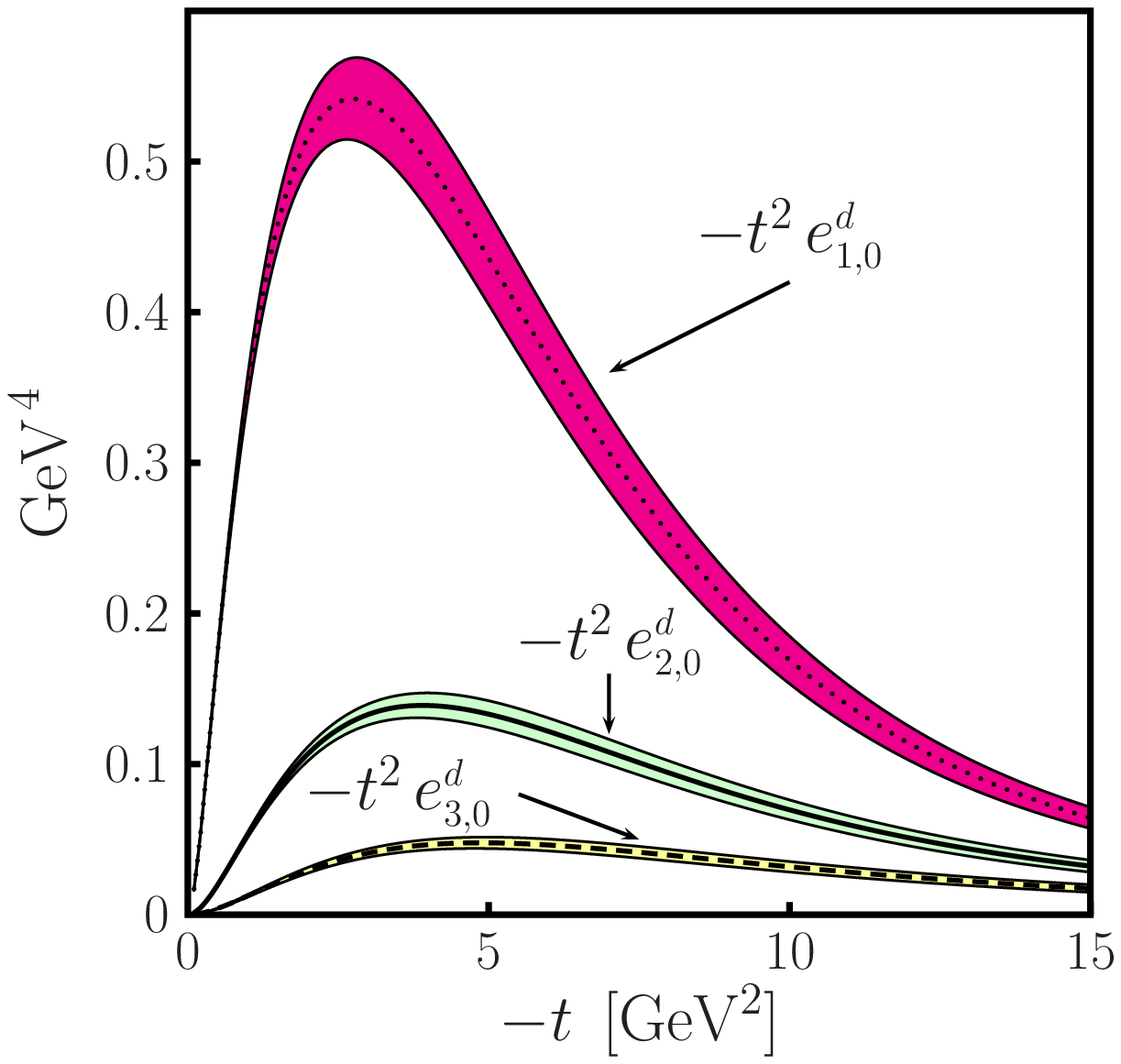}
\\[1em]\
\end{center}
\caption{\label{fig:E-moments} The first three moments of the valence
GPDs $E_v^u$ (left) and $E_v^d$ (right), scaled with $t^2$.  The error
bands denote the parametric uncertainty resulting from our fit to the
Pauli form factors $F_2^{p}$ and~$F_2^{n}$.}
\label{fig:Emom}
\end{figure}

The parametric errors on $h_{n,0}^u$ are rather small, reflecting the
quality of the experimental data on the proton form factor. On the
other hand the uncertainties on $h_{n,0}^d$ become large for $|t|$
above, say, $10~{\rm GeV}^2$, which is due to the lack of data on the
neutron form factor in that region.  In addition, we have to be aware
of the systematic effects related to different choices or constraints
for the profile functions.  In Fig.~\ref{fig:Hmom_sys} we study these
effects by comparing the results for the moments $h_{2,0}^q$ from
different fits to the Dirac form factor data, as we did for the
distance function $d_q(x)$ in Fig.~\ref{fig:H-syst}.  We see that the
systematic error for $u$ quark moments is reasonably small, whereas
for $d$ quarks the uncertainty on individual moments above $|t|=3
\gev^2$ becomes very large.  We should mention that in \emph{ratios}
of moments for the same quark flavor the systematic effects drop out
to some extent.

For $\tilde h_{n,0}^q$ and $e_{n,0}^q$, which are shown in
Figs.~\ref{fig:Htmom} and \ref{fig:Emom}, similar comments concerning
the $t$ dependence apply.  We should point out that the moments
$\tilde{h}_{n,0}$ are subject to additional uncertainties due to the
polarized parton densities $\Delta q_v(x)$, which are not shown for
simplicity.  In the case of $e_{n,0}^q$ the rather large systematic
uncertainties discussed in Section~\ref{sec:ansatz-e} should be kept
in mind.  We remark that for our default fit to the Pauli form
factors, the powers governing the large-$x$ behavior of the forward
distributions $e_v^q(x)$ are $\beta_u \simeq 4$ and $\beta_d \simeq
5.6$.  Via the Drell-Yan relation this translates into a power
behavior $e_{n,0}^u \sim |t|^{-2.5}$ and $e_{n,0}^d \sim |t|^{-3.3}$
in the region where the soft Feynman mechanism applies.  As we can see
from Fig.~\ref{fig:Emom}, this behavior sets in for values of $|t|$
above $5$~GeV$^2$, which is at the border of the region presently
covered by experiment. Measurements of $F_2^{p}$ and $F_2^{n}$ up to
momentum transfers of, say, $10$~GeV$^2$ could rather directly probe
the region of the Feynman mechanism and thus provide valuable
constraints on the exponents $\beta_u$ and $\beta_d$.  The $t$
dependence of the presently available data probes the distributions in
the transition region between small and large $x$.

\begin{figure}[t]
\begin{center}
\includegraphics[width=.45\textwidth,
  bb=115 240 430 590]{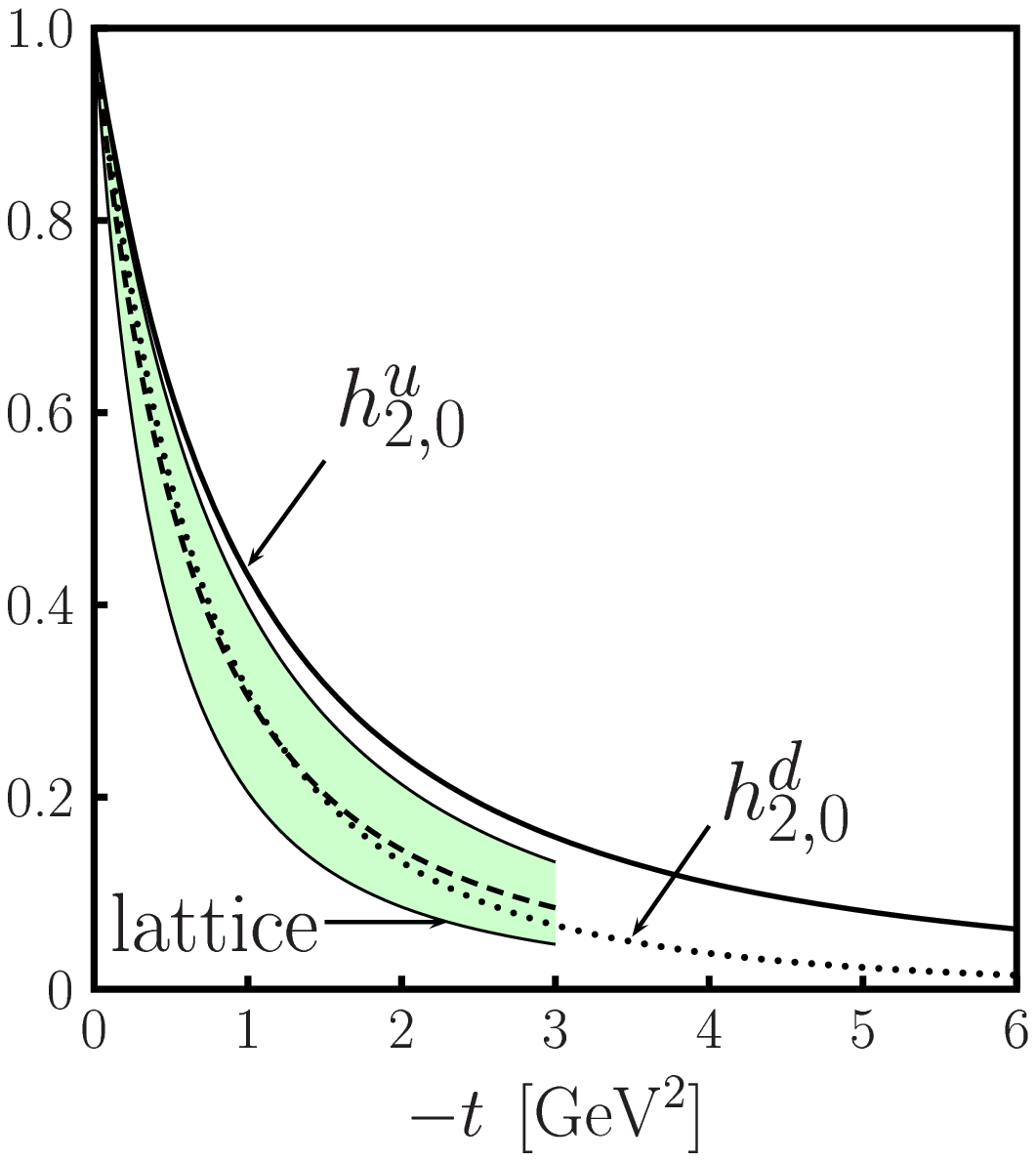}
\hspace{3em}
\includegraphics[width=.45\textwidth,
  bb=115 240 430 590]{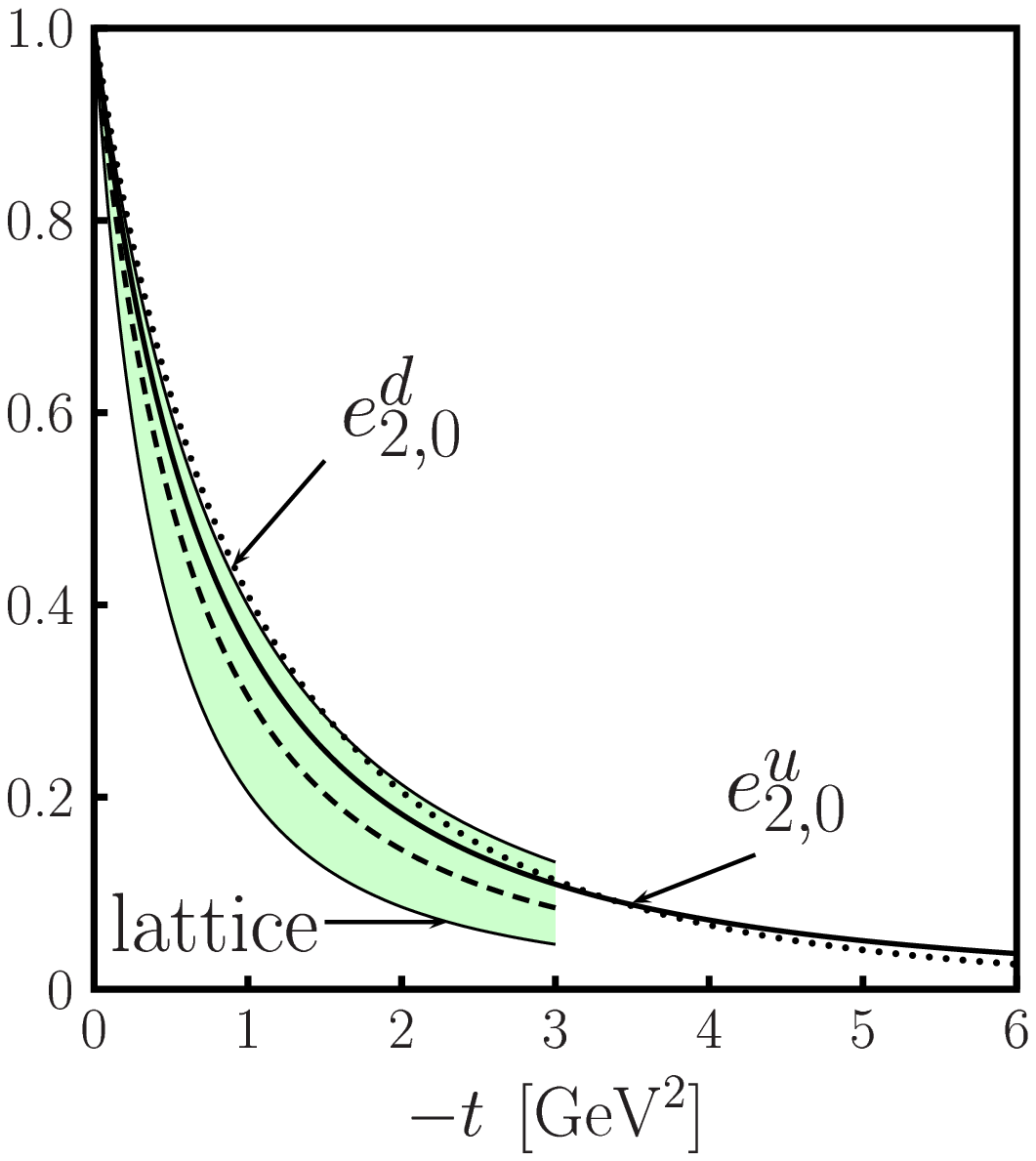}
\end{center}
\caption{\label{fig:mom-latt} Comparison between lattice QCD results
\protect\cite{goe:03} and our default fit for the normalized moments
$h_{2,0}^q(t)/h_{2,0}^q(0)$ and $e_{2,0}^q(t)/e_{2,0}^q(0)$. The
dashed line with the band represents a fit to the lattice data with a
common dipole mass $M=1.11\pm 0.20~{\rm GeV}$ for all $n=2$ moments.}
\end{figure}

The lowest moments of GPDs have been calculated in lattice QCD.  In
particular, the moments $A_{2,0}^q$ and $B_{2,0}^q$ for $u$ and $d$
quarks have been obtained in \cite{goe:03} for $-t$ up to $3 \gev^2$.
The result has been parameterized in terms of a simple dipole fit with
a \emph{common} dipole mass $M=1.11 \pm 0.20 \gev$ for all $n=2$
moments.  One should keep in mind that the value of $M$ has been
obtained by linear extrapolation to the chiral limit of results
obtained with rather high pion (or equivalently quark) masses.
Furthermore, we have evaluated the valence quark contributions to
$A_{2,0}^q$ and $B_{2,0}^q$, whereas the lattice calculation includes
sea quark effects to the extent that they are included in connected
diagrams and in the quenched approximation of QCD.  In
Fig.~\ref{fig:mom-latt} we compare our result for the normalized
moments $h_{2,0}^q(t)/h_{2,0}^q(0)$ and $e_{2,0}^q(t)/e_{2,0}^q(0)$
with 
the dipole fit to the lattice results.  Within their uncertainties the
results are in remarkable agreement.  As already pointed out, our
analysis leads to quite different behavior of $u$ and $d$ quarks.  In
particular, $h_{2,0}^d$ falls of significantly faster than
$h_{2,0}^u$.  It will be interesting to see whether this effect is
also observed in the improved lattice calculations currently under
way.

\subsection{Valence contribution to Ji's sum rule}
\label{sec:ji}

Our determination of the GPDs $E_v^q(x,t)$ from the fit to the Pauli
form factors of proton and neutron has enabled us to estimate the
forward distributions $e_v^q(x)$, which are not accessible in
inclusive processes but play an essential role in understanding how
the total spin of the nucleon is made up from quarks and gluons.  The
forward distributions $e_v^q(x)$ enter the angular momentum sum rule
\cite{ji:96} in the form
\begin{equation}
  \label{valence-ji}
  2 \langle L_v^q \rangle = \int_0^1 \d x \Big[
    x e_v^q(x) + x q_v(x) - \Delta q_v(x) \Big] \, ,
\end{equation}
where we have only given the contribution from valence quarks.  The
second moments of unpolarized parton densities $q_v(x)$ are
well-determined from DIS and other inclusive processes.  The first
moments of polarized parton densities can be obtained from the
axial-vector couplings of nucleons and hyperons under the assumption
that the couplings satisfy flavor SU(3) symmetry and that the
polarized quark sea is also SU(3) symmetric.  For a recent analysis we
refer to \cite{Goto:1999by}, whose results were also taken as an input
for the polarized parton distributions of \cite{blum:03} used in this
work.  The contribution to (\ref{valence-ji}) from the second moments
of $e_v^q(x)$ can be computed from our default fit.  For this
particular quantity the systematic uncertainties are dominated by our
insufficient knowledge on the power $\beta_q$ used in the ansatz
(\ref{good-old-ans}).  In comparison to this parameter, the power
$\alpha$ in (\ref{good-old-ans}) is rather well determined.

\begin{figure}
\begin{center}
\includegraphics[width=.45\textwidth,
  bb=40 385 400 730]{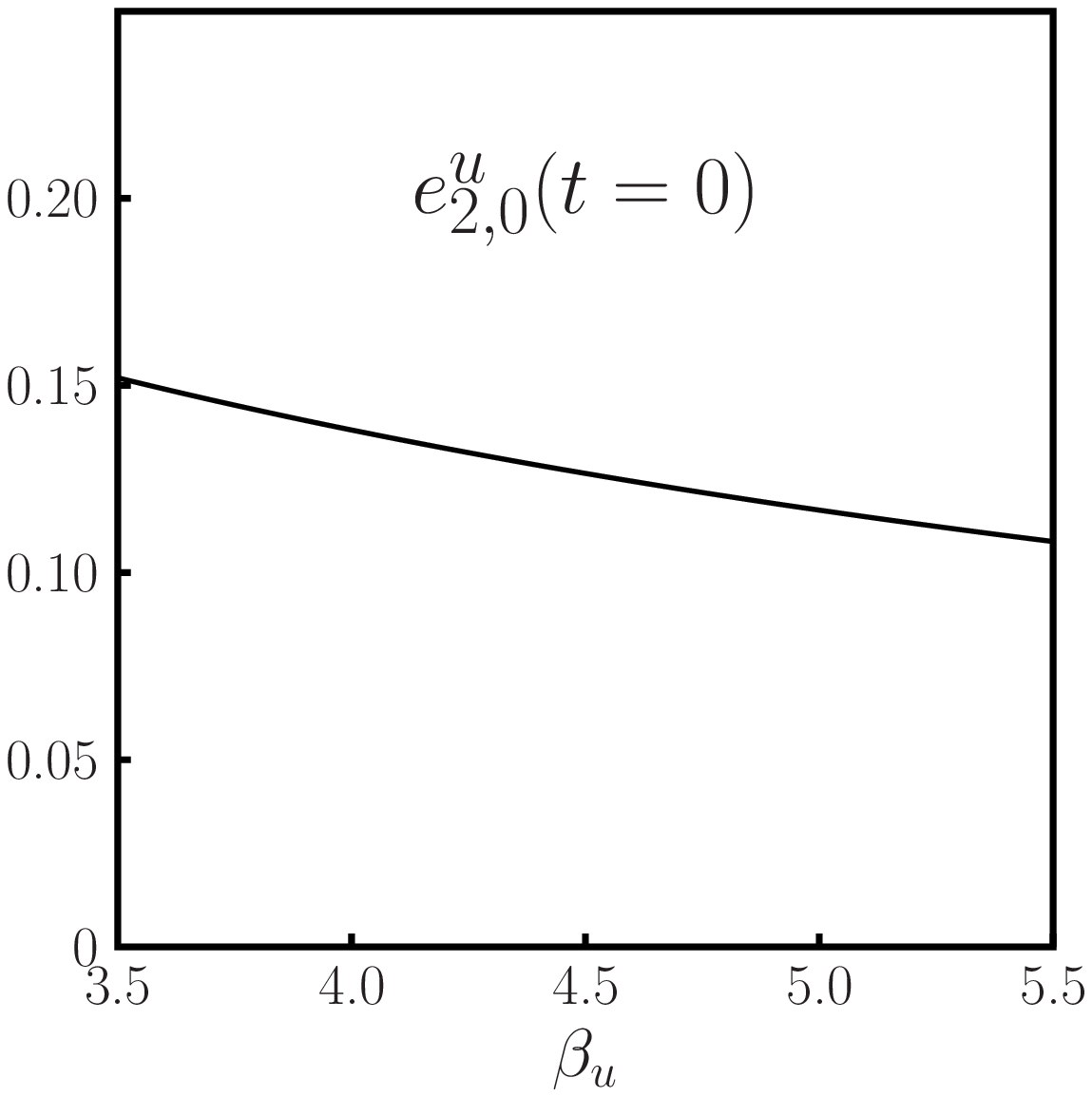}
\hspace{2em}
\includegraphics[width=.45\textwidth,
  bb=40 385 400 730]{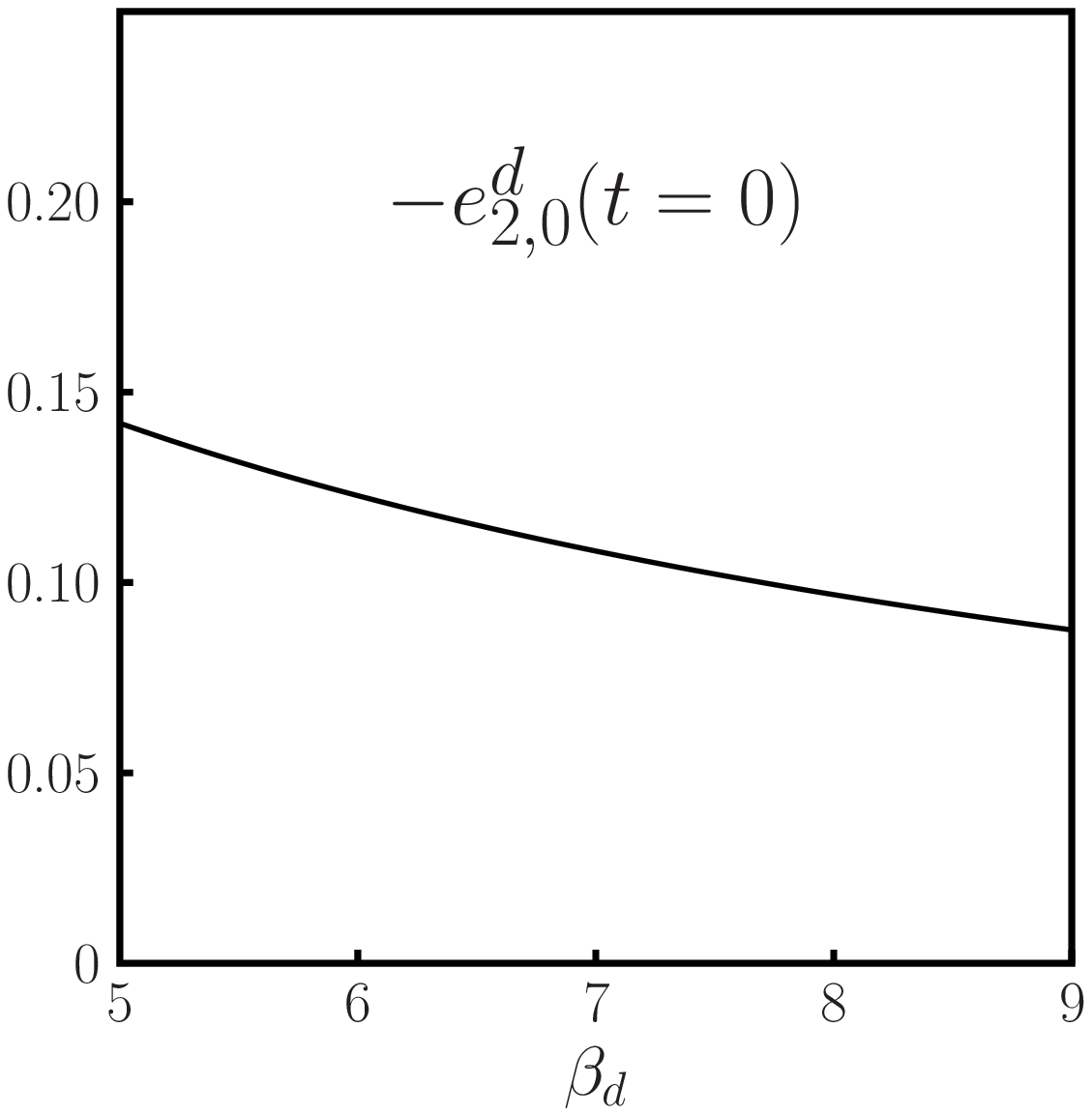}
\end{center}
\caption{\label{fig:xEsys} Variation with $\beta_q$ of the second
moment of $e_v^q(x)$ at $\mu= 2 \gev$.}
\end{figure}

\begin{figure}
\begin{center}
\includegraphics[width=.45\textwidth,
  bb=40 385 400 730]{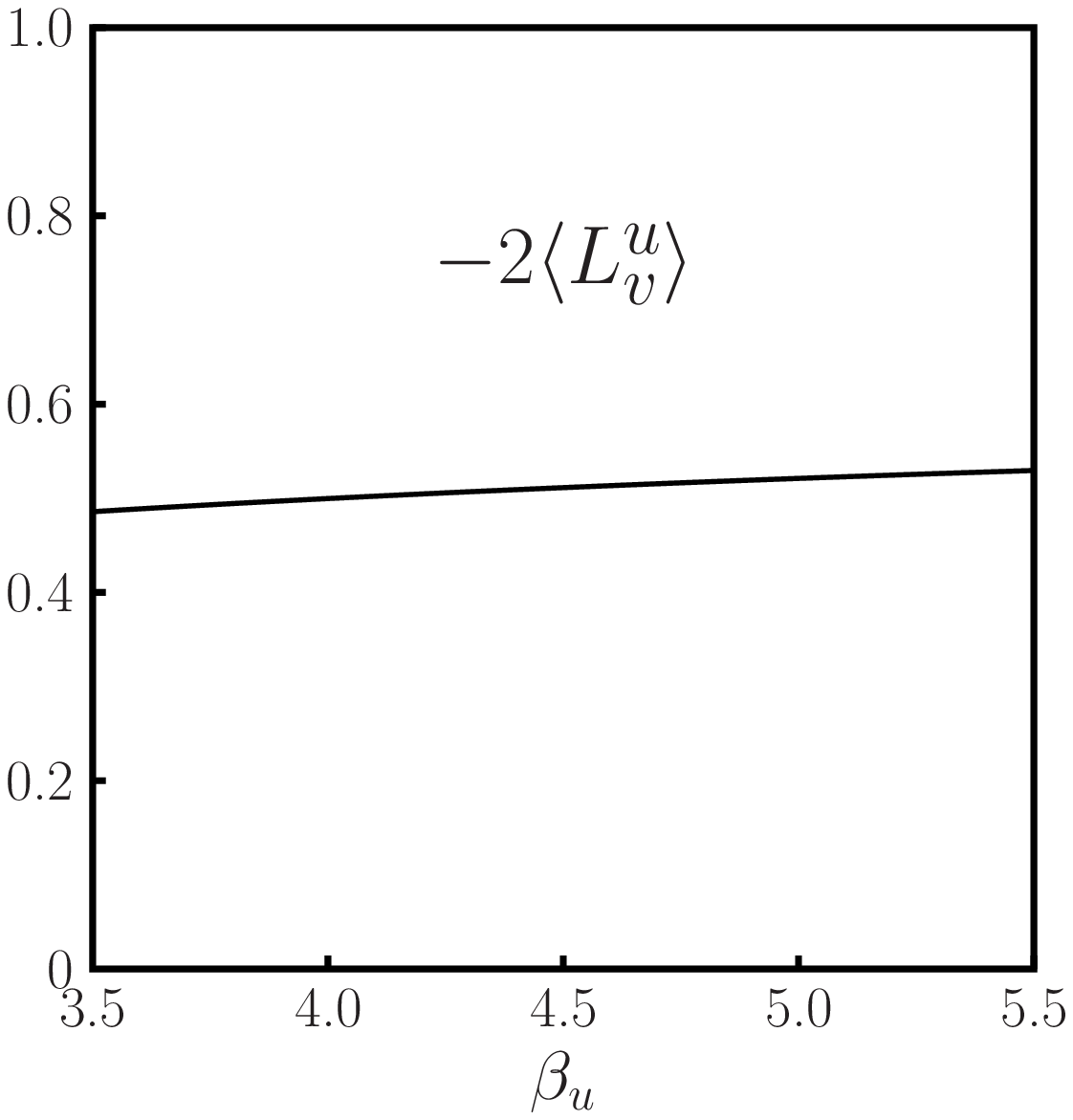}
\hspace{2em}
\includegraphics[width=.45\textwidth,
  bb=40 385 400 730]{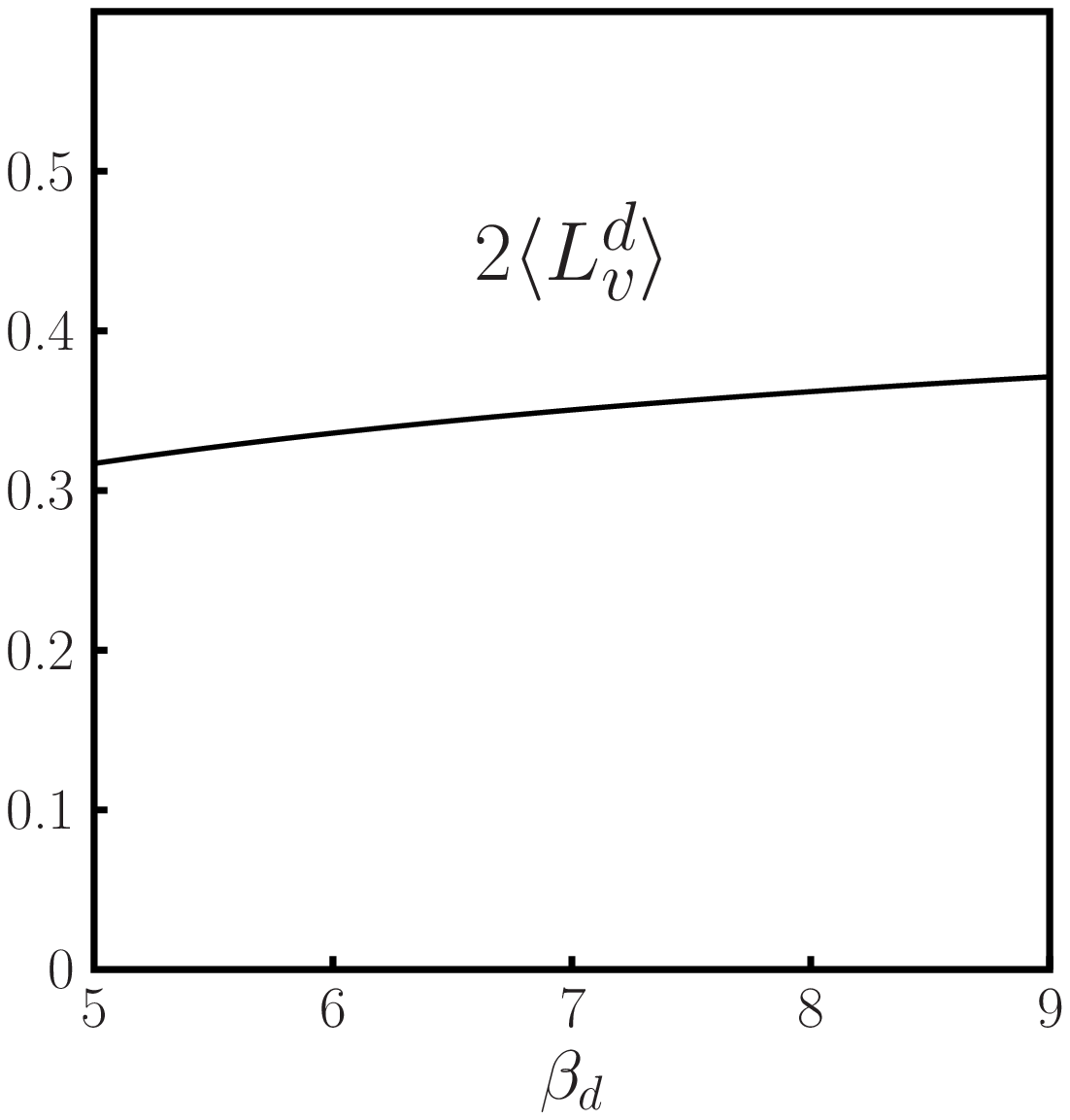}
\end{center}
\caption{\label{fig:Lsys} As in Fig.~\ref{fig:xEsys} but for the
valence contribution to $2 \langle L^q \rangle$ at
$\mu = 2 \gev$.}
\end{figure}

In Fig.~\ref{fig:xEsys} we study the variation of the second moments
$e_{2,0}^q$ at $t=0$ within the wide range of $\beta_q$ values for
which we found consistent fits to the Pauli form factors.  We observe
that the contributions of $u$ and $d$ quarks are almost equal in
magnitude (varying between $0.10$ and $0.15$) and opposite in sign.
The corresponding relative uncertainties on the valence contributions
to the orbital angular momentum are rather small.  This is shown in
Fig.~\ref{fig:Lsys}, where we have used the unpolarized CTEQ6M
distributions at $\mu=2\gev$ and the results of \cite{Goto:1999by} to
evaluate the last two terms in (\ref{valence-ji}).  In the sum
$\langle L_v\rangle = \langle L_v^u\rangle + \langle L_v^d\rangle$ the
contributions from $u$ and $d$ quarks cancel to a large extent, so
that the relative error on this quantity is bigger again.  With the
central values of $\beta_u$ and $\beta_d$ from our default fit we
obtain $2 \langle L_v\rangle \approx - 0.17$ at $\mu=2\gev$.

As we already mentioned, sea quarks do not enter the Pauli form factor
and are thus not accessible in our fit for the GPDs.  For completeness
we note that the sea quark contributions to $2 \langle L^u \rangle$ and
$2 \langle L^d \rangle$ at $\mu=2\gev$ are $2 \int \d x\, (x \bar{u} -
\Delta\bar{u}) \approx -0.09$ and $2 \int \d x\, (x \bar{d} -
\Delta\bar{d}) \approx -0.08$ if we take the distributions from CTEQ6M
\cite{CTEQ} and the NLO distributions in scenario 1 of \cite{blum:03}.
Since we have no information on the sea quark contributions
corresponding to the first term in (\ref{valence-ji}), we refrain from
comparing our results with evaluations of $\langle L^q \rangle$ in
lattice QCD \cite{goe:03,hae:03a,Mathur:1999uf}.

\subsection{Proton tomography}
\label{sec:tomo}

\begin{figure}[p]
\begin{center}
\includegraphics[height=.37\textwidth,
  bb=120 75 370 280,clip=true]{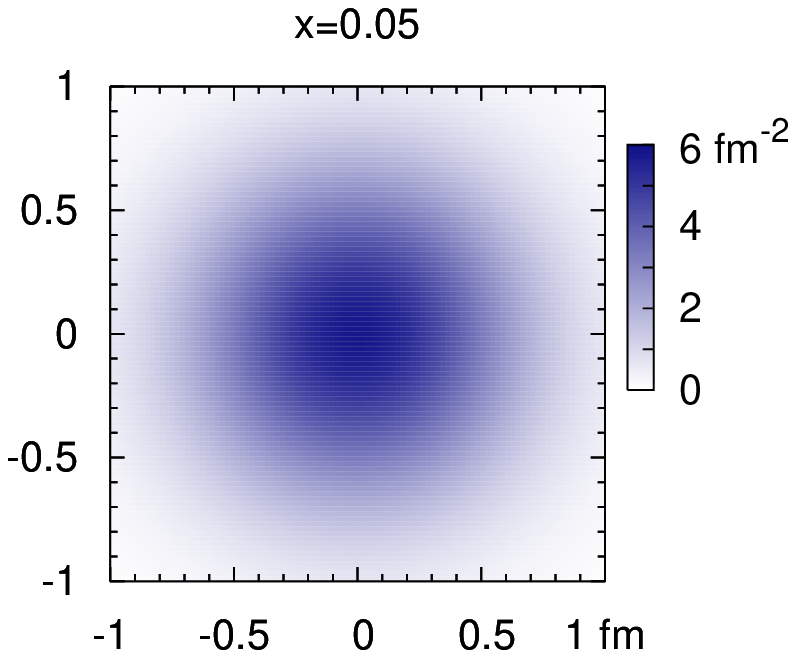}
\includegraphics[height=.37\textwidth,
  bb=120 75 320 280,clip=true]{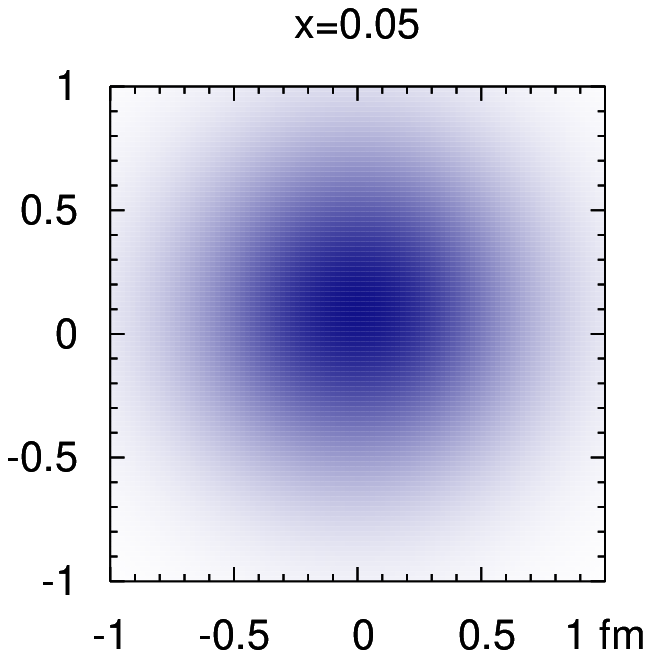}
\\[2em]
\includegraphics[height=.37\textwidth,
  bb=120 75 370 280,clip=true]{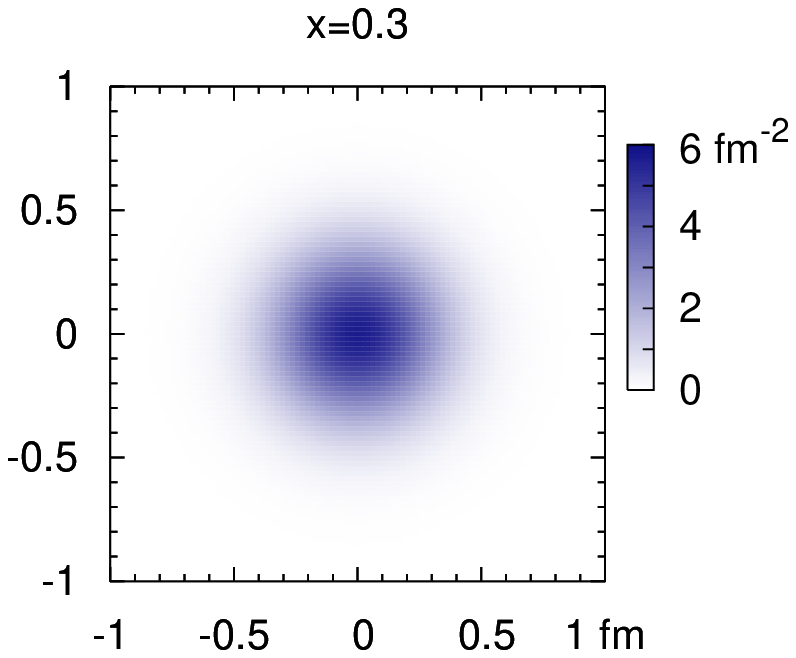}
\includegraphics[height=.37\textwidth,
  bb=120 75 320 280,clip=true]{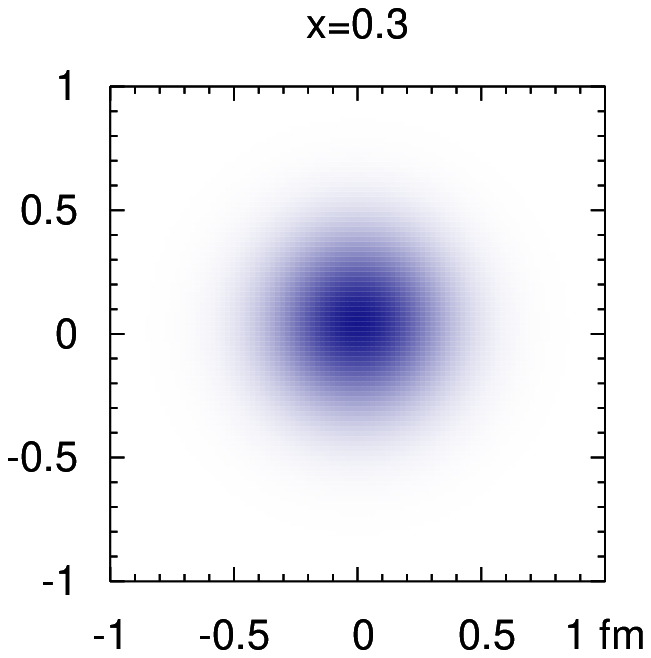}
\\[2em]
\includegraphics[height=.37\textwidth,
  bb=120 75 370 280,clip=true]{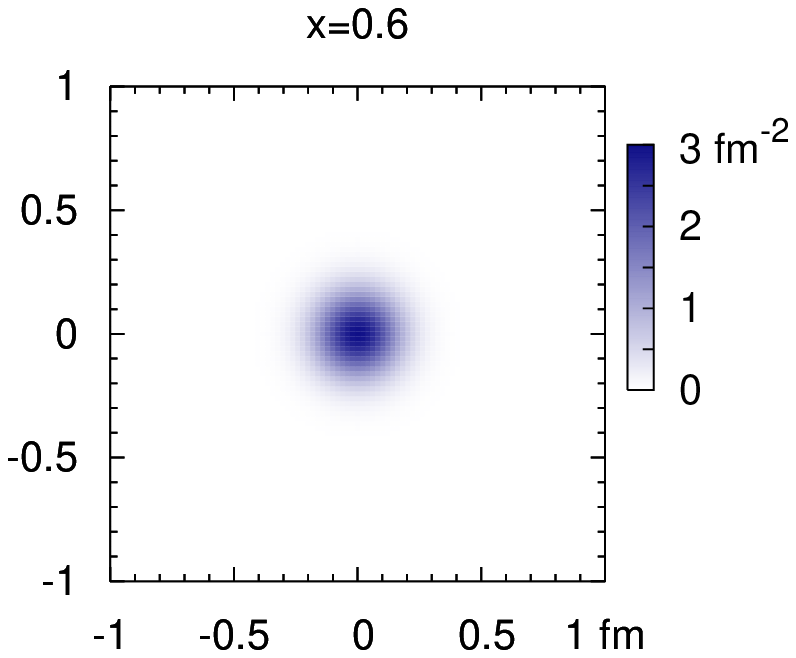}
\includegraphics[height=.37\textwidth,
  bb=120 75 320 280,clip=true]{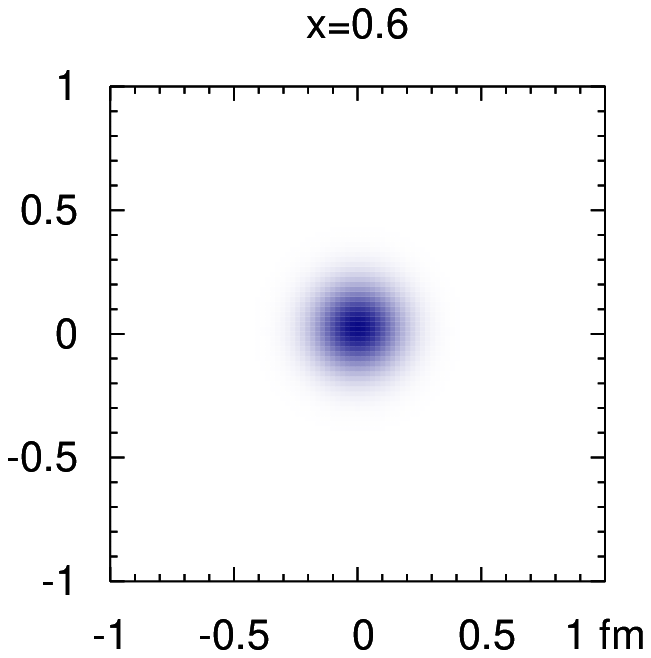}
\end{center}
\caption{\label{fig:tomo_u} Tomography plots of $u_v(x,\tvec{b})$
(left) and $u_v^X(x,\tvec{b})$ (right) in the transverse $b^x$--$b^y$
plane.  Note that the scale of intensity for longitudinal momentum
fraction $x=0.6$ differs from the one for $x=0.3$ and $x=0.05$.}
\end{figure}

\begin{figure}[p]
\begin{center}
\includegraphics[height=.37\textwidth,
  bb=120 75 370 280,clip=true]{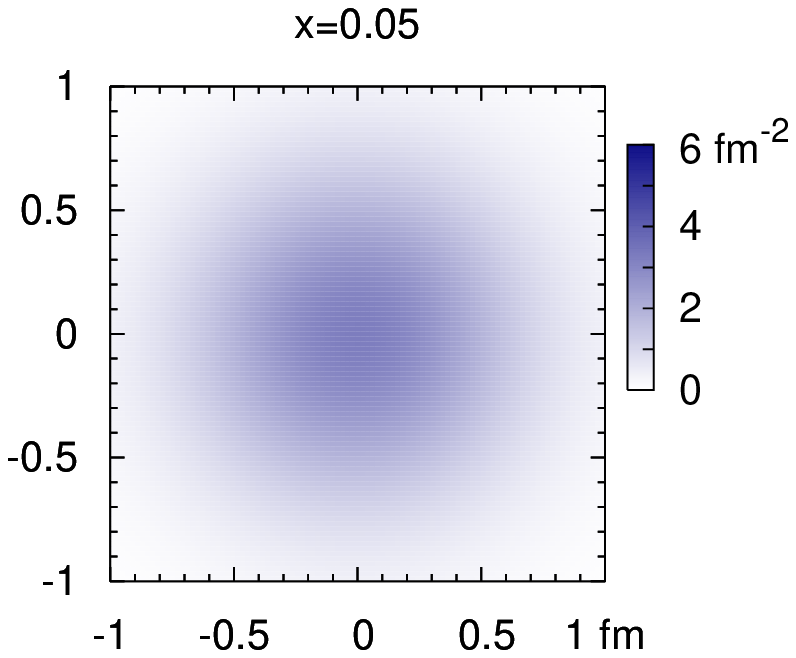}
\includegraphics[height=.37\textwidth,
  bb=120 75 320 280,clip=true]{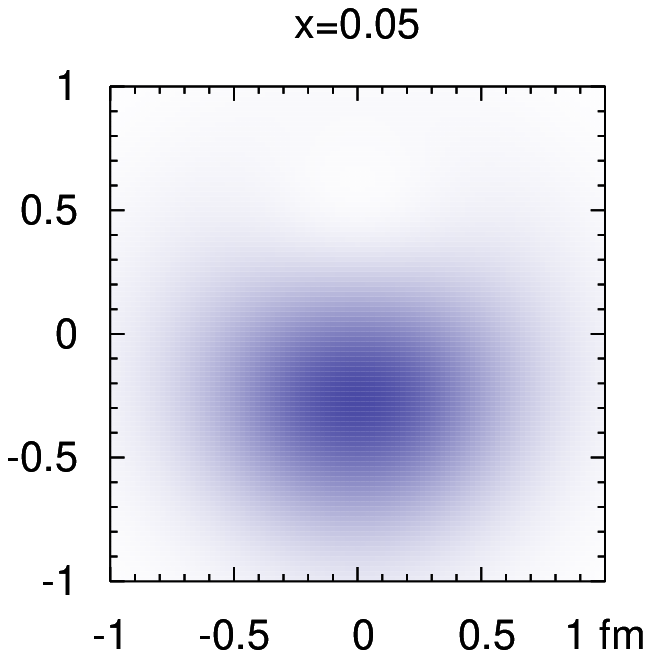}
\\[2em]
\includegraphics[height=.37\textwidth,
  bb=120 75 370 280,clip=true]{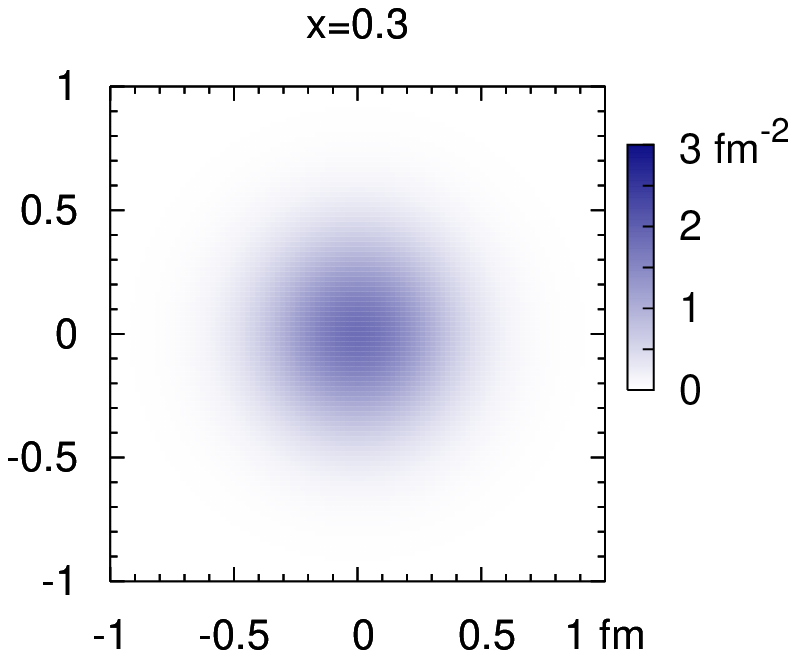}
\includegraphics[height=.37\textwidth,
  bb=120 75 320 280,clip=true]{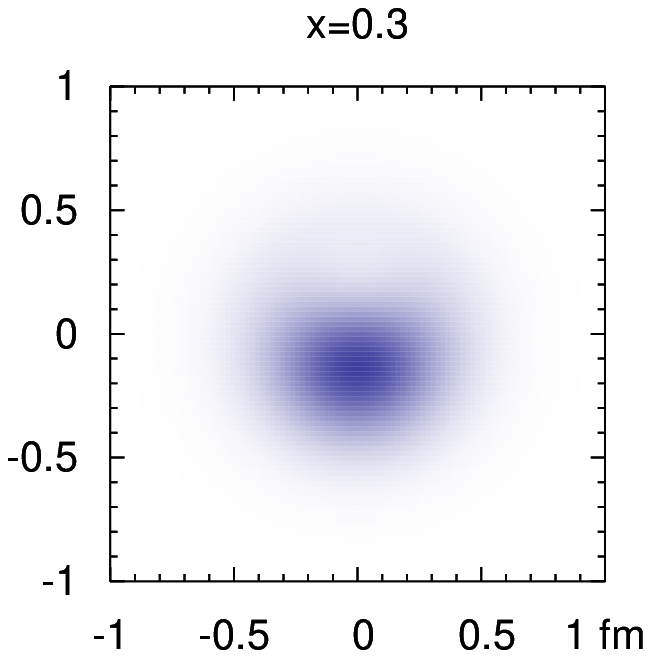}
\\[2em]
\includegraphics[height=.37\textwidth,
  bb=120 75 370 280,clip=true]{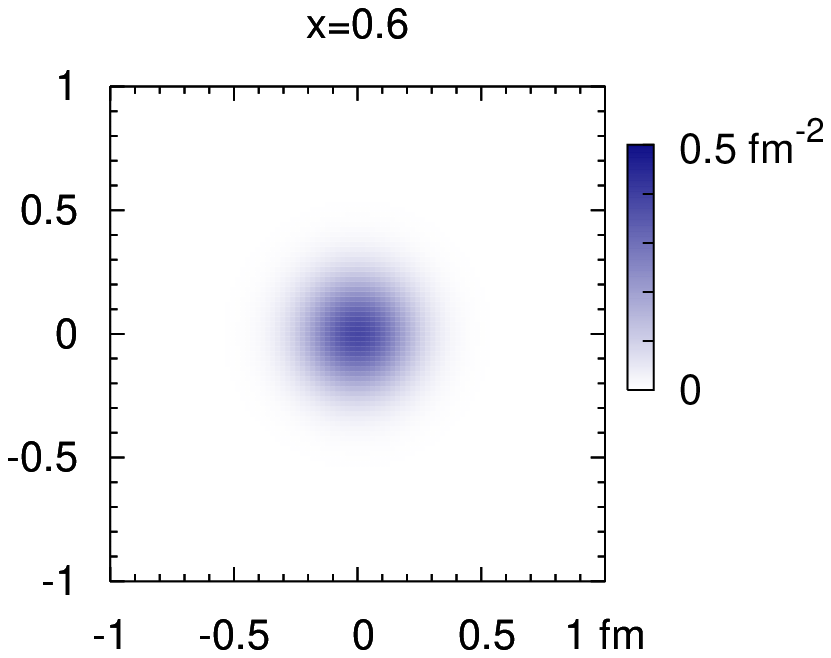}
\includegraphics[height=.37\textwidth,
  bb=120 75 320 280,clip=true]{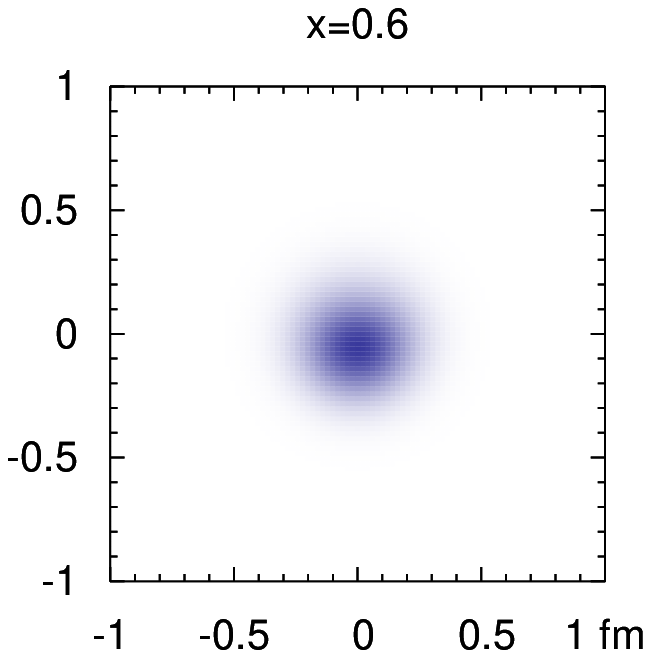}
\end{center}
\caption{\label{fig:tomo_d} Tomography plots of $d_v(x,\tvec{b})$
(left) and $d_v^X(x,\tvec{b})$ (right) in the transverse $b^x$--$b^y$
plane.  Note that the scale of intensity is different for all three
longitudinal momentum fractions $x$.}
\end{figure}

In Figs.~\ref{fig:tomo_u} and \ref{fig:tomo_d} we illustrate our
default results for GPDs as tomography plots in impact parameter space
for fixed longitudinal momentum fraction $x$. We show the unpolarized
density $q_v(x,\tvec{b})$ as well as its analog $q_v^X(x,\tvec{b})$
for a proton polarized in the $x$ direction, see (\ref{impact-gpd})
and (\ref{impact-pol}).  Notice that in general $q_v^X(x,\tvec{b})$
has no rotational symmetry in the impact parameter plane, as is
readily seen from (\ref{impact-pol}).  One can nicely see that the
displacement of
the center of the distribution along the $b^y$ axis is different for
$u$ and $d$ quarks.  According to our discussion in
Section~\ref{sec:gpd-e} the average displacement is described by the
quantity $s_q(x)$ defined in (\ref{sq-def}).  Indeed we read off from
Fig.~\ref{fig:s-syst} that for small values of $x$ the shift
$|s_d(x)|$ in our default fit is significantly larger than $s_u(x)$.
This explains the pronounced difference between the polarized and
unpolarized densities $d_v^X(x,\tvec{b})$ and $d_v(x,\tvec{b})$ at
small $x$ compared to the corresponding $u$ quark densities. The
different signs of $s_u(x)$ and $s_d(x)$, corresponding to the
different signs of the anomalous magnetic moments $\kappa_u$ and
$\kappa_d$, imply that for $d$ quarks the center of the polarized
density is shifted toward negative $b_y$, whereas for $u$ quarks it is
found at positive $b_y$.

\section{Handbag approach to wide-angle Compton scattering}
\label{sec:compton}

Our analysis in Sect.~\ref{sec:feynman} has shown that at momentum
transfer $-t$ around $10 \gev^2$ the proton Dirac form factor can be
described in terms of the soft Feynman mechanism.  Dominance of this
mechanism is the basis of the so-called handbag approach to wide-angle
Compton scattering \cite{rad:98,DFJK1} and related processes
\cite{Huang:2000kd,Huang:2003jd}.  In this approach the relevant
process amplitudes factorize into a hard scattering on a single quark
in the target and moments of GPDs at skewness $\xi=0$.  The handbag
approach is restricted to kinematical situations where all Mandelstam
variables $s$, $t$, $u$ are sufficiently large compared to a hadronic
scale $\Lambda^2$, but not yet in the asymptotic regime where the
standard hard-scattering approach \cite{Lepage:1980fj} is applicable.
Before turning to a quantitative analysis of the Compton form factors
on the basis of our results from Section~\ref{sec:ansatz}, we wish to
clarify some important theoretical aspects, concerning in particular
the inclusion and interpretation of radiative QCD corrections in the
handbag formalism.  A formal proof for handbag factorization would
require a systematic treatment of these corrections, an issue which
has not been addressed in too much detail yet.  As a first step,
one-loop radiative corrections to the partonic scattering amplitude in
Compton scattering have been evaluated in
\cite{Huang:2001ej,Huang:2003uy}.

\subsection{Compton scattering and Compton form factors}
\label{sec:compton-basics}

Let us recall some important features of the handbag approximation to
Compton scattering, $\gamma p \to \gamma p$, at large angles.  (The
discussion in this and the following subsection can be generalized to
the case where the incoming photon is off-shell.)  The approximation
assumes that the soft configurations discussed in
Sect.~\ref{sec:feynman} are dominant in the scattering process.  The
process amplitude can then be written in terms of hard-scattering
amplitudes for Compton scattering on a free quark, $\gamma q \to
\gamma q$, multiplied with Compton form factors describing the
emission and reabsorption of the struck quark by the proton target.
The form factors are given in terms of GPDs as
\begin{eqnarray}
\label{compton-ffs}
  R_V(t,\mu^2) &=&   \sum_q e_q^2 \, \int_{-1}^1 \frac{\d x}{x} 
                     H^{q}(x,t,\mu^2)  \ , 
\qquad \qquad
  R_T(t,\mu^2) \:=\: \sum_q e_q^2 \, \int_{-1}^1 \frac{\d x}{x}
                     E^{q}(x,t,\mu^2)  \ ,
\nonumber \\
  R_A(t,\mu^2) &=&   \sum_q e_q^2 \, \int_{-1}^1 \frac{\d x}{x} \,
                     \mbox{sgn}(x) \, \widetilde H^{q}(x,t,\mu^2) \ ,
\end{eqnarray}
with the momentum fraction $x$ defined in a frame where the
plus-momentum of the proton is equal before and after the scattering,
see Sect.~\ref{sec:feynman} and \cite{DFJK1}.  The derivation of the
factorization formula requires approximations appropriate for the soft
kinematics specified in (\ref{soft-classif-t}), where terms suppressed
by $\Lambda /\sqrt{|t|}$ are neglected.\footnote{We recall that the
power counting given in \cite{DFJK1} refers to the ultrasoft region in
(\protect\ref{soft-classif-t}) and should be amended accordingly.
This affects the parametric estimate of corrections to the
approximations made, but not their final result.}
In particular, the amplitudes for the partonic subprocess $\gamma q
\to \gamma q$ are evaluated with the four-momentum $k$ of the struck
quark approximated by the full momentum $p$ of the proton (see
Fig.~\ref{fig:feynman}a for the kinematics).  The Mandelstam variables
$\hat{s}$, $\hat{t}$, $\hat{u}$ of the subprocess $\gamma q \to \gamma
q$ are then equal to the Mandelstam variables $s$, $t$, $u$ of the
overall process $\gamma p \to \gamma p$.  The longitudinal momentum
fraction $x$ of the struck quark is thus set to $1$ in the
hard-scattering amplitude, leaving the integration over $x$ to be
performed at the level of the proton matrix elements as given in
(\ref{compton-ffs}).  The results of our phenomenological fit for
$H_v^q(x,t)$ show that at $-t$ of several $\gev^2$ the proton matrix
elements are not dominated by very large values of $x$.  In
Fig.~\ref{fig:Hgpd} we see for instance that at $-t= 10 \gev^2$ the
maximum of $H_v^u(x,t)$ is around $x\approx 0.62$, with half the
maximum attained at $x\approx 0.47$ and $x\approx 0.77$.  This is well
within the parametric estimate $(1-x) \sim \Lambda/\sqrt{|t|}$ if we
take for $\Lambda$ a value around $1 \gev$, but one may worry that the
resulting approximation of the hard-scattering amplitude is very bad.
This need not necessarily be the case.  To see this, let us consider
more sophisticated relations between $\hat{s}$, $\hat{t}$, $\hat{u}$
and $s$, $t$, $u$.  The ansatz $\hat{s} = x s$ and $\hat{u} = x u$ for
instance can be argued to give a better first approximation than the
naive relation with $x=1$.  Since the
leading-order scattering kernels for $\gamma q \to \gamma q$ depend
only on $\hat{u} / \hat{s}$, the value of $x$ cancels in this
approximation.\footnote{We remark that such a cancellation does not
take place with the expressions for $\hat{s}, \hat{t}, \hat{u}$
recently used in \protect\cite{Chen:2004tw}.  These expressions cannot
be used for wide-angle Compton scattering, since they imply a zero of
$\hat{u}$ at a particular value of $x$, see Sect.~4.1 of
\protect\cite{DFJK1}.}
A consistent treatment of the kinematics in the handbag graphs beyond
the $x=1$ approximation must however be complemented by the inclusion
of graphs that are not described by the proton matrix elements
corresponding to the twist-two GPDs in (\ref{compton-ffs}).  Such
graphs involve for instance operators with two quark fields and an
additional gluon, or four-quark operators
associated with the so-called cat's ears diagrams.  The inclusion of
higher-twist operators is already required to preserve electromagnetic
gauge invariance of the Compton amplitude, as has been seen in the
treatment of deeply virtual Compton scattering
\cite{Anikin:2000em,Radyushkin:2000ap}.  A corresponding analysis for
wide-angle scattering would be very important but has remained elusive
so far.

A related theoretical uncertainty concerns the factor $1/x$ in the
integrals (\ref{compton-ffs}).  In the derivation of \cite{DFJK1} it
is somewhat ambiguous whether to associate this factor with the soft
matrix elements or with the hard-scattering amplitude, where it would
be approximated by 1.  The difference between $1/x$ and $1$ is among
the effects which are presently beyond theory control.  Numerically it
is quite 
large in the $t$ range of Compton scattering experiments,
given that the expansion parameter $\Lambda /\sqrt{|t|}$ relevant
for the soft Feynman mechanism only scales with the square root of the
large momentum transfer.  The expressions (\ref{compton-ffs}) should
only be used for values of $t$ that are large enough for the
underlying approximations---at small $t$ the corresponding integrals
even diverge.  As we observed at the end of
Sect.~\ref{sec:power-law-t}, the valence GPDs from our ansatz with an
exponential $t$ dependence vanish like a power of $x$ for $x\to 0$
provided that $|t| \gsim 0.7 \gev^2$, so that their $1/x$ integrals
then converge.  However, sea quarks also contribute to the Compton
form factors, which describe exchange of positive $C$ parity in the
$t$-channel.  Since in the distributions at $t=0$ the small-$x$ rise
is faster for sea quarks than for valence quarks, one expects that
larger $|t|$ is necessary to make (\ref{compton-ffs}) well defined.
We shall return to the issue of sea quarks in
Sect.~\ref{sec:compton-pheno}.

The $1/x$ factors in (\ref{compton-ffs}), as well as the
$\mbox{sgn}(x)$ in the case of $R_A$, give rise to a $\mu$ dependence
of the Compton form factors.  Arising from DGLAP evolution of the GPDs
under the integral, this scale dependence is an effect of order
$\alpha_s$.  We shall see in the next subsection that at present NLO
corrections to wide-angle Compton scattering are only understood in
the approximation where $x$ is set to $1$ in the hard-scattering
amplitude and where the form factors (\ref{compton-ffs}) are replaced
with their scale independent counterparts
\begin{eqnarray}
\label{simple-ffs}
  R_V^{(0)}(t) &=&   \sum_q e_q^2 \, \int_{-1}^1 \d x \,
                     H^{q}(x,t,\mu^2)  \ , 
\qquad \qquad
  R_T^{(0)}(t) \:=\: \sum_q e_q^2 \, \int_{-1}^1 \d x \,
                     E^{q}(x,t,\mu^2)  \ ,
\nonumber \\
  R_A^{(0)}(t) &=& \sum_q e_q^2 \, \int_{-1}^1 \d x \,
                     \widetilde H^{q}(x,t,\mu^2) \, ,
\end{eqnarray}
which apart from charge factors are just the usual Dirac, Pauli, and
axial form factors.  In addition to neglecting the difference between
$1/x$ and $1$, we have counted antiquarks with the ``wrong'' sign in
the integral over $\widetilde{H}^q$, which should again be a valid
approximation when large enough momentum fractions dominate in the
integral.


\subsection{NLO corrections to wide-angle Compton scattering}
\label{sec:compton-nlo}

Next-to-leading order corrections to the partonic subprocess in
Compton scattering have been calculated in \cite{Huang:2001ej}.  This
means to consider virtual corrections to $\gamma q \to \gamma q$ where
the virtualities of the loop momenta are hard, i.e., where they scale
with the large momentum transfer $t$.  The unrenormalized result for
the hard-scattering amplitudes has the schematic form\footnote{Our
discussion refers to the case of photon helicity conserving
amplitudes.  Photon helicity-flip amplitudes are zero at LO and finite
at NLO.}
\begin{equation}
    {\cal H}^{\rm LO}(s,t) 
    \left(1 + \frac{\alpha_s C_F}{4\pi} \, C_{\rm IR}(t,\mu^2)\right)
    + \frac{\alpha_s C_F}{4\pi} \, {\cal H}^{\rm NLO}(s,t) \ ,
\label{krolletal}
\end{equation} 
where the NLO contribution has been decomposed into a divergent part
$C_{\rm IR}(\mu,t)$ multiplied with the LO amplitude and a finite
piece $ {\cal H}^{\rm NLO}(s,t)$.  Using dimensional regularization
with $D=4+\epsilon$, one has
\begin{equation}
  C_{\rm IR}(t,\mu^2) = -\frac{8}{\epsilon^2} + \frac{4}{\epsilon}
       \ln \frac{\mu^2}{|t|} + \frac{6}{\epsilon}  
      - \ln^2 \frac{\mu^2}{|t|} - 3 \ln \frac{\mu^2}{|t|}
      + \mbox{constant terms} \, ,
\label{CIR}
\end{equation}
which contains a collinear and a soft infrared divergence, and the
well-known Sudakov double logarithms.  It has been pointed out in
\cite{Huang:2001ej} that the soft and collinear divergences and the
associated double logarithms are universal: not only do they appear in
the corrections to $\gamma q\to \gamma q$ but also in the corrections
to $\gamma^* q \to q$, which are relevant for the elastic proton form
factors.  In a ``physical'' factorization scheme one considers the
radiative corrections to the quark-photon vertex as part of the form
factor itself.  To renormalize the Compton form factors and the
corresponding hard-scattering amplitudes one then subtracts the full
expression $C_{\rm IR}$, where the constant terms are fixed by the
condition that in this scheme the elastic proton form factor is
unchanged.\footnote{\label{const} The relation between our notation in
(\protect\ref{compton_final}) and the one in
\protect\cite{Huang:2001ej} is ${\cal H}^{\rm
NLO}_{\mbox{\tiny\protect\cite{Huang:2001ej}}} = \frac{\alpha_s
C_F}{4\pi} \left({\cal H}^{\rm NLO}_{\mbox{\tiny here}} +
\frac{\pi^2}{3} - 8\right)$.}
Writing the overall Compton amplitudes as $M_i$, where $i$ corresponds
to the subscripts $V$, $A$, $T$ of the Compton form factors, we then
have
\begin{eqnarray}
  M_i(s,t) &=& 
  \left( {\cal H}_{i}^{\rm LO}(s,t) 
   + \frac{\alpha_s C_F}{4\pi} \,{\cal H}_{i}^{\rm NLO}(s,t) \right)
   R_i(t,\mu^2) + \ldots 
\nonumber \\[0.2em]
 &=& 
  \left( {\cal H}_{i}^{\rm LO}(s,t) 
   + \frac{\alpha_s C_F}{4\pi} \,{\cal H}_{i}^{\rm NLO}(s,t) \right)
   R_i^{(0)}(t) 
+ {\cal H}_{i}^{\rm LO}(s,t) \left[ R_i^{\phantom{()\!\!}}(t,\mu^2) 
                                  - R_i^{(0)}(t) \right]
   + \ldots .  \hspace{2.5em}
 \label{compton_final}
\end{eqnarray}
Note that the NLO calculations for $\gamma q\to \gamma q$ and
$\gamma^* q \to q$ with on-shell quarks enabled us to find a
renormalization scheme at the level of the electromagnetic and weak
nucleon form factors and of their flavor combinations
(\ref{simple-ffs}) relevant for Compton scattering.  They do
\emph{not} permit an analogous renormalization of the Compton form
factors (\ref{compton-ffs}) or of the GPDs themselves.  This is
manifest in the first line of (\ref{compton_final}), where we have an
uncanceled $\mu$ dependence.  In the second line of
(\ref{compton_final}) we have therefore indicated that, for internal
consistency of the presently available results, the NLO calculation
should be restricted to the part of the Compton form factors which
involves the same local operators as $F_1$, $F_2$ and $F_A$.  The
parametric uncertainties of the second line in (\ref{compton_final})
are of order $\alpha_s^2$ or of order $\alpha_s\, \Lambda
/\sqrt{|t|}$.  Up to this accuracy the result is $\mu$ independent as
it must be.  Controlling effects of order $\alpha_s\, \Lambda
/\sqrt{|t|}$ would require a more complete analysis of
power-suppressed terms already at Born level.

Following an alternative point of view, one could use a minimal
subtraction scheme in (\ref{krolletal}) to render the resummation of
Sudakov logarithms explicit in hard coefficient functions.  The
Compton amplitudes would then be written as
\begin{equation}
  M_i(s,t) = C_{i}(s,t,\mu^2) \, R_i^{\rm eff}(t,\mu^2) + \ldots ,
\label{Mi}
\end{equation}
where $C_{i}(s,t,\mu^2)$ obeys the renormalization group equation
\begin{equation}
  \mu^2 \frac{\d}{\d \mu^2} \, C_{i}(s,t,\mu^2)  = 
   \gamma(t,\mu^2) \, C_{i}(s,t,\mu^2) 
\end{equation}
whose anomalous dimension follows from (\ref{CIR}),
\begin{equation}
  \gamma(t,\mu^2) = - \frac{\alpha_s C_F}{2 \pi} 
  \left( \ln \frac{\mu^2}{|t|} 
   + \gamma_0 \right) + {\cal O}(\alpha_s^2) \ .
\end{equation}
The value of the constant $\gamma_0$ depends on the renormalization
scheme.  Renormalization group evolution from $\mu^2 \sim |t|$ down to
$\mu_0^2 \sim \Lambda \sqrt{|t|}$ corresponds to the summation of
Sudakov logarithms between those two scales, which are then
explicitly contained in $C_{i}(s,t,\mu_0^2)$.  In this sense, the new
Compton form factors $R_i^{\rm eff}(t,\mu_0^2)$ only contain dynamics
below the factorization scale $\mu_0^2$, which is the virtuality of
the struck quark in soft kinematics.  They can be considered as being
defined in soft-collinear effective theory
\cite{Bauer:2000yr,Beneke:2002ph}, which is an effective field theory
where the hard degrees of freedom are integrated out, leaving degrees
of freedom with virtualities smaller than $|t|$.  The scale dependence
of $R_i^{\rm eff}(t,\mu^2)$ compensates that of $C_{i}(s,t,\mu^2)$.

Similarly, in the effective theory the Dirac form factor can be
written as
\begin{equation}
  \label{eff-ff}
  F_1(t) = C_1(t,\mu^2) \, F_1^{\rm eff}(t,\mu^2) + \ldots
\end{equation}
with a hard coefficient function that obeys the same evolution
equation as in the case of Compton form factors,
\begin{equation}
  \mu^2 \frac{\d}{\d \mu^2} \, C_1(t,\mu^2)  = 
   \gamma(t,\mu^2) \, C_1(t,\mu^2)  .
\end{equation}
In other words, the function $C_1(t,\mu^2)$ resums the same (leading)
Sudakov logarithms as $C_{i}(t,\mu^2)$.  The physical scheme where the
coefficient function in (\ref{eff-ff}) is equal to 1 is related to a
general scheme by a finite renormalization.  Applying the same
renormalization to the Compton form factors one would write
\begin{equation}
  R_i^{(0)}(t) = C_1(t,\mu^2) R_i^{\rm eff}(t,\mu^2) + \ldots,
\end{equation}
which together with (\ref{Mi}) gives
\begin{equation}
  M_i(s,t) = \frac{C_{i}(s,t,\mu^2)}{C_1(t,\mu^2)} \, R_i^{(0)}(t)
+ \ldots .
 \label{compton_eff}
\end{equation}
Sudakov effects cancel in the ratio of hard-coefficient
functions,\footnote{The situation is similar to the case of
heavy-to-light form factors, where the issues related to the summation
of Sudakov logarithms in the effective theory are irrelevant for
\emph{ratios} of soft form factors \protect\cite{Beneke:2000wa}.}
which can therefore be calculated in fixed-order perturbation theory
as
\begin{equation}
  \label{coeff-ratio}
  \frac{C_{i}(s,t,\mu^2)}{C_1(t,\mu^2)} =
   {\cal H}_{i}^{\rm LO}(s,t) 
   + \frac{\alpha_s C_F}{4\pi} \,{\cal H}_{i}^{\rm NLO}(s,t) 
   + \ldots ,
\end{equation}
where we recover (\ref{compton_final}) to leading accuracy in the
expansion parameter $\Lambda /\sqrt{|t|}$ of the effective theory.

Returning to the level of GPDs, we have so far dealt with
distributions $H_v^q(x,t,\mu^2)$ defined by standard collinear
renormalization in QCD.  Note that even for $\mu^2 \ll |t|$ such
distributions can be defined, since they can be obtained from
distributions with $\mu^2 \gsim |t|$ by DGLAP evolution, which may be
understood as a finite renormalization.  The result does however not
correspond to matrix elements with internal virtualities of at most
$\Lambda \sqrt{|t|}$, which would be appropriate for the soft Feynman
mechanism.  One may speculate that such a ``soft GPD'' would be
obtained in the effective theory as
\begin{equation}
  \label{H-eff}
H_v^{q\, \rm eff}(x,t,\mu^2) =
\frac{1}{C_1(t,\mu^2)}\,  H_v^q(x,t,\mu^2) 
\end{equation}
for $|t| \gg \Lambda^2$, with associated sum rules
\begin{equation}
F_1^{\rm eff}(t,\mu^2) 
  = \sum_q e_q   \int_0^1 \d x \, H_v^{q\, \rm eff}(x,t,\mu^2) , 
\qquad
R_V^{\rm eff}(t,\mu^2)
  = \sum_q e_q^2 \int_0^1 \d x \, H_v^{q\, \rm eff}(x,t,\mu^2) .
\end{equation}
Notice that $H_v^{q\, \rm eff}$ obeys a \emph{different} evolution
equation than $H_v^q$, since its renormalization includes the Sudakov
logarithms from the endpoint region $x\to 1$.  It is however not clear
whether the $x$-independent renormalization factor in (\ref{H-eff})
corresponds indeed to the structure emerging in the effective theory.

In the absence of a proper definition for ``soft GPDs'', we understand
the distributions extracted in this work as ``standard GPDs'' obeying
DGLAP evolution.  This is also convenient in connection with our
phenomenological ansatz (\ref{master-ansatz}), which interpolates
between the standard $\overline{\mbox{MS}}$ parton densities at $t=0$
and distributions at large $t$.  We point out that even for the
highest values of $-t \sim 30 \gev^2$ we consider in our applications,
Sudakov double logarithms are not very large, with $\ln^2 (\mu^2/|t|)
\leq 4$ for $\mu=2 \gev$, so that one may expect the difference
between ``soft'' and ``standard'' GPDs to be reasonably small.

In our calculations for wide-angle Compton scattering we will use the
form factors (\ref{compton-ffs}), regarding their $1/x$ factors as a
phenomenological estimate of effects beyond a strict $\Lambda
/\sqrt{|t|}$ expansion.  To keep the analysis simple we have refrained
from using the expression (\ref{compton_final}) and instead take the
full form factors (\ref{compton-ffs}) multiplied with the
hard-scattering amplitudes at NLO.  Since the $\mu^2$ dependence
inherent in the $1/x$ moments is an effect beyond present theoretical
control, we will require that the variation of the form factors
obtained with GPDs fitted for $\mu=1$, $2$ and $4 \gev$ remains
reasonably small.  This will provide a criterion of how large $t$ is
required to obtain stable results within our present approach.


\subsection{Results for form factors and observables}
\label{sec:compton-pheno}

\begin{figure}
\begin{center}
\includegraphics[width=.45\textwidth, 
  bb=50 385 395 735]{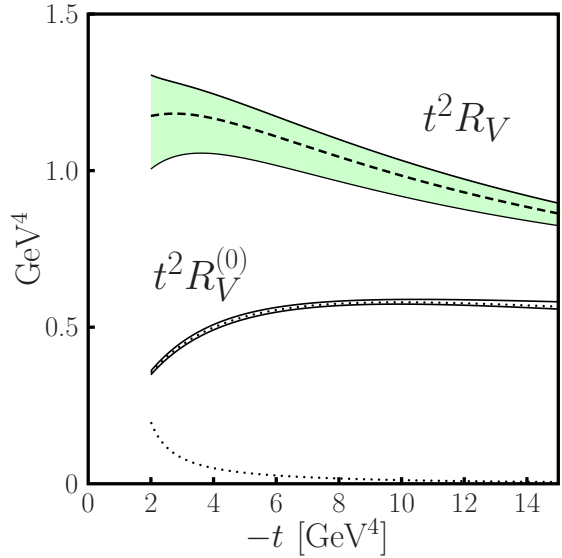}
\end{center}
\caption{\label{fig:RV-mu} Study of systematic uncertainties regarding
the Compton form factor $R_V$.  Shown are $t^2 R_V$ and $t^2
R_V^{(0)}$ from (\protect\ref{compton-ffs}) and
(\protect\ref{simple-ffs}) evaluated from our default fit for $H_v^q$
at $\mu=2 \gev$.  The limits of the error bands correspond to the fits
for $H_v^q$ at $\mu=1$ and $4 \gev$ in
Table~\protect\ref{tab:alpha-mu-fits}.  The dotted line shows the sea
quark contribution to $t^2 R_V$ as specified before and in
(\protect\ref{sea-quarks-RV}).}
\end{figure}

After our discussion of theoretical uncertainties let us now see how
they quantitatively look like with the GPDs we have obtained from our
fits to elastic form factors.  We restrict this discussion to $R_V$,
which dominates the unpolarized Compton cross section.  In
Fig.~\ref{fig:RV-mu} we compare the full form factor $R_V$ and its
analog $R_V^{(0)}$ without a factor of $1/x$ under the integral, both
evaluated with the result of our default fit for $H_v^q(x,t)$.  As is
to be expected from the shape of $H_v^q$ in Fig.~\ref{fig:Hgpd}, the
difference is considerable in the $t$ range relevant for existing or
planned measurements.  The error bands on the curves for $R_V$ and
$R_V^{(0)}$ indicate the variation of $H_v^q$ with $\mu$, with the
limits of the bands corresponding to the fits with $\mu=1$ and $4
\gev$ listed in Table~\ref{tab:alpha-mu-fits}.  The scale variation
for $R_V$ becomes rather large for $-t$ below about $3 \gev^2$, which
gives a first indication of when effects of order $\alpha_s\, \Lambda
/\sqrt{|t|}$ can no longer be neglected.  The tiny scale dependence of
$R_V^{(0)}$ confirms that our fits to $F_1^p$ and $F_1^n$ with input
parton densities at different $\mu$ indeed give a scale invariant
result to a very good approximation.

As we already mentioned, sea quark contributions do not cancel in the
Compton form factors defined in (\ref{compton-ffs}).  Contrary to the
isovector axial form factor $F_A(t)$, the quark flavor combination in
Compton scattering is also sensitive to the flavor singlet
distribution, which is more difficult to model due to its mixing with
gluons as explained at the end of Sect.~\ref{sec:phys-mot}.  For a
naive estimate we have taken the same impact parameter profile for sea
quarks and valence quarks.  We thus define ${H}^{\bar{q}}(x,t) = -
H^q(-x,t)$ and set ${H}^{\bar{q}}(x,t) = \bar{q}(x) \exp[ t f_q(x) ]$
with the CTEQ6M \cite{CTEQ} antiquark densities $\bar{q}(x)$ at $\mu=2
\gev$ and with $f_q(x)$ determined in our default fit for $H_v^q$.
The resulting sea quark contribution
\begin{equation}
  \label{sea-quarks-RV}
R_V^{\rm sea}(t,\mu^2) = 
  2 \sum_q e_q^2 \, \int_{0}^1  \frac{\d x}{x} H^{\bar{q}}(x,t,\mu^2) 
\end{equation}
to $R_V$ is shown as the dotted curve in Fig.~\ref{fig:RV-mu}, where
we have restricted the sum over quark flavors $u$ and $d$.  When the
sea quark contribution becomes important, we are clearly outside the
region where the physical picture and the approximations of the soft
Feynman mechanism are applicable.  We see that with our estimate one
need not worry about sea quarks for $-t \ge 4 \gev^2$.  Below $-t =2
\gev^2$ their contribution quickly grows out of control, and we will
therefore not show Compton form factors below this value.

We have seen in Fig.~\ref{fig:neff} that our default fit for $H_v^q$
exhibits the scaling properties characteristic of the soft Feynman
mechanism in $F_1^p(t)$ for $-t$ above $10\gev^2$.  Below this value
one enters a transition region, where the scaling becomes more and
more approximate.  Figure~\ref{fig:RV-mu} indicates that current
theoretical control over the Feynman mechanism in wide-angle Compton
scattering does not extend below $-t = 2 \gev^2$.

\begin{figure}[p]
\begin{center}
\includegraphics[width=.35\textwidth, height=.35\textwidth,
  bb= 158 391 447 676]{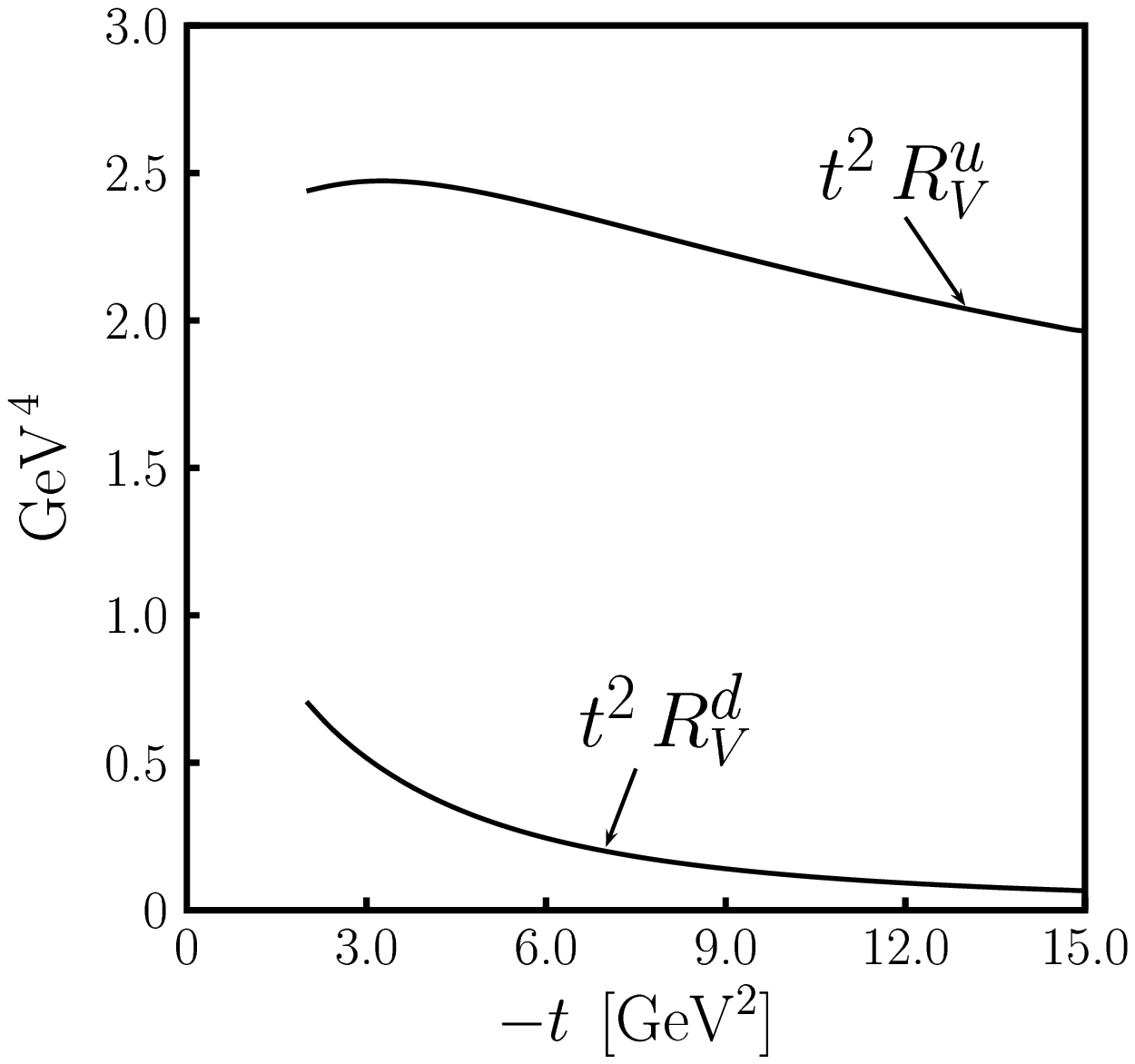}
\hspace{4em}
\includegraphics[width=.35\textwidth, height=.35\textwidth,
  bb= 159 423 453 711]{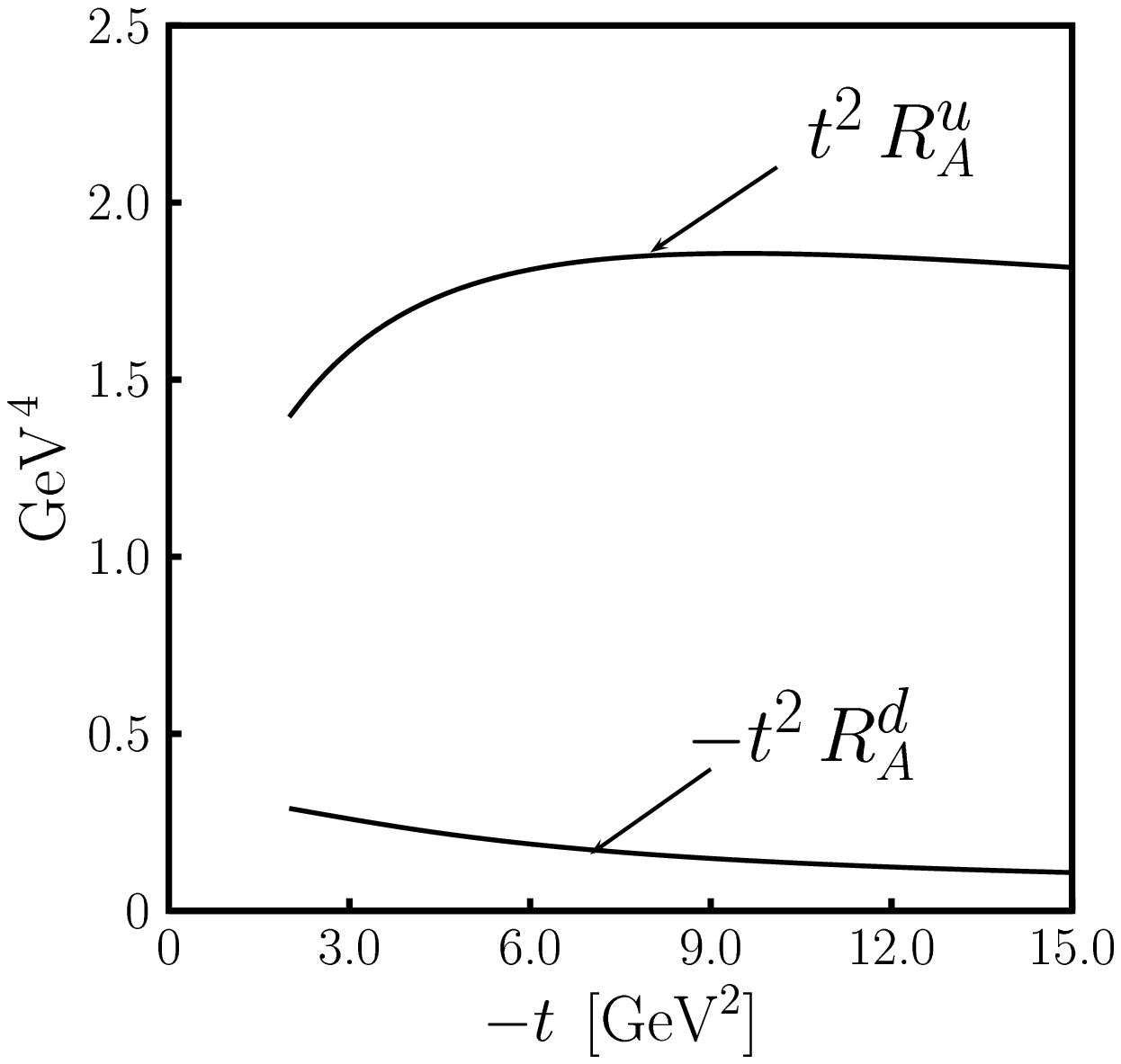}
\\[4em]
\includegraphics[width=.35\textwidth, height=.35\textwidth,
  bb= 149 421 443 708]{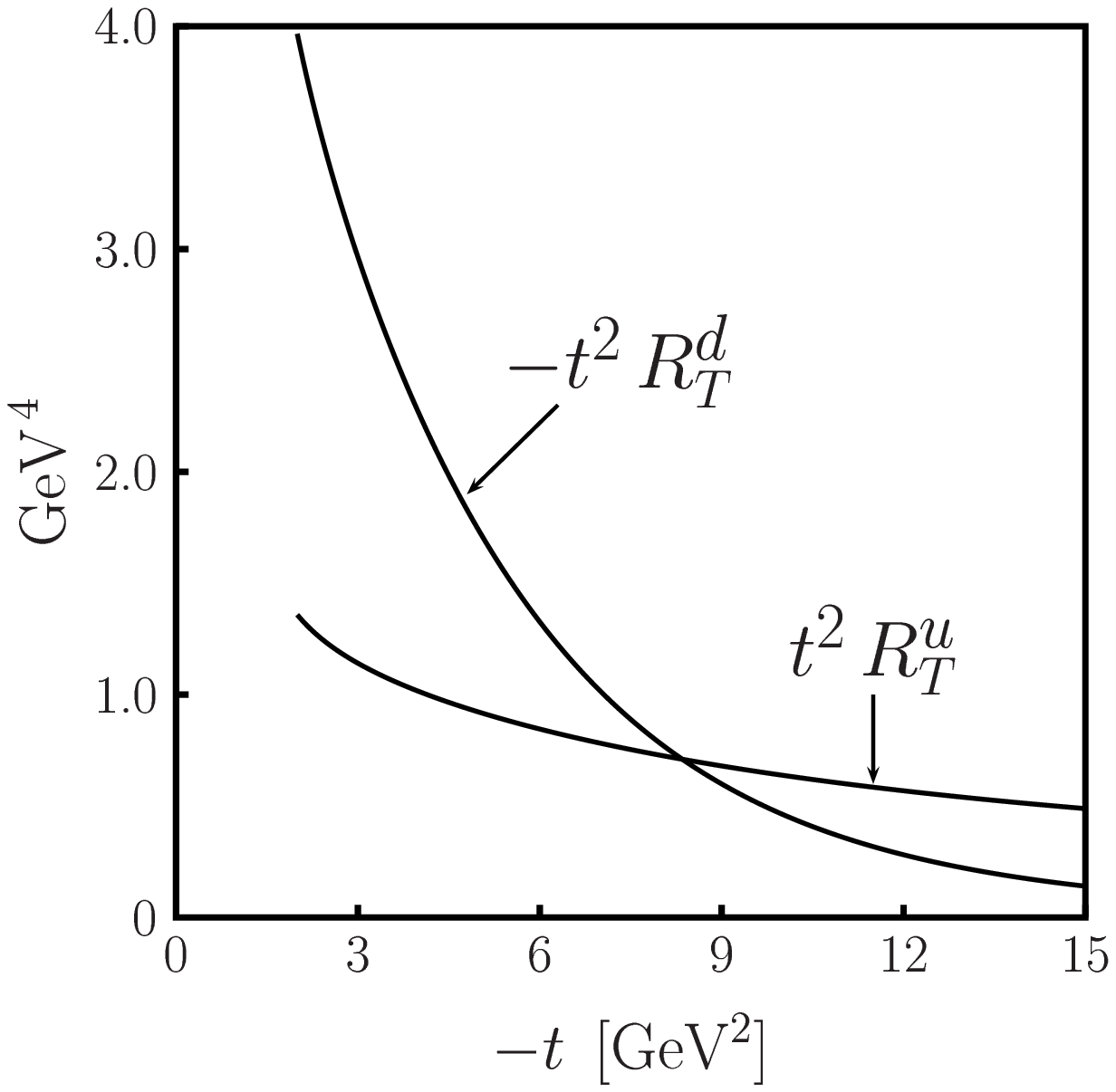}
\hspace{4em}
\includegraphics[width=.35\textwidth, height=.35\textwidth,
  bb= 101 157 379 430]{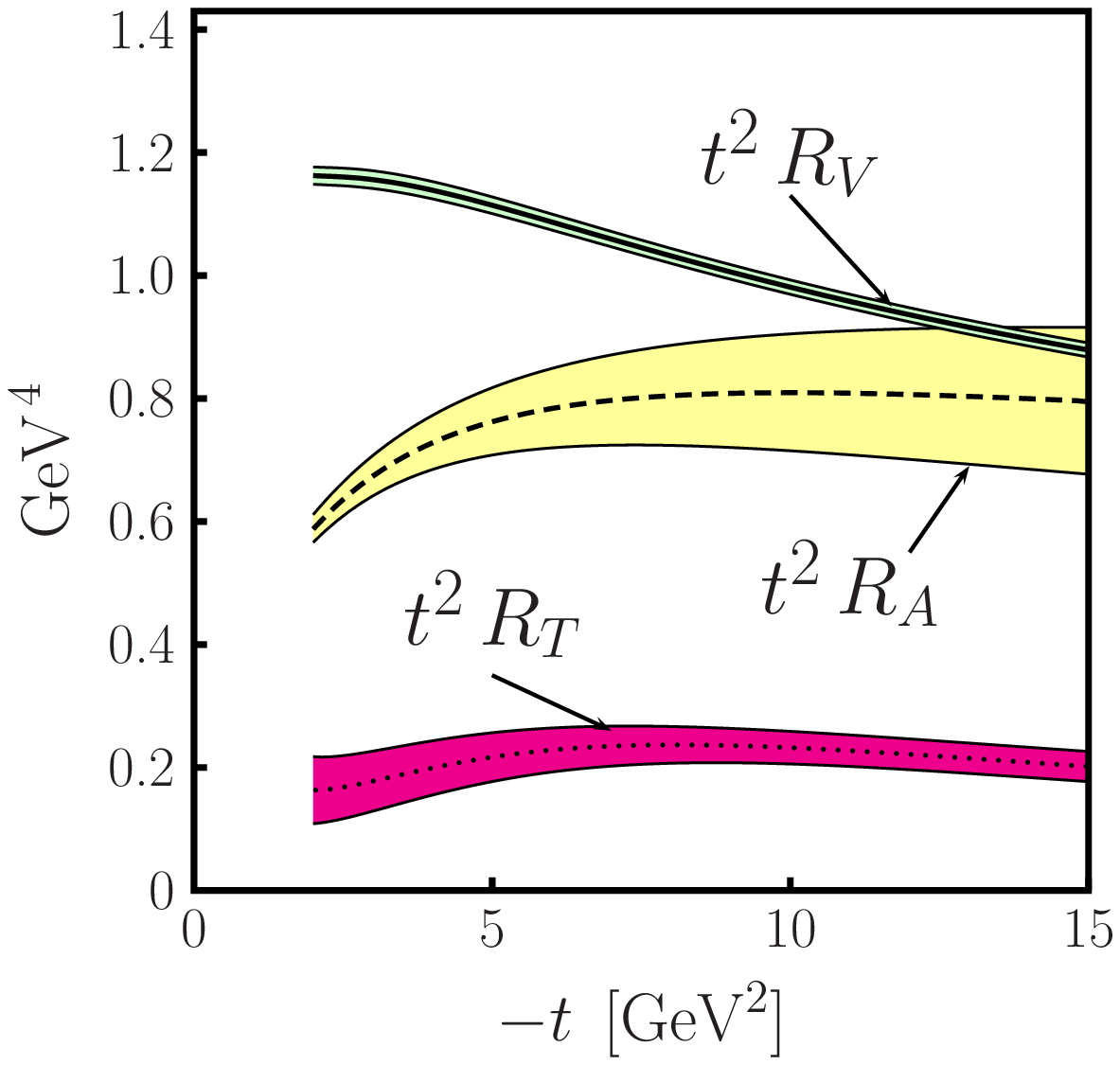}
\\[1em]\
\end{center}
\caption{\label{fig:compton-ffs} Scaled Compton form factors for
individual quark flavors and summed with the appropriate squared
charge factors (bottom right).  The error bands shown for $R_V$ and
for $R_T$ correspond to the $1\, \sigma$ uncertainties of our default
fits for $H_v^q$ and $E_v^q$.  The band for $R_A$ corresponds to the
$1\, \sigma$ uncertainties on our profile function added in quadrature
to the errors on the polarized parton densities in
\protect\cite{blum:03}, which are by far dominant for this quantity.}
\end{figure}

In Fig.~\ref{fig:compton-ffs} we show the Compton form factors
obtained from our default fits for $H_v^q$, $E_v^q$ and from the
ansatz (\ref{Htilde-ansatz}) for $\widetilde{H}_v^q$ with $\tilde{f}_q
= f_q$.  We plot both individual flavor contributions and the flavor
combinations $R_i = \frac{4}{9} R^u_i + \frac{1}{9} R^d_i$ for Compton
scattering on the proton.  In the case of $R_V^q$ and $R_A^q$ we see
the clear dominance of $u$ quarks over $d$ quarks even without the
charge factors, as we already did for the corresponding moments in
Sect.~\ref{sec:moments}.  The full form factors $R_V$ and $R_A$ turn
out to be somewhat smaller than those modeled in \cite{DFJK1}.

For $R_T^q$ the relative weight of flavors is different than for
$R_V^q$ and $R_A^q$.  Even more than for the moments $e_{1,0}^q$ in
Fig.~\ref{fig:E-moments}, $d$ quarks dominate for smaller and $u$
quarks for larger values of $t$.  Given the opposite signs of $R_T^u$
and $R_T^d$ and the squared charge factors, the resulting $R_T$ is
significantly smaller than $R_V$ and $R_A$ at lower values of $t$.
The ratio $R_T /R_V$ behaves quite differently from the analogous
ratio $F_2^p /F_1^p$ of electromagnetic form factors.  With
\begin{equation}
\kappa_{\rm C}(t)  = \frac{\sqrt{-t}}{2m}\, 
                     \frac{R_T(t)}{R_V(t)} \, ,
\qquad\qquad
\kappa_{\rm em}(t) = \frac{\sqrt{-t}}{2m}\, 
                     \frac{F_2^p(t)}{F_1^p(t)} \, ,
\end{equation}
the measurement of \cite{Gayou:2001qd} gives an approximately constant
$\kappa_{\rm em} \approx 0.37$ for $2 \gev^2 \le -t \le 5.5 \gev^2$,
whereas we find that with our default fits $\kappa_{\rm C}$ rises from
0.1 to 0.25 in the same interval and continues to grow to 0.4 at $-t =
10 \gev^2$.  Our result for $R_T$ is however subject to rather large
uncertainties, given the considerable freedom we encountered in
extracting $E_v^q$ from the Pauli form factors alone.  The fits shown
in Table~\ref{tab:scan}, which all give a good description of $F_2^p$
and $F_2^n$, produce form factors $R_T$ which at $-t = 5 \gev^2$ are
up to a factor of 1.5 larger than $R_T$ shown in
Fig.~\ref{fig:compton-ffs}, with even larger discrepancies at lower
$t$.  Within this set of fits, the curve in the figure acts more like
a lower bound.

\begin{figure}
\begin{center}
\includegraphics[width=.35\textwidth, height=.35\textwidth,
  bb= 192 263 468 548]{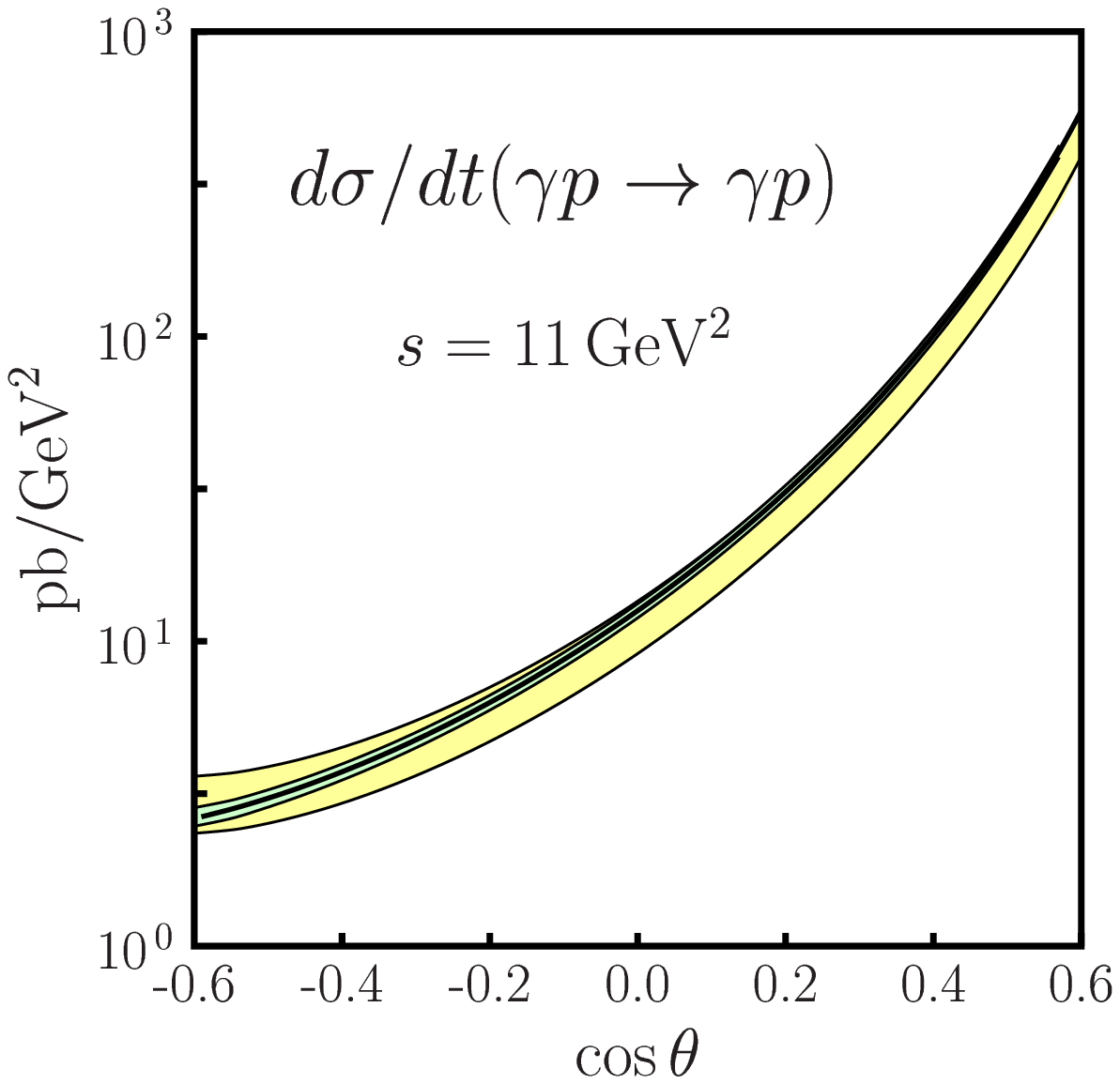}
\hspace{4em}
\includegraphics[width=.35\textwidth, height=.35\textwidth,
  bb= 146 391 423 678]{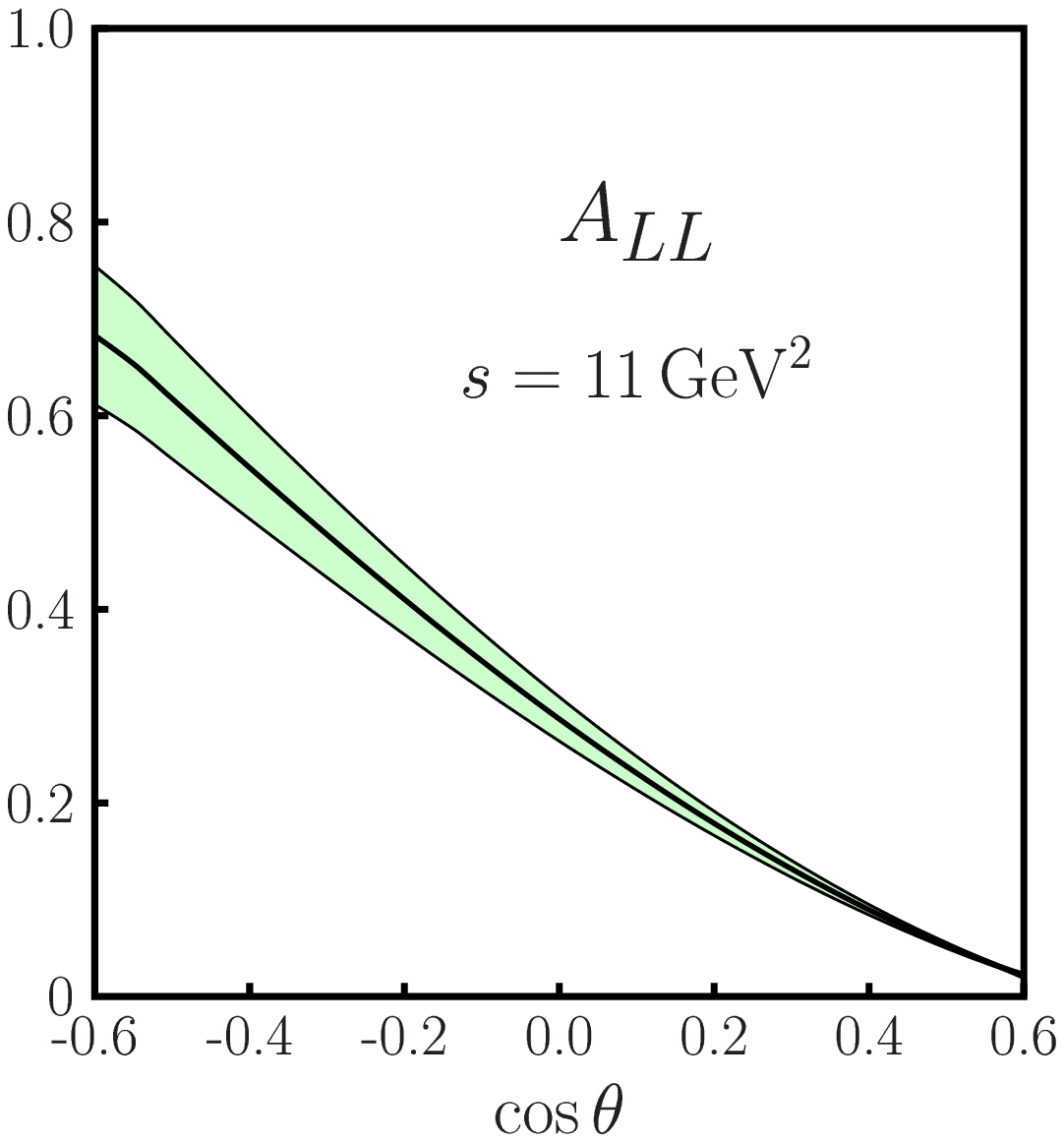}
\\[1em]\
\end{center}
\caption{\label{fig:wacs-obs} The unpolarized cross section (left) and
the helicity correlation parameter $A_{LL}$ (right) for wide-angle
Compton scattering at $s= 11 \gev^2$ as a function of the scattering
angle $\theta$ in the c.m.  Both observables are evaluated at NLO with
the Compton form factors shown in Fig.~\protect\ref{fig:compton-ffs}.
The error bands are explained in the text.}
\end{figure}

With the form factors in Fig.~\ref{fig:compton-ffs} we obtain the
unpolarized cross section $d\sigma /dt$ shown in
Fig.~\ref{fig:wacs-obs} for a squared c.m.\ energy of $s = 11 \gev^2$.
We have included the NLO corrections in the physical scheme described
in Sect.~\ref{sec:compton-nlo}, subtracting a constant term
$\frac{\alpha_s C_F}{4\pi} \left( \frac{\pi^2}{3} - 8\right)$ from the
quark scattering kernels of \cite{Huang:2001ej} as mentioned in
footnote~\ref{const}.  For the form factor $R_V^g$ describing gluon
exchange at NLO we have taken the model result of \cite{Huang:2001ej}
divided by a factor of 1.3, which is the typical discrepancy at $-t
\sim 5 \gev^2$ between $R_V$ obtained in the present study and in
\cite{Huang:2001ej}.  The resulting gluonic contribution to $d\sigma
/dt$ is only a few percent.

The inner band for the curve of $d\sigma /dt$ reflects the parametric
errors on the form factors shown in Fig.~\ref{fig:compton-ffs}.  The
corresponding uncertainty is very small.  This is because $R_V$ is
strongly dominated by $H_v^u$, which is the distribution best
constrained by the abundant data for $F_1^p$, whereas the individual
contributions of both $R_A$ and $R_T$ to $d\sigma /dt$ are between 1\%
and 10\% for the kinematics shown in Fig.~\ref{fig:wacs-obs}.  The
central curve in the figure corresponds to scenario 2 in
\cite{Diehl:2002ee}, where the Mandelstam variables of $\gamma q\to
\gamma q$ and of $\gamma p\to \gamma p$ are related by $\hat{s} = s -
m^2$, $\hat{t} = t$ and $\hat{u} = u - m^2$.  The limits of the outer
band correspond to scenarios 1 and 3 in the same study and reflect the
uncertainty in the evaluation of $d\sigma /dt$ that is due to the
finite proton mass.

An observable with greater sensitivity to $R_A$ is the correlation
parameter $A_{LL}$ between the helicities of the incoming photon and
the incoming proton \cite{Diehl:1999tr,Huang:2001ej}, which to a good
approximation is given by the ratio $R_A /R_V$ times a known
kinematical factor.  Within our approach $A_{LL}$ is equal to the
correlation parameter $K_{LL}$ between the helicities of the incoming
photon and the outgoing proton, which is a consequence of neglecting
the quark mass in the hard-scattering subprocess (see also
\cite{Miller:2004rc}).  Our result for $s= 11 \gev^2$ is shown in
Fig.~\ref{fig:wacs-obs}, with the error band reflecting the parametric
uncertainties on $R_A$, $R_V$ and $R_T$.

The correlation parameter $K_{LS}$ between the helicity of the
incoming photon and the transverse polarization of the outgoing proton
is sensitive to the tensor form factor $R_T$.  Assuming $\kappa_{\rm
C} = \kappa_{\rm em} \approx 0.37$ the study \cite{Huang:2001ej} found
values well in the range of 10\%.  With our default fit we find
$K_{LS}$ at $s= 11 \gev^2$ as small as $1\%$ for backward angles
$\theta$ and yet smaller in the forward hemisphere.  This can be
traced back to the approximate cancellation of two terms in the
expression for $K_{LS}$, one proportional to $R_T$ and the other to
$R_V$, see eq.~(397) in \cite{die:03}.  For this particular
kinematics, the uncertainties on $R_T$ from our fits hence greatly
amplify in the observable $K_{LS}$, which unlike $A_{LL}$ turns out to
be strongly energy dependent.

\section{Summary and outlook}
\label{sec:summary}

We have developed a parameterization of valence quark GPDs at zero
skewness which takes into account dynamically motivated correlations
between their $x$ and $t$ dependence.  At small $x$ their form is
consistent with dominance of the leading meson trajectories known from
Regge phenomenology.  For large $t$ they incorporate the soft Feynman
mechanism, where the struck quark has a virtuality of order $\Lambda
\sqrt{|t|}$.  Both analytically and numerically our distributions
satisfy the Drell-Yan relation between the large-$x$ power behavior of
parton distributions at $t=0$ and the large-$t$ power behavior of the
associated elastic form factors.  This holds for both $H_v^q$ and
$E_v^q$ and is to be understood in the sense of effective powers
describing the $x$ or $t$ dependence in a finite range of these
variables.  Key features of our functional ansatz for the GPDs are
approximately invariant under DGLAP evolution, so that with an
appropriate change of the free parameters this ansatz can be used for
a reasonable range of scales $\mu$.

Good fits of the GPDs can be obtained to describe both the Dirac and
Pauli form factors of proton and neutron, with an accuracy better than
20\% for almost all data points and better than 5\% in wide ranges of
$t$.  {}From $F_1^p$, where most data is available, we get a good
determination of $H_v^u(x,t)$ up to $x \lsim 0.8$.  It is stable under
a change of parameterization, provided we assume an exponential $t$
dependence as in (\ref{master-ansatz}).  Under this assumption we have
thus well determined the average squared impact parameter $\bsq{u}{x}$
of $u$ quarks in the $x$ range just quoted.  We have however seen that
this does not suffice to fix the limiting behavior of $\bsq{u}{x}$ for
$x\to 1$.  Given the paucity of data on $F_1^n$, the distribution
$H_v^d(x,t)$ is less well determined, but we find clear indications
that $\bsq{d}{x} > \bsq{u}{x}$ for larger values of $x$.  This implies
that with increasing $|t|$ the suppression of $d$ quarks compared with
$u$ quarks becomes even stronger than in the forward parton densities
at large $x$.

With the fit results for $H_v^q$ we also achieve a rather good
description of the isovector axial form factor $F_A$ under the
simplifying assumptions that sea quarks can be neglected and that the
impact parameter distribution of valence quarks is independent of the
quark helicity.  With our ansatz for $\widetilde{H}_v^q$ we find that
the largest uncertainty on the resulting form factor $F_A$ is due to
the current errors on the polarized quark densities $\Delta d$ and
$\Delta u$.

Using our ansatz for the proton helicity flip distributions we get
good fits to the Pauli form factors $F_2^p$ and $F_2^n$.  We find
however considerable ambiguities in determining $E_v^q(x,t)$ since its
forward limit is not known.  The form factor data disfavor an
identical shape in $x$ and $t$ of $E_v^u(x,t)$ and $E_v^d(x,t)$, but
they have only little preference concerning the details of this flavor
dependence.  In particular, we cannot determine whether the forward
limit $E_v^u(x,0)$ has a harder or a softer $x$ dependence than
$E_v^d(x,0)$.  Positivity constraints prove to be extremely valuable
in constraining $E_v^u$ and $E_v^d$, as they severely restrict their
size for $x\gsim 0.5$ as well as their $t$ dependence.  We find that,
despite the uncertainties on these distributions, their moments $\int
\d x\, x E_v^q(x,t)$ are reasonably well determined within our ansatz.
With Ji's angular momentum sum rule, the resulting valence quark
contribution to the orbital angular momenta $\langle L^u \rangle$ and
$\langle L^d \rangle$ have even smaller relative errors for the
allowed range of our parameters.

The $x$ moments of our extracted GPDs are in reasonable agreement with
results from lattice QCD, given the uncertainties of both our
phenomenological determination and of the lattice calculations.  With
our results we have also calculated the distribution in the transverse
plane of valence quarks with momentum fraction $x$, both in an
unpolarized and in a transversely polarized proton.  We have thus
shown that ``proton tomography'' as proposed in \cite{ral:02,bur:02}
is feasible on the basis of experimental data.  Finally, we have used
our GPDs to evaluate the form factors needed in the soft handbag
approach to wide-angle Compton scattering, as well as the associated
experimental observables.  From a number of indicators we conclude
that consistency of this approach requires momentum transfers $-t$
above $3 \gev^2$ or so.

Where can further progress be made?  High-quality data on the neutron
form factors in a wide $t$ range would be highly valuable for pinning
down the differences in the spatial distribution of $u$ and $d$
quarks.  Our fit results indicate in fact drastic differences in the
behavior of $u$ and $d$ contributions to the form factors, which
should be tested experimentally.  The fit to the Pauli form factors
suggests that the $t$ behavior of $F_2^p$ will continue to change
beyond $5.5 \gev^2$, and it would be good to see whether this is
indeed the case.  Since Dirac and Pauli form factors are most natural
to connect with the physics of parton distributions, we encourage
experimenters to provide data on the Sachs form factors $G_M$ and
$G_E$ at equal values of $t$ and with correlated errors wherever
possible.  Only then can one transform to $F_1$ and $F_2$ without
introducing further theoretical bias or having to guess errors.  With
the available data on the axial form factor $F_A$ we could only take a
glimpse at the role of quark helicity.  More data on this fundamental
quantity (in the form of data points rather than dipole
parameterizations whenever possible) should lead to important progress
in the study of $\widetilde{H}^q$.  An improved determination of the
polarized forward densities $\Delta q$ is under way at several
experimental facilities and will also be of benefit to this study.
For a more precise study of wide-angle Compton scattering, progress is
required in the theory of power-suppressed contributions.  Together
with ongoing and planned measurements this should lead to an improved
understanding of the dynamics and, if the soft handbag approach is
further validated, eventually to pinning down the relevant form
factors.

The fits to the electromagnetic form factors in this work have
restricted us to zero skewness and to the valence quark sector.  The
ansatz we have developed for GPDs at $\xi=0$ can be used as an input
for modeling distributions at nonzero $\xi$, using the concept of
double distributions as shown e.g.\ in \cite{Musatov:1999xp}.  A
nontrivial interplay between $t$ and the longitudinal variables can
readily be implemented in this framework
\cite{Mankiewicz:1998kg,goe:01}.  Interesting observations concerning
this formalism for proton distributions have recently been made in
\cite{Tiburzi:2004qr}, building on earlier work in
\cite{Polyakov:1999gs,Teryaev:2001qm}.  Most of the exclusive
processes where GPDs at nonzero $\xi$ occur are sensitive to sea
quarks and also to gluons, in particular the important channels of
deeply virtual Compton scattering and of neutral vector meson
production.  It will be interesting to investigate the dynamical
interplay between the quark singlet and gluons in the impact parameter
distributions, similarly to what has been seen for the distributions
in momentum fraction $x$.  In some channels, like the production of
pseudoscalar mesons or pion pairs, one can however selectively probe
the valence quark combinations of GPDs, see Sect.~1 of \cite{goe:01}
and Sect.~5 of \cite{die:03}.  This opens the possibility to
investigate the transverse distribution of valence quarks at small
$x$, where elastic form factors have only a limited reach.  The
versatility of the GPD formalism and vigorous experimental activity
let us hope that important progress is still to come.


\section*{Acknowledgments}

It is our pleasure to thank H. B\"ottcher, P. H\"agler, F. Schrempp
and C. Weiss for discussions, B. Wojtsekhowski for information on
Ref.~\cite{bogdan}, and E. Brash and J. Roche for helpful
correspondence.

We acknowledge the partial support by the Department of Energy and the
hospitality of the Institute for Nuclear Studies at the University of
Washington, where part of this work was done.  M.D. and
P.K. acknowledge partial support by the Integrated Infrastructure
Initiative ``Hadron Physics'' of the European Union, contract
No.~506078.


\appendix

\section{Details on form factor data}
\label{app:A}

Experimental data are usually given for the magnetic and electric form
factors $G_E$ and $G_M$, or for their ratio.  Natural quantities for
our parameterization of GPDs and their physical interpretation are the
Dirac and Pauli form factors, so that we must perform the conversion
\begin{equation}
  \label{ff-conversion}
F_1 = \frac{G_E + \tau G_M}{1+\tau} , \qquad\qquad\qquad
F_2 = \frac{G_M - G_E}{1+\tau} ,
\end{equation}
where $\tau = -t /(4 m^2)$.  To obtain central values for $F_1$ and
$F_2$ is rather straightforward, but we could not derive errors on
these quantities in a rigorous way, lacking information on the
correlation between the errors on $G_E$ and $G_M$.  We encourage
experimentalists to provide correlated errors on the two form factors
(or derived quantities) whenever possible, since depending on the
physical context one or the other set of form factors is more
adequate.

For the proton form factors we have used the results from
\cite{Brash:2001qq}, where the original data from
\cite{Janssens:1966,Berger:1971kr,Sill:1992qw} was reanalyzed using
the additional information on the ratio $G_E /G_M$ measured by the
recoil polarization method in \cite{Jones:1999rz,Gayou:2001qd}.  With
increasing $t$ the measurements using this method deviate from those
obtained with a Rosenbluth separation, which is affected by QED
radiative corrections growing with $t$ (see \cite{Chen:2004tw} and
references therein).

To obtain $F_1^p$ we have used the central values and errors on
$G_M^p$ together with the analytic parameterization for $G_E^p /G_M^p$
given in \cite{Brash:2001qq}.  In calculating errors on $F_1$ we have
only used the errors on $G_M$ but not those on the parameterization of
$G_E /G_M$.  The uncertainty on $G_E /G_M$ is included in the errors
on the reanalyzed $G_M$ data given in \cite{Brash:2001qq}, and
counting it again in the conversion (\ref{ff-conversion}) would not be
consistent.  In discarding them we obtain the same relative errors on
$F_1^p$ as those on $G_M$.  They are between $0.8\%$ and $2.4\%$ for
$-t \le 10\gev^2$ and then grow up to $9\%$ at the highest $t$ of the
data.  If we had added the errors on the parameterization of $G_E
/G_M$ in quadrature, the resulting errors on $F_1^p$ would have been
increased by at most a factor of 1.5 for $-t \le 7.7 \gev^2$.  (Beyond
this value the ratio $G_E/G_M$ is taken as zero in \cite{Brash:2001qq}
and it is unclear which error one should assign to it.)

For $F_2^p$ we have used the analytic parameterization of $G_M^p$
given in \cite{Brash:2001qq} together with the data and errors on
$G_E^p /G_M^p$ measured with the recoil polarization method in
\cite{Jones:1999rz,Gayou:2001qd,Milbrath:1997de}.  For calculating the
errors on $F_2$ we have used those on $G_E /G_M$ but not those on the
parameterization of $G_M$ (which overestimate the experimental errors
on $G_M$ in the $t$ range we need).  If we were to add in quadrature
an error on $G_M$ of $1.5\%$ for $-t< 1 \gev^2$ and of $2\%$
otherwise, our errors on $F_2^p$ would increase by at most a factor of
1.6.  The recoil data only go down to $-t= 0.32 \gev^2$, and in order
to have some data for lower $-t$ we have also included the results of
\cite{Price:1971zk} for $G_M$ and $G_E$ obtained by a Rosenbluth
separation, using (\ref{ff-conversion}) and adding the errors on $G_M$
and $G_E$ in quadrature.  The values for $F_2^p$ we obtain from these
data match rather well with those from the recoil measurements in the
$t$ region where they overlap (see Fig.~\ref{fig:F2data2}), indicating
that in the kinematics of interest the two methods are reasonably
consistent.  The errors we estimate for $F_2^p$ are between $1\%$ and
$4\%$, with the exception of two data points with $5\%$ and one point
with $7\%$.

For the neutron form factors we have restricted ourselves to results
published after 1990 and only considered values of $t$ where data for
both $G_M^n$ and $G_E^n$ are available, adding their errors in
quadrature (as well as statistical, systematic and theoretical errors
when they are given separately).  The measurements in
\cite{War:2003ma,Madey:2003av} cover a range $0.5 \gev^2 \le -t \le
1.5 \gev^2$, and we obtain $F_1^n$ with errors between 6\% and 17\%
and $F_2^n$ with errors between $0.8\%$ and $2.4\%$.  At $-t= 0.255
\gev^2$ we have combined the measurement of $G_M^n$ from
\cite{Mar:1993hx} and the one of $G_E^n$ from
\cite{Eden:1994ji}.  The resulting error on $F_2^n$ is 5\%, whereas we
obtain $F_1^n= -0.010 \pm 0.035$ with a huge error.  This is because
the individual terms $G_E^n = 0.066 \pm 0.037$ and $\tau G_M^n= -0.077
\pm 0.003$ in (\ref{ff-conversion}) nearly cancel.  We nevertheless
keep this data point, since a $\chi^2$ fit to the data will correctly
take into account its large uncertainty.  At the high-$t$ end we have
taken data from \cite{Lung:1992bu}.  We do not use their results at
$-t= 1.75 \gev^2$ and $2.5 \gev^2$ since the central values of
$(G_E^n)^2$ came out negative.  {}From the remaining points $-t=3.25
\gev^2$ and $4 \gev^2$ we obtained $F_1^n$ with errors around 30\% and
$F_2^n$ with errors around 20\%.

To have an indication of how well our fits describe proton data in the
lower and upper range of available $t$, we give in App.~\ref{app:B}
separate values of $\chi^2$ for data sets labeled ``low'' and ``high''
as specified in Table~\ref{tab:data-sets}.

\begin{table}
\caption{\label{tab:data-sets} Details of the data sets for which
separate values of $\chi^2$ are given in the tables of
App.~\protect\ref{app:B}.}
\begin{center}
\renewcommand{\arraystretch}{1.2}
\begin{tabular}{llcl} \hline\hline
data set & references & data points & range of $-t /\gev^2$ \\
\hline
$F_1^p$ low  & \protect\cite{Janssens:1966,Berger:1971kr} &
               49 & $0.156$ to $7$ \\
$F_1^p$ high & \protect\cite{Sill:1992qw} &
               13 & $2.862$ to $31.2$ \\
$F_1^n$ & 
\protect\cite{Mar:1993hx,Eden:1994ji,War:2003ma,Madey:2003av,Lung:1992bu} &
               ~8 & $0.255$ to $4$ \\
$F_1$ total & & 70 & \\
\hline
$F_2^p$ low  & \protect\cite{Price:1971zk} &
               18 & $0.0389$ to $1.75$ \\
$F_2^p$ high & \protect\cite{Gayou:2001qd,Jones:1999rz,Milbrath:1997de} &
               28 & $0.32$ to $5.54$ \\
$F_2^n$ & 
\protect\cite{Mar:1993hx,Eden:1994ji,War:2003ma,Madey:2003av,Lung:1992bu} &
               ~8 & $0.255$ to $4$ \\
$F_2$ total & & 54 & \\
\hline\hline
\end{tabular}
\end{center}
\end{table}


\section{Details of fit results}
\label{app:B}

In the Tables of this section we list the results of the various fits
discussed in this paper.  We have performed $\chi^2$ fits to the form
factor data using the minimization package Minuit~\cite{James:1975dr}.
The $1\, \sigma$ errors on the parameters are defined by the contour in
parameter space for which
\begin{equation}
\chi^2 = \chi^2_{\rm min} + \Delta \chi^2 ,
\end{equation}
where $\Delta \chi^2$ corresponds to a confidence level of 68\%.  One
respectively has $\Delta\chi^2 = 2.30$, $3.53$, $4.72$ for 2, 3, 4
free parameters in the fit.  The error bands shown in our figures have
been derived from the $1\, \sigma$ errors on the fitted parameters by
standard error propagation (see e.g.\ Sect.~31.2 of \cite{PDG}).  This
takes properly into account correlations between errors on different
parameters, which are typically quite large in our fits.

Our default fit for $H_v^q$ is specified in Sect.~\ref{sec:select} and
listed in the first row of Table~\ref{tab:n2-fits}, and our default
fit for $E_v^q$ is explained in Sect.~\ref{sec:fit-e} and shown in
Table \ref{tab:def2}.  The covariance and error correlation matrices
of these fits are given in Tables~\ref{tab:covar-H} and
\ref{tab:covar-E}, respectively.  All our fits for $E_v^q$ use the
exponential ansatz (\ref{master-e}) with the profile function
(\ref{profile-e}) and the forward limit (\ref{good-old-ans}).  Except
for Table~\ref{tab:flavor-e} we have fixed $\alpha=0.55$ in all fits.
No error is given on $C_u$ when a fit has chosen its maximum value
$1.22 \gev^{-2}$ allowed by the positivity constraint
(\ref{profile-bound}).

\begin{table}[b]
\caption{\label{tab:covar-H} The covariance matrix (left) and
correlation matrix (right) of the default fit to $H_v^q$ described in
Sect.~\protect\ref{sec:select} and listed in the first row of
Table~\protect\ref{tab:n2-fits}.  The entries of the covariance matrix
are in units of $\gev^{-4}$ and are normalized such that diagonal
elements give the squared $1\, \sigma$ errors on a parameter.}
\begin{center}
\renewcommand{\arraystretch}{1.2}
\begin{tabular}{cccc} \hline\hline
      & $B_q$ & $A_u$ & $A_d$ \\ \hline
$B_q$ & $\phantom{-}6.95 \cdot 10^{-4}$ & $-4.84 \cdot 10^{-4}$ &
        $\phantom{-}5.69 \cdot 10^{-3}$ \\
$A_u$ & $-4.84 \cdot 10^{-4}$ & $\phantom{-}6.11 \cdot 10^{-4}$ &
        $-1.38 \cdot 10^{-3}$ \\
$A_d$ & $\phantom{-}5.69 \cdot 10^{-3}$ & $-1.38 \cdot 10^{-3}$ &
        $\phantom{-}8.22 \cdot 10^{-2}$ \\
\hline\hline
\end{tabular}
\hspace{2em}
\begin{tabular}{cccc} \hline\hline
      & $B_q$              & $A_u$          & $A_d$ \\ \hline
$B_q$ & $\phantom{-}1$     & $-0.743$       & $\phantom{-}0.750$ \\
$A_u$ & $-0.743$           & $\phantom{-}1$ & $-0.194$ \\
$A_d$ & $\phantom{-}0.750$ & $-0.194$       & $\phantom{-}1$  \\
\hline\hline
\end{tabular}
\end{center}
\end{table}

\begin{table}
\caption{\label{tab:covar-E} The covariance matrix (left) and
correlation matrix (right) of the default fit to $E_v^q$ described in
Sect.~\protect\ref{sec:fit-e} and listed in
Table~\protect\ref{tab:def2}.  The entries of the covariance matrix
are in units of $\gev^{-4}$ and are normalized such that diagonal
elements give the squared $1\, \sigma$ errors on a parameter.}
\begin{center}
\renewcommand{\arraystretch}{1.2}
\begin{tabular}{cccc} \hline\hline
          & $\beta_u$ & $D_u$ & $D_d$ \\ \hline
$\beta_u$ & $\phantom{-}5.01 \cdot 10^{-2}$ & $-2.15 \cdot 10^{-2}$ &
            $-8.24 \cdot 10^{-3}$ \\
$D_u$   & $-2.15 \cdot 10^{-2}$ & $\phantom{-}1.17 \cdot 10^{-2}$ &
            $\phantom{-}1.80 \cdot 10^{-3}$ \\
$D_d$   & $-8.24 \cdot 10^{-3}$ & $\phantom{-}1.80 \cdot 10^{-3}$ &
            $\phantom{-}2.84 \cdot 10^{-3}$ \\
\hline\hline
\end{tabular}
\hspace{2em}
\begin{tabular}{cccc} \hline\hline
          & $\beta_u$ & $D_u$ & $D_d$ \\ \hline
$\beta_u$ & $\phantom{-}1$ & $-0.884$           & $-0.693$ \\
$D_u$   & $-0.884$       & $\phantom{-}1$     & $\phantom{-}0.311$ \\
$D_d$   & $-0.693$       & $\phantom{-}0.311$ & $\phantom{-}1$ \\
\hline\hline
\end{tabular}
\end{center}
\end{table}

In Tables~\ref{tab:old-fits} to \ref{tab:constr} we give the fitted
parameters and their $1\, \sigma$ errors.  For the sake of legibility
we have omitted the units of $A_q$, $B_q$, $C_q$, $D_q$ and
$\alpha'$, which are always taken as $\gev^{-2}$.  The columns labeled
``low'' and ``high'' give the $\chi^2$ per number of data points for
the separate proton data sets specified in Table~\ref{tab:data-sets},
and the columns ``$p$'' and ``$n$'' give the corresponding numbers for
the combined proton data and for the neutron data.  The column
``total'' gives the $\chi^2$ per degrees of freedom, which is the
quantity minimized in the fit.  Note that if different fits in the
same table have different numbers of free parameters, then the
normalization of $\chi^2 /\dof$ differs slightly from one row to
another.  The precise $\chi^2$ values should be interpreted with care
because of the inadequacies affecting our errors on the Dirac and
Pauli form factors (see App.~\ref{app:A}).  This applies in particular
to the low-$t$ data on $F_1^p$, whose errors are particularly small.
To assess the quality of a fit it is hence useful to consider in
addition to its $\chi^2$ also the ``pull'' as shown in
Figures~\ref{fig:n2pull} and \ref{fig:F2pull}.

{}From Table~\ref{tab:data-sets} it is clear that the low-$t$ data on
$F_1^p$ tend to dominate the $\chi^2$ minimization in the fit to the
Dirac form factor because of the large number of data points.  This
could be circumvented by reweighting the $\chi^2$ for different data
points, but we have opted for the more straightforward procedure
without weights, given that our best fit turns out to describe the
large-$t$ data on $F_1^p$ and the data on $F_1^n$ quite well.  In a
similar fashion, the proton data tend to dominate over those for the
neutron in the fits to both the Dirac and the Pauli form factors due
to the paucity of neutron data.


\begin{table}[p]
\caption{\label{tab:old-fits} Fits to $H_v^q(x,t) = q_v(x) \exp[t
f_q(x)]$ using the CTEQ6M parton densities at scale $\mu=2 \gev$ and
$f_q(x) = \alpha' (1-x)^{n-1} \log(1/x) + (A_q - \alpha') (1-x)^n$
from (\protect\ref{fit:var1}) and (\protect\ref{fit:var2}).  More
information on the different columns is given in the text.}
\begin{center}
\renewcommand{\arraystretch}{1.2}
\begin{tabular}{ccccccccc} \hline\hline
 & \multicolumn{3}{l}{fit parameters} & 
   \multicolumn{4}{l}{$\chi^2 /$data points} & $\chi^2 /\dof$ \\
$n$ & $\alpha'$ & $A_u$ & $A_d$ &
      low & high & $p$ & $n$ & total \\ \hline
1 & $  1.38\pm  0.01$ & $  0.070\pm  0.006$ & $  0.25\pm  0.04$ &
       1.42&   0.60&   1.25&   1.55&  1.34 \\ 
2 & $  1.08\pm  0.02$ & $  0.78\pm 0.01 $ & $  1.05\pm  0.05$ &
       4.48&   8.79&   5.39&   2.22&  5.25 \\
\hline\hline
\end{tabular}
\end{center}
\end{table}

\begin{table}
\caption{\label{tab:n1-fits} Fits to $H_v^q(x,t) = q_v(x) \exp[t
f_q(x)]$ using the CTEQ6M parton densities at $\mu=2 \gev$ and $f_q(x)
= \alpha' (1-x)^2 \log(1/x) + B_q (1-x)^2 + A_q\, x (1-x)$ from
(\protect\ref{fit:n1}) with fixed $\alpha' = 0.9 \gev^{-2}$.}
\begin{center}
\renewcommand{\arraystretch}{1.2}
\begin{tabular}{cccccccc} \hline\hline
 \multicolumn{3}{l}{fit parameters} &
 \multicolumn{4}{l}{$\chi^2 /$data points} & $\chi^2 /\dof$ \\
 $B_u = B_d$ & $A_u$ & $A_d$ &
      low & high & $p$ & $n$ & total \\ \hline
    $0.43\pm  0.02$ & $  0.164\pm  0.014$ & $  1.35\pm  0.21$ &
        3.05&   1.75&   2.77&   2.15&  2.82 \\
    $0.32\pm  0.01$ & \multicolumn{2}{c}{$  0.190\pm  0.009$} &
        4.23&   1.64&   3.69&   9.46&  4.47 \\
\hline\hline
\end{tabular}
\end{center}
\end{table}

\begin{table}
\caption{\label{tab:n2-fits} Fits to $H_v^q(x,t) = q_v(x) \exp[t
f_q(x)]$ using the CTEQ6M parton densities at $\mu=2 \gev$ and $f_q(x)
= \alpha' (1-x)^3 \log(1/x) + B_q (1-x)^3 + A_q\, x (1-x)^2$ from
(\protect\ref{fit:n2}) with fixed $\alpha' = 0.9 \gev^{-2}$.}
\begin{center}
\renewcommand{\arraystretch}{1.2}
\begin{tabular}{ccccccccc} \hline\hline
 \multicolumn{4}{l}{fit parameters} &
 \multicolumn{4}{l}{$\chi^2 /$data points} & $\chi^2 /\dof$ \\
 $B_u$ & $B_d$ & $A_u$ & $A_d$ &
      low & high & $p$ & $n$ & total \\ \hline
 \multicolumn{2}{c}{$0.59\pm  0.03$} &
     $  1.22\pm  0.02$ & $  2.59\pm  0.29$ &
        2.27&   0.70&   1.94&   1.12&  1.93 \\
 $  0.59\pm  0.03$ & $  0.32\pm  0.14$ &
     $  1.26\pm  0.04$ & $  3.82\pm  0.78$ &
        1.78&   1.32&   1.68&   0.73&  1.67 \\
 $  0.56\pm  0.02$ & $  0.75\pm  0.08$ &
 \multicolumn{2}{c}{$  1.17\pm  0.03$} &
        3.22&   0.76&   2.70&   2.36&  2.78 \\
 \multicolumn{2}{c}{$  0.50\pm  0.01$} &
 \multicolumn{2}{c}{$  1.20\pm  0.02$} &
        2.98&   0.69&   2.50&   9.24&  3.36 \\
\hline\hline
\end{tabular}
\end{center}
\end{table}

\begin{table}
\caption{\label{tab:error-fits} Fits to $H_v^q(x,t) = q_v(x) \exp[t
f_q(x)]$ with $f_q(x)$ as in Table~\protect\ref{tab:n2-fits} and
different sets of CTEQ6M parton distributions at $\mu=2 \gev$.}
\begin{center}
\renewcommand{\arraystretch}{1.2}
\begin{tabular}{ccccccccc} \hline\hline
 & \multicolumn{3}{l}{fit parameters} &
   \multicolumn{4}{l}{$\chi^2 /$data points} & $\chi^2 /\dof$ \\
 set & $B_u=B_d$ & $A_u$ & $A_d$ &
      low & high & $p$ & $n$ & total \\ \hline
 17 & $  0.55\pm  0.03$ & $  1.21\pm  0.02$ & $  2.52\pm  0.30$ &
      1.79&   1.23&   1.68&   1.09&  1.68 \\
 18 & $  0.64\pm  0.03$ & $  1.24\pm  0.02$ & $  2.60\pm  0.27$ &
      2.96&   0.58&   2.46&   1.30&  2.44 \\
 35 & $  0.58\pm  0.03$ & $  1.25\pm  0.03$ & $  2.88\pm  0.28$ &
      2.20&   0.72&   1.89&   0.98&  1.86 \\
 36 & $  0.59\pm  0.03$ & $  1.22\pm  0.03$ & $  2.85\pm  0.29$ &
      1.99&   0.91&   1.77&   0.81&  1.73 \\
\hline\hline
\end{tabular}
\end{center}
\end{table}

\begin{table}
\caption{\label{tab:alpha-mu-fits} Fits to $H_v^q(x,t) = q_v(x) \exp[t
f_q(x)]$ with $f_q(x) = \alpha' (1-x)^3 \log(1/x) + B_q (1-x)^3 +
A_q\, x (1-x)^2$ from (\protect\ref{fit:n2}) and CTEQ6M parton
densities taken at different scales $\mu$.}
\begin{center}
\renewcommand{\arraystretch}{1.2}
\begin{tabular}{cccccccccc} \hline\hline
 & \multicolumn{4}{l}{fit parameters} &
   \multicolumn{4}{l}{$\chi^2 /$data points} & $\chi^2 /\dof$ \\
 $\mu$ & $\alpha'$ & $B_u=B_d$ & $A_u$ & $A_d$ &
      low & high & $p$ & $n$ & total \\ \hline
 1\gev & $  1.17\pm  0.05$ & $  0.50\pm  0.12$ &
         $  1.81\pm  0.07$ & $  2.62\pm  0.31$ &
     1.16&   0.70&   1.06&   1.11&  1.13 \\
 2\gev & $  0.97\pm  0.04$ & $  0.43\pm  0.09$ &
         $  1.32\pm  0.06$ & $  2.51\pm  0.35$ &
     1.77&   1.53&   1.72&   0.86&  1.72 \\
 4\gev & $  0.89\pm  0.03$ & $  0.36\pm  0.08$ &
         $  1.10\pm  0.06$ & $  2.58\pm  0.38$ &
      2.23&   2.09&   2.20&   0.89&  2.17 \\
 8\gev & $  0.84\pm  0.03$ & $  0.29\pm  0.07$ &
         $  0.96\pm  0.06$ & $  2.71\pm  0.41$ &
      2.56&   2.42&   2.53&   1.10&  2.51 \\
\hline\hline
\end{tabular}
\end{center}
\end{table}

\begin{table}
\caption{\label{tab:power-fits} Fits to $H_v^q(x,t) = q_v(x) \, (1 - t
f_q(x) /p)^{-p}$ with $f_q(x)$ as in Table~\ref{tab:alpha-mu-fits} and
the CTEQ6M parton densities at $\mu= 2\gev$.}
\begin{center}
\renewcommand{\arraystretch}{1.2}
\begin{tabular}{cccccccccc} \hline\hline
 & \multicolumn{4}{l}{fit parameters} &
   \multicolumn{4}{l}{$\chi^2 /$data points} & $\chi^2 /\dof$ \\
 $p$ & $\alpha'$ & $B_u=B_d$ & $A_u$ & $A_d$ &
      low & high & $p$ & $n$ & total \\ \hline
 $\infty$ & $  0.97\pm  0.04$ & $\phantom{-}0.43\pm  0.09$ &
     $  1.32\pm  0.06$ & $  2.51\pm  0.35$ &
     1.77&   1.53&   1.72&   0.86&  1.72 \\
 4~ & $  0.91\pm  0.06$ & $  \phantom{-}0.84\pm  0.17$ &
      $  2.28\pm  0.14$ & $  4.48\pm  0.50$ &
      1.06&   0.85&   1.02&   1.25&  1.11 \\
 3~ & $  0.96\pm  0.09$ & $  \phantom{-}0.66\pm  0.25$ &
      $  3.34\pm  0.22$ & $  6.11\pm  0.59$ &
      0.92&   0.70&   0.87&   1.64&  1.02 \\
 2.5 & $  1.21\pm  0.14$ & $ -0.12\pm  0.39$ &
       $  5.38\pm  0.42$ & $  8.75\pm  0.73$ &
      0.81&   0.44&   0.73&   2.18&  0.95 \\
 2~ & $  2.73\pm  0.18$ & $ -6.44\pm 0.74 $ &
      $ 21.0 \pm  2.3 $ & $ 27.3 \pm 2.7 $ &
      1.34&   6.19&   2.36&   4.14&  2.72 \\
\hline\hline
\end{tabular}
\end{center}
\end{table}


\begin{table}
\caption{\label{tab:scan} Fits for $E_v^q$ with free parameters
$D_u$, $D_d$ and $C_u$, $C_d$ subject to the constraint
(\protect\ref{profile-bound}).  In all fits $C_d$ takes it maximum
allowed value $2.59 \gev^{-2}$.  Further explanations on the fits for
$E_v^q$ are given in the text.}
\begin{center}
\renewcommand{\arraystretch}{1.2}
\begin{tabular}{cccccccccc} \hline\hline
 & & \multicolumn{3}{l}{fit parameters} &
     \multicolumn{4}{l}{$\chi^2 /$data points} & $\chi^2 /\dof$ \\
 $\beta_u$ & $\beta_d$ & $D_u$ & $D_d$ & $C_u$ &
      low & high & $p$ & $n$ & total \\ \hline
 4 & 5 & $\phantom{-}0.38\pm 0.12$ & $-0.65\pm 0.13$ &  $1.22$ &
        1.89&   0.89&   1.28&   0.83&   1.32 \\
 5 & 5 & $\phantom{-}0.14\pm 0.07$ & $-0.65\pm 0.05$ &  $0.85\pm 0.14$ &
        2.03&   0.90&   1.34&   0.79&   1.36 \\
 6 & 5 & $-0.06\pm 0.06$ & $-0.65\pm 0.05$ &   $0.60\pm 0.13$ &
        2.13&   0.94&   1.40&   0.76&   1.41 \\
\hline
 4 & 6 & $\phantom{-}0.38\pm 0.07$  & $-0.82\pm 0.04$ &  $1.22$ &
        1.94&   0.87&   1.29&   0.79&   1.31 \\
 5 & 6 & $\phantom{-}0.14\pm 0.07$  & $-0.82\pm 0.04$ &  $0.82\pm 0.14$ &
        2.03&   0.91&   1.35&   0.76&   1.36 \\
 6 & 6 & $-0.05\pm 0.10$ & $-0.82\pm 0.12$ &   $0.58\pm 0.46$ &
        2.14&   0.94&   1.41&   0.74&   1.41 \\
\hline\hline
\end{tabular}
\end{center}
\end{table}

\begin{table}
\caption{\label{tab:def2} Fit for $E_v^q$ with fixed parameters $C_u=
1.22 \gev^{-2}$, $C_d= 2.59 \gev^{-2}$ and $\beta_d-\beta_u = 1.60$.}
\begin{center}
\renewcommand{\arraystretch}{1.2}
\begin{tabular}{cccccccc} \hline\hline
 \multicolumn{3}{l}{fit parameters} &
 \multicolumn{4}{l}{$\chi^2 /$data points} & $\chi^2 /\dof$ \\
 $\beta_u$ & $D_u$ & $D_d$ &
      low & high & $p$ & $n$ & total \\ \hline
 $3.99\pm 0.22$ & $0.38\pm 0.11$ & $-0.75\pm 0.05$ &
   1.92 & 0.88 & 1.29 &  0.80 & 1.31 \\
\hline\hline
\end{tabular}
\end{center}
\end{table}

\begin{table}
\caption{\label{tab:flavor-e} Four-parameter fits for $E_v^q$ with
equal values of the parameters for $u$ and $d$ quarks.  In the fit of
the first row the constraint $C_q \le 1.22 \gev^{-2}$ from
positivity is imposed, whereas in the fit of the second row $C_q$ is
left completely free.}
\begin{center}
\renewcommand{\arraystretch}{1.2}
\begin{tabular}{ccccccccc} \hline\hline
 \multicolumn{4}{l}{fit parameters} &
 \multicolumn{4}{l}{$\chi^2 /$data points} & $\chi^2 /\dof$ \\
 $\alpha$ & $\beta_u=\beta_d$ & $D_u=D_d$ & $C_u=C_d$ &
      low & high & $p$ & $n$ & total \\ \hline
 $0.53 \pm 0.02$ & $7.56 \pm 1.95 $ &
 $-0.59 \pm 0.31$ & $1.22$ &
      4.48&   1.01&   2.37&  13.26&   4.30 \\
 $0.61 \pm 0.05$ & $2.24\pm 0.71$ &
 $\phantom{-}0.04 \pm 0.17$ & $3.50 \pm 1.04$ &
      3.87&   0.80&   2.00&  13.27&   3.97 \\
\hline\hline
\end{tabular}
\end{center}
\end{table}

\begin{table}
\caption{\label{tab:constr-scan} Fits for $E_v^q$ with free parameters
$D_u=D_d$ and $C_u$, $C_d$ subject to the constraint
(\protect\ref{profile-bound}).}
\begin{center}
\renewcommand{\arraystretch}{1.2}
\begin{tabular}{cccccccccc} \hline\hline
 & & \multicolumn{3}{l}{fit parameters} &
 \multicolumn{4}{l}{$\chi^2 /$data points} & $\chi^2 /\dof$ \\
 $\beta_u$ & $\beta_d$ & $D_u=D_d$ & $C_u$ & $C_d$ &
      low & high & $p$ & $n$ & total \\ \hline
 7 & 5 & $-0.41\pm 0.02$ & $1.22$ & $1.61\pm 0.25$ &
         2.79&   0.84&   1.61&   3.38&   1.98 \\
 8 & 5 & $-0.49\pm 0.02$ & $1.00\pm 0.25$ & $1.78\pm 0.28$ &
         2.63&   0.91&   1.58&   1.88&   1.72 \\
 9 & 5 & $-0.56\pm 0.02$ & $0.60\pm 0.20$ & $2.16\pm 0.31$ &
         2.58&   0.96&   1.60&   1.15&   1.62 \\
10 & 5 & $-0.63\pm 0.02$ & $0.29\pm 0.16$ & $2.48\pm 0.32$ &
         2.55&   1.06&   1.64&   0.76&   1.60 \\
\hline\hline
\end{tabular}
\end{center}
\end{table}

\begin{table}
\caption{\label{tab:constr} Fits for $E_v^q$ with free parameters
$\beta_u$, $\beta_d$, $D_u$, $D_d$ and $C_u = C_d$.  In all
fits $C_q$ takes its maximum value $1.22 \gev^{-2}$ allowed by the
constraint (\protect\ref{profile-bound}).}
\begin{center}
\renewcommand{\arraystretch}{1.2}
\begin{tabular}{ccccccccc} \hline\hline
 \multicolumn{4}{l}{fit parameters} &
 \multicolumn{4}{l}{$\chi^2 /$data points} & $\chi^2 /\dof$ \\
 $\beta_u$ & $\beta_d$ & $D_u$ & $D_d$ &
      low & high & $p$ & $n$ & total \\ \hline
 $8.97\pm 2.86$ & $6.75\pm 1.82$ &
 $-0.62\pm 0.34$ & $-0.72\pm  0.27$ &
     2.34 & 1.36 & 1.75 & 2.93 & 2.12 \\
 $9.54\pm 1.79$ & $6.49\pm 1.27$ &
 \multicolumn{2}{c}{$-0.69\pm 0.20$} &
     2.38 & 1.35 & 1.75 & 2.93 & 2.08 \\
 \multicolumn{2}{c}{$7.76\pm 1.23$} &
 $-0.47\pm 0.18$ & $-0.85\pm 0.16$ &
   2.24 & 1.47 & 1.77 & 3.15 & 2.13 \\
 \multicolumn{2}{c}{$8.42\pm 1.30$} &
 \multicolumn{2}{c}{$-0.74\pm 0.16$} &
   3.57 & 1.11 & 2.08 & 15.9 & 4.36 \\
\hline\hline
\end{tabular}
\end{center}
\end{table}


\end{document}